\documentclass[12pt,fleqn]{iiscthes}
\usepackage[utf8x]{inputenc}
\usepackage{graphicx}
\usepackage{amsmath,amssymb,bm}
\usepackage{slashed}
 \usepackage{caption}
\usepackage{leftidx,cite,subfig,epsfig,float}

\begin{document}

\begin{frontmatter}
%
%

\title{ Spin-Statistics Correlations in Various Noncommutative Field Theories }
\author{Rahul Srivastava}
\submitdate{July 2012}
\dept{Centre for High Energy Physics}
\sciencefaculty
\iisclogotrue 
\tablespagetrue 
\maketitle
%

\begin{dedication}
\begin{center}
\large\it Dedicated to my Grandparents
\end{center}
\end{dedication}


\declaration

I hereby declare that the work presented in this thesis entitled
``Spin-Statistics Correlations in Various Noncommutative Field Theories''
is the result of the investigations carried out by me under the 
supervision of Prof. Sachindeo Vaidya at the Centre for High Energy Physics,
Indian Institute of Science, Bangalore, India, and that it has not been
submitted elsewhere for the conferment of any degree or diploma
of any Institute or University. Keeping with the
general practice, due acknowledgements have been made wherever the work
described is based on other investigations. \\

\bigskip

Dated: 26/07/2012 \quad \quad \quad \quad \quad \quad  \quad \quad \quad \quad \quad \quad 
\quad \quad \quad \quad \quad \quad \quad  Rahul Srivastava

\bigskip
\bigskip
\bigskip
\bigskip
\bigskip
\bigskip
{\bf Certified} \\
 Sachindeo Vaidya \\
\quad Professor \\
\quad Centre for High Energy Physics,\\
\quad Indian Institute of Science,\\
\quad Bangalore - 560012,\\
\quad Karnataka, India\\


\acknowledgements
First and foremost I would  like to thank my research supervisor Prof. Sachindeo Vaidya. Working with him was a wonderfully enlightening experience. Over the years I have
benefited a lot from his clarity of thought, his deep insight and understanding of the subject. He always took great pains to make sure that I understood
the various intricate concepts as well as calculational details of a given work. Whatever I have learned during the five years of my Ph.D is in large parts due to and from him.
He has always been quite supportive and encouraging. It was a pleasure to work with him and I must say that 
I was quite fortunate to get such a wonderful supervisor. I owe my deepest gratitude to him.
 
I would also like to take this opportunity to thank Prof. Manu Paranjape. Thanks to the Canadian Commonwealth Fellowship, I got an opportunity to work with him for five months. Working
with him was quite a pleasure. The sight of him working out pages and pages of calculations in his head was quite impressive and will last in my memory forever. Needless
to say, I have ever since been trying to do the same without ever succeeding in going past two steps. His friendly and easy going nature was in large parts responsible for
making my stay in Montr\'eal such a wonderful experience. 

I would also like to thank my collaborators Prof. Amilcar R. de Queiroz and Dr. Prasad Basu from whom I learned a lot. I am especially grateful to Amilcar for putting in a lot 
of effort in
helping me finish a part of my work in this thesis. He was always helpful and always available for discussion and I want to thank him for his efforts which lead to the timely
completion of this thesis. Prasad is not just my collaborator but also a very close friend. I still cherish all the evenings we spent together discussing a wide range of topics
from our work in physics to world politics to Arab-Israel crisis to what not. I also benefited from his deep insights and clarity of thought. As a young student just entering
research, he helped me a lot by clarifying quite a few misconceptions of mine and by always motivating me when things were not going well. I want to thank him from the deepest
corners of my heart.

I would also take this opportunity to thank Prof. A.P. Balachandran and Prof. T.R. Govindarajan for various discussions I had with them. They were always encouraging and
helped me get a better understanding of my field by patiently clarifying the various doubts I had.

I also thank all the faculty members of CHEP for their support and encouragement. Many thanks to my teachers who taught me various courses at CHEP, namely, Prof. B. Ananthanarayan,
Prof. Apoorva Patel, Prof. Rohini Godbole and Prof. Sudhir Vempati.

My sincere thanks to the office staff, Mr. Keshava and Mr Shankar. Their efforts and support for the students over the years are highly appreciable. They were always readily 
available for help.
 
I would also like to thank my other collaborators Gauhar Abbas, Nitin Chandra and Sandeep Chatterjee. Apart from being my close friends, I owe them a lot for clarifying various
silly doubts and misconceptions in my understanding of physics. I will always cherish the many friendly discussions I had with them. I would especially like to thank Gauhar
for all he has done for me over the years. We joined the department together and right from the first day we formed a great friendship. Despite our different natures and
ideologies, our friendship has grown stronger and stronger over time. Over these years he has always helped me and stood by me at every hour of need, from helping me do my assignments
in time to providing me the latex template for this thesis. I owe him a lot (monetarily as well as metaphorically) and would like to take this opportunity to thank him for being 
such a great friend and colleague.

I also offer my gratitude to my close friends Monalisa Patra and Nirmalendu Acharyya with whom I have also shared my office for many a years. They both, along with Gauhar have
suffered me and my idiosyncrasies in silence for all these years without ever complaining. I am also grateful to Monalisa for sometimes inviting me for dinner and helping me
get out of the monotony of mess food. Nirmalendu has been a very close friend of mine for many years and I have spent some great times with him. I would like to take this opportunity
to thank him for all the fun times.
 
I would also take this opportunity to thank my other friends of IISc namely Sumit, Senti, Suropriya, Chandni, Marsha, Semonti, Ranjan, Ketan, Nitin, Bidya, Amit, Thakur, Abhay,
Akshaya and Debangsu. They helped add colour to my life and also helped me in times of need. My sincere apologies to anyone I am missing out here.

I would also like to thank my childhood friends Rohit, Deepak, Durgesh and Rajeev for all the lovely memories of my childhood. They were the main reason why I failed in some
exams and I was the reason for their failure. We literally wrote each others destiny by making sure we do every thing other than study. Thanks to them for whatever I have achieved
and my sincere apologies for not letting them achieve to their potential. My gratitude to my other childhood friends as well for all the sweet memories. 

Special thanks are also due to Rati for being a constant companion. I have always found her by my side in times of need. She has helped me have a somewhat more disciplined approach
towards work and life. Her caring and affectionate nature has worked as a healer for me in various difficult times. Ever since coming into my life, she has been a constant source
of encouragement for me. Like my close friends, she also has put up with all my shortcomings without ever complaining. I owe her a lot (again monetarily as well as metaphorically)
in life. I am looking forward to a great future with her.

Finally I would like to thank my whole family for the support and encouragement I have received from them. My mother and father have gone through a lot of hardships to make sure
I can go ahead and chase my dreams. I would like to thank them for their support. My aunts and uncles have always been encouraging and have also helped me a lot in times of need.
I would also like to thank all my brothers and sisters for the lovely memories of childhood and for their support. Special thanks are due to my grandparents. Right from the
time when they taught me how to walk and speak to this day, they have been an integral part of my life and it is because of them that I have finally managed to complete my thesis.
They both were quite excited about the prospect of me completing my Ph.D and were really looking forward to this day. Unfortunately my grandmother passed away a month
  ago and couldn't see me reach the finish line. I am not sure whether she is still there somewhere in this or some other world but I am quite sure that if she is there somewhere 
at all, she will take great pleasure from the fact that I have finally managed to submit my Ph.D thesis. I would like to dedicate this thesis to my grandfather and to the memories of
my grandmother.

The work done in this thesis was supported by C.S.I.R under the award no: F. No 10 – 2(5)/2007(1) E.U. II.


\vspace{0.0cm}
\publications{
\vspace{2.0cm}
\begin{enumerate}
\item {\it Thermal Correlation Functions of Twisted Quantum Fields. }\\
      Prasad Basu, Rahul Srivastava and Sachindeo Vaidya. \\
       Phys. Rev. D, 82, 025005, (2010),  arXiv: 1003.4069 [hep-th].

 \item {\it 	
Signatures of New Physics from HBT Correlations in UHECRs. }\\
   Rahul Srivastava.\\ 
   Mod. Phys. Lett. A, 27, 1250160, (2012),  arXiv: 1201.2380 [hep-th].

\item {\it Renormalization of Noncommutative Quantum Field Theories. } \\
   Amilcar R. de Queiroz, Rahul Srivastava and Sachindeo Vaidya.\\
    Phys. Rev. D, 87, 064014, (2013),   arXiv: 1207.2358 [hep-th].

 \item {\it Poincaré Invariant Quantum Field Theories With Twisted Internal Symmetries.}\\
    Rahul Srivastava and Sachindeo Vaidya.\\
   J.H.E.P, 01, 019, (2013),  arXiv: 1207.5669 [hep-th].\\

\end{enumerate}

}
%

\abstract{

Intuitive arguments involving standard quantum mechanics uncertainty relations suggest that at length scales close to Planck length, strong gravity effects will limit the 
spatial as well  as temporal resolution smaller than fundamental length scale ($l_p \approx$ Planck Length), leading to space-space as well as space-time uncertainties. 
Spacetime cannot be probed with a resolution beyond this scale i.e. spacetime becomes “fuzzy” below this scale, resulting into noncommutative spacetime. Hence it becomes important 
and interesting to study in detail the structure of such noncommutative spacetimes and the properties of quantum fields on such spaces, because it not only helps us improve our 
understanding of the Planck scale physics but also helps in bridging standard particle physics with physics at Planck scale. 

In this thesis we study field theories written on a particular model of noncommutative spacetime, the Groenewold-Moyal (GM) plane. We start with briefly reviewing the novel features
of field theories on GM plane e.g. the $\ast$-product, restoration of Poincar\'e-Hopf symmetry and twisted commutation relations. We then discuss our work on renormalization of
field theories on GM plane. We show that any generic noncommutative theory involving pure matter fields is a renormalizable theory if the analogous commutative theory is renormalizable.
We further show that all such noncommutative theories will have same fixed points and $\beta$-functions for the couplings, as that of the analogous commutative theory. 
The unique feature of these field theories is the twisted statistics obeyed by the particles. Motivated by it, we look at the possibility of twisted statistics by deforming
internal symmetries instead of spacetime symmetries. We construct two different twisted theories which can be viewed as internal symmetry analogue of the GM plane and dipole 
field theories which arise in the low energy limit of certain string configurations. We further study their various properties like the issue of causality and the scattering formalism. 
Having studied the mathematical properties 
of noncommutative and twisted internal symmetries we move on to discuss their potential phenomenological signatures. We first discuss the noncommutative thermal correlation functions   
and show that because of the twisted statistics, all correlation functions except two-point function get modified.  
Finally we discuss the modifications in Hanbury-Brown Twiss (HBT) correlation functions due to twisted statistics on GM plane and the potential of observing
signatures of noncommutativity by doing a HBT correlation experiment with Ultra High Energy Cosmic Rays (UHECRs). 

The plan of the thesis is as follows : 

In the first chapter we review the basic concepts and well-known results of field theories written on GM plane. In the second chapter we start with reviewing the formalism 
of noncommutative interaction picture and the noncommutative scattering theory. We then show the equivalence of the interaction picture and Lehmann-Symanzik-Zimmermann
(LSZ) approach for such theories. We
then discuss the issue of renormalizability. We show that any generic noncommutative field theory having only matter fields is a renormalizable theory, provided the corresponding 
commutative theory is also renormalizable. All such theories are free from UV/IR mixing and have identical fixed points and $\beta$-functions for the various couplings as 
the analogous commutative theory.

In the third chapter we discuss the possibility of constructing Poincar\'e invariant field theories having twisted statistics. For the sake of concreteness we construct such theories 
by deforming the transformation
properties of the fields under a global $SU(N)$ group. We construct two such twisted field theories. We further study the issue of causality and the scattering formalism.

In the fourth chapter we discuss the formalism of Green's functions to compute correlation functions and adapt it to the noncommutative case. We show that due to twisted commutation 
relations satisfied by the fields on GM plane, all correlation functions apart from two-point correlation functions get modified. 

In the fifth chapter we look at probable signatures of noncommutativity in UHECRs. We look at the modifications in noncommutative HBT correlation function
due to the twisted statistics. We show that the commutative and noncommutative HBT correlation functions differ from each other and the difference gets more and more pronounced as 
we go to higher and higher energies. Hence an HBT experiment with UHECRs can provide us potential signatures of noncommutativity. We finally conclude our work in the last chapter.

}
%

\makecontents


\vspace{10MM}

\end{frontmatter}

\chapter{Introduction}
\label{chap:chap1}
The idea that spacetime geometry at very short distances may be noncommutative is quite old. It goes back to Schr\"odinger and Heisenberg who raised this possibility to obtain an effective cut-off 
in quantum field theories by introducing fundamental length scale below which spacetime becomes noncommutative. The earliest work along this approach is due to Snyder 
who used the noncommutative structure of spacetime to introduce a small length scale cut-off in field theory without breaking Lorentz invariance \cite{Snyder}. His work was 
further extended by Yang who used it to describe a general geometry where the algebra of noncommuting linear operators is replaced by the algebra of functions \cite{yang}. 
With the successful development of the renormalization program Snyder's idea was forgotten. Later, Connes \cite{connesA} and Woronowicz \cite{woronowicz} 
revived the idea of noncommutative geometry by introducing a differential structure in the noncommutative framework. 

It is now widely believed that the picture of spacetime as a manifold of points will break down at length scales close to Planck length and spacetime events cannot be 
localized with an accuracy higher than Planck length. This is expected to happen because in order to probe physics at a fundamental length scale $l_p$ close to the Planck scale, 
the Compton wavelength $\frac{\hbar}{mc}$ of the probe must fulfill \cite{dop}
\begin{equation}
\frac{\hbar}{mc}\,\leq\, l_p\ \ \textrm{or}\ \
m\,\geq\,\frac{\hbar}{l_p \, c} \,\simeq\, \textrm{Planck mass}.
\end{equation}
Such a high mass in a small volume $l_p^3$ will strongly affect gravity and can cause formation of black holes and their horizons. This suggests a fundamental length limiting spatial 
localization and space-space uncertainty,
\begin{equation}
\Delta x_{1}\Delta x_{2} + \Delta x_{2}\Delta x_{3} + \Delta x_{3}\Delta x_{1} \ \gtrsim l_p^{2}
\end{equation}
Similar arguments can be made about time localization as observation of very short time scales requires very high energies. Such high energies can again produce black holes and 
their horizons which will then limit spatial resolution indicating 
\begin{equation}
\Delta x_{0}(\Delta x_{1} + \Delta x_{2} + \Delta x_{3}) \geq l_p^{2}.
\end{equation}
Just as replacing classical phase space by noncommutative phase space in quantum physics leads to Heisenberg's uncertainty relations, 
the above uncertainty relations suggest that spacetime ought to be described as a noncommutative manifold.  The points on the classical commutative manifold should then be 
replaced by states on a noncommutative algebra. In this thesis we study a particular model for such noncommutative spacetime and its implications on field theories written on it.


\section{The Groenewold-Moyal Plane}


Of the various approaches to model the noncommutative structure of spacetime, the simplest one is noncommutative Groenewold-Moyal spacetime, usually referred as Groenewold-Moyal (GM)
plane. The Groenewold-Moyal plane is a deformation of ordinary spacetime in which the spacetime coordinate functions $\widehat{x}_\mu$ satisfy a commutation relation of the form 
\cite{ConnesC, Madore, Landi, Bondia}: 
\begin{equation} 
[\hat{x}_{\mu},\hat{x}_{\nu}] = i \theta_{\mu\nu} \; ; \quad  \mu, \nu \, = \, 0,1,2,3,  \quad \theta \: \; \text{a real, constant, antisymmetric matrix.}
\label{gm}
\end{equation} 
where the coordinate representation of the operators $\widehat{x}_\mu$ gives Cartesian coordinates $x_{\mu}$ of (flat) spacetime:
\begin{eqnarray} 
\widehat{x}_{\mu}(x)=x_{\mu}.
\end{eqnarray} 
The elements of the $\theta$ matrix have the dimension of $\text{(length)}^2$ and set the scale for the area of the smallest possible localization in the 
$\mu - \nu$ plane, giving a measure for the strength of noncommutativity \cite{dopi1}. One cannot probe spacetime with a resolution below this scale i.e. spacetime is ``fuzzy'' 
\cite{Ydri} below this scale. In the limit $\theta_{\mu \nu} \to 0$, one recovers ordinary spacetime.


\subsection{Star Product and Deformed Algebra }


Let us look at some details of the GM plane which will be useful for us. On GM plane, the noncommutative nature of spacetime can be incorporated by replacing point-by-point multiplication of two 
fields by a type of ``smeared'' product called a star product. In general there exists a way of deforming the algebra of functions on a manifold $M$ \cite{queiroz}.  As we show, 
the GM plane, ${\cal A}_{\theta}({\mathbb R}^{d+1})$, associated with spacetime ${\mathbb R}^{d+1}$ is an example of such a deformed algebra. 

Let us consider a Riemannian manifold $(M,g)$ with metric $g$. If the group ${\mathbb R}^N\;(N \geq 2)$ acts as a group of isometries on $M$, then it acts on the Hilbert space 
$L^2(M,d\mu_g)$ of square integrable functions on $M$. Also $g$ induce the volume form $d\mu_g$ for the scalar product on $L^2(M,d\mu_g)$.
If $\Big\{\lambda = (\lambda_1, \ldots, \lambda_N)\Big\}$ stands for the unitary irreducible representations (UIR's) of ${\mathbb R}^N$, then we have
\begin{equation}\label{eq:1}
L^2(M, d\mu_g) = \bigoplus_\lambda {\cal H}^{(\lambda)} \;,
\end{equation}
where ${\mathbb R}^N$ acts by the UIR $\lambda$ on ${\cal H}^{(\lambda)}$. 

For $a=(a_{1}, a_{2}, \cdots, a_{N}) \in {\mathbb R}^{N}$, the $\lambda$ is chosen such that
\begin{eqnarray}
\lambda : a \longrightarrow e^{i \lambda a}
\end{eqnarray}
If $f_{\lambda}$ and $f_{\lambda '}$ are two smooth functions in ${\cal H}^{(\lambda)}$ and ${\cal H}^{(\lambda ')}$, then under pointwise multiplication we have
\begin{eqnarray}
f_{\lambda} \otimes f_{\lambda '} \rightarrow f_{\lambda} f_{\lambda '}.
\end{eqnarray}
Also, if $p$ is a point on $M$ then
\begin{eqnarray}
(f_{\lambda} f_{\lambda '})(p) = f_{\lambda}(p) f_{\lambda '}(p).
\end{eqnarray}
and
\begin{eqnarray} \label{eq:2}
f_{\lambda} f_{\lambda '} \in {\cal H}^{(\lambda + \lambda ')}
\end{eqnarray}
where the group law is taken as addition. 

Let $\theta^{\mu \nu}$ be a constant antisymmetric matrix in the space of UIR's of ${\mathbb R}^N$. Then we can define a new deformed algebra where the pointwise product 
becomes a $\theta$ dependent ``smeared" $\ast$-product given by
\begin{eqnarray} 
\label{eq:3}
f_\lambda \ast f_{\lambda'} =  f_\lambda \;  f_{\lambda'} \; e^{-\frac{i}{2} \lambda_\mu \theta^{\mu \nu} \lambda'_\nu}\;.
\end{eqnarray}
The GM plane, ${\cal A}_\theta({\mathbb R}^{d+1})$, is a special case of this algebra. This deformed algebra can also be shown to be associative. 

In the case of the GM plane, the group ${\mathbb R}^{d+1}$ acts on ${\cal A}_{\theta}({\mathbb R}^{d+1})$ $\{={\cal C}^\infty({\mathbb R}^{d+1}) \; \textrm{as a set} \}$ by 
translations. The UIR's are labeled by the ``momenta" $\lambda = p = (p^{0}, p^{1}, \ldots, p^{d})$. It leaves the flat Euclidean metric invariant. Plane waves $e_{p}$
form a basis for the Hilbert space ${\cal H}^{(p)}$ with $e_{p}(x) = e^{-ip_{\mu}  x^{\mu}}$, $x = (x^0,x^1, \ldots, x^d)$ being a point on ${\mathbb R}^{d+1}$. 
From (\ref{eq:3}) we have
\begin{eqnarray} 
\label{eq:4}
e_{p} \ast e_{q} = e_{p} \; e_{q} \; e^{-\frac{i}{2}p_\mu \theta^{\mu \nu}q_\nu}\;.
\end{eqnarray}
This $*$-product defines the Moyal plane ${\cal A}_\theta({\mathbb R}^{d+1})$.
%


\section{Spacetime Symmetries on GM Plane}


In this section we discuss how to implement spacetime symmetries on GM plane. The GM plane is characterized by the commutation relations (\ref{gm}).
These relations are not invariant under naive Lorentz transformations and hence our model of noncommutativity breaks usual Lorentz symmetry.
In this section we show how one can interpret these relations in a Lorentz-invariant way by implementing a deformed Lorentz group action \cite{chaichian,chaichian-2}.


\subsection{The Deformed Poincar\'e Group Action}


In quantum mechanics the single particle states are identified with the carrier space of the one-particle unitary irreducible representations (UIR's) of the  two-fold cover of identity 
component of the Poincar\'e group, $\bar{P}^{\uparrow}_{+}$. If $U(g)$, $g \in \bar{P}^{\uparrow}_{+}$, is the UIR for a spinless particle of mass $m$ on a Hilbert space ${\cal H}$, 
then ${\cal H}$ has the basis $\{ |k \rangle\}$ of momentum eigenstates, where $k=(k_{0}, \vec{k})$, $k_{0}=|\sqrt{\vec{k}^{2}+m^{2}}|$. 

If $|k\rangle$ is a single-particle state, $U(g)$ transforms $|k\rangle$ as
\begin{eqnarray}
U(g) |k\rangle = |g k\rangle.
\end{eqnarray}
The usual action of $\bar{P}^{\uparrow}_{+}$ on the two-particle Hilbert space ${\cal H} \otimes {\cal H}$ is then given by
\begin{eqnarray}
U(g) \otimes U(g) \,|k_1\rangle \otimes |k_2\rangle & \equiv & [U \otimes U](g \times g) \,|k_1\rangle \otimes |k_2\rangle \, = \, |g k_1\rangle \otimes |g k_2\rangle.
\end{eqnarray}
Similar relations hold for all multi-particle states. 

Thus in defining the group action on multi-particle states, we made use of the isomorphism $G \rightarrow G \times G$ defined by $g \rightarrow g \times g$. 
This map is essential for defining the group action on multi-particle states. It is said to be a coproduct on $G$ and denoted by $\Delta$. The usual Poincar\'e group  
has associated with it a canonical coproduct $\Delta_0$
\begin{eqnarray}
\Delta_0 : G & \rightarrow & G \times G, \nonumber \\
\Delta_0(g) & = & g \times g.
\end{eqnarray}
The coproduct also exists at the level of the algebra. Tensor products of representations of an algebra are determined by the coproduct \cite{mack1, mack2,mack3}. It is a homomorphism 
from the group algebra $G^*$ to $G^* \otimes G^*$. Let $u$ be an element of the Lie algebra $\bar{P}^{\uparrow}_{+}$ of Poincar\'e group. Then
\begin{equation}
\Delta_0(u)\ =\ u\,\otimes\,\bold{1} \,+ \, \bold{1} \,\otimes\,u.
\end{equation}
The coproduct map need not be unique and all choices of coproduct are not equivalent. In particular, the Clebsch-Gordan coefficients, 
occurring in the reduction of group representations depend upon choice of coproduct. Any choice of coproduct must satisfy
\begin{eqnarray}
\label{eq:g1g2Action}
\Delta(g_{1})\Delta(g_{2}) & = & \Delta (g_{1}g_{2}), \; \; g_{1}, g_{2} \in G.
\end{eqnarray}
These definitions extend to the group algebra $G^{*}$ by linearity. If $\alpha, \beta : G \rightarrow {\mathbb C}$ are smooth compact
functions on $G$, then the group algebra $G^{*}$ contains the generating elements
$\int d\mu(g) \alpha(g) g$ and $\int d\mu(g') \alpha(g') g'$, where $d\mu$ is the measure in $G$. The coproduct action on $G^{*}$ is
\begin{eqnarray}
\Delta : G^{*} &\rightarrow& G^{*} \otimes G^{*}, \nonumber \\
\int d\mu(g) \alpha(g) g &\rightarrow&  \int d\mu(g) \alpha(g) \Delta(g).
\end{eqnarray}
If $U_{k}$ are representations of $G^{*}$ on ${\cal H}_{k}$, $k = i, j$ then
\begin{eqnarray}
U_{k} : \int d\mu(g) \alpha(g) g &\rightarrow&  \int d\mu(g) \alpha(g) U_{k}(g).
\end{eqnarray}
It also extends to the representation $U_{i} \otimes U_{j}$ on ${\cal H}_{i} \otimes {\cal H}_{j}$ as
\begin{eqnarray}
U_{i} \otimes U_{j} : \int d\mu(g) \alpha(g) g &\rightarrow&  \int d\mu(g) \alpha(g) (U_{i} \otimes U_{j})\Delta(g).
\end{eqnarray}
Thus it is the coproduct $\Delta$ which determines the action of the symmetry group on the tensor product of two representations $\rho_1$ and $\rho_2$,
\begin{eqnarray}
g \triangleright (\alpha \otimes \beta) = (\rho_1 \otimes \rho_2)\Delta(g)(\alpha \otimes \beta).
\end{eqnarray}
If the representation space is itself an algebra ${\cal A}$ and $\alpha, \beta \in {\cal A}$, then there exists a rule for taking products of elements of ${\cal A}$ 
called as multiplication map $m$ given by
\begin{eqnarray}
&& m : {\cal A} \otimes {\cal A} \rightarrow {\cal A}, \nonumber \\
&& \alpha \otimes \beta \rightarrow m(\alpha \otimes \beta)=\alpha \beta.
\end{eqnarray}
It is essential that the choice for $\Delta$ be compatible with a given multiplication map $m$ i.e. 
\begin{eqnarray}
\label{eq:compatibility}
m\Big[(\rho \otimes \rho)\Delta(g)(\alpha \otimes \beta)\Big]=\rho(g)m(\alpha \otimes \beta),
\end{eqnarray}
where $\rho$ is a representation of the group acting on the algebra.

The compatibility condition (\ref{eq:compatibility}) can be better illustrated by the diagram
\begin{equation}
\begin{array}{ccc}
\alpha \otimes \beta & \longrightarrow & ( \rho \otimes \rho ) \Delta
(g) \alpha \otimes \beta \\
& &  \\ m \,\, \downarrow  & & \downarrow \,\, m \\ &  & \\ m(\alpha
\otimes
\beta) &
\longrightarrow & \rho(g) m (\alpha \otimes \beta)
\end{array}
\end{equation}
It implies that, multiplying two functions and then transforming them under the group action should be same as first transforming the individual functions and then multiplying them.
If such a $\Delta$ compatible with $m$ can be found, $G$ is said to be an automorphism of ${\cal A}$. In the absence of such a $\Delta$, $G$ does not act on ${\cal A}$ as a symmetry 
group.

With the above defined compatibility condition, let us consider the action of $P_{+}^{\uparrow}$ on the noncommutative spacetime algebra (GM plane)  ${\cal A}_{\theta}({\mathbb R}^{d+1})$. 
The algebra ${\cal A}_{\theta}({\mathbb R}^{d+1})$ consists of smooth functions on ${\mathbb R}^{d+1}$ with the multiplication map
\begin{equation}
m_{\theta}: {\cal A}_{\theta}({\mathbb R}^{d+1}) \otimes {\cal A}_{\theta}({\mathbb R}^{d+1}) \rightarrow {\cal A}_{\theta}({\mathbb R}^{d+1}).
\end{equation}
For two functions $\alpha$ and $\beta$ in the algebra ${\cal A}_{\theta}$, the multiplication map is not the point-wise multiplication but the $*$-multiplication
\begin{eqnarray}
m_{\theta} (\alpha \otimes \beta) (x) \, = \, (\alpha \ast \beta)(x) \, = \, \exp \Big(\frac{i}{2}\theta^{\mu \nu}\frac{\partial}{\partial x^{\mu}}
\frac{\partial}{\partial y^{\nu}}\Big)\alpha(x)\beta(y)\Big|_{x=y}. 
\label{starproduct}
\end{eqnarray}
We can write $m_{\theta}$ in terms of the commutative multiplication map $m_{0}$ as
\begin{eqnarray}
m_{\theta} = m_{0} {\cal F}_{\theta}, \qquad 
{\cal F}_{\theta} = \textrm{exp}(-\frac{i}{2}\theta^{\mu \nu} P_{\mu} \otimes P_{\nu}), \qquad P_{\mu} = -i \partial_{\mu}
\label{dtwist}
\end{eqnarray}
where ${\cal F}_{\theta}$ is the ``Drinfel'd twist". Also, in writing (\ref{dtwist}), we have used Minkowski metric with signature ($+, -, -, -$) for raising or lowering the indices.

It can be shown that on the noncommutative algebra of functions, the Poincar\'e group action through the usual coproduct $\Delta_0(g)$  is not compatible with the
 $\ast$-product in the sense of (\ref{eq:compatibility}). Thus, $P_{+}^{\uparrow}$ does not act on ${\cal A}_{\theta}({\mathbb R}^{d+1})$ in the usual way. 

But inspite of the incompatibility of usual coproduct $\Delta_0(g)$ with $\ast$-product, there is a way to implement Poincar\'e symmetry on noncommutative algebra. As mentioned 
before the choice of coproduct is not unique. Using the twist element, the coproduct of the universal enveloping algebra ${\cal U}(\bar{P}^{\uparrow}_{+})$ of the Poincar\'e algebra can be 
appropriately deformed in such a way that the deformed coproduct $\Delta_{\theta}$ is compatible with $\ast$-multiplication. The deformed coproduct is given by
\begin{eqnarray}
\Delta_{\theta} = {\cal F}^{-1}_{\theta} \Delta_0 {\cal F}_{\theta}.
\label{tcop}
\end{eqnarray}
It is easy to check that the twisted coproduct $\Delta_{\theta}$ is compatible with the twisted multiplication $m_{\theta}$ as
\begin{eqnarray}
m_{\theta} \left( (\rho \otimes \rho) \Delta_{\theta}(g) ( \alpha \otimes
\beta ) \right) &=&
m_{0} \left( {\cal F}_{\theta} ({\cal F}_{\theta}^{-1} \rho(g) \otimes
\rho(g) {\cal F}_{\theta}) \alpha \otimes \beta \right) \nonumber \\
&=& \rho(g) \left( \alpha * \beta \right), \quad
 \alpha,\beta \in {\cal A}_\theta(\mathbb{R}^{d+1}).
\label{proofcomp}
\end{eqnarray}
The compatibility between $\Delta_{\theta}$ and $m_{\theta}$ given by (\ref{proofcomp}), can be illustrated by the diagram
\begin{equation}
\begin{array}{ccc}
\alpha \otimes \beta & \longrightarrow & ( \rho \otimes \rho ) \Delta_{\theta}
(g) \alpha \otimes \beta \\
& &  \\ m_{\theta} \,\, \downarrow  & & \downarrow \,\, m_{\theta} \\ &  & \\ \alpha * \beta &
\longrightarrow & \rho(g) (\alpha * \beta)
\end{array}
\end{equation}
Thus with the choice of twisted coproduct $\Delta_{\theta}$, $G$ acts as an automorphism of ${\cal A}_{\theta}$.

From (\ref{tcop}) it is easy to see that the coproduct for the generators of the Lie algebra of the Lorentz group $M_{\mu \nu}$ gets deformed but that for the generators 
of the translation group $P_{\alpha}$ remains unchanged i.e.
\begin{eqnarray}
\Delta_{\theta}(M_{\mu \nu}) & = & 1 \otimes M_{\mu \nu} + M_{\mu \nu} \otimes 1 - \frac{1}{2}\Big[(P\cdot \theta)_{\mu}\otimes P_{\nu}-P_{\nu}\otimes (P\cdot \theta)_{\mu} 
- (\mu \leftrightarrow \nu) \Big], \nonumber \\
\Delta_{\theta}(P_{\alpha}) & = & \Delta_0 (P_{\alpha}),
\label{tgenrators}
\end{eqnarray}
where $(P \cdot \theta)_{\mu} = P_{\sigma} \theta^{\sigma}_{\mu}$. 

The idea of twisted coproduct can be traced back to Drinfel'd's work in mathematics \cite{drinfeld}. The Drinfel'd twist leads naturally to deformed $R$-matrices and statistics 
for quantum groups \cite{majid}. The idea of twisting the coproduct in noncommutative spacetime algebra is due to \cite{chaichian, drinfeld, majid, fiore2, 
fiore1, fioresolo1, fioresolo2, watts1, Oeckl:2000eg, watts2, gms, Dimitrijevic:2004rf, matlock, aschieri3}.


\section{The Twisted Statistics}


In this section we discuss the implications of twisted Poincar\'e symmetry in quantum theory \cite{sachin-twist}. As we shall see, in quantum theory, the twisting of the coproduct 
implies a new type of statistics called as ``twisted statistics''. We start with first discussing the origin of twisted statistics in quantum mechanics and then generalize 
our discussion to the quantum field theory.


\subsection{In Quantum Mechanics}


In quantum mechanics the physical wave functions describing identical particles are required to be either symmetric (bosons) or antisymmetric (fermions). This requires us to work 
with either the symmetrized or antisymmetrized tensor product i.e. if $\phi$ and $\psi$ are single particle wavefunctions of two identical particles then the two-particle 
(anti-)symmetrized wavefunction is given by 
 \begin{eqnarray} 
  \phi \otimes_S \psi & \equiv & \frac{1}{2} \left[ \phi \otimes \psi \, + \, \psi \otimes \phi \right]  \quad = \quad \left( \frac{1 \, + \, \tau_{0}}{2} \right) ( \phi \otimes \psi ),
\nonumber \\
\phi \otimes_A \psi & \equiv & \frac{1}{2} \left[ \phi \otimes \psi \, - \, \psi \otimes \phi \right]  \quad = \quad \left( \frac{1 \, - \, \tau_{0}}{2} \right) ( \phi \otimes \psi ),
\label{sym}
 \end{eqnarray}
where $\tau_0$ is the statistics (flip) operator associated with exchange 
\begin{eqnarray}
 \tau_0 \, ( \phi \otimes \psi ) & = & \psi \otimes \phi, \qquad \tau^2_0 \, = \, \bold{1} \otimes \bold{1}.
 \label{flip}
\end{eqnarray}
The (anti-)symmetrized wavefunction (\ref{sym}) satisfy
\begin{eqnarray} 
  \phi \otimes_S \psi & = & + \, \psi \otimes_S \phi,  \nonumber \\
\phi \otimes_A \psi & = & - \, \psi \otimes_A \phi.
\label{sym1}
 \end{eqnarray}
In a theory with Lorentz invariance, these relations should hold in all frames of reference. In other words, performing a Lorentz transformation on $\phi \otimes \psi $ and then
(anti-) symmetrizing has to be the same as (anti-)symmetrization followed by the Lorentz transformation. In the commutative case, since  $\Delta_0(g)=g \times g$, we have 
\begin{eqnarray} 
\tau_0 (\rho \otimes \rho) \Delta_0(g) = (\rho \times \rho)\Delta_0(g) \tau_0, \; \; g \in P_{+}^{\uparrow}.
\label{mcomsym}
\end{eqnarray}
Hence in commutative case, due to (\ref{mcomsym}) the Lorentz transformations preserve (anti-) symmetrization. 

As shown in \cite{sachin-twist}, the twisted coproduct action of the Lorentz group is not compatible with usual symmetrization/antisymmetrization.
The origin of this non-compatibility can be traced to the fact that the coproduct is not cocommutative except when $\theta^{\mu \nu}={0}$. That is,
\begin{eqnarray} 
\tau_0 (\rho \otimes \rho) \Delta_\theta (g) \neq (\rho \times \rho)\Delta_\theta (g) \tau_0. 
\label{mncomsym}
\end{eqnarray}
Hence the usual statistics is not compatible with the twisted coproduct and one has to construct an appropriate twisted statistics operator $\tau_{\theta}$ compatible with 
twisted coproduct. Such a twisted statistics operator can be easily constructed upon noticing that $\Delta_{\theta}(g)={\cal F}_{\theta}^{-1}\Delta_0(g){\cal F}_{\theta}$.
The new statistics operator $\tau_{\theta}$ compatible with $\Delta_\theta$ is given by
\begin{eqnarray}
\tau_{\theta} &=& {\cal F}_{\theta}^{-1} \tau_{0} {\cal F}_{\theta}, \qquad \tau^2_\theta \, = \, \bold{1} \otimes \bold{1}.
\end{eqnarray}
So the appropriate physical two-particle wave functions are twisted (anti-)symmetrized ones and are given by
\begin{eqnarray} 
  \phi \otimes_{S_{\theta}} \psi & \equiv &  \left( \frac{1 \, + \, \tau_{\theta}}{2} \right) ( \phi \otimes \psi ), \nonumber \\
\phi \otimes_{A_{\theta}} \psi & \equiv &  \left( \frac{1 \, - \, \tau_{\theta}}{2} \right) ( \phi \otimes \psi ).
\label{tsym}
 \end{eqnarray}
For plane waves states $e_p(x) = e^{-i p \cdot x}$ we get
\begin{eqnarray} 
\left(\frac{\bold{1} \, \pm \, \tau_\theta}{2}\right)(e_p \otimes e_q) & \equiv & e_p \otimes_{S_\theta,A_\theta} e_q \, = \, \pm\, e^{-i p_\mu \theta^{\mu \nu} q_\nu} e_q 
\otimes_{S_\theta,A_\theta}  e_p \, ,\\
(e_p \otimes_{S_\theta,A_\theta} e_q)(x_1,x_2) & = & \pm \, e^{-i \frac{\partial}{\partial x^\mu_1} \theta^{\mu \nu} \frac{\partial}{\partial x^\nu_2}} (e_p \otimes_{S_\theta,A_\theta}
  e_q)(x_2,x_1),
\label{tplane}
\end{eqnarray} 
where $+$ sign should be taken for bosons and $-$ sign should be taken for fermions.

A similar analysis can be done for all multi-particle wavefunctions and one can easily see that the correct multi-particle wavefunctions on GM plane have to be twisted (anti-) symmetrized.
As we show in next section, the above analysis can also be extended to field theories on GM plane.


\subsection{Twisted Quantum Fields}


Having studied the implications of twisted Poincar\'e symmetry on statistics of particles in a quantum mechanics context, we now look at its implications in a quantum field theory.

In quantum field theory, a quantum field when evaluated at a spacetime point gives us an operator-valued distribution acting on a Hilbert space. For example, a quantum field at a 
spacetime point $x_{1}$ acting on the vacuum gives us one-particle state centered at $x_{1}$. 
Similarly, the product of two quantum fields at spacetime points $x_{1}$ and $x_{2}$ act on the vacuum and generate a two-particle state where one particle is centered at $x_{1}$ 
and the other at $x_{2}$. 

In the commutative case, a free real scalar quantum field $\phi_{0}(x)$ has the mode expansion
\begin{eqnarray} 
\phi_{0}(x) & = & \int \frac{d^3 p}{(2\pi)^3 2 E_p}\, \left[ c_ p \, \textrm{e}_{p}(x) \, + \, c^\dagger_ p \; \textrm{e}_{-p}(x) \right]
\end{eqnarray}
where $\textrm{e}_{p}(x) = \textrm{e}^{-i\; p\cdot x}$. The creation/annihilation operators satisfy the standard commutation relations
\begin{eqnarray}
  c_{p_{1}} c_{p_{2}} & = & \eta \, c_{p_{2}} c_{p_{1}}, \nonumber \\
c^{\dagger}_{p_{1}} c^{\dagger}_{p_{2}} & = & \eta \, c^{\dagger}_{p_{2}} c^{\dagger}_{p_{1}}, \nonumber \\
c_{p_{1}} c^{\dagger}_{p_{2}} & = & \eta \, c^{\dagger}_{p_{2}} c_{p_{1}} + (2\pi)^3 \, 2 E_p \,\delta^{3}(p_{1} \, - \, p_{2}),
\label{ccom}
\end{eqnarray}
where $\eta = 1$ for bosons and $\eta = -1$ for fermions. The creation operator must satisfy
\begin{eqnarray}
\langle 0 |\phi_{0}(x) c^\dagger_ p |0\rangle & = & e_p(x) \, = \, e^{-i p \cdot x}, \\
\frac{1}{2} \langle 0 | \phi_{0}(x_1) \phi_{0}(x_2) c^\dagger_ q c^\dagger_ p |0\rangle & = &  \left(\frac{\bold{1} \pm \tau_0 }{2}\right) (e_p \otimes e_q)(x_1,x_2) \nonumber \\
& \equiv & (e_p \otimes_{S_0,A_0} e_q)(x_1,x_2).
\end{eqnarray}
The above equations give us the standard prescription to establish the connection between quantum field operators and (multi-)particle wavefunctions.

In the noncommutative case let $a^\dagger_p$ be the creation operator of the noncommutative real scalar field $\phi_\theta(x)$. Then as in standard quantum field theory, we should have
\begin{eqnarray} 
\langle 0 |\phi_\theta(x) a^\dagger_p |0\rangle & = & e_p(x), \\
\frac{1}{2}\langle 0 |\phi_\theta(x_1) \phi_\theta(x_2) a^\dagger_q a^\dagger_p |0\rangle & = & \left(\frac{{\bf 1} \pm \tau_\theta}{2}\right)(e_p \otimes e_q)(x_1,x_2) \nonumber \\
& \equiv & (e_p \otimes_{S_\theta,A_\theta} e_q)(x_1,x_2) .
\label{tbasis} 
\end{eqnarray} 
If $|p, q\rangle_{S_\theta,A_\theta} $ represents a two-particle state labeled by momenta, then from (\ref{tbasis}) we see that upon exchanging $p$ and $q$  we get 
\begin{equation}
|p, q\rangle_{S_\theta,A_\theta} = \eta \, e^{ -i p_\mu \theta^{\mu\nu} q_\nu}\,|q,p \rangle_{S_\theta,A_\theta} .
\label{tistate}
\end{equation}
If we define $a^\dagger_k$ to be an operator which adds a particle to the right of the particle list,
\begin{eqnarray}
 a^\dagger_k | p_1,p_2 \dots p_n \rangle_\theta & = & | p_1,p_2 \dots p_n, k \rangle_\theta.
\label{tiaanni}
\end{eqnarray}
Then the two particle state can be written as 
\begin{eqnarray}
 |p, q  \rangle_{\theta} & = & a^\dagger_q\, a^\dagger_p \, | 0 \rangle.
\label{iint2state}
\end{eqnarray} 
Hence from (\ref{tistate}) and (\ref{iint2state}) we have 
\begin{eqnarray}
  a_{p} a_{q} & = & \eta \, e^{ip \wedge q} \,a_{q} a_{p}, \nonumber \\
a^{\dagger}_{p} a^{\dagger}_{q} & = & \eta \, e^{ip \wedge q} \, a^{\dagger}_{q} a^{\dagger}_{p}, \nonumber \\
a_{p} a^{\dagger}_{q} & = & \eta \, e^{-ip \wedge q} \, a^{\dagger}_{q} a_{p} + (2\pi)^3 \, 2 E_p \,\delta^{3}(p \, - \, q),
\label{tcom}
\end{eqnarray}
where $E^2_p = \vec{p}^2 + m^2 $, $p \wedge q \, = \, p_\mu \theta^{\mu\nu} q_\nu$ and $\eta = \pm 1$ depending on whether the particles are ``twisted bosons'' $(+1)$ or 
``twisted fermions'' $(-1)$. 

Thus we have a new type of commutation relation reflecting the deformed quantum symmetry. Therefore, while constructing a quantum field theory on noncommutative spacetime, 
we should twist the creation and annihilation operators in addition to the $\ast$-multiplication between the fields. Noncommutative field theories without twisted commutation 
relations do not preserve the classical twisted Poincar\'e invariance at quantum level and suffer from UV/IR mixing 
\cite{pinzul-uvir}. The twisted statistics is a novel feature of fields on GM plane. It leads to interesting new effects like Pauli forbidden transitions \cite{bal-pauli,pramod-pauli} and 
changes in certain thermodynamic quantities \cite{basu-th, basu1-th}. It can be used to search for signals of noncommutativity in certain experiments involving 
Ultra High Energy Cosmic Rays (UHECRs) \cite{rahul-hbt} and Cosmic Microwave Background (CMB) \cite{anosh-cmb}.

At this point we like to remark that, in literature there exist another approach to quantization of noncommutative fields \cite{chaic}. In this approach, the quantization 
is done according to the usual rules and the 
quantum fields follow usual bosonic/fermionic statistics. However, such a quantization scheme does not preserve the classical twisted Poincar\'e invariance and suffers
from UV/IR mixing \cite{pinzul-uvir}. In this thesis we only discuss the twisted quantization as it preserves twisted Poincar\'e invariance in noncommutative quantum field theories.

The twisted creation/annihilation operators ($a^{\dagger}_p, a_p $) are related to ordinary creation/ 
annihilation operators ($ c^{\dagger}_p, c_p $) 
satisfying usual statistics by the ``dressing transformation'' \cite{Grosse, Zamolodchikov,Faddeev}: 
\begin{eqnarray}
a_p & = &  c_p \, e^{-\frac{i}{2} \, p \wedge P}, \nonumber \\
a^\dagger_p & = &  c^\dagger_p \, e^{\frac{i}{2} \, p \wedge P}, 
\label{dresstransform}
\end{eqnarray}
where 
\begin{eqnarray}
 P_\mu \, = \, \int \frac{d^3 p}{(2\pi)^3 2 E_p} \, p_\mu \, c^\dagger_p c_p \, = \, \int  \frac{d^3 p}{(2\pi)^3 2 E_p} \, p_\mu a^\dagger_p a_p 
\label{momfop}
\end{eqnarray}
is the Fock space momentum operator. The antisymmetry of $\theta^{\mu \nu}$ allows us to write
\begin{eqnarray}
c_{ p} e^{-\frac{i}{2} p \wedge P} = e^{-\frac{i}{2} p \wedge P} c_{ p}, 
\end{eqnarray}
\begin{eqnarray}
c^{\dagger}_{ p} e^{\frac{i}{2} p \wedge P} = e^{\frac{i}{2} p \wedge P} c^{\dagger}_{ p}. 
\end{eqnarray}
Hence the ordering of factors here is immaterial. 

It should also be noted that the map from the $c$- to the $a$-operators is invertible,
\begin{eqnarray}
c_{ p} & = & a_{ p} \; e^{\frac{i}{2}p \wedge P }, \nonumber \\
c^\dagger_{p} & = & a^\dagger_{ p}\; e^{-\frac{i}{2} p \wedge P}.
\end{eqnarray}
where $P_{\mu}$ is written as in eqn.~(\ref{momfop}).
Using the ``dressing transformation'' of (\ref{dresstransform}), one can relate $\phi_{\theta}$ with the commutative real scalar field $\phi_{0} $ as
\begin{eqnarray}
 \phi_{\theta} (x) & = & \phi_{0} (x) \, e^{\frac{1}{2} \, \overleftarrow{\partial} \wedge P}.
\label{fielddress}
\end{eqnarray}
This is an important identity and helps us to relate noncommutative expressions with their analogous commutative ones. In the later chapters of this thesis we will repeatedly 
make use of the relations (\ref{tcom})-(\ref{fielddress}) to simplify our computations. \\  
Thus the twisted quantum field $\phi_{\theta}$ differs from the untwisted quantum field $\phi_{0}$ in two ways:\\
\textbf{(1)} \quad $e_{p} \in {\cal A}_\theta(\mathbb{R}^{d+1})$: When evaluating the product of $\phi_\theta$'s at the same point, we must multiply them by $\ast$-product. \\
and \\
\textbf{(2)} \quad $a_{{ p}}$ satisfy twisted commutation relations: The field $\phi_\theta$ follows a twisted statistics. \\
The $\ast$-product between the twisted quantum fields is
\begin{eqnarray}
(\phi_{\theta} \ast \phi_{\theta})(x) = \phi_{\theta}(x) e^{\frac{i}{2} \overleftarrow{\partial} \wedge \overrightarrow{\partial}} \phi_{\theta}(y) |_{x=y}, \qquad
\overleftarrow{\partial} \wedge \overrightarrow{\partial} = \overleftarrow{\partial}_{\mu} \theta^{\mu \nu} \overrightarrow{\partial}_{\nu}.
\end{eqnarray}
From the dressing transformation (\ref{fielddress}) between noncommutative $\phi_\theta$ and commutative $\phi_0$ fields, it follows that the $\ast$-product of an 
arbitrary number of fields $\phi_{\theta}^{(i)}$ ($i$ = 1, 2, 3, $\cdots$ n) can be written as
\begin{eqnarray}
\phi_{\theta}^{(1)} \ast \phi_{\theta}^{(2)} \ast \cdots \ast \phi_{\theta}^{(n)} & = & (\phi^{(1)}_{0}\phi^{(2)}_{0} \cdots \phi^{(n)}_{0}) \; 
e^{\frac{i}{2} \overleftarrow{\partial} \wedge P}.
\label{eq:productfields}
\end{eqnarray}
This is a very useful relation and in the following chapters we will make extensive use of it to simplify computations.

Although, for sake of simplicity we have confined our discussion only to the real scalar fields on GM plane but it can be easily generalized to all tensorial 
and spinorial quantum fields. In case of spinor fields we have to consider anti-commuting creation/annihilation operators which again satisfy the twisted commutation relations
(\ref{tcom}) but with $\eta = -1$. The rest of the discussion follows closely to the scalar case we discussed. 

Using the twisted fields one can write field theories on GM plane. For example, a generic interaction Hamiltonian density built out of the twisted scalar quantum fields is given by
\begin{eqnarray}
{\cal H}_{I}(x) = \phi(x) \ast \phi(x) \ast \cdots \ast \phi(x).
\end{eqnarray}
It transforms like a scalar in the noncommutative theory also. (This is the case only when we choose a $\ast$-product between the fields to write down the 
Hamiltonian density.) This form of the Hamiltonian and the twisted statistics of the fields is all that is required to show that there is no UV-IR mixing in this theory. 
Twisted field theories involving real scalar field $\phi_{\theta}$ and having a $\phi^4_{\theta,\ast}$ interactions are discussed in \cite{bal-uvir} and 
are shown to be free from UV/IR mixing. The issue of locality and causality of twisted fields is discussed in \cite{sachin-locality}. The discussion regarding CPT properties
can be found in \cite{bal-cpt}.

Gauge field theories with nonabelian gauge groups are constructed in \cite{amilcar-pinzul, sachin-pinzul}. Noncommutative field theories involving nonabelian gauge 
fields violate twisted Poincar\'e invariance and are know to suffer from UV/IR mixing. Construction of thermal field theories is done in \cite{amilcar-th1, amilcar-th2, trg }.
Some interesting aspects of quantum mechanics in noncommutative spacetime can be found in \cite{nitin, nitint},
while \cite{asrarul} discusses the twisted bosonization in two dimensional noncommutative spacetime. A comprehensive review of twisted field theories can be found in \cite{balreview}.


\section{Plan of the Thesis}


In this thesis we study field theories on the GM plane. In this chapter we briefly reviewed the novel features
of field theories on GM plane e.g. the $\ast$-product, restoration of Poincar\'e-Hopf symmetry and twisted commutation relations. After this brief review, in next chapter we discuss 
our work on renormalization of field theories on GM plane. We first give a review of the noncommutative interaction picture, the Lehmann-Symanzik-Zimmermann
(LSZ) reduction formula and show their equivalence. 
We then take up the problem of renormalization of noncommutative theories involving only matter fields. We show that any generic noncommutative theory involving pure matter fields is 
a renormalizable theory if the analogous commutative theory is renormalizable.
We further show that all such noncommutative theories will have same fixed points and $\beta$-functions for the couplings, as that of the analogous commutative theory. 
As discussed in this chapter, the unique feature of the these field theories is the twisted statistics of the particles. Motivated by it, in the third chapter, we look at 
the possibility of twisted statistics by deforming internal symmetries instead of spacetime symmetries. We construct two different twisted theories which can be viewed as internal 
symmetry analogue of the GM plane and dipole field theories. We further study their various properties like the issue of causality and the scattering formalism. 
Having studied the mathematical properties of noncommutative and twisted internal symmetries we move on to discuss their potential phenomenological signatures. We first discuss the 
noncommutative thermal correlation functions and show that because of the twisted statistics, all correlation functions except two-point function get modified. 
Finally we discuss the modifications in Hanbury-Brown Twiss (HBT) correlation functions due to twisted statistics on GM plane 
and the potential of observing signatures of noncommutativity by doing a HBT correlation experiment with UHECRs. 

The plan of the thesis is as follows : 

In the first chapter we have reviewed the basic concepts and well-known results of field theories written on GM plane. 

In the second chapter we start with reviewing the formalism 
of noncommutative interaction picture and the noncommutative scattering theory. After that we discuss the noncommutative LSZ formalism and the reduction formula. 
We also show the equivalence of the interaction picture and LSZ approach for such theories. We then discuss the issue of renormalizability. For sake of simplicity, we start 
with discussing the renormalization and computation of $\beta$-function for a noncommutative real scalar field theory with $\phi^4_{\theta,\ast}$ self interactions. We show 
that the $\phi^4_{\theta,\ast}$ theory is renormalizable, is free from UV/IR mixing and the $\beta$-function is same as that of the analogous commutative theory. 
We then look at a generic pure matter theory without gauge fields. We show that any generic noncommutative field theory having only matter fields is a renormalizable theory, 
provided the corresponding commutative theory is also renormalizable. All such theories are free from UV/IR mixing and have identical fixed points and $\beta$-functions for 
the various couplings as the analogous commutative theory. 

In the third chapter we discuss the possibility of constructing Poincar\'e invariant field theories having twisted statistics. In other words, we construct field theories where 
fields transform in standard way under Poincar\'e transformation (as opposed to the twisted transformation on GM plane) but nonetheless have twisted statistics. Such theories 
are constructed by deforming the transformation
properties of the fields under a global $SU(N)$ group. We construct two such twisted field theories. We further study the issue of causality, the scattering formalism.
We begin with briefly reviewing the treatment of global symmetries, in particular $SU(N)$ group, in the usual untwisted case. 
We then discuss a specific type of twist called ``antisymmetric twist''. This kind of twist is quite similar in spirit to the twisted noncommutative field theories. 
The formalism developed here will closely resemble (with generalizations and modifications which we will elaborate on) the formalism of twisted noncommutative theories. 
We then go on to construct
more general twisted statistics which can be viewed as internal symmetry analogue of dipole theories. We also discuss the construction of interaction terms and scattering formalism
for both types of twists. We end the chapter with discussion of causality of such twisted field theories.

In the fourth chapter we discuss the formalism of Green's functions to compute correlation functions and adapt it to the noncommutative case. We show that due to twisted commutation 
relations satisfied by the fields on GM plane, all correlation functions apart from two-point correlation function get modified.

In the fifth chapter we look at probable signatures of noncommutativity in UHECRs. We look at the modifications in noncommutative HBT correlation function
due to the twisted statistics. We show that the commutative and noncommutative HBT correlation functions differ from each other and the difference gets more and more pronounced as 
we go to higher and higher energies. Hence an HBT experiment with UHECRs can provide us potential signatures of noncommutativity. 

We finally conclude our work in the last chapter.


\chapter{Renormalization and $\beta$-Function}
\label{chap:chap2}

In this chapter, we discuss the issue of renormalization of noncommutative field theories on GM plane. We use the formalism of twisted field theories as outlined in chapter one
of the thesis. We show that any generic noncommutative field theory with only matter fields is a renormalizable theory, provided the corresponding commutative theory is also 
renormalizable. Moreover, we show that all such theories are free of UV/IR mixing. We further argue that they have identical fixed points as analogous commutative theory. 
We also obtain the $\beta$-functions for the various couplings in analogy with commutative theory.

It should be noted that the results obtained in this chapter hold true only for twisted field theories on GM plane. As remarked in the first chapter, in literature there exist 
other approaches to noncommutative theories on GM plane \cite{chaic, gurau, grosse-2, gayral, shiraz-uvir, grosse-nc, sachin-nc-matrix, blaschke}. 
Typically, in such models the quantization is done according to the commutative rules 
and the quantum fields follow usual bosonic/fermionic statistics. As a result, the quantum fields in these models are same as their commutative counterparts. 
The noncommutativity is encoded only in the $\ast$-product with which the fields are multiplied. However, such a quantization scheme does not preserve the classical 
twisted Poincar\'e invariance \cite{pinzul-uvir}. Moreover, some of the models also suffer from UV/IR mixing \cite{pinzul-uvir, shiraz-uvir, sachin-nc-matrix}. 
In this chapter we will restrict only to the discussion of twisted field theories and will not discuss any of these models. Interested readers can look at \cite{blaschke}
for a recent work and a brief review of these models. 

The plan of the chapter is as follows. We first start reviewing the formalism of noncommutative interaction picture and the noncommutative scattering theory. For the sake of simplicity 
we choose a specific model of the noncommutative real scalar fields having a $ \phi^4_{\theta,\ast}$ self interaction. We compute the $S$-matrix elements and show that the $S$-matrix 
elements have only an overall noncommutative phase and hence absence of UV/IR mixing in this theory. We also show the equivalence of noncommutative $S$-matrix with the commutative 
$S$-matrix. We then review the noncommutative Lehmann-Symanzik-Zimmermann
(LSZ) formalism (again for simplicity we will restrict only to real scalar fields) for computing $S$-matrix elements and show the equivalence 
of the two approaches. We then present our work on renormalization of this theory and show that it is renormalizable. We further compute the fixed point and $\beta$-function for 
the coupling. We show that this noncommutative theory shares the same fixed point and $\beta$-function as the analogous commutative $\phi^4_0$ theory. We also show the absence of 
UV/IR mixing in the renormalized theory. We then conclude with comments about more complicated and generic noncommutative theories involving only matter fields. We finally argue 
that our analysis although explicitly done only for a specific model holds true for all such theories.

This chapter is based on the work published in \cite{renor}.


\section{Noncommutative Interaction Picture}  


For the sake of completeness, in this section, we start reviewing the formalism of scattering theory for a generic noncommutative theory using the 
``noncommutative interaction picture''. For the sake of simplicity and definiteness, we choose a specific type of interaction hamiltonian 
$H_{\theta, \rm{Int}} = \phi^4_{\theta,\ast}$. We will compute the $S$-matrix $\hat{S}_\theta$ and $S$-matrix elements for a generic scattering problem. We also show the relation of 
these quantities with the commutative $S$-matrix $\hat{S}_0$ and $S$-matrix elements. The results discussed in this section are due to the work of \cite{bal-uvir} and the interested 
reader is referred to it for further details.


\subsection{General Formalism}


Many questions in field theory, especially those related to the scattering problem, are best discussed in the interaction (also known as Dirac) picture. Using interaction picture 
for calculations has many obvious advantages, making the calculations 
much easier. Hence, it is desirable, for the work done here, to have a noncommutative interaction picture. With this in mind, we briefly review the noncommutative interaction picture.
 The formalism developed here is quite similar to that of ordinary commutative field theories, for which any good book on field theory \cite{weinberg,greiner} can be consulted.

Let $\hat{H}_{\theta}$ be the full Hamiltonian for the system of interest and we assume that it can be split into two parts, the free part $\hat{H}_{\theta,F}$ and the interaction 
part $\hat{H}_{\theta, \rm{Int}} $ i.e.
\begin{eqnarray}
\hat{H}_{\theta} = \hat{H}_{\theta, F} + \hat{H}_{\theta, \rm{Int}}.
 \label{ham}
\end{eqnarray}
Let $\hat{O}^{H}_{\theta}(t)$ be a noncommutative operator in the Heisenberg Picture satisfying the Heisenberg equation of motion
\begin{eqnarray}
i\partial_{t}\hat{O}^{H}_{\theta}(t) = \left[\hat{O}^{H}_{\theta}(t) , \hat{H}_{\theta} \right].
 \label{heq}
\end{eqnarray}  
The formal solution of (\ref{heq}) is given by
\begin{eqnarray}
\hat{O}^{H}_{\theta}(t) = e^{i \hat{H}_{\theta}(t-t_{0})} \hat{O}^{H}_{\theta}(t_{0}) e^{-i \hat{H}_{\theta}(t-t_{0})}.
 \label{htev}
\end{eqnarray}
Furthermore, like in the commutative case, the state vectors $\left| \alpha, t \right> ^{H}_{\theta} $ are time independent, i.e.
\begin{eqnarray}
 \left| \alpha, t \right> ^{H}_{\theta} = \left| \alpha, t_{0} \right> ^{H}_{\theta} \equiv \left| \alpha \right> ^{H}_{\theta}.
\label{hstate}
\end{eqnarray}
Now, we define the noncommutative interaction picture operator $\hat{O}^{I}_{\theta}(t)$ and state vector $\left| \alpha, t \right> ^{I}_{\theta} $ as
\begin{eqnarray}
\hat{O}^{I}_{\theta}(t)  =  e^{i \hat{H}_{\theta,F}t} e^{-i \hat{H}_{\theta}t} \hat{O}^{H}_{\theta}(0) e^{i \hat{H}_{\theta}t} e^{-i \hat{H}_{\theta,F}t}   
\label{iop}
\end{eqnarray}
and
\begin{eqnarray}
 \left| \alpha, t \right> ^{I}_{\theta} =  e^{i \hat{H}_{\theta,F}t} e^{-i \hat{H}_{\theta}t} \left| \alpha \right> ^{H}_{\theta}.
\label{istate}
\end{eqnarray}
In writing (\ref{iop}) and (\ref{istate}) we have assumed that the two pictures agree at the (arbitrarily chosen) time $t_0$.

The interaction picture operator $\hat{O}^{I}_{\theta}(t)$ defined by (\ref{iop}) satisfies the equation of motion
\begin{eqnarray}
i\partial_{t}\hat{O}^{I}_{\theta}(t) = \left[\hat{O}^{I}_{\theta}(t) , \hat{H}_{\theta,F} \right]
 \label{ieq}
\end{eqnarray} 
with formal solution written as
\begin{eqnarray}
\hat{O}^{I}_{\theta}(t) = e^{i \hat{H}_{\theta,F}(t-t_{0})} \hat{O}^{I}_{\theta}(t_{0}) e^{-i \hat{H}_{\theta,F}(t-t_{0})}.
 \label{itev}
\end{eqnarray}

Also, the state vectors $\left| \alpha, t \right> ^{I}_{\theta} $ defined by (\ref{istate}) satisfy
\begin{eqnarray}
 i \partial_t \left| \alpha, t \right> ^{I}_{\theta} =  \hat{H}^{I}_{\theta, \rm{Int}} \left| \alpha,t \right> ^{I}_{\theta}.
\label{iseveq}
\end{eqnarray}
The formal solution of (\ref{iseveq}) is given by 
\begin{eqnarray}
\left| \alpha, t \right> ^{I}_{\theta}  &  =  &  \hat{U}_{\theta}(t , t_{0})  \left| \alpha,t_{0} \right> ^{I}_{\theta}  \nonumber \\ 
& = &  e^{i \hat{H}_{\theta,F}t}  e^{-i \hat{H}_{\theta}(t - t_{0})}  e^{-i \hat{H}_{\theta,F}t_{0}}  \left| \alpha,t_{0} \right> ^{I}_{\theta}
\label{istev}
\end{eqnarray}

The operator $ \hat{U}_{\theta}(t , t_{0}) $ is the ``noncommutative time evolution operator''.  Just like its commutative counterpart it also satisfies certain properties :

\begin{enumerate}
\item  Group Law:
\begin{eqnarray}
 \hat{U}_{\theta}(t_{2} , t_{1}) \hat{U}_{\theta}(t_{1} , t_{0}) =  \hat{U}_{\theta}(t_{2} , t_{0}).
\label{tg}
\end{eqnarray}

\item Identity:
\begin{eqnarray}
 \hat{U}_{\theta}(t_{0} , t_{0})  =  \mathbb{I}.
\label{tid}
\end{eqnarray}

\item Inverse Operator:
\begin{eqnarray}
 \hat{U}^{-1}_{\theta}(t_{1} , t_{0}) =  \hat{U}_{\theta}(t_{0} , t_{1}).
\label{tin}
\end{eqnarray}

\item Unitarity:
\begin{eqnarray}
 \hat{U}^{\dagger}_{\theta}(t_{1} , t_{0}) =  \hat{U}^{-1}_{\theta}(t_{1} , t_{0}).
\label{tun}
\end{eqnarray}

\item Relation between Heisenberg and interaction pictures: If the two pictures agree at (an arbitrarily chosen) time $t = t_0$ , then we have
\begin{eqnarray}
 \hat{O}^{I}_{\theta}(t) =  \hat{U}_{\theta}(t , t_0) \hat{O}^{H}_{\theta}(t) \hat{U}^{\dagger}_{\theta}(t , t_0)
\label{tevop}
\end{eqnarray}
and
\begin{eqnarray}
  \left| \alpha, t \right> ^{I}_{\theta} =  \hat{U}_{\theta}(t , t_0)   \left| \alpha\right> ^{H}_{\theta},
\label{tevs}
\end{eqnarray}
so that $ \hat{U}_{\theta}(t , t_{0}) $ satisfies the differential equation
 \begin{eqnarray}
 i \partial_{t} \hat{U}_{\theta}(t , t_{0}) =   \hat{H}^{I}_{\theta, \rm{Int}} (t) \hat{U}_{\theta}(t , t_{0})
\label{tdfeq}
 \end{eqnarray}
with the boundary condition given by (\ref{tid}). This differential equation can be transformed into an equivalent integral equation, in  exactly the same manner 
as done in commutative field theory and we have 
\begin{eqnarray}
 \hat{U}_{\theta}(t , t_{0}) \quad = \quad \mathbb{I} \quad + \quad  (-i) \int_{t_{0}}^{t} dt' \hat{H}^{I}_{\theta, \rm{Int}} (t') \hat{U}_{\theta}(t' , t_{0}).
\label{tinteq}
\end{eqnarray}
The formal solution of  (\ref{tinteq}) can be written in terms of ``time ordered exponential function'' as
\begin{eqnarray}
  \hat{U}_{\theta}(t , t_{0}) = \mathcal{T} \exp\left[-i\int^{t}_{t_{0}} dt' \hat{H}^{I}_{\theta, \rm{Int}} (t')\right]
\label{tev}
\end{eqnarray}  
where the time ordering operator $\mathcal{T}$ is defined in the same way as in standard commutative case. 
\end{enumerate}


\subsection{Computation of $S$-matrix}


In the previous section we have described the noncommutative interaction picture. In this section we use it to compute $S$-matrix elements for a typical scattering process. 
We use a particular model of real scalar fields having quartic self-interactions. The commutative interaction Hamiltonian density $\mathcal{\hat{H}}_{0, \rm{Int}}(x)$ that we 
consider is given by
\begin{eqnarray}
\mathcal{\hat{H}}_{0, \rm{Int}}(x) & = & \frac{\lambda}{4!} \, \phi_0 (x) \cdot \phi_0 (x) \cdot \phi_0 (x) \cdot \phi_0 (x) \, = \,  \frac{\lambda}{4!} \, \phi^4_0 (x)
\label{chint}
\end{eqnarray}
and the analogous noncommutative interaction hamiltonian density $\mathcal{\hat{H}}_{\theta, \rm{Int}}(x)$ is 
\begin{eqnarray}
 \mathcal{\hat{H}}_{\theta, \rm{Int}} (x) \, = \, \frac{\lambda}{4!} \, \phi_{\theta} (x) \ast \phi_{\theta}(x) \ast \phi_{\theta}(x) \ast \phi_{\theta} (x) 
\, = \, \frac{\lambda}{4!} \, \phi^4_{\theta,\ast} (x) \, = \, \frac{\lambda}{4!} \, \phi^4_0 (x) \, e^{\frac{1}{2} \overleftarrow{\partial} \wedge P},
\label{nchint}
\end{eqnarray}
where in writing the last equality we have used the dressing transformation (\ref{fielddress}) and the expression for the star product (\ref{starproduct}). 

Our aim is to compute the noncommutative $S$-matrix elements for a typical scattering process. We do that by first finding a relation between noncommutative $S$-matrix elements and 
their commutative counterparts by making use of the dressing transformations (\ref{dresstransform}) and (\ref{fielddress}). We briefly review the standard treatment in commutative 
case before discussing the noncommutative case and establishing its relation with commutative case.


\subsubsection{Commutative Case}


Let us restrict ourself to two particle scattering processes $ p_{1}, p_{2} \rightarrow  p'_{1}, p'_{2} $. The case of two-to-many and many-to-many will be 
taken up later. For a typical two-to-two particle scattering, the $S$-matrix element is given by
\begin{eqnarray}
 S_{0}[p_{2}, p_{1} \rightarrow  p'_{1}, p'_{2}] & \equiv &  S_{0}[p'_{2}, p'_{1} ; p_{2}, p_{1}]
 \; = \;  \leftidx{_{\rm{out},0}}{\left \langle p'_{2}, p'_{1} | p_{2}, p_{1} \right \rangle }{_{0,\rm{in}}}
\label{csm}
\end{eqnarray}
where $ |p'_{1}, p'_{2} \rangle_{0,\rm{out}} $ is the two particle out-state measured in the far future and $ |p_{2}, p_{1} \rangle_{0,\rm{in}} $ is the two particle in-state 
prepared in the far past. The in- and out-states can be related with each other using $S$-matrix $\hat{S}_{0}$. Therefore we have
\begin{eqnarray}
S_{0}[p'_{2}, p'_{1} ; p_{2}, p_{1}] & = &  \leftidx{_{\rm{out},0}}{\left \langle  p'_{2}, p'_{1}| \hat{S}_{0} | p_{2}, p_{1}  \right \rangle}{_{\rm{out},0}}
\; = \; \leftidx{_{\rm{in},0}}{\left \langle  p'_{2}, p'_{1}| \hat{S}_{0} | p_{2}, p_{1}  \right \rangle}{_{\rm{in},0}},
\label{csmat}
\end{eqnarray}
where $\hat{S}_{0}$ can be written in interaction picture as 
\begin{eqnarray}
 \hat{S}_{0} & = & \lim_{t_{1} \rightarrow \infty} \lim_{t_{2} \rightarrow -\infty} U_{0}(t_{1},t_{2}) \nonumber \\
& = & \mathcal{T} \exp\left[-i\int^{\infty}_{-\infty} d^{4}z \frac{\lambda}{4!} \phi^{4}_{0}(z) \right].
\label{csoperator}
\end{eqnarray}
In the last line we have used the form (\ref{chint}) for the interaction Hamiltonian density.

The two particle states are defined as
\begin{eqnarray}
  |p, q \rangle_{0} & = & c^{\dagger}_{q} c^{\dagger}_{p}  |0 \rangle
\label{c2state}
\end{eqnarray}
where $ c^{\dagger}_{p} $ is the creation operator for the commutative theory with the usual commutation relations.

Using (\ref{csoperator}) and (\ref{c2state}) in (\ref{csmat}) we obtain 
\begin{eqnarray}
 S_{0}[p'_{2}, p'_{1} ; p_{2}, p_{1}] & = & \left \langle 0 \left | c_{p'_{1}} c_{p'_{2}} \mathcal{T} \exp\left[-i\int^{\infty}_{-\infty} d^{4}z \frac{\lambda}{4!} 
\phi^{4}_{0}(z) \right]
c^{\dagger}_{p_{1}} c^{\dagger}_{p_{2}} \right |0 \right \rangle.
\label{csmatrix}
\end{eqnarray}
Now, to calculate any specific process, $\hat{S}_{0}$ is expanded in power series of coupling constant $\lambda$ (provided $\lambda$ is small enough to allow
 perturbative expansion) up to some desired order of coupling constant. It is evaluated using standard techniques, e.g. Wick's theorem and Feynman diagrams. 

The two-to-many ($ 2 \rightarrow N $) or many-to-many particle ($ M \rightarrow N $)  scattering cases can be similarly discussed. For instance, 
for ($ M \rightarrow N $) scattering we have
\begin{eqnarray}
 S_{0}[ p'_{N}, ... p'_{1} ; p_{M}, ... p_{1} ] & = &  \leftidx{_{\rm{out},0}}{\left \langle p'_{N}, ... p'_{1} | p_{M} ... p_{1} \right \rangle }{_{0,\rm{in}}}
\label{cmsm}
\end{eqnarray}
where $ |p'_{1}, ... p'_{N} \rangle_{0,\rm{out}} $ is the N-particle out-state and $ | p_{M} ... p_{1} \rangle_{0,\rm{in}} $ is the M-particle in-state.
 As before, the in- and out-states can be related with each other using $S$-matrix $\hat{S}_{0}$. Therefore we have
\begin{eqnarray}
 S_{0}[ p'_{N}, ... p'_{1} ; p_{M}, ... p_{1} ] & = & \leftidx{_{\rm{out},0}}{\left \langle p'_{N}, ... p'_{1} \left| \hat{S}_{0} \right| p_{M} ... p_{1}
 \right \rangle}{_{\rm{out},0}} \nonumber \\
& = & \leftidx{_{\rm{in},0}}{\left \langle p'_{N}, ... p'_{1} \left| \hat{S}_{0} \right| p_{M} ... p_{1} \right \rangle}{_{\rm{in},0}}
\label{cmsmat}
\end{eqnarray}
where $\hat{S}_{0}$ is given by $(\ref{csoperator})$.

These multiple-particle states can be written as 
\begin{eqnarray}
  |p_{M} ... p_{1} \rangle_{0} & = & c^{\dagger}_{p_{1}} ... c^{\dagger}_{p_{M}} |0 \rangle.
\label{cmstate}
\end{eqnarray}
Using (\ref{csoperator}) and (\ref{cmstate}) in (\ref{cmsmat}) we obtain
\begin{equation}
  S_{0}[ p'_{N}, ... p'_{1} ; p_{M}, ... p_{1} ]  =  \left \langle 0 \left | c_{p'_{1}} ... c_{p'_{N}} \mathcal{T} \exp\left[-i\int^{\infty}_{-\infty} d^{4}z \frac{\lambda}{4!}
\phi^{4}_{0}(z) \right] c^{\dagger}_{p_{1}} ... c^{\dagger}_{p_{M}} \right |0 \right \rangle.
\label{cmsmatrix}
\end{equation}
Again, any specific process can be calculated using perturbative expansion in $\lambda$ (if possible) and invoking standard tools like Wick's theorem and Feynman diagrams.

 
\subsubsection{Noncommutative Case}


Our treatment of the noncommutative case follows closely the formalism of commutative case. Therefore, as in the commutative case, for a two-to-two particle scattering processes 
the $S$-matrix elements are given by
\begin{eqnarray}
 S_{\theta}[p_{2}, p_{1} \rightarrow  p'_{1}, p'_{2}] & \equiv &  S_{\theta}[p'_{2}, p'_{1} ; p_{2}, p_{1}]
\; = \;  \leftidx{_{\rm{out},\theta}}{\left \langle  p'_{2}, p'_{1} | p_{2}, p_{1} \right \rangle }{_{ \theta, \rm{in}}}
\label{ncsm}
\end{eqnarray}
where $ |p'_{1}, p'_{2} \rangle_{\theta,\rm{out}} $ is the noncommutative two particle out-state which is measured in the far future and $ |p_{2}, p_{1}  \rangle_{\theta,\rm{in}} $ 
is the noncommutative two particle in-state prepared in the far past. Now, because of the twisted statistics (\ref{tcom}) there is an ambiguity in defining the action of the 
twisted creation and annihilation operators on the Fock space of states. Following \cite{bal-pinzul} we choose to define $a^\dagger_k$ to be an operator which adds a particle to 
the right of the particle list,
\begin{eqnarray}
 a^\dagger_k | p_1,p_2 \dots p_n \rangle_\theta & = & | p_1,p_2 \dots p_n, k \rangle_\theta.
\label{taanni}
\end{eqnarray}
Hence the two particle in-state can be written as 
\begin{eqnarray}
 |p_{2}, p_{1}  \rangle_{\theta,\rm{in}} & = & a^\dagger_{p_1}\, a^\dagger_{p_2} \, | 0 \rangle.
\label{int2state}
\end{eqnarray}
Since the noncommutative vacuum is the same as that of the commutative theory, no extra label is needed for $| 0 \rangle$.

Just like in the commutative case, the noncommutative in- and out-states can be related with each other using $S$-matrix $\hat{S}_{\theta}$. Therefore we have
\begin{eqnarray}
S_{\theta}[p'_{2}, p'_{1} ; p_{2}, p_{1}] & = & \leftidx{_{\rm{out},\theta}}{\left \langle p'_{2}, p'_{1} \left|\hat{S}_{\theta} \right| p_{2}, p_{1} \right \rangle}{_{\rm{out},\theta}}
\; = \;  \leftidx{_{\rm{in},\theta}}{\left \langle  p'_{2}, p'_{1} \left | \hat{S}_{\theta} \right| p_{2}, p_{1}  \right \rangle}{_{\rm{in},\theta}}
\label{rncsmat}
\end{eqnarray}
where the noncommutative $S$-matrix $\hat{S}_{\theta}$ in interaction picture can be written as 
\begin{eqnarray}
 \hat{S}_{\theta} & = & \lim_{t_{1} \rightarrow \infty} \lim_{t_{2} \rightarrow -\infty} U_{\theta}(t_{1},t_{2})
\label{ncsop}
\end{eqnarray} 
where $U_{\theta}(t_{1},t_{2}) $ is given by (\ref{tev}). For the interaction Hamiltonian density given in (\ref{nchint}) we obtain
\begin{eqnarray}
\hat{S}_{\theta} & = & \mathcal{T}  \exp \left[ -i\int^{\infty}_{-\infty} d^{4}z \frac{\lambda}{4!} \phi^{4}_{0}(z) e^{\frac{1}{2} \overleftarrow{\partial_{z}} \wedge P }\right]. 
\label{ncsint}
\end{eqnarray}
One can formally expand the exponential and write $\hat{S}_{\theta}$ as a time-ordered power series like
\begin{eqnarray}
\hat{S}_{\theta} & = &   \mathbb{I} \, + \, -i\int^{\infty}_{-\infty} d^{4}z   \frac{\lambda}{4!} \phi^{4}_{0}(z) \, e^{\frac{1}{2} \overleftarrow{\partial_{z}} \wedge P } \nonumber \\
& + &  \mathcal{T}\, \frac{(-i )^2}{2!} \int^{\infty}_{-\infty} d^{4}z \int^{\infty}_{-\infty} d^{4}z'\,  \frac{\lambda}{4!} \phi^{4}_{0}(z) \, e^{\frac{1}{2} \overleftarrow{\partial_{z}} \wedge P } 
\,  \frac{\lambda}{4!} \phi^{4}_{0}(z') \, e^{\frac{1}{2} \overleftarrow{\partial_{z'}} \wedge P } \, + \, \cdots
\label{rspower}
\end{eqnarray} 
As done in \cite{bal-uvir}, each term in the power series in (\ref{rspower}) can be further simplified by expanding the exponential $ e^{\frac{1}{2} \overleftarrow{\partial} \wedge P } $, 
integrating by parts and discarding the surface terms. For instance, the second term in (\ref{rspower}) becomes
\begin{eqnarray}
 -i\int^{\infty}_{-\infty} d^{4}z    \frac{\lambda}{4!} \phi^{4}_{0}(z) \, e^{\frac{1}{2} \overleftarrow{\partial_{z}} \wedge P } 
& = &  -i\int^{\infty}_{-\infty} d^{4}z \left [   \frac{\lambda}{4!} \phi^{4}_{0}(z) 
\, + \, \partial_\mu \left( \frac{\lambda}{4!} \phi^{4}_{0}(z) \right)\, \theta^{\mu \nu}\, P_\nu \, + \, \dots \right] \nonumber \\
& = & -i\int^{\infty}_{-\infty} d^{4}z \frac{\lambda}{4!} \phi^{4}_{0}(z) .
\label{rsecpower}
\end{eqnarray}
One can similarly show that all the higher order terms in the power series of (\ref{rspower}) are also free 
of any $\theta$ dependence. We refer to \cite{bal-uvir} for more details.

We then have
\begin{eqnarray}
 \hat{S}_{\theta} & = & \mathcal{T} \exp\left[-i\int^{\infty}_{-\infty} d^{4}z \frac{\lambda}{4!} \phi^{4}_{0}(z) \right] \; = \; \hat{S}_{0}.
\label{rncsoperator}
\end{eqnarray}
Using (\ref{rncsoperator}) and (\ref{int2state}) in (\ref{rncsmat}) we obtain 
\begin{eqnarray}
 S_{\theta}[p'_{2}, p'_{1} ; p_{2}, p_{1}] & = & \left \langle 0 \left | a_{p'_{1}} a_{p'_{2}} \mathcal{T} \exp\left[-i\int^{\infty}_{-\infty} d^{4}z \frac{\lambda}{4!}
 \phi^{4}_{0}(z) \right] a^{\dagger}_{p_{1}} a^{\dagger}_{p_{2}} \right |0 \right \rangle.
\label{ncsmatr}
\end{eqnarray}
But the noncommutative creation/annihilation operators are related with those of commutative theory by dressing transformation (\ref{dresstransform}), so that
\begin{eqnarray}
 S_{\theta}[p'_{2}, p'_{1} ; p_{2}, p_{1}] & = & \left \langle 0 \left | c_{p'_{1}} e^{\frac{-i}{2} p'_{1} \wedge P} c_{p'_{2}} e^{\frac{-i}{2} p'_{2} \wedge P}
 \mathcal{T} \exp\left[-i\int^{\infty}_{-\infty} d^{4}z \frac{\lambda}{4!} \phi^{4}_{0}(z) \right] \right. \right. \nonumber \\
& & \left. \left. c^{\dagger}_{p_{1}} e^{\frac{i}{2} p_{1} \wedge P} c^{\dagger}_{p_{2}} e^{\frac{i}{2} p_{2} \wedge P}\right |0 \right \rangle \nonumber \\
& = &  e^{\frac{-i}{2} p'_{2} \wedge p'_{1}} \, e^{\frac{i}{2} p_{1} \wedge p_{2}} \left \langle 0 \left | c_{p'_{1}} c_{p'_{2}} \mathcal{T} \exp\left[-i\int^{\infty}_{-\infty} d^{4}z 
\frac{\lambda}{4!} \phi^{4}_{0}(z) \right]   c^{\dagger}_{p_{1}} c^{\dagger}_{p_{2}} \right |0 \right \rangle \nonumber \\
& = &  e^{\frac{-i}{2} p'_{2} \wedge p'_{1}} \, e^{\frac{i}{2} p_{1} \wedge p_{2}} \; S_{0}[p'_{2}, p'_{1} ; p_{2}, p_{1}].
\label{ncsmatrix}
\end{eqnarray}
The expression (\ref{ncsmatrix}) relates the noncommutative $S$-matrix element for a two-to-two particle scattering process with its commutative counterpart. We remark that this 
correspondence is a nonperturbative one and it is true to all orders in perturbation of the coupling constant. Also, the only noncommutative dependence of 
$S_{\theta}[p'_{2}, p'_{1} ; p_{2}, p_{1}]$ is by an overall phase. Therefore the model is essentially free from any UV/IR mixing. 

An analogous relation between noncommutative and commutative $S$-matrix for two-to-many ($ 2 \rightarrow N $) and many-to-many ($ M \rightarrow N $) particle scattering processes 
can be established in a similar way. For instance, for  ($ M \rightarrow N $) scattering we have
\begin{eqnarray}
 S_{\theta}[ p'_{N}, ... p'_{1} ; p_{M}, ... p_{1} ] & = &  \leftidx{_{\rm{out},\theta}}{\left \langle  p'_{N}, ... p'_{1} | p_{M} ... p_{1} \right \rangle }{_{\theta,\rm{in}}},
\label{ncmsm} 
\end{eqnarray}
where $ |p'_{1}, ... p'_{N} \rangle_{\theta,\rm{out}} $ is the noncommutative N-particle out-state and $ | p_{M} ... p_{1} \rangle_{\theta,\rm{in}} $ is the noncommutative 
N-particle in-state. As before, the in- and out-states can be related with each other using $S$-matrix $\hat{S}_{\theta}$. Therefore we have
\begin{eqnarray}
 S_{\theta}[ p'_{N}, ... p'_{1} ; p_{M}, ... p_{1} ] & = &  \leftidx{_{\rm{out},\theta}}{\left \langle p'_{N}, ... p'_{1} \left| \hat{S}_{0} \right| p_{M} ... p_{1}
 \right \rangle}{_{\rm{out},\theta}} \nonumber \\
& = &  \leftidx{_{\rm{in},\theta}}{\left \langle p'_{N}, ... p'_{1} \left| \hat{S}_{0} \right| p_{M} ... p_{1} \right \rangle}{_{\rm{in},\theta}}
\label{ncmsmat}
\end{eqnarray}
where $\hat{S}_{\theta}$ is given by $(\ref{rncsoperator})$.

Just like the two-particle states, the noncommutative multiple-particle states can be written as 
\begin{eqnarray}
  | p_{M} ... p_{1} \rangle_{\theta} & = & a^{\dagger}_{p_{1}} ... a^{\dagger}_{p_{M}}  |0 \rangle.
\label{ncmstate}
\end{eqnarray}
Using (\ref{rncsoperator}) and (\ref{ncmstate}) in (\ref{ncmsmat}) we obtain 
\begin{equation}
S_{\theta}[ p'_{N}, ... p'_{1} ; p_{M}, ... p_{1} ]  =  \left \langle 0 \left | a_{p'_{1}} ... a_{p'_{N}} \mathcal{T} \exp\left[-i\int^{\infty}_{-\infty} d^{4}z \frac{\lambda}{4!}
\phi^{4}_{0}(z) \right] a^{\dagger}_{p_{1}} ... a^{\dagger}_{p_{M}} \right |0 \right \rangle.
\label{ncmsmatr}
\end{equation}
Using the dressing transformation (\ref{dresstransform}) in (\ref{ncmsmatr}) we obtain
\begin{eqnarray}
\hspace{-.5cm} 
S_{\theta}[p'_{N}, ... p'_{1} ; p_{M}, ... p_{1}]  & = & \left \langle 0 \left | c_{p'_{1}} e^{\frac{-i}{2} p'_{1} \wedge P} ... c_{p'_{N}} e^{\frac{-i}{2} p'_{N} \wedge P}
\, \mathcal{T} \exp\left[-i\int^{\infty}_{-\infty} d^{4}z \frac{\lambda}{4!} \phi^{4}_{0}(z) \right]  \right. \right. \nonumber \\
& & \left. \left. c^{\dagger}_{p_{1}} e^{\frac{i}{2} p_{1} \wedge P} ... 
c^{\dagger}_{p_{M}} e^{\frac{i}{2} p_{M} \wedge P} \right |0 \right \rangle \nonumber \\
& = &   e^{\frac{i}{2} \left( \sum^{M}_{i,j=1,j>i}p_{i} \wedge p_{j}  -   \sum^{N}_{i,j= N, j< i} p'_{i} \wedge p'_{j}\right)}  S_{0}[p'_{N}, ... p'_{1} ; p_{M}, ... p_{1}]. 
\label{ncmsmatrix}
\end{eqnarray}
This is the generic result relating the noncommutative many-to-many particle $S$-matrix with its commutative analogue. Again, it should be noted that the proof is completely 
nonperturbative and hence valid to all orders in the coupling constant. Also, as argued before, the phenomena of UV/IR mixing is completely absent.


\section{Noncommutative LSZ Formalism}


In this section we review the noncommutative LSZ formalism and calculate the noncommutative $S$-matrix elements via the reduction formula.
 The noncommutative $S$-matrix computed via LSZ will be shown to be completely equivalent to that computed in the previous section using interaction picture.  
This establishes the equivalence  of the two approaches. Also, this second method brings out the difference between scattering amplitudes and off-shell Green's functions.

We consider as an example the time ordered product of four real scalar fields with $\phi^4$ type self-interactions representing a process of two particles going 
into two other particles. This is described by the correlation function 
\begin{eqnarray}
G_{2+2}(x_1', x_2';~x_1,x_2)=\langle\Omega|\mathcal{T}\left(\phi(x_1')\phi(x_2')\phi(x_1)\phi(x_2)\right)|\Omega\rangle
\label{gengreen}
\end{eqnarray}
where $|\Omega\rangle$ is the vacuum of the full interacting theory.

The Green's function $G_{2+2}^0$ in the commutative case is given by the time ordered product of four commutative fields $\phi_0$. The corresponding Green's function 
$G^{\theta}_{2+2}$
in the noncommutative case is obtained by replacing the commutative fields $\phi_0$ by the noncommutative ones  $\phi_\theta$ in the time ordered product in (\ref{gengreen}). 
The case of many particle scattering will be taken up later.

As done in previous section, we start first by briefly reviewing the derivation of commutative LSZ reduction formula before going on to the noncommutative case. 
The derivation presented in this section is originally due to \cite{amilcar} which can be consulted for further details.


\subsection{ Commutative Case }


In this section we use the following notations: 
\begin{eqnarray}
& & \hat{p}~ \textrm{is an on-shell momentum}= (E_{\vec{p}}=\sqrt{\vec{p}^{~2}+m^2},~ \vec{p}), \nonumber \\
& & p~ \textrm{is a generic 4-momentum, with}~ p^0>0.
\end{eqnarray}

Let us consider the time ordered product of four commutative fields $\phi_0 (x)$ given by
\begin{eqnarray}
G^0_{2+2}(x_1', x_2';~x_1,x_2) & = & \langle\Omega|\mathcal{T}\left(\phi_0(x_1')\phi_0(x_2')\phi_0(x_1)\phi_0(x_2)\right)|\Omega\rangle.
\label{com2green}
\end{eqnarray}
As mentioned before, $G^0_{2+2}(x_1', x_2';~x_1,x_2) $ can be related to a process of two particles scattering/decaying into two other particles.

We Fourier transform $G_{2+2}^0(x_1', x_2';~x_1, x_2)$ only in $x_1'$. Without loss of generality, we can assume that $x_1'$ is associated with an outgoing particle. 
We can split the $x_1'^0$-integral into three time intervals as
 \begin{eqnarray}
\left( \int_{-\infty}^{T_-}dx_1'^0 + \int_{T_-}^{T_+}dx_1'^0 + \int_{T_+}^{\infty}dx_1'^0 \right)d^3x_1'~e^{ip_1'^0x_1'^0 - i \vec{p}_1'\cdot\vec{x}_1'} \, G_{2+2}^0(x_1', x_2';~x_1, x_2)
\end{eqnarray}
Here $T_{+} \geq \textrm{max}(x_2'^0, x_1^0, x_2^0)$ and $T_{-} \leq \textrm{min}(x_2'^0, x_1^0, x_2^0)$. Since $T_+ \geq x_1'^0 \geq T_-$ is a finite interval, the corresponding integral 
gives no pole. A pole comes from a single particle insertion in the integral over $x_1'^0\geq T_+$ in $G_{2+2}^0$. In the integration between the limits 
$T_+$ and $+\infty$, $\phi(x_1')$ stands to the extreme left inside the time-ordering so that
 \begin{equation}
\hspace{-.5cm}
G_{2+2}^0(x_1', x_2';~x_1, x_2)= \int\frac{d^3q_1}{(2\pi)^3}\frac{1}{2E_{\vec{q}_1}} \langle\Omega|\phi_0(x_1')|q_1\rangle\langle q_1|
T\left(\phi_0(x_2')\phi_0(x_1)\phi_0(x_2)\right)|\Omega\rangle  +  \textrm{OT}
\end{equation}
where $\textrm{OT}$ stands for the other terms. The matrix element of the field $\phi_0(x_1')$ can be written as
\begin{eqnarray}
 \langle\Omega|e^{iP\cdot x_1'}\phi_0(0)e^{-iP\cdot x_1'}|E_{\vec{q}_1}, \vec{q_1}\rangle & = & \langle\Omega|\phi_0(0)|E_{\vec{q_1}}, 
\vec{q_1}\rangle e^{-iq_1\cdot x_1'}|_{q_1^0=E_{\vec{q}_1}}  \nonumber \\ 
& = & \langle\Omega|\phi_0(0)|q^0_1,\vec{q}_1=0\rangle e^{-iq_1\cdot x_1'}|_{q_1^0=E_{\vec{q}_1}}
\end{eqnarray} 
where $E^2_{\vec{q_1}} = \vec{q_1}^2+m^2 $.  We have used the Lorentz invariance of the vacuum $|\Omega\rangle$ and $\phi_0(0)$ in above. We then have
\begin{eqnarray}
 \langle\Omega|\phi_0(x_1')|E_{\vec{q_1}}, \vec{q}_1\rangle = \sqrt{Z} e^{-i\left(E_{\vec{q_1}}x_1'^0-\vec{q}_1\cdot\vec{x}_1'\right)}
\end{eqnarray}
where the field-strength renormalization factor $\sqrt{Z}$ is defined by
\begin{eqnarray}
\sqrt{Z} = \langle\Omega|\phi_0(0)|q^0_1,\vec{q}_1=0\rangle
\end{eqnarray}
and $q_1^0 > 0$. Hence the integral between $T_+$ and $+\infty$ becomes
\begin{eqnarray}
\sqrt{Z}\frac{1}{2E_{\vec{p'}_1}}\int_{T_+}^{\infty}dx_1'^0 ~e^{i\left(p_1'^0 - E_{\vec{p}_1'} +i\epsilon\right)x_1'^0}~\langle p_1'|
T\left(\phi_{2'}\phi_1\phi_2\right)|\Omega\rangle \, + \, \textrm{OT}
\end{eqnarray}
where $\epsilon > 0$ is a cut-off and $\phi_i = \phi_0(x_i)$. After the $x_1'^0$ integral we obtain
\begin{eqnarray}
\widetilde{G}_0^{(1)}(p_1', x_2', x_1, x_2)= \sqrt{Z}\frac{i}{2E_{\vec{p'}_1}} \frac{e^{i\left(p_1'^0-E_{\vec{p}_1'}+i\epsilon\right)T_+}}{\left(p_1'^0-E_{\vec{p}_1'}+i\epsilon\right)}\langle
 p_1'|T\left(\phi_{2'}\phi_1\phi_2\right)|\Omega\rangle  \, + \, \textrm{OT}.
\end{eqnarray}
As $p_1'^0 \rightarrow E_{\vec{p}_1'}$, it becomes
\begin{eqnarray}
\label{pole}
\widetilde{G}_0^{(1)}(p_1', x_2', x_1, x_2) = \sqrt{Z}  \frac{i}{p_1'^2-m^2-i\epsilon}~\langle p_1'|T\left(\phi_{2'}\phi_1\phi_2\right)|\Omega\rangle \, + \, \textrm{OT}.
\end{eqnarray}
Now in the case of integration over $(-\infty, T_-)$, $\phi_0(x_1')$ stands to the extreme right in the time ordered product, so the one-particle state contribution comes from
\begin{eqnarray}
 \langle q_1|\phi_0(x_1')|\Omega\rangle = \sqrt{Z}e^{i\left(E_{\vec{q}_1}x_1'^0-\vec{q}_1\cdot\vec{x}_1'\right)}.
\end{eqnarray}
The energy denominator is thus $\frac{1}{p_1'^0+E_{\vec{p}_1'}-i\epsilon}$ and has no pole for $p_1'^0>0$. The only pole comes from the single particle insertion in the integral 
over $x_1'^0\geq T_+$. It is given by (\ref{pole}).

Similarly, for the two-particle scattering $p_1, p_2\rightarrow p_1', p_2'$, the poles appear in both $p_1'^0$ and $p_2'^0$ when both $x_1'^0$ and $x_2'^0$ integrations are large, 
that is
\begin{eqnarray}
  x_1'^0,~ x_2'^0~ >>~ T_1~>>~x_1^0,~ x_2^0.
\end{eqnarray}
So for these poles, we obtain
\begin{eqnarray}
 \widetilde{G}_0^{(2)}(p_1', p_2', x_1, x_2) & = & \int_{T_+}^{\infty} dx_1'^0 dx_2'^0 d^3x_1' d^3x_2'~ e^{ip_1'\cdot x_1' + ip_2'\cdot x_2'}\frac{1}{2!}\left(\frac{1}{(2\pi)^3}\right)^2
\frac{d^3q_1 d^3q_2}{(2E_{\vec{q}_1})(2E_{\vec{q}_2})} \nonumber \\ 
& & \langle\Omega|\phi_0(x_1')\phi_0(x_2')|\vec{q}_1\vec{q}_2\rangle\langle\vec{q}_1\vec{q}_2|T\left(\phi_1\phi_2\right)|\Omega\rangle \, + \, \textrm{OT}.
\end{eqnarray}
Here $T_+$ is supposed to be very large. We take $\phi_0(x_1')$, $\phi_0(x_2')$ to be out fields. As we set $|\vec{q}_2\vec{q}_1\rangle$ to $|\vec{q}_2\vec{q}_1\rangle_{\textrm{out}}$
 for large $T_+$, only $\langle\Omega|\phi^{\textrm{out}+}_0(x_1')\phi^{\textrm{out}+}_0(x_2')|\vec{q}_2\vec{q}_1\rangle_{\textrm{out}}$, where $\phi_0^{\textrm{out}+}$ is 
the positive frequency part of the out-field, contributes. Thus we do not need any time-ordering between these out-fields. So we have
\begin{eqnarray}
\widetilde{G}_0^{(2)}(p_1',p_2',x_1,x_2) & = & \int_{T_+}^{\infty}d^4x_1'd^4x_2' ~e^{ip_1'\cdot x_1'+ip_2'\cdot x_2'}\frac{1}{2!}\left(\frac{1}{(2\pi)^3}\right)^2
\left(\frac{d^3q_1}{2E_{\vec{q}_1}}\right)\left(\frac{d^3q_2}{2E_{\vec{q}_2}}\right) \nonumber \\ 
& & \langle\Omega|\phi_0^{\textrm{out}}(x_1')\phi_0^{\textrm{out}}(x_2')|\vec{q}_2\vec{q}_1\rangle_{\textrm{out}}~_{\textrm{out}}\langle\vec{q}_2\vec{q}_1|T
\left(\phi_1\phi_2\right)|\Omega\rangle.
\end{eqnarray}
Now, 
\begin{eqnarray}
\langle\Omega|\phi_0^{\textrm{out}}(x_1')\phi_0^{\textrm{out}}(x_2')|\vec{q}_2\vec{q}_1\rangle_{\textrm{out}}=\langle\Omega|\phi_0^{\textrm{out}}(x_1')|\vec{q}_1
\rangle\langle\Omega|\phi_0^{\textrm{out}}(x_2')|\vec{q}_2\rangle + \vec{q}_2\leftrightarrow\vec{q}_1.
\end{eqnarray}
Thus we can generalize (\ref{pole}) to
\begin{eqnarray}
\widetilde{G}_0^{(2)}(p_1',p_2', x_1,x_2) & = & \left[\sqrt{Z} \left(\frac{i}{p^{'2}_1-m^2-i\epsilon}\right)\right]\left[\sqrt{Z} \left(\frac{i}{p^{'2}_2-m^2-i\epsilon}\right)\right] 
\nonumber \\
& &  _{\textrm{out}}\langle p_1' p_2'|T\left(\phi_1\phi_2\right)|\Omega\rangle \, + \, \textrm{OT}.
\end{eqnarray}
Similar calculations for incoming poles, with $x_1^0, x_2^0 << T_- << x_1'^0, x_2'^0$, leads to
\begin{eqnarray}
\label{commutativeLSZ}
\widetilde{G}_0^{(4)}(p_1', p_2', p_1, p_2) & = & \prod_{i=1}^2\prod_{j=1}^2 \left[\sqrt{Z}  \left(\frac{1}{p_i^{'2}-m^2-i\epsilon}\right)\right]\left[\sqrt{Z} 
\left(\frac{1}{p_j^2-m^2-i\epsilon}\right)\right] \nonumber \\
& &  _{\textrm{out}}\langle p_1'~ p_2'~|~ p_1~p_2\rangle_{\textrm{in}}.
\end{eqnarray}


\subsection{Noncommutative Case}


Our treatment of the noncommutative case is quite similar to that of the commutative case just discussed. Our aim is to arrive at the noncommutative 
version of (\ref{commutativeLSZ}). However, instead of considering a 2-particle scattering process first and then generalizing, as done in the commutative case, we directly 
start with the generic process where $M$ particles go into $N$ particles. 

Before discussing the noncommutative LSZ formalism we list down a few relations:

\begin{enumerate}
\item \textbf{The completeness relations :} These remain same for the twisted in- and out-states like in the commutative case. Recall that the noncommutative phases arising because 
of the twisted statistics (\ref{tcom}) followed by $a_p$ and $ a^\dagger_p$,  cancel each other. Therefore
 \begin{equation}
\label{5.25}
 a^{\dag\textrm{in, out}}_{p_N}\cdots a^{\dag\textrm{in, out}}_{p_1}|\Omega\rangle\langle\Omega| a^{\textrm{in, out}}_{p_1}\cdots a^{\textrm{in, out}}_{p_N}  
 =   c^{\dag\textrm{in, out}}_{p_N}\cdots c^{\dag\textrm{in, out}}_{p_1}| \Omega\rangle\langle\Omega|c^{\textrm{in, out}}_{p_1}\cdots c^{\textrm{in, out}}_{p_N}.
\end{equation} 
Using (\ref{5.25}) one can also check the resolution of identity (given below) as well as the completeness for the twisted in- and out-states.

\item \textbf{ Resolution of identity:} 
\begin{equation}
I' = \sum_N\frac{1}{N!}\left(\int \prod_{i=1}^N \frac{d^3p_i }{(2\pi)^3}\frac{1}{2E_{\vec{p_i}}}\right) a^{\dag\textrm{in, out}}_{p_N}\cdots a^{\dag\textrm{in, out}}_{p_1}
|\Omega\rangle\langle\Omega|a^{\textrm{in, out}}_{p_1}\cdots a^{\textrm{in, out}}_{p_N}.
\label{reid}
\end{equation} 
This turns out to be independent of $\theta_{\mu\nu}$ due to (\ref{5.25}). Hence we have
\begin{equation}
\label{id}
I' = \sum_N\frac{1}{N!}\left(\int \prod_{i=1}^N \frac{d^3p_i}{(2\pi)^3}\frac{1}{2E_{\vec{p_i}}}\right) c^{\dag\textrm{in, out}}_{p_N}
\cdots c^{\dag\textrm{in, out}}_{p_1}|\Omega\rangle\langle\Omega|c^{\textrm{in, out}}_{p_1}\cdots c^{\textrm{in, out}}_{p_N}.
\end{equation}
\end{enumerate}

We are interested in the scattering process of $M$ particles going to $N$ particles. We then consider the twisted $N+M$-point Green's function
\begin{equation}
G^{\theta}_{N+M}(x_1',..., x_N';~x_1,..., x_M) = \langle\Omega|T\left(\phi_{\theta}(x_1')\cdots\phi_{\theta}(x_N')\phi_{\theta}(x_1)\cdots\phi_{\theta}(x_M)\right)|\Omega\rangle.
\label{nmncgreenf}
\end{equation}
As mentioned before, the twisted $N+M$-point Green's function is obtained by replacing the commutative fields $\phi_0$ with noncommutative fields $\phi_\theta$ in the time-ordered 
product of fields. Also, the Fourier transform of (\ref{nmncgreenf})  can be obtained by integrating with respect to the measure 
$ \left(\prod_i d^4x_i'\right)\left(\prod_j d^4x_j\right) e^{i\left(\sum_{i\leq N}p_i'\cdot x_i' - \sum_{j\leq M}p_j\cdot x_j\right)}. $

Integration over $x_i$, $x_i'$ gives us $\widetilde{G}_{\theta}^{N+M}(p_1'\cdots , p_N'; p_1 \cdots ,p_M)$. The residue at the poles in all the momenta multiplied together gives 
the scattering amplitude. This is just the noncommutative version of the LSZ reduction formula. We now show that it gives the same expression for the $S$-matrix elements, as the 
one obtained in previous section using interaction picture.

As done in the commutative case, the pole in $p_1'$ can be obtained by Fourier transforming in just $x_1'$, i.e.
\begin{eqnarray}
\widetilde{G}_{\theta}^{(1)}(p_1',\cdots ,x_N', x_1, \cdots ,x_M) & = & \int d^4x_1' ~e^{i\left(p_1'^0x_1'^0 - \vec{p}_1'\cdot\vec{x}_1'\right)} \, \langle\Omega|
\mathcal{T} \left(\phi_{\theta}(x_1')\cdots\phi_{\theta}(x_N')\phi_{\theta}(x_1) \right. \nonumber \\
 & & \left. \cdots\phi_{\theta}(x_M)\right)|\Omega\rangle.
\end{eqnarray}
Taking $T_+ >> x_N'^0\cdots x_2'^0, x_M^0, \cdots ,x_1^0$, we can isolate the term with pole in $\widetilde{G}_{\theta}^{(1)}$. Hence
\begin{eqnarray}
 \widetilde{G}_{\theta}^{(1)}(p_1',\cdots ,x_N', x_1\cdots x_M) & = & \sqrt{Z}\int_{T_+}^{\infty} dx_1'^0 d^3x_1'~ e^{i\left(p_1'^0x_1'^0 - \vec{p}_1'\cdot\vec{x}_1'\right)} 
\langle\Omega|\phi_{\theta}^{\textrm{out}}(x_1') \mathcal{T} \left(\phi_{\theta}(x_2')\cdots \right.  \nonumber \\
& &  \left. \phi_{\theta}(x_N')\phi_{\theta}(x_1)\cdots\phi_{\theta}(x_M)\right)|\Omega\rangle 
\, + \, \textrm{OT} \nonumber \\
& = & \sqrt{Z}\int_{T_+}^{\infty}dx_1'^0d^3x_1'\frac{1}{(2\pi)^3} \frac{d^3q_1}{2E_{\vec{q}_1}}e^{i\left(p_1'^0x_1'^0 - \vec{p}_1'\cdot\vec{x}_1'\right)} 
\langle\Omega|\phi_{\theta}^{\textrm{out}}(x_1')|\hat{q}_1\rangle \nonumber \\ 
& &  \langle\hat{q}_1| \mathcal{T} \left(\phi_{\theta}(x_2')\cdots\phi_{\theta}(x_N')\phi_{\theta}(x_1)
\cdots\phi_{\theta}(x_M)\right)|\Omega\rangle  +  \textrm{OT} \nonumber \\ 
\end{eqnarray}
where
\begin{eqnarray}
{\langle\Omega|\phi_{\theta}^{\textrm{out}}(x_1')|\hat{q}_1\rangle = \langle\Omega|\phi_0^{\textrm{out}}(x_1')|\hat{q}_1\rangle }
\end{eqnarray}
because the twist gives just 1 in this case. This can be seen by using the dressing transformation (\ref{fielddress}), i.e. writing $\phi_{\theta}^{\textrm{out}}$ 
as $e^{\frac{1}{2}\partial \wedge P} \phi_{0}^{\textrm{out}}$ and acting with $P_{\nu}$ on $\langle\Omega|$. 

Repeating essentially the same procedure as in the commutative case, one can extract the pole $\frac{1}{p_1'^2-m^2-i\epsilon}$ and its coefficient. 

For poles at $p'_1$, $p'_2$, we have
 \begin{eqnarray}
& & \widetilde{G}_{\theta}^{(2)}(p_1', p_2', x_3', \cdots , x_N', x_1, \cdots ,x_M) \, = \, \int_{T_+}^{\infty} d^4x_1' d^4x_2'~ e^{ip_1'\cdot x_1' + ip_2'\cdot x_2'} 
(\sqrt{Z})^2\frac{d^3\hat{q}_1d^3\hat{q}_2}{2! (2E_{\vec{q}_1})(2E_{\vec{q}_2})}  \nonumber \\
& & \langle\Omega|\phi_{\theta}^{\textrm{out}}(x_1')\phi_{\theta}^{\textrm{out}}(x_2')|\hat{q}_1,
\hat{q}_2\rangle\langle\hat{q}_1, \hat{q}_2| \mathcal{T} \left(\phi_{\theta}(x_3')\cdots\phi_{\theta}(x_N')\phi_{\theta}(x_1)\cdots\phi_{\theta}(x_M)\right)|\Omega\rangle  
+ \textrm{OT}. \nonumber \\
\end{eqnarray}
 Because of (\ref{id}) there is no twist factor in $|\hat{q}_1, \hat{q}_2\rangle$ and $\langle\hat{q}_2, \hat{q}_1|$. 

Now we compute the matrix element of the two out-fields:
\begin{eqnarray}
\langle\Omega|\phi_{\theta}^{\textrm{out}}(x_1')\phi_{\theta}^{\textrm{out}}(x_2')|\hat{q}_1, \hat{q}_2\rangle  
& = & \int \left(\frac{1}{(2\pi)^3}\right)^2 \frac{d^3p_1''}{E_{\vec{p''}_1}}\frac{d^3p_2''}{2E_{\vec{p''}_2}}e^{-i\hat{p}_1''\cdot x_1' - 
i\hat{p}_2''\cdot x_2'} e^{-\frac{i}{2}\hat{p}_1''\wedge\left(-\hat{p}_2''+\hat{q}_1 + \hat{q}_2\right)} 
\nonumber \\ 
& & e^{-\frac{i}{2}\hat{p}_2''\wedge\left(\hat{q}_1 + \hat{q}_2\right)} \, \langle\Omega|c^{\textrm{out}}_{p_1''}c^{\textrm{out}}_{p_2''}c^{\dag\textrm{out}}_{q_2}
c^{\dag\textrm{out}}_{q_1}|\Omega\rangle,
\label{mel}
\end{eqnarray}
where the matrix element is
\begin{eqnarray}
\langle\Omega|c^{\textrm{out}}_{p_1''}c^{\textrm{out}}_{p_2''}c^{\dag\textrm{out}}_{q_2}c^{\dag\textrm{out}}_{q_1}|\Omega\rangle  
& = & \left(2\pi\right)^3\left(2\pi\right)^3 \,2E_{\vec{p''}_1} 2E_{\vec{p''}_2} \, \left[\delta^3(\vec{p''}_1-\vec{q}_1)\delta^3(\vec{p''}_2-\vec{q}_2) \right. \nonumber \\
& + & \left. \delta^3(\vec{p''}_1-\vec{q}_2)\delta^3(\vec{p''}_2-\vec{q}_1)\right].
\label{momcon}
\end{eqnarray}
It is then clear that the whole matrix element in (\ref{mel}) vanishes unless 
\begin{eqnarray}
\hat{p}_1'' + \hat{p}_2'' = \hat{q}_1 + \hat{q}_2 ,
\label{momcond}
\end{eqnarray}
so that
\begin{eqnarray}
{e^{-\frac{i}{2}\hat{p}_1''\wedge\left(-\hat{p}_2'' + \hat{p}_1'' + \hat{p}_2''\right) -\frac{i}{2}\hat{p}_2''\wedge\left(\hat{p}_1'' + \hat{p}_2''\right)} =
 e^{-\frac{i}{2}\hat{p}_2'' \wedge\hat{p}_1''}.}
\end{eqnarray}

Now, integrations over $\vec{x}_1'$, $\vec{x}_2'$ give us further $\delta$-functions which imply that
\begin{eqnarray}
{\vec{p''}_1=\vec{p'}_1~,~ \vec{p''}_2 = \vec{p'}_2}
\end{eqnarray}
 and hence
\begin{eqnarray}
{\hat{p}_1''=\hat{p}_1'~,~ \hat{p}_2'' = \hat{p}_2'.}
\end{eqnarray}
Thus we finally obtain the noncommutative phase $e^{-\frac{i}{2}\hat{p}_2'\wedge\hat{p}_1'}$. 

Moreover, since
\begin{eqnarray}
{_{\textrm{out}}\langle\hat{q}_1, \hat{q}_2| \rightarrow~ _{\textrm{out}}\langle\hat{p}_1', \hat{p}_2'|}
\end{eqnarray}
 and due to the identity
\begin{eqnarray}
{_{\textrm{out}}\langle\Omega|c^{\textrm{out}}_{q_1}c^{\textrm{out}}_{q_2} = ~_{\textrm{out}}\langle\Omega|c^{\textrm{out}}_{q_2}c^{\textrm{out}}_{q_1}},
\end{eqnarray}
 we finally obtain
\begin{eqnarray}
& & \widetilde{G}_{\theta}^{(2)}(p_1', p_2', \cdots , x_N', x_1, \cdots , x_M) = \frac{\sqrt{Z}}{p_1'^2 - m^2 - i\epsilon}\frac{\sqrt{Z}}{p_2'^2 - m^2 - 
i\epsilon} e^{-\frac{i}{2}\hat{p}_2'\wedge\hat{p}_1'} \nonumber \\ 
& &  _{\textrm{out}}\langle\hat{p}_1' \hat{p}_2'|\mathcal{T}\left(\phi_{\theta}(x_3')\cdots\phi_{\theta}(x_N')\phi_{\theta}(x_1)\cdots\phi_{\theta}(x_M)\right)|\Omega\rangle 
\, + \, \textrm{OT}.
\end{eqnarray}
 The phase can be absorbed so that the twisted out-state becomes
\begin{eqnarray}
{\langle\Omega|a_{\theta}^{\textrm{out}}(\hat{p}_2')a_{\theta}^{\textrm{out}}(\hat{p}_1').}
\end{eqnarray}
Hence the two-particle residue gives us the same expression as obtained in (\ref{ncmsmatrix}). 

As shown in \cite{amilcar} the above analysis can be easily generalized to $N$ outgoing particles. For this purpose it is enough to analyze the phases associated with the outgoing 
fields. Indeed, let us look at
\begin{eqnarray}
\label{25.1}
\langle\Omega|a^{\textrm{out}}_{\hat{p}_1'}a^{\textrm{out}}_{\hat{p}_2'}\cdots a^{\textrm{out}}_{\hat{p}_N'}|\hat{q}_1\cdots\hat{q}_N\rangle
\qquad \text{and} \qquad
\langle \hat{q}_1\cdots\hat{q}_N|a^{\dag}_{\hat{p}_N'}\cdots a^{\dag}_{\hat{p}_1'} |\Omega\rangle.
\end{eqnarray}
 The above two matrix elements have phases related with each other by complex conjugation.  One can easily calculate them by using (\ref{tcom}) and moving the twist of $a_{\hat{p}'}$ 
in the first term to the left and in the second term to the right. This will give the appropriate phase seen in (\ref{ncmsmatrix}).

 One can similarly do a computation for incoming particles as well, where the conjugates of (\ref{25.1}) will appear. Putting all this together, the final answer can
easily be seen to be the same as the one obtained in (\ref{ncmsmatrix}).


\section{Renormalization and $\beta$-function}


In this section, we carry out the renormalization of twisted $\phi^4_{\theta,\ast}$ scalar field theory on the Moyal plane. We argue that the twisted theory is renormalizable,
with the renormalization prescription being similar to that of commutative $\phi^4_{0}$ theory. In particular, we explicitly check the above claim by carrying 
out renormalization to one loop, computing the beta-function upto one loop and analyzing the RG flow of coupling. We show that the twisted-$\beta$ function
is essentially the same as the $\beta$ function of the commutative theory. The case of more general pure matter theories will be considered in the next section. 

In this section, we follow the treatment of \cite{ramond} and \cite{srednicki} for the computations in the commutative $\phi^4_{0}$ theory.


 \subsection{Superficial Degree of Divergence}


We begin by analyzing superficial degree of divergence of a generic Feynman diagram for a $\phi^n_{\theta,\ast}$ scalar field theory in d-dimensions. It is easy to see that the 
criterion for superficial degree of divergence will be the same as that for a generic Feynman diagram for a $\phi^n_{0}$ scalar field theory in d-dimensions. The reason is that 
the noncommutative $S$-matrix (and Feynman diagrams) differ from their commutative counterparts only by an overall noncommutative phase
which does not contribute to the superficial degree of divergence of a diagram. For a generic noncommutative Feynman diagram (involving only scalars) in d-dimensions 
with E external lines, I internal lines and $V_N$ vertices having N-legs (internal or external) attached to them, the superficial degree of divergence D is
\begin{eqnarray}
 D & = & d - \frac{1}{2}(d - 2)E  + V_N \left( \frac{N-2}{2}d - N \right) .
\label{sdd}
\end{eqnarray}
In $d=4$ dimensions this reduces to
\begin{eqnarray}
 D & = & 4 - E  + V_N \left( N - 4 \right). 
\label{sdd4}
\end{eqnarray}
Furthermore, for  $\phi^4_{\theta,\ast}$ theory in $d=4$ dimensions we have 
\begin{eqnarray}
 D & = & 4 - E  .
\label{ncsdd4}
\end{eqnarray}
We notice that, as expected, the superficial degree of divergences in (\ref{sdd}), (\ref{sdd4}) and (\ref{ncsdd4}) are all the same as that for commutative case. 
So the criterion for determining which of the diagrams will be divergent, remains the same, i.e. the diagrams with $D  \geq 0$ 
are the divergent ones.  Thus, it follows immediately from (\ref{ncsdd4}), that for  $\phi^4_{\theta,\ast}$ theory in $d=4$ dimensions, which is the model we are presently 
interested in, there are divergences for $E=2$ and $E=4$. These correspond to the one particle irreducible (1PI) 2-point function $\Gamma^{(2)}_{\theta}$ and 4-point 
function $\Gamma^{(4)}_{\theta}$ respectively, implying that, $\Gamma^{(2)}_{\theta}$ and $\Gamma^{(4)}_{\theta}$ will be divergent. We need to renormalize them, resulting in 
corrections to propagators and vertices. Furthermore, like in commutative case, by making 1PI two-point function and four-point functions finite, we can make the whole theory finite, 
as these two functions are the only source of divergences.

We further remark that, like in commutative case, just because the superficial degree of divergence of a given diagram is less than zero does not mean that it is divergence free, 
as it can have divergent sub-diagrams. But if we renormalize $\Gamma^{(2)}_{\theta}$ and $\Gamma^{(4)}_{\theta}$, all these sub-divergences 
will be taken into account, resulting in the renormalized theory being divergence free.


\subsection{Dimensional Regularization and Renormalization using the Minimal Subtraction Scheme}


In this section, we carry out the renormalization of $\phi^4_{\theta,\ast}$ scalar field theory on Moyal Plane, using dimensional regularization and minimal subtraction scheme. 
We use $\overline{MS}$ scheme and dimensional regularization by working in d = 4 - $\epsilon$ dimensions. In  d = 4 - $\epsilon$ dimensions the coupling $\lambda$ is no longer 
dimensionless, so we change it to $\lambda \rightarrow \lambda \tilde{\mu}^{\epsilon}$, where $\tilde{\mu}$ is a mass parameter. 

The bare ($\tilde{\phi}_{\theta}$, $m_B$ and $\lambda_B$) and renormalized ($\phi_{\theta}$, $m$ and $\lambda$) fields and parameters are related with each other as 
\begin{eqnarray}
\tilde{\phi}_{\theta} & = & Z^{1/2}_{\phi} \, \phi_{\theta} , \nonumber \\
m_B & = & Z^{-1/2}_{\phi} \,  Z_{m} \, m , \nonumber \\
\lambda_B & = & Z^{-2}_{\phi} \, Z_{\lambda} \, \lambda \, \tilde{\mu}^{\epsilon} .
\label{b2r}
\end{eqnarray}
where $Z_{\Phi}$ is the wavefunction renormalization constant, $Z_m$ is the mass renormalization constant and $Z_{\lambda}$ is the coupling renormalization constant. 
The Zs are as of yet unknown constants and are to be evaluated perturbatively. It should also be noted that the functional form of the Zs depends on the renormalization scheme. 
Moreover, it turns out that in $\overline{MS}$ renormalization scheme, the Zs will have a generic form like
\begin{eqnarray}
 Z_{\Phi} & = & 1 \, + \, \sum_{n=1}^{\infty}\, \frac{a_n(\lambda)}{\epsilon^n}, \nonumber \\
 Z_{m} & = & 1 \, + \, \sum_{n=1}^{\infty}\, \frac{b_n(\lambda)}{\epsilon^n}, \nonumber \\
 Z_{\lambda} & = & 1 \, + \, \sum_{n=1}^{\infty}\, \frac{c_n(\lambda)}{\epsilon^n}.
\label{zstrt}
\end{eqnarray}
From (\ref{zstrt}) and as we will argue later in this section, the Zs are all independent of $\theta$ to all orders in perturbation theory. This implies that the $\beta$-function, 
the anomalous dimensions of mass and n-point Green's functions will be the same as that for commutative $\phi^4_0$ theory.

\vspace{4cm}

\subsection{2-Point Function}


The Feynman diagrams contributing at one loop to the two-point function are seen in Figure 2.1.

\begin{figure}[H]
        \centerline{
               \mbox{\includegraphics*[angle=0,width=7in]{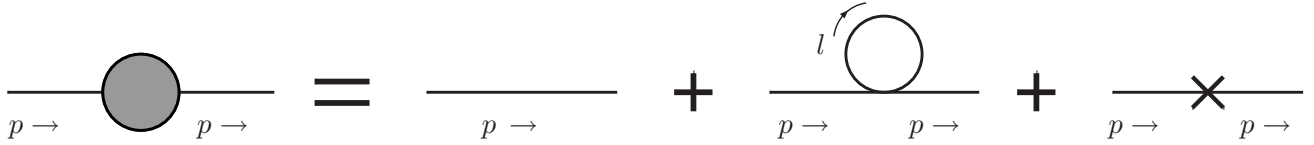}}
               }
 \caption{ Feynman diagrams for the 2-point function at one loop.  }
               \end{figure}
               
So the loop contribution to the 2-point function is given by
\begin{eqnarray}
 -i \Pi(k^2) & = &  \frac{1}{2}(-i Z_{ \lambda}\,  \lambda \, \tilde{\mu}^{\epsilon}) \int \frac{d^d l}{(2\pi)^d}\, \frac{i}{l^2 - m^2} \, + \, i (A \, k^2 - Bm^2),
\label{twopointfunction}
\end{eqnarray}
where $A = Z_{\phi} - 1$ and $B = Z_m - 1$.

Now, let us consider the integral
\begin{eqnarray}
\xi & = &  \tilde{\mu}^{\epsilon} \int \frac{d^d l}{(2\pi)^d}\, \frac{i}{l^2 - m^2}.
\label{twopointintegral}
\end{eqnarray}
Substituting $l^0 = i l^0_E$ and going to Euclidean plane we have 
\begin{eqnarray}
\xi & = &  \tilde{\mu}^{\epsilon} \int \frac{d^d l_E}{(2\pi)^d}\, \frac{1}{l_E^2 + m^2},
\label{2pointintegral}
\end{eqnarray}
where $l_E^2 = (l_E^0)^2 + \vec{l}^2 $. The integral evaluates to (d = 4 - $\epsilon$) \cite{srednicki}
\begin{eqnarray}
 \xi & = & \frac{\Gamma(-1 + \frac{\epsilon}{2})}{(4\pi)^2} \, m^2 \, \left( \frac{4\pi\tilde{\mu}^2}{m^2} \right)^{\frac{\epsilon}{2}}
\label{2pointinteg}
\end{eqnarray}
Now, we use the identity
\begin{eqnarray}
 \Gamma (-n + x) & = & \frac{(-1)^n}{n!} \left[ \frac{1}{x} - \gamma + \sum^n_{k=1} k^{-1} + O(x)\right],
\label{gammaidentity}
\end{eqnarray}
where $\gamma$ is the Euler-Mascheroni constant. Using (\ref{gammaidentity}) in (\ref{2pointinteg}) we obtain
\begin{eqnarray}
\xi & = & \frac{-m^2}{(4\pi)^2} \, \left( \frac{2}{\epsilon} - \gamma + 1 \right) \, \left( \frac{4\pi\tilde{\mu}^2}{m^2} \right)^{\frac{\epsilon}{2}} \nonumber \\
& = & \frac{-m^2}{(4\pi)^2} \, \left[ \left( \frac{2}{\epsilon} - \gamma + 1 \right) \, + \,\ln \left( \frac{4\pi\tilde{\mu}^2}{m^2} \right) \, + \, (- \gamma + 1) \frac{\epsilon}{2}
 \ln \left( \frac{4\pi\tilde{\mu}^2}{m^2}\right) \right],
\label{2pointint}
\end{eqnarray}
where we have used the relation $X^{\frac{\epsilon}{2}} = 1 + \frac{\epsilon}{2} \ln X + O(\epsilon^2),\, \text{for} \, \epsilon << 1$. Since we are interested 
in the d = 4 case, we take the limit $\epsilon \rightarrow 0$ in (\ref{2pointint}), so that 
\begin{eqnarray}
\lim_{\epsilon \rightarrow 0} \xi & = & \frac{-m^2}{(4\pi)^2} \, \left[  \frac{2}{\epsilon}  + 1 \, + \,\ln \left(\frac{\mu^2}{m^2} \right) \right],
\label{lim2pointint}
\end{eqnarray}
where we have $\mu^2 = 4\pi\tilde{\mu}^2 \, e^{-\gamma}$. Using (\ref{lim2pointint}) in (\ref{twopointfunction}) we obtain
\begin{eqnarray}
 \lim_{\epsilon \rightarrow 0} -i \Pi(k^2) & = &  \frac{(-i  \lambda)}{2} \, \frac{-m^2}{(4\pi)^2} \, \left[  \frac{2}{\epsilon}  + 1 \, + \,\ln \left(\frac{\mu^2}{m^2} \right) \right]
 \, + \, i (A \, k^2 - B \, m^2) .
\label{limtwopointfunction}
 \end{eqnarray}
As can be seen from (\ref{limtwopointfunction}), the singularities due to loop contribution manifest themselves as certain terms developing singularities in the 
limit $\epsilon \rightarrow 0$. Since we are interested in only the singular terms we may split (\ref{limtwopointfunction}) as
\begin{eqnarray}
 \lim_{\epsilon \rightarrow 0} \Pi(k^2) & = & - \frac{ \lambda \, m^2}{(4\pi)^2} \, \frac{1}{\epsilon} \, - \, A \, k^2  +  B \, m^2 \, + \, \text{Terms of finite order}.
\label{limtwopointfn}
\end{eqnarray}

Now, according to the $\overline{MS}$ scheme, the constants $A$ and B are to be chosen in such a way as to cancel all the singular terms in (\ref{limtwopointfn}). So we have
\begin{eqnarray}
  A & = & Z_{\phi} - 1 \; = \; O( \lambda^2) \qquad \qquad \qquad \; \; \; \;  \Rightarrow \, Z_{\phi} \; = \; 1 \, + \, O( \lambda^2) \nonumber \\
B & = & Z_{m} - 1 \; = \; \frac{ \lambda}{16 \pi^2} \, \frac{1}{\epsilon} \, + \, O( \lambda^2 ) \qquad \Rightarrow \, Z_{m} \; = \; 1 \, + \,  \frac{ \lambda}{16 \pi^2}
 \, \frac{1}{\epsilon} \, + \, O( \lambda^2 )
\label{ab}
\end{eqnarray}

\vspace{2.5cm}
\subsection{4-Point Function}


The Feynman diagrams up to one loop for the four-point function are depicted in Figure 2.2.

\begin{figure}[H]
        \centerline{
               \mbox{\includegraphics*[angle=0,width=7in]{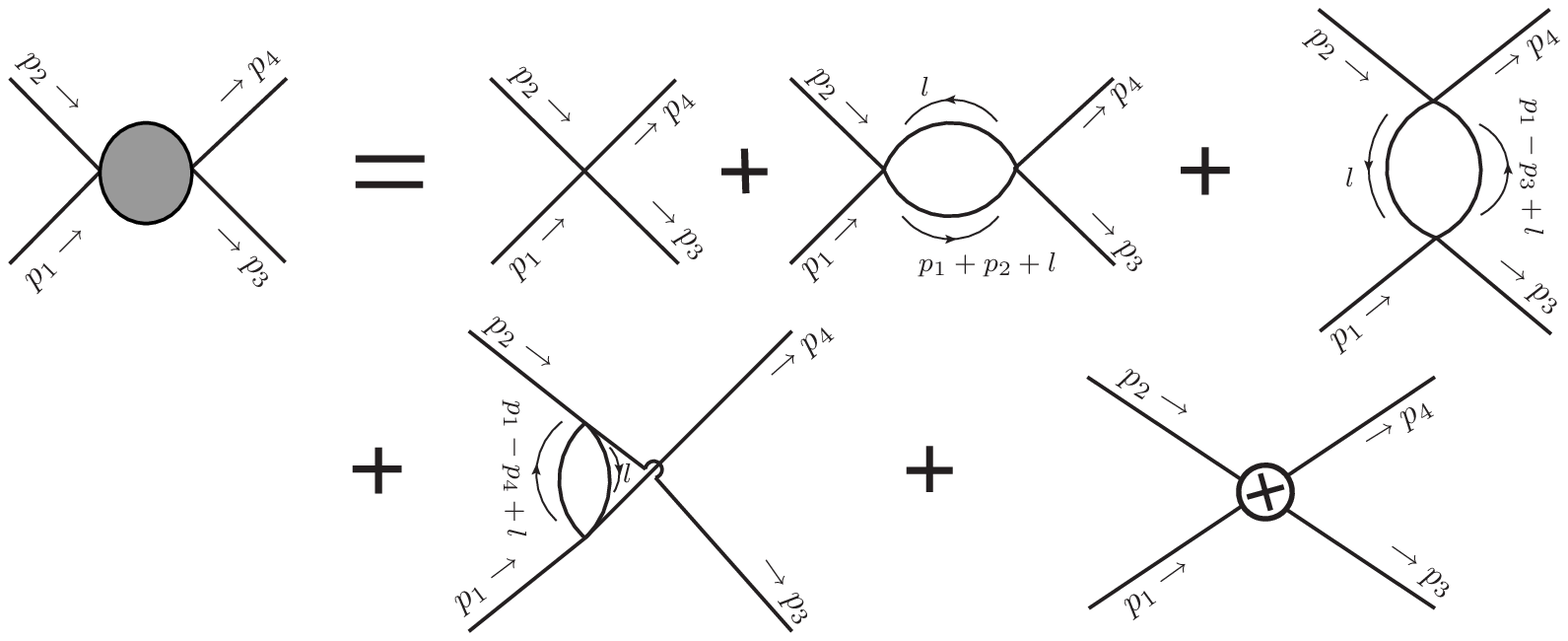}}
               }
 \caption{ Feynman diagrams for the 4-point function at one loop. }
               \end{figure}

The 4-point function is given by
\begin{equation}
i \Gamma^{(4)}_{\theta}  =  e^{\frac{i}{2}\left(p_1\wedge p_2 - p_3\wedge p_4\right)}  \left[-i Z_{\lambda} \lambda  \tilde{\mu}^{\epsilon}  
  + \frac{1}{2} \left( -i Z_{\lambda} \lambda  \tilde{\mu}^{\epsilon} \right)^2  \left \{ i V(s) + iV(t) + iV(u) \right \}  +  O(\lambda^3) \right]
\label{fourpointfunction}
\end{equation}
where s, t, u are the Mandelstam variables defined as $ s = (p_1 + p_2)^2$, $ t = (p_3 - p_1)^2$ and $ u = (p_4 - p_1)^2$ and 
\begin{eqnarray}
 iV(p^2) & = & \int \frac{d^d l}{(2\pi)^d} \, \frac{i}{(l + p)^2 - m^2} \, \frac{i}{l^2 - m^2}.
\label{4pointintegral}
\end{eqnarray}
The appearance of noncommutative phases in (\ref{fourpointfunction}) is an attribute of the twisted statistics followed by the particles. Moreover, these phases insure that 
$\Gamma^{(4)}_{\theta} $ has right symmetries vis-a-vis twisted Poincar\'e invariance.

Now, consider the integral
\begin{eqnarray}
\tilde{\mu}^{\epsilon} \, iV(p^2) & = &  \tilde{\mu}^{\epsilon} \, \int \frac{d^d l}{(2\pi)^d} \, \frac{i}{(l + p)^2 - m^2} \, \frac{i}{l^2 - m^2},
\label{div4pointintegral}
\end{eqnarray}
which evaluates after Wick rotation $q^0 \rightarrow iq^0_E$ to \cite{srednicki}
\begin{eqnarray}
 \tilde{\mu}^{\epsilon} \, iV(p^2) & = &  - i \tilde{\mu}^{\epsilon} \, \int_0^1 dx \, \int \frac{ d^d q_E}{(2\pi)^d} \, \frac{1}{[q_E^2 + D]^2}
\label{d4ptintegral}
\end{eqnarray}
where $D = m^2 - x(1-x)p^2 $, with $x$ being a Feynman parameter. 

Using the standard integral
\begin{eqnarray}
 \int \frac{ d^d q_E}{(2\pi)^d} \, \frac{(q_E^2)^a}{[q_E^2 + D]^b} & = & \frac{\Gamma(b - a - \frac{d}{2}) \,
 \Gamma(a + \frac{d}{2})}{(4\pi)^{\frac{d}{2}}\,\Gamma(b)\,\Gamma(\frac{d}{2})},
\, D^{-(b - a - \frac{d}{2})}
\label{stdint}
\end{eqnarray}
we have
\begin{eqnarray}
 \tilde{\mu}^{\epsilon} \, iV(p^2) & = &  - i \tilde{\mu}^{\epsilon} \, \int_0^1 dx \, \frac{\Gamma(2 - \frac{d}{2})}{(4\pi)^{\frac{d}{2}}} \, D^{-(2 - \frac{d}{2})} .
\label{d4pintegral}
\end{eqnarray}
Putting $ d = 4 - \epsilon $, we have
\begin{eqnarray}
 \tilde{\mu}^{\epsilon} \, iV(p^2) 
& = &  \frac{-i}{(4\pi)^2} \, \Gamma \left( \frac{ \epsilon}{2} \right ) \, \int_0^1 dx \, \left( \frac{4\pi \, \tilde{\mu}^{2}}{D} \right )^{\frac{ \epsilon}{2}}.
\label{d4pinteg}
\end{eqnarray}
Using the identity
\begin{eqnarray}
 \Gamma(-n + x) & = & \frac{(-1)^n}{n!}\, \left[ \frac{1}{x} - \gamma + \sum_{k=1}^n \, k^{-1} + O(x) \right],
\label{gammafn}
\end{eqnarray}
we have 
\begin{eqnarray}
 \Gamma \left( \frac{ \epsilon}{2} \right ) & = &  \frac{2}{\epsilon} - \gamma +  O(\epsilon) .
\label{gamma}
\end{eqnarray}
Using (\ref{gamma}) in (\ref{d4pinteg}) we have
\begin{eqnarray}
 \tilde{\mu}^{\epsilon} \, iV(p^2) & = & \frac{-i}{(4\pi)^2} \, \left( \frac{2 }{\epsilon} - \gamma \right) \, \int_0^1 dx \, 
\left( \frac{4\pi \, \tilde{\mu}^{2}}{D} \right )^{\frac{ \epsilon}{2}}.
\label{d4pinte}
\end{eqnarray}
In the limit $\epsilon \rightarrow 0$, we have
\begin{eqnarray}
 \lim_{\epsilon \rightarrow 0} \tilde{\mu}^{\epsilon} \, iV(p^2) & = &  \frac{-i}{(4\pi)^2} \, \left[ \frac{2 }{\epsilon} +   \int_0^1 dx \, \ln \left( \frac{\mu^{2}}{ D} \right) \right],
\label{limd4pint}
\end{eqnarray}
Using (\ref{limd4pint}) into (\ref{fourpointfunction}) we obtain
\begin{eqnarray}
& & \lim_{\epsilon \rightarrow 0} i \Gamma^{(4)}_{\theta} \, = \, e^{\frac{i}{2}\left(p_1\wedge p_2 - p_3\wedge p_4\right)} \left[-i Z_{\lambda} \lambda  
  + \frac{1}{2} \left( -i Z_{\lambda} \lambda \right)^2 \left( \frac{-i}{(4\pi)^2} \right) \left \{ \frac{6 }{\epsilon} +  \int_0^1 dx 
 \left(  \ln \left( \frac{\mu^{2}}{ D(s)} \right)  \right. \right. \right.  \nonumber \\
& & +  \left.\left. \left. \ln \left( \frac{\mu^{2}}{ D(t)} \right)  \, + \, \ln \left( \frac{\mu^{2}}{ D(u)} \right) \right) \right \} + O(\lambda^3) \right]  \nonumber \\
& & \approx \, e^{\frac{i}{2}\left(p_1\wedge p_2 - p_3\wedge p_4\right)} \left[-i Z_{\lambda} \lambda  
  +  \left( -i \lambda \right)^2 \left( \frac{-i}{32\pi^2} \right)  \left \{ \frac{6 }{\epsilon} +  \text{Finite Terms} \right \} + O(\lambda^3) \right] 
\label{4pointfunction}
\end{eqnarray}
where in writing last line we have neglected higher powers of $Z_{\lambda}$.

Now, in accordance with $\overline{MS}$ scheme, matching the divergent parts in (\ref{4pointfunction}), we obtain
\begin{eqnarray}
 Z_{\lambda} & = & 1 + \frac{3\lambda}{16 \pi^2}\, \frac{1}{\epsilon} 
\label{zlambda}
\end{eqnarray}
which is the same as that for the commutative theory. Note that, as remarked in the beginning of this section, the $Z_{\lambda}$, $Z_{\phi}$ and $Z_{m}$ are all completely 
independent of $\theta$. This is what we naively expected from our analysis of the tree level theory in previous sections. The noncommutative corrections are just phases. 
Hence they do not result in any new source of divergence. Moreover, since the form of Zs is completely fixed (within a given renormalization scheme) by the demand that the 
renormalized theory 
should be divergence free, if we try to put an implicit dependence of $\theta$ in Zs, then (\ref{zlambda}) and (\ref{ab}) will not be satisfied, implying that the renormalized 
theory is still not completely free from divergences. So the demand that renormalized theory be completely free of any divergence, forces us to choose Zs of the form (\ref{zlambda}) 
and (\ref{ab}) and hence no dependence of Zs on $\theta$, whether implicit or explicit, is allowed. Moreover, although we have done calculations with a particular 
renormalization scheme, it is easy to see that whatever renormalization scheme one chooses to use, the source and form of divergences always remains the same. The noncommutative 
phases will never result in any new divergence or contribute to any divergence and hence the demand to cancel all the divergences will always imply that at least the divergent 
part of Zs is completely independent of $\theta$. As it does in commutative theory, the prescription dependence of renormalization scheme will only effect 
the finite terms. Hence, even changing renormalization scheme or for that matter, even the regularization technique, does not change the essential result that the divergent part 
of Zs have no dependence, implicit or explicit, on $\theta$. 

\textbf{Higher Loop Corrections to 2-Point and 4-Point Functions :} 

Although in this chapter we restrict ourself only to one loop corrections to 2-point and 4-point functions, higher loop effects can similarly be computed. 
The noncommutative corrections are always a phase (to all orders of perturbation). They never give rise to new sources of divergences. So the Zs to all orders in perturbation will 
be always independent (implicitly as well as explicitly) of $\theta$ and will have the same form as that of commutative theory. So, like in commutative case the generic form of Zs 
are   
\begin{eqnarray}
 Z_{\phi} & = & 1 \, + \, \sum_{n=1}^{\infty}\, \frac{a_n(\lambda)}{\epsilon^n} \nonumber \\
 Z_{m} & = & 1 \, + \, \sum_{n=1}^{\infty}\, \frac{b_n(\lambda)}{\epsilon^n} \nonumber \\
 Z_{\lambda} & = & 1 \, + \, \sum_{n=1}^{\infty}\, \frac{c_n(\lambda)}{\epsilon^n}
\label{1zstrt}
\end{eqnarray}
where $a_n(\lambda)$, $b_n(\lambda)$ and $c_n(\lambda)$ are unknown functions which are evaluated perturbatively by demanding that the renormalized theory be independent of 
divergences at all orders of perturbation. Also note that $a_n(\lambda)$, $b_n(\lambda)$ and $c_n(\lambda)$ are all independent of $\theta$ and as argued before they have the 
same form as for 
commutative $\phi^4_0$ theory.


\subsection{Renormalization Group and $\beta$-Function}


In previous section we showed that all Zs are independent of $\theta$ and are the same as in the commutative case. In view of this, we expect and will show by explicit computations 
that it is indeed the case. The $\beta$-function and R.G equation are also independent of $\theta$. They are the same as in the commutative case.

For $\beta$-function computation we start with noticing the fact that the bare and renormalized couplings are related with each other via
\begin{eqnarray}
 \lambda_B & = & Z^{-2}_{\phi} \, Z_{\lambda} \, \lambda \, \tilde{\mu}^{\epsilon},
\label{btor}
\end{eqnarray}
or
\begin{eqnarray}
\ln \lambda_B & = & \ln (Z^{-2}_{\phi} \, Z_{\lambda}) \, + \, \ln \lambda \, + \, \epsilon \, \ln \tilde{\mu}.
\label{lnbtor}
\end{eqnarray}
Differentiating (\ref{lnbtor}) with respect to $\ln \mu$, we obtain
\begin{eqnarray}
\frac{\partial (\ln \lambda_B)}{\partial(\ln \mu)} & = & \frac{\partial (\ln (Z^{-2}_{\phi} \, Z_{\lambda}))}{\partial (\ln \mu)} \, + \, \frac{\partial(\ln \lambda)}{\partial (\ln \mu)}
 \, + \, \frac{\partial (\epsilon \, \ln \tilde{\mu})}{\partial (\ln \mu)} \nonumber \\
& = & \frac{\partial (\ln (Z^{-2}_{\phi} \, Z_{\lambda}))}{\partial (\ln \mu)} \, + \, \frac{\partial (\ln \lambda)}{\partial (\ln \mu)}
 \, + \, \epsilon .
\label{dlnbtor}
\end{eqnarray}
where $\mu$ is a mass scale, $\mu^{2} = 4\pi \, e^{-\gamma}\, \tilde{\mu}^{2} $. 
Now, we demand that the bare coupling be independent of $\mu$, i.e. $\frac{\partial (\ln \lambda_B)}{\partial (\ln \mu)} = 0 $. Then
\begin{eqnarray}
 0 & = & \frac{\partial (\ln (Z^{-2}_{\phi} \, Z_{\lambda}))}{\partial (\ln \mu)} \, + \, \frac{\partial (\ln \lambda)}{\partial (\ln \mu)}
 \, + \, \epsilon \nonumber \\
& = & \frac{\partial (\ln (Z^{-2}_{\phi} \, Z_{\lambda}))}{\partial\, \lambda} \, \frac{\partial\, \lambda}{\partial (\ln \mu)} 
\, + \, \frac{1}{\lambda} \, \frac{\partial \, \lambda}{\partial (\ln \mu)} \, + \, \epsilon .
\label{dlnbtorb0}
\end{eqnarray}
From (\ref{ab}) and (\ref{zlambda}) we have 
\begin{eqnarray}
 \ln (Z^{-2}_{\phi} \, Z_{\lambda}) & = & \ln \left( 1 + \frac{3\lambda}{16 \pi^2}\, \frac{1}{\epsilon} \right) \; = \; \frac{3\lambda}{16 \pi^2}\, \frac{1}{\epsilon} 
\, + \, O(\lambda^2) .
\label{lnzform}
\end{eqnarray}
Using (\ref{lnzform}) in (\ref{dlnbtorb0}) we obtain 
\begin{eqnarray}
\frac{3}{16 \pi^2}\,\frac{1}{\epsilon}\,\frac{\partial\,\lambda}{\partial (\ln \mu)} \, + \, \frac{1}{\lambda}\, \frac{\partial\, \lambda}{\partial (\ln \mu)}\, + \, \epsilon & = & 0,
\label{gbetafn1}
\end{eqnarray}
or
\begin{eqnarray}
\frac{\partial\,\lambda}{\partial (\ln \mu)} & = & - \epsilon \lambda \, + \, \frac{3 \lambda^2}{16 \pi^2} \, + \, O(\lambda^3) .
\label{gbetafn}
\end{eqnarray}
Therefore the $\beta$-function is given by
\begin{eqnarray}
 \beta (\lambda) & = & \lim_{\epsilon \rightarrow 0} \, \frac{\partial\,\lambda}{\partial (\ln \mu)} \; = \; \frac{3 \lambda^2}{16 \pi^2} \, + \, O(\lambda^3). 
\label{beta} 
\end{eqnarray}
We note that, as expected, (\ref{beta}) is completely independent of $\theta$ and is the same as the commutative $\beta$-function.

Integrating (\ref{beta}) we can immediately calculate the running of coupling constant with respect to variation in the scale $\mu$, which turns out to be the same as in the 
commutative theory. It is given by
\begin{eqnarray}
 \lambda_2  & = & \frac{\lambda_1}{1 \, - \,\frac{3\lambda_1 }{16 \pi^2} \, \ln \left( \frac{\mu_2}{\mu_1} \right) }.
\label{runningcoupling}
\end{eqnarray}

Now we calculate the R.G equation for a generic n-point 1PI function $\Gamma^{(n)}_{\theta}$. 

The bare n-point 1PI functions $\Gamma^{(n)}_{\theta, B}$ and renormalized n-point 1PI functions  $\Gamma^{(n)}_{\theta, R}$ are related with each other as
\begin{eqnarray}
 \Gamma^{(n)}_{\theta, B} \, (p_1, \dots, p_n, \theta; \, \lambda_B, m_B, \epsilon) & = &  Z^{\frac{-n}{2}}_{\phi} 
\, \Gamma^{(n)}_{\theta, R} \, (p_1, \dots, p_n, \theta; \, \lambda, m, \epsilon, \mu)
\label{n1pifn}
\end{eqnarray}
where all the $\Gamma^{(n)}_{\theta, R}$ are finite as $\epsilon \rightarrow 0$. From (\ref{n1pifn}) we see that the left hand side does not depend on the arbitrary scale $\mu$ 
but the right hand side has explicit as well as implicit (through the mass $m$ and coupling $\lambda$) dependence on $\mu$ \footnote{ Its worth noting that the functional 
dependence of  both $\Gamma^{(n)}_{\theta, B}$ and  $\Gamma^{(n)}_{\theta, R}$ on the noncommutative phases (like on momenta) is same, so the noncommutative phases will not 
affect the R.G. equations}.  So if (\ref{n1pifn}) is correct then the explicit and implicit dependence of $\Gamma^{(n)}_{\theta, R}$ on $\mu$ should cancel each other, i.e.
\begin{eqnarray}
 \frac{ \partial \ln \left \{\Gamma^{(n)}_{\theta, B} \, (p_1, \dots, p_n, \theta; \, \lambda_B, m_B, \epsilon) \right \}}{ \partial \ln \mu}  \, = \, 
\frac{ \partial \ln \left \{ Z^{\frac{-n}{2}}_{\phi} \, \Gamma^{(n)}_{\theta, R} \, (p_1, \dots, p_n, \theta; \, \lambda, m, \epsilon, \mu) \right \}}{ \partial \ln \mu} & = & 0 
\nonumber \\
\left[ \mu \, \frac{\partial }{\partial \mu} \, + \, \mu \, \frac{\partial \lambda}{\partial \mu} \, \frac{\partial}{\partial \lambda} 
\, + \,\mu \, \frac{\partial m }{\partial  \mu} \, \frac{\partial}{\partial m} \, - \, \mu \, \frac{n}{2} \, \frac{\partial \ln Z_{\phi}}{\partial \mu}  \right]
\, \Gamma^{(n)}_{\theta, R} \, (p_1, \dots, p_n, \theta; \, \lambda, m, \epsilon, \mu) = 0 \nonumber \\
\left[ \mu \, \frac{\partial }{\partial \mu} \, + \, \beta \, \frac{\partial}{\partial \lambda} \, + \, \frac{1}{m} \,\gamma_m \, \frac{\partial}{\partial m} \, - \, n \gamma_d  \right]
\, \Gamma^{(n)}_{\theta, R} \, (p_1, \dots, p_n, \theta; \, \lambda, m, \epsilon, \mu) = 0
\label{nrg}
\end{eqnarray}
where $\beta$ is the $\beta$-function, $\gamma_m$ is the anomalous mass dimension and $\gamma_d$ is the anomalous scaling dimension of $\Gamma^{(n)}_{\theta, R}$.

The equations (\ref{nrg}) are the R.G equations for a noncommutative n-point 1PI function. The noncommutativity does not give rise to any new divergences. Since the functional 
dependence of bare and renormalized n-point 1PI functions on noncommutative parameters are same, the R.G equations are essentially the same as that for commutative theory. 
Hence the noncommutative phases in (\ref{nrg}) sit more like spectators and do not affect the R.G. equations.


\section{Generic Pure Matter Theories}


So far we have restricted ourselves to the case of noncommutative real scalar fields having a $\phi^4_{\theta,\ast}$ self interaction. In this section, we consider general 
noncommutative theories involving only matter fields. As we show in this section, the formalism developed and discussed in previous sections, for
real scalar fields, goes through (with appropriate generalizations) for all such theories. Noncommutative theories involving gauge fields need a separate treatment and will 
not be discussed in this work.


\subsection{Complex Scalar Fields}
 

Let $\phi_{\theta}$ be a noncommutative complex scalar field having a normal mode expansion
\begin{eqnarray}
 \phi_{\theta} (x) & = & \int d^3 \tilde{k} \left[ a_k \, e^{-ikx} \, + \, b^\dagger_k e^{ikx}  \right] \nonumber \\
\phi^\dagger_{\theta} (x) & = & \int d^3 \tilde{k} \left[ b_k \, e^{-ikx} \, + \, a^\dagger_k e^{ikx}  \right] 
\label{csncfields}
\end{eqnarray}
where $ d^3 \tilde{k} \, = \, \frac{d^3 k}{(2\pi)^3 2 E_k}$, and $a_k$, $b_k$ are noncommutative annihilation operators satisfying the twisted algebra:
\begin{eqnarray}
  a^\#_{p_{1}} a^\#_{p_{2}} & = & \eta \, e^{ip_{1}\wedge p_{2}} \,a^\#_{p_{2}} a^\#_{p_{1}} \nonumber \\
(a^{\dagger}_{p_{1}})^\# (a^{\dagger}_{p_{2}})^\# & = & \eta \, e^{ip_{1}\wedge p_{2}} \, (a^{\dagger}_{p_{2}})^\# (a^{\dagger}_{p_{1}})^\# \nonumber \\
a_{p_{1}} a^{\dagger}_{p_{2}} & = & \eta \, e^{-ip_{1}\wedge p_{2}} \, a^{\dagger}_{p_{2}} a_{p_{1}} + (2\pi)^3 \, 2 E_p \,\delta^{3}(p_{1} \, - \, p_{2}) \nonumber \\
b_{p_{1}} b^{\dagger}_{p_{2}} & = & \eta \, e^{-ip_{1}\wedge p_{2}} \, b^{\dagger}_{p_{2}} b_{p_{1}} + (2\pi)^3 \, 2 E_p \,\delta^{3}(p_{1} \, - \, p_{2}) \nonumber \\
a_{p_{1}} b^{\dagger}_{p_{2}} & = & \eta \, e^{-ip_{1}\wedge p_{2}} \, b^{\dagger}_{p_{2}} a_{p_{1}}  
\label{nccstcom}
\end{eqnarray}
where $a^\#_{p}$ and $(a^{\dagger}_{p})^\# $ stands for either of the operators $a_p, b_p$ and $a^\dagger_p, b^\dagger_p$ respectively. For $\eta = 1$, these are bosonic operators. 
For $\eta = -1$, these are fermionic operators. We consider in the following $\eta = 1$. 

The  noncommutative creation/annihilation operators are related with their commutative counterparts (denoted by $c_k$, $d_k$ respectively) by the dressing transformations
\begin{eqnarray}
 a_k & = & c_k \, e^{-\frac{i}{2} \, k \wedge P}, \nonumber \\
b_k & = & d_k \, e^{-\frac{i}{2} \, k \wedge P}, \nonumber \\
 a^\dagger_k & = & c^\dagger_k \, e^{\frac{i}{2} \, k \wedge P}, \nonumber \\
b^\dagger_k & = & d^\dagger_k \, e^{\frac{i}{2} \, k \wedge P} 
\label{csdressingtransformations}
\end{eqnarray}
where $P_\mu$ is the Fock space momentum operator
\begin{eqnarray}
 P_\mu & = & \int d^3 \tilde{p} \, p_\mu [c^\dagger_p c_p \, + \, d^\dagger_p d_p] \nonumber \\
& = & \int d^3 \tilde{p} \, p_\mu [a^\dagger_p a_p \, + \, b^\dagger_p b_p]
\label{csmomentumoperator}
\end{eqnarray}
Using (\ref{csncfields}) and (\ref{csdressingtransformations}) one can easily check that the noncommutative fields are also related with commutative fields by the dressing 
transformation
\begin{eqnarray}
 \phi_{\theta} (x) & = & \phi_{0} (x) \, e^{\frac{1}{2} \, \overleftarrow{\partial} \wedge P} \nonumber \\
 \phi^\dagger_{\theta} (x) & = & \phi^\dagger_{0} (x) \, e^{\frac{1}{2} \, \overleftarrow{\partial} \wedge P}
\label{nccsfdress}
\end{eqnarray}
 where $\phi_{0}$ and $\phi^\dagger_{0}$ are the commutative complex scalar fields having the mode expansion
\begin{eqnarray}
 \phi_{0} (x) & = & \int d^3 \tilde{k} \left[ c_k \, e^{-ikx} \, + \, d^\dagger_k e^{ikx}  \right] \nonumber \\
\phi^\dagger_{0} (x) & = & \int d^3 \tilde{k} \left[ d_k \, e^{-ikx} \, + \, c^\dagger_k e^{ikx}  \right] 
\label{cscomfields}
\end{eqnarray}

Since $\phi_\theta$ is composed of the operators $a_p$ and $b_p$ following twisted statistics, unlike commutative fields, the commutator of $\phi_\theta$ 
and $\phi^\dagger_\theta$ evaluated at same spacetime points does not vanish, i.e.
\begin{eqnarray}
 [\phi_\theta (x), \phi^\dagger_\theta (x)] & \neq & 0
\label{op}
\end{eqnarray}

In view of (\ref{op}), one can in principle write six different quartic self interaction terms which naively seem inequivalent to each other. Hence, a generic 
interaction hamiltonian density with quartic self interactions can be written as
\begin{eqnarray}
& &  \mathcal{H}^{\theta}_{\rm{Int}} (x)  \, = \, \frac{\lambda_1}{4} \, \phi^\dagger_{\theta} \ast \phi^\dagger_{\theta} \ast \phi_{\theta} \ast \phi_{\theta} (x) 
\, + \, \frac{\lambda_2}{4} \,\phi^\dagger_{\theta} \ast \phi_{\theta} \ast \phi^\dagger_{\theta} \ast \phi_{\theta} (x)
  \, + \,\frac{\lambda_3}{4} \, \phi^\dagger_{\theta} \ast \phi_{\theta} \ast \phi_{\theta} \ast \phi^\dagger_{\theta} (x)  \nonumber \\
 & & + \, \frac{\lambda_4}{4} \, \phi_{\theta} \ast \phi^\dagger_{\theta} \ast \phi^\dagger_{\theta} \ast \phi_{\theta} (x)
\, + \, \frac{\lambda_5}{4} \,\phi_{\theta} \ast \phi^\dagger_{\theta} \ast \phi_{\theta} \ast \phi^\dagger_{\theta} (x)
 \, + \,\frac{\lambda_6}{4} \,\phi_{\theta} \ast \phi_{\theta} \ast \phi^\dagger_{\theta} \ast \phi^\dagger_{\theta} (x),
\label{csnchamiltonian}
\end{eqnarray}
where the $\lambda_i$ are the six coupling constants and in general they need not be equal to each other.


\subsection*{Some Identities}


We now list some identities that the noncommutative fields satisfy.
\begin{eqnarray}
\textbf{ 1)} \qquad  &  \phi^\dagger_\theta (x) \, \ast \, \phi_\theta (x)  \, = \,  \phi_\theta (x) \, \ast \, \phi^\dagger_\theta (x)  &  \qquad
\qquad \qquad \qquad \qquad \qquad \qquad \qquad \qquad 
\label{identity1}
\end{eqnarray}
\textbf{Proof : } Using (\ref{csncfields}) we have
\begin{eqnarray}
 \phi^\dagger_\theta (x) \, \ast \, \phi_\theta (x) & = &  \int d^3 \tilde{k}_1 \left[ b_{k_1} \, e^{-ik_1x} \, + \, a^\dagger_{k_1} e^{ik_1x} \right] 
\, e^{\frac{i}{2} \,\overleftarrow{\partial} \wedge \overrightarrow{\partial}}  \, \int d^3 \tilde{k}_2 \left[ a_{k_2} \, e^{-ik_2x} \, + \, b^\dagger_{k_2} e^{ik_2x}  \right]
\nonumber \\
& = & \int d^3 \tilde{k}_1 \, d^3 \tilde{k}_2 \,\left[ b_{k_1} \, e^{-ik_1x} \, e^{\frac{i}{2} (-ik_1) \wedge (-ik_2)} \, a_{k_2} \, e^{-ik_2x}
 \, + \, b_{k_1} \, e^{-ik_1x} \, e^{\frac{i}{2} (-ik_1) \wedge (ik_2)}  \right. \nonumber \\
& + & \left.  b^\dagger_{k_2} e^{ik_2x} a^\dagger_{k_1} e^{ik_1x}  e^{\frac{i}{2} (ik_1) \wedge (-ik_2)}  a_{k_2}  e^{-ik_2x} 
 +   a^\dagger_{k_1} e^{ik_1x} e^{\frac{i}{2} (ik_1) \wedge (ik_2)}  b^\dagger_{k_2} e^{ik_2x} \right]
\label{proof11}
\end{eqnarray}
The operators $a_p$ and $b_p$ satisfy twisted commutation relations, so using (\ref{nccstcom}) in (\ref{identity1}) we have 
\begin{eqnarray}
\phi^\dagger_\theta (x) \ast \phi_\theta (x) & = & \int d^3 \tilde{k}_1 d^3 \tilde{k}_2 \left[ a_{k_2}  b_{k_1}  e^{\frac{i}{2} (k_1) \wedge (k_2)}  e^{-ik_1x} \,e^{-ik_2x}
  +  b^\dagger_{k_2} \, b_{k_1} e^{\frac{-i}{2} (k_1) \wedge (k_2)} e^{-ik_1x} e^{ik_2x} \right. \nonumber \\
& - & (2\pi)^3 \, 2 E_{k_1} \, \delta^3 (k_1 - k_2)\, e^{-ik_1x} \,e^{ik_2x} \, + \,  a_{k_2} \, a^\dagger_{k_1}\, e^{\frac{-i}{2} (k_1) \wedge (ik_2)} \,e^{ik_1x} \, e^{-ik_2x}
\nonumber \\
 & + & \left. (2\pi)^3 \, 2 E_{k_1} \, \delta^3 (k_1 - k_2)\,e^{ik_1x} \, e^{-ik_2x} 
\, + \,   b^\dagger_{k_2}\,  a^\dagger_{k_1} \, e^{\frac{i}{2} (k_1) \wedge (k_2)}\, e^{ik_1x}\,e^{ik_2x}  \right] \nonumber \\
& = &  \int d^3 \tilde{k}_2 \left[ a_{k_2} \, e^{-ik_2x} \, + \, b^\dagger_{k_2} e^{ik_2x}  \right] \, e^{\frac{i}{2} \,\overleftarrow{\partial} \wedge \overrightarrow{\partial}}  \,
 \int d^3 \tilde{k}_1 \left[ b_{k_1} \, e^{-ik_1x} \, + \, a^\dagger_{k_1} e^{ik_1x} \right] \nonumber \\
& = &  \phi_\theta (x) \, \ast \, \phi^\dagger_\theta (x) 
\label{proof1}
\end{eqnarray}
\begin{eqnarray}
 \textbf{2)} \quad & \left [ \phi^\dagger_0 (x) \, \phi_0 (x) \right] \, e^{\frac{1}{2}\, \overleftarrow{\partial} \wedge P}
\,  = \, \left [ \phi_0 (x) \, \phi^\dagger_0 (x) \right] \, e^{\frac{1}{2}\, \overleftarrow{\partial} \wedge P} & \qquad \qquad \qquad 
\qquad \qquad \qquad 
\label{identity2}
\end{eqnarray}

One can check this identity by explicit calculations. But in view of (\ref{identity1}), this is easily checked to be true. Indeed this is nothing but (\ref{identity1}) rewritten 
in terms of commutative fields using dressing transformations (\ref{nccsfdress}). 

These two identities can be generalized to a product of arbitrary number of fields. Hence for a string of fields we have 
\begin{eqnarray}
 \textbf{3)} \qquad \quad \phi^\dagger_\theta (x) \, \ast \, \phi_\theta \,  \dots \, \phi^\dagger_\theta (x) \, \ast \, \phi_\theta  
\, & = &\,  \phi^\dagger_\theta (x) \, \ast \, \phi_\theta \, \dots  \,\phi_\theta (x) \, \ast \, \phi^\dagger_\theta (x) \qquad \qquad \qquad \qquad  
\nonumber \\
\qquad & = & \phi_\theta (x) \, \ast \, \phi^\dagger_\theta (x) \, \dots \, \phi_\theta (x) \, \ast \, \phi^\dagger_\theta (x)  \nonumber \\
\qquad & = & \text{Other Permutations.}
\label{identity3}
\end{eqnarray}
Using dressing transformation (\ref{nccsfdress}),  (\ref{identity3}) can be rewritten in terms of the commutative fields, so that 
\begin{eqnarray}
 \textbf{4)} \qquad \quad \left [ \phi^\dagger_0 (x) \, \phi_0 (x) \, \dots \phi^\dagger_0 (x) \, \phi_0 (x) \right] \, e^{\frac{1}{2}\, \overleftarrow{\partial} \wedge P}
& = &  \left [ \phi^\dagger_0 (x) \, \phi_0 (x) \, \dots \phi_0 (x) \, \phi^\dagger_0 (x) \right] \, e^{\frac{1}{2}\, \overleftarrow{\partial} \wedge P} 
  \nonumber \\
& = &  \left [\phi_0 (x) \, \phi^\dagger_0 (x) \dots \phi^\dagger_0 (x) \, \phi_0 (x) \,\right] \, e^{\frac{1}{2}\, \overleftarrow{\partial} \wedge P} \nonumber \\
& = & \text{Other Permutations.} 
\label{identity4}
\end{eqnarray}

From (\ref{identity3}) it is clear that inspite of $\phi_\theta$ not satisfying usual commutation relation (\ref{op}), the six possible apparently different terms in 
(\ref{csnchamiltonian}) are one and the same. Hence (\ref{csnchamiltonian}) simplifies to
 \begin{eqnarray}
 \mathcal{H}^{\theta}_{\rm{Int}} & = & \left\{\frac{\lambda_1}{4} \, + \, \frac{\lambda_2}{4} \, + \, \frac{\lambda_3}{4}  \, + \, \frac{\lambda_4}{4}  \, + \, \frac{\lambda_5}{4}  
\, + \,\frac{\lambda_6}{4}\right\} \, \phi^\dagger_{\theta} \ast \phi^\dagger_{\theta} \ast \phi_{\theta} \ast \phi_{\theta} (x)  \nonumber \\
& = &  \frac{\lambda}{4} \, \phi^\dagger_{\theta} \ast \phi^\dagger_{\theta} \ast \phi_{\theta} \ast \phi_{\theta} (x),
\label{csncham}
\end{eqnarray}
where $ \lambda = \lambda_1 \, + \, \lambda_2 \, + \,\lambda_3 \, + \,\lambda_4 \, + \, \lambda_5 \, + \, \lambda_6 $.

One can further simplify (\ref{csncham}) using the dressing transformation, so that
\begin{eqnarray}
  \mathcal{H}^{\theta}_{\rm{Int}} & = & \frac{\lambda}{4} \,\phi^\dagger_{\theta} \ast \phi^\dagger_{\theta} \ast \phi_{\theta} \ast \phi_{\theta} (x) \nonumber \\
& = & \frac{\lambda}{4} \,\int d^3 \tilde{k}_{1} \left[ b_{k_{1}} \, e^{-ik_{1}x} \, + \, a^\dagger_{k_{1}} e^{ik_{1}x} \right] 
\,  e^{\frac{i}{2}\,\overleftarrow{\partial}\wedge\overrightarrow{\partial}} \left\{ \,
\int d^3 \tilde{k}_{2} \left[ b_{k_{2}} \, e^{-ik_{2}x} \, + \, a^\dagger_{k_{2}} e^{ik_{2}x} \right] \right.\nonumber \\
& &  \left.e^{\frac{i}{2} \overleftarrow{\partial}\wedge\overrightarrow{\partial}} \left\{
 \int d^3 \tilde{k}_3 \left[ a_{k_3}  e^{-ik_3x}  +  b^\dagger_{k_3} e^{ik_3x}  \right]  e^{\frac{i}{2}\,\overleftarrow{\partial}\wedge\overrightarrow{\partial}} 
\int d^3 \tilde{k}_4 \left[ a_{k_4}  e^{-ik_4x} \right.  \right.  \right. \nonumber \\
& & + \, \left. \left. \left. b^\dagger_{k_4} e^{ik_4x}  \right] \right\} \right\}.
\label{i1}
 \end{eqnarray}
Now, let us take a generic term like
 \begin{eqnarray}
 & &  b_{k_{1}} \, e^{-ik_{1}x} \, e^{\frac{i}{2} \,\overleftarrow{\partial} \wedge \overrightarrow{\partial}}  \, \left\{  a^\dagger_{k_{2}} \, e^{ik_{2}x} 
\,  e^{\frac{i}{2}\,\overleftarrow{\partial} \wedge \overrightarrow{\partial}}  \, \left\{ b^\dagger_{k_3} \, e^{ik_3x} 
\,  e^{\frac{i}{2}\,\overleftarrow{\partial} \wedge \overrightarrow{\partial}} \,  a_{k_4} \, e^{-ik_4x}  \right\} \right\} \nonumber \\
& = &  d_{k_{1}} \, e^{-\frac{i}{2} \, k_1 \wedge P}  \, e^{-ik_{1}x} \, e^{\frac{i}{2} \,\overleftarrow{\partial} \wedge \overrightarrow{\partial}}  \,
\left\{ c^\dagger_{k_{2}} \, e^{\frac{i}{2} \, k_2 \wedge P} \, e^{ik_{2}x} \, e^{\frac{i}{2} \,\overleftarrow{\partial} \wedge \overrightarrow{\partial}}  \,
\left\{ d^\dagger_{k_{3}} \, e^{\frac{i}{2} \, k_3 \wedge P} \, e^{ik_3x} \right. \right. \nonumber \\
& & \left. \left. e^{\frac{i}{2} \,\overleftarrow{\partial} \wedge \overrightarrow{\partial}}  \,  c_{k_{4}} \, e^{-\frac{i}{2} \, k_4 \wedge P} \, e^{-ik_4x} \right\} \right\}
\nonumber \\
& = &  d_{k_{1}} \, e^{-ik_{1}x} \, e^{-\frac{i}{2} \, k_1 \wedge P} \, e^{\frac{i}{2} \, (-ik_1) \wedge (ik_2 + ik_3 -ik_4)} 
\, c^\dagger_{k_{2}} \, e^{ik_{2}x} \, e^{\frac{i}{2} \, k_2 \wedge P} \, e^{\frac{i}{2} \, ( ik_2) \wedge ( ik_3 -ik_4)} \nonumber \\
& & d^\dagger_{k_{3}} \, e^{ik_3x}\, e^{\frac{i}{2} \, k_3 \wedge P} \, e^{\frac{i}{2} \, (ik_3) \wedge (-ik_4)}\, c_{k_{4}} \, e^{-ik_4x}\, e^{-\frac{i}{2}\,k_4 \wedge P}
 \nonumber \\
& = &  d_{k_{1}} \, e^{-\frac{i}{2} \, k_1 \wedge P}  \, c^\dagger_{k_{2}} \, e^{\frac{i}{2} \, k_2 \wedge P} \, d^\dagger_{k_{3}} \, e^{\frac{i}{2} \, k_3 \wedge P} \,
c_{k_{4}} \, e^{-\frac{i}{2} \, k_4 \wedge P} \, e^{-ik_{1}x} \, e^{ik_{2}x} \, e^{ik_3x}\, e^{-ik_4x}\, \nonumber \\
& &  e^{\frac{i}{2} \, (-ik_1) \wedge (ik_2 + ik_3 -ik_4)} \, e^{\frac{i}{2} \, ( ik_2) \wedge ( ik_3 -ik_4)}\, e^{\frac{i}{2} \, ( ik_3) \wedge (-ik_4)}
\label{1genterm}
 \end{eqnarray}
To simplify it further, we need the identities 
 \begin{eqnarray}
e^{\frac{i}{2} q \wedge P} \, c_p \, e^{\frac{-i}{2} q \wedge P} & = & e^{\frac{-i}{2} q \wedge p} \, c_p \nonumber \\
e^{\frac{i}{2} q \wedge P} \, d_p \, e^{\frac{-i}{2} q \wedge P} & = & e^{\frac{-i}{2} q \wedge p} \, c_p \nonumber \\
e^{\frac{i}{2} q \wedge P} \, c^\dagger_p \, e^{\frac{-i}{2} q \wedge P} & = & e^{\frac{i}{2} q \wedge p} \, c^\dagger_p \nonumber \\
e^{\frac{i}{2} q \wedge P} \, d^\dagger_p \, e^{\frac{-i}{2} q \wedge P} & = & e^{\frac{i}{2} q \wedge p} \, d^\dagger_p.
\label{ncidentities}
 \end{eqnarray} 
Using (\ref{ncidentities}) in (\ref{1genterm}) we obtain
\begin{eqnarray}
 & & b_{k_{1}} \, e^{-ik_{1}x} \, e^{\frac{i}{2} \,\overleftarrow{\partial} \wedge \overrightarrow{\partial}}  \, \left\{  a^\dagger_{k_{2}} \, e^{ik_{2}x} 
\,  e^{\frac{i}{2}\,\overleftarrow{\partial} \wedge \overrightarrow{\partial}}  \, \left\{ b^\dagger_{k_3} \, e^{ik_3x} 
\,  e^{\frac{i}{2}\,\overleftarrow{\partial} \wedge \overrightarrow{\partial}} \,  a_{k_4} \, e^{-ik_4x}  \right\} \right\} \nonumber \\
& = & d_{k_{1}} \, c^\dagger_{k_{2}} \, d^\dagger_{k_{3}} \, c_{k_{4}} \, e^{-ik_{1}x} \, e^{ik_{2}x} \, e^{ik_3x}\, e^{-ik_4x}\, e^{\frac{1}{2} (-ik_1 + ik_2 + ik_3 - ik_4) \wedge P}
\nonumber \\
& = & d_{k_{1}} \, c^\dagger_{k_{2}} \, d^\dagger_{k_{3}} \, c_{k_{4}} \, e^{-ik_{1}x} \, e^{ik_{2}x} \, e^{ik_3x}\, e^{-ik_4x}\, e^{\frac{1}{2} \overleftarrow{\partial} \wedge P}
\label{11genterm}
\end{eqnarray}
One can check by similar computations that each and every term in (\ref{i1}) can be similarly simplified. Hence for a generic string of creation/annihilation operators we have
\begin{equation}
  (a_1)^{\#}_{k_1}\, (a_2)^\#_{k_2} \dots (a_4)^\#_{k_4} \, e^{i(\pm k_1 \pm k_2 \dots \pm k_4)x} 
 =   (c_1)^\#_{k_1}\, (c_2)^\#_{k_2} \dots (c_4)^\#_{k_4} 
\, e^{i(\pm k_1 \pm k_2 \dots \pm k_4)x} \, e^{\frac{1}{2} \overleftarrow{\partial} \wedge P}
\label{genncop}
\end{equation}
where $a^\#$ represents any of the twisted creation/annihilation operators and $c^\#$ is the analogous commutative operator.

Therefore using (\ref{11genterm}) and its generalized form (\ref{genncop}), (\ref{i1}) can be simplified to
\begin{eqnarray}
 \mathcal{H}^{\theta}_{\rm{Int}} & = & \frac{\lambda}{4} \, \phi^\dagger_{\theta} \ast \left(\phi^\dagger_{\theta} \ast \left(\phi_{\theta} \ast \phi_{\theta} \right) \right) \nonumber \\
&= & \left[\frac{\lambda}{4} \, \phi^\dagger_{0} \phi^\dagger_{0} \phi_{0} \phi_{0} \right] \, e^{\frac{1}{2} \overleftarrow{\partial} \wedge P} \nonumber \\
& = & \mathcal{H}^{0}_{\rm{Int}}\, e^{\frac{1}{2} \overleftarrow{\partial} \wedge P}
\label{simp1l}
\end{eqnarray}
where $\mathcal{H}^{0}_{\rm{Int}}  =  \frac{\lambda}{4} \, \phi^\dagger_{0} \phi^\dagger_{0} \phi_{0} \phi_{0} $ is the analogous commutative hamiltonian density.


\subsection*{The $S$-matrix}


The computation of $S$-matrix in this case is quite similar to that of real scalar fields discussed in earlier sections.

For a process of two-to-two particle scattering, the $S$-matrix elements are given by  
\begin{eqnarray}
 S_{\theta}[p_{2}, p_{1} \rightarrow  p'_{1}, p'_{2}] & \equiv &  S_{\theta}[p'_{2}, p'_{1} ; p_{2}, p_{1}]
\; = \;  \leftidx{_{\rm{out},\theta}}{\left \langle  p'_{2}, p'_{1} | p_{2}, p_{1} \right \rangle }{_{ \theta,\rm{in}}}
\label{csncsm}
\end{eqnarray}
where $ |p'_{1}, p'_{2} \rangle_{\theta,\rm{out}} $ is the noncommutative two particle out-state which is measured in the far future and $ |p_{2}, p_{1}  \rangle_{\theta,\rm{in}} $ 
is the noncommutative two particle in-state prepared in the far past.

Just like the case of real scalar fields, the noncommutative in- and out-states can be related with each other using $S$-matrix $\hat{S}_{\theta}$. Therefore we have
\begin{eqnarray}
 S_{\theta}[p'_{2}, p'_{1} ; p_{2}, p_{1}] & = &  \leftidx{_{\rm{out},\theta}}{\left \langle  p'_{2}, p'_{1} | \hat{S}_{\theta} | p_{2}, p_{1}  \right \rangle}{_{\rm{out},\theta}}
\; = \;  \leftidx{_{\rm{in},\theta}}{\left \langle  p'_{2}, p'_{1} | \hat{S}_{\theta} | p_{2}, p_{1}  \right \rangle}{_{\rm{in},\theta}}
\label{csncsmat}
\end{eqnarray}
where the noncommutative $S$-matrix $\hat{S}_{\theta}$, in interaction picture, can be written as 
\begin{eqnarray}
 \hat{S}_{\theta} & = & \lim_{t_{1} \rightarrow \infty} \lim_{t_{2} \rightarrow -\infty} U_{\theta}(t_{1},t_{2}) \nonumber \\
& = & \mathcal{T}  \exp \left[-i\int^{\infty}_{-\infty} d^{4}z \mathcal{H}^{\theta}_{\rm{Int}} (z) \right] \nonumber \\
& = & \mathcal{T}  \exp \left[ -i\int^{\infty}_{-\infty} d^{4}z \mathcal{H}^{0}_{\rm{Int}} (z)\, e^{\frac{1}{2} \overleftarrow{\partial_{z}} \wedge P }\right] 
\label{csncsint}
\end{eqnarray} 
where $\mathcal{H}^{\theta}_{\rm{Int}}$ is given by (\ref{simp1l}) and $\mathcal{H}^{0}_{\rm{Int}} (z)$ is its commutative analogue.

We can formally expand the exponential and write the $\hat{S}$ as a time-ordered power series given by
\begin{eqnarray}
 \hat{S}_{\theta} & = &   \mathbb{I} \, + \, -i\int^{\infty}_{-\infty} d^{4}z  \mathcal{H}^{0}_{\rm{Int}} (z)\, e^{\frac{1}{2} \overleftarrow{\partial_{z}} \wedge P } 
\nonumber \\
& + &  \mathcal{T}\, \frac{(-i )^2}{2!} \int^{\infty}_{-\infty} d^{4}z \int^{\infty}_{-\infty} d^{4}z'\, \mathcal{H}^{0}_{\rm{Int}} (z)\, e^{\frac{1}{2} \overleftarrow{\partial_{z}} \wedge P } 
\,  \mathcal{H}^{0}_{\rm{Int}} (z')\, e^{\frac{1}{2} \overleftarrow{\partial_{z'}} \wedge P } \, + \, \dots
\label{csspower}
\end{eqnarray}
Now let us take the second term and simplify it to 
\begin{eqnarray}
 -i\int^{\infty}_{-\infty} d^{4}z  \mathcal{H}^{0}_{\rm{Int}} (z)\, e^{\frac{1}{2} \overleftarrow{\partial_{z}} \wedge P } 
& = &   -i\int^{\infty}_{-\infty} d^{4}z  \mathcal{H}^{0}_{\rm{Int}} (z) 
\label{cssecpower}
\end{eqnarray}
where as done in \cite{bal-uvir} we have expanded the exponential, integrated and discarded all the surface terms. With computations analogous to that done in \cite{bal-uvir} one 
can similarly show that all the the higher order terms in power series of (\ref{csspower}) will be free of any $\theta$ dependence. We refer to \cite{bal-uvir} for more details. 

Hence we have
\begin{eqnarray}
 \hat{S}_{\theta} & = & \mathcal{T} \exp\left[-i\int^{\infty}_{-\infty} d^{4}z  \mathcal{H}^{0}_{\rm{Int}} (z)  e^{\frac{1}{2} \overleftarrow{\partial_{z}} \wedge P } \right]
  \; = \; \hat{S}_{0}
\label{csncsoperator}
\end{eqnarray}

Like the previously discussed real scalar field case,  here also the $\hat{S}_{\theta} $ turns out to be completely equivalent to $\hat{S}_{0}$.  
The noncommutative $S$-matrix elements have only overall noncommutative phases in them. This implies that there is no UV/IR mixing and the physical observables e.g 
scattering cross-section and decay rates are independent of $\theta$. 


\subsection{Yukawa Interactions}


Like the scalar fields, a noncommutative spinor field $\psi_\theta$ is composed of twisted fermionic creation/annihilation operators and has a normal mode expansion 
\begin{eqnarray}
 \psi_\theta (x) & = & \int d^3 \tilde{k} \sum_s \left[ a_{s,k} \, u_{s,k}\, e^{-ikx} \, + \, b^\dagger_{s,k} \, v_{s.k}\,e^{ikx} \right], \nonumber \\
\overline{\psi}_\theta (x) & = & \int d^3 \tilde{k} \sum_s \left[ b_{s,k} \, \overline{v}_{s,k}\, e^{-ikx} \, + \, a^\dagger_{s,k} \,\overline{u}_{s.k}\,e^{ikx} \right]
\label{ncspinor}
\end{eqnarray}
where  $u_{s,k}$ and $v_{s,k}$ are four component spinors (same as commutative case), $ d^3 \tilde{k} \, = \, \frac{d^3 k}{(2\pi)^3 2 E_k}$ and $a_{s,p}$, $b_{s,p}$ are twisted 
fermionic operators satisfying relations similar to (\ref{nccstcom}) but with $\eta = - 1$. 

The operators $a_{s,p}$, $b_{s,p}$ can again be related with their commutative counterparts $c_{s,p}$, $d_{s,p}$ by dressing transformations similar to (\ref{nccsfdress}). Hence, 
$\psi_\theta$ can also be related with the commutative spinor field $\psi_0$ by
\begin{eqnarray}
 \psi_{\theta} (x) & = & \psi_{0} (x) \, e^{\frac{1}{2} \, \overleftarrow{\partial} \wedge P}, \nonumber \\
 \overline{\psi}_{\theta} (x) & = & \overline{\psi}_{0} (x) \, e^{\frac{1}{2} \, \overleftarrow{\partial} \wedge P}
\label{spfdress}
\end{eqnarray}
Using $\psi_{\theta}$ and $\phi_{\theta}$ we can construct a Yukawa interaction term given by
\begin{eqnarray}
  \mathcal{H}^{\theta}_{Yuk} & = & \eta_1 \, \overline{\psi}_{\theta} \ast \phi_{\theta} \ast \psi_{\theta}
\, + \, \eta_2 \, \overline{\psi}_{\theta} \ast \psi_{\theta} \ast \phi_{\theta} 
\label{yukawa}
\end{eqnarray}
 Using identities similar to (\ref{identity1}) - (\ref{identity4}) one can show that the two terms in (\ref{yukawa}) are the same, so that
\begin{eqnarray}
  \mathcal{H}^{\theta}_{Yuk} & = & \eta \, \overline{\psi}_{\theta} \ast \phi_{\theta} \ast \psi_{\theta}
\label{yukawa1}
\end{eqnarray}
with $\eta = \eta_1 + \eta_2$. Using the dressing transformation (\ref{spfdress}) one can see that
\begin{eqnarray}
\mathcal{H}^{\theta}_{Yuk} & = & \left[ \eta \, \overline{\psi}_{0} \phi_{0} \psi_{0} \right] \, e^{\frac{1}{2} \, \overleftarrow{\partial} \wedge P} 
\; = \; \mathcal{H}^{0}_{Yuk} \, e^{\frac{1}{2} \, \overleftarrow{\partial} \wedge P}
\label{yukawa2}  
\end{eqnarray}
where $\mathcal{H}^{0}_{Yuk} = \eta \, \overline{\psi}_{0} \phi_{0} \psi_{0} $ is the commutative Yukawa interaction term.

We again find that the noncommutative interaction hamiltonian density is (analogous commutative hamiltonian density) $\times$ $e^{\frac{1}{2} \overleftarrow{\partial} \wedge P}$. 
By computations similar to that done before we have 

\begin{eqnarray}
 \hat{S}_{\theta} & = & \mathcal{T}  \exp \left[-i\int^{\infty}_{-\infty} d^{4}z \mathcal{H}^{\theta}_{Yuk} (z) \right] \nonumber \\
& = & \mathcal{T}  \exp \left[ -i\int^{\infty}_{-\infty} d^{4}z  \mathcal{H}^{0}_{Yuk} (z)\, e^{\frac{1}{2} \overleftarrow{\partial_{z}} \wedge P }\right] \nonumber \\
& = &  \hat{S}_{0}
\label{yuksop}
\end{eqnarray} 
Since $\hat{S}_{\theta} = \hat{S}_{0}$, the $S$-matrix elements for any process have only overall noncommutative phases coming due to the twisted statistics 
of the in- and out-states. 

The equivalence between $\hat{S}_{\theta}$ and $\hat{S}_{0}$ and the fact that only an overall noncommutative phase appears in $S$-matrix elements is a generic result. It holds true 
for any noncommutative field theory involving only matter fields \cite{ssb}. 


\subsection{Renormalization}


Since for any pure matter theory, the $\hat{S}_{\theta}$ always turns out to be the same as $\hat{S}_{0}$ and the noncommutative $S$-matrix elements have only overall noncommutative 
phase dependences, the noncommutative 1PI functions also have only overall noncommutative phase dependences. Apart from the divergences already present in analogous commutative 
theories, there are no new source of divergences, in any noncommutative theory involving only matter fields. So all such theories are renormalizable provided the analogous 
commutative theory is itself renormalizable. Moreover, as we saw from explicit calculations for the case of $\phi^4_{\theta,\ast}$ theory, the essential techniques of 
renormalization remains the same as the commutative ones and these noncommutative theories can always be renormalized in a way very similar to the commutative theories.

Also, as in case of $\phi^4_{\theta,\ast}$ theory, the $\theta$ dependent phases present in 1PI functions for all such theories will sit more like spectators and will not 
change the $\beta$-functions, RG flow of couplings or the fixed points, from those of the analogous commutative theory.


\section{Conclusions}


In this chapter we have presented a complete and comprehensive treatment of noncommutative theories involving only matter fields. We have shown first for real scalar fields having a 
$\phi^4_{\theta,\ast}$ interaction and then for generic theories that the noncommutative $\hat{S}_{\theta}$ is the same as $\hat{S}_{0}$ and that the $S$-matrix elements only have 
an overall phase dependence on the noncommutativity scale $\theta$. We have also argued that since there is only an overall phase dependence on the noncommutativity scale $\theta$, 
there is no UV/IR mixing in any such theory. 

We have further showed that all such theories are renormalizable if and only if the corresponding commutative theories are renormalizable. The usual commutative techniques for 
renormalization can be used to renormalize such theories. Moreover, we showed by explicit calculations for $\phi^4_{\theta,\ast}$ case and argued for generic case, that for 
all such theories the $\beta$-functions, RG flow of couplings or the fixed points are all same as those of the analogous commutative theory. 

The equivalence of $S$-matrix and also that of $\beta$-functions, RG flow and fixed points with those of the corresponding 
commutative theories does not mean that the all such noncommutative theories are one and the same as their commutative counterparts. One can always construct, even for free theories, 
appropriate observables, which unambiguously distinguish between a noncommutative and commutative theory \cite{rahul-hbt}.

The discussion of this chapter was limited only to matter fields and interaction terms constructed out of only matter fields. Noncommutative field theories involving nonabelian gauge 
fields violate twisted Poincar\'e invariance and are know to suffer from UV/IR mixing \cite{amilcar-pinzul}. They require special treatment which is outside the scope of present 
work.


\chapter{Twisted Internal Symmetries}
\label{chap:chap3}
In previous chapters we saw that deformation of spacetime symmetries leads to twisting of statistics for quantum fields written on such spacetime. Motivated by this, 
in this chapter we discuss the possibility of having twisted statistics by deforming global internal symmetries. Following up the work of \cite{queiroz} on deformed algebras, 
we present a class of Poincaré invariant quantum field theories with twisted bosonic/fermionic particles having
deformed internal symmetries. The twisted quantum fields discussed in this work, satisfy commutation relations different from the usual bosonic/fermionic 
commutation relations. Such twisted fields by construction (and in view of CPT theorem) are nonlocal in nature. We show that inspite of the basic ingredient fields being nonlocal,  
it is possible to construct local interaction Hamiltonians which satisfy cluster decomposition principle and are Poincaré invariant. Although the formalism developed here can be
adapted to the discussion of a generic global internal symmetry group but for sake of concreteness we restrict ourself only to the discussion of global $SU(N)$ symmetries. 
As a specific example of interesting application of these ideas we show that twisted internal symmetries can significantly simplify the discussion of the marginal deformations
(also known as $\beta$-deformations) of the scalar sector of $N=4$ supersymmetric (SUSY) theories.

The plan of this chapter is as follows. We begin with briefly reviewing the treatment of global symmetries, in particular $SU(N)$ group, in the usual untwisted case. 
We then discuss a specific type of twist called ``antisymmetric twist''. This kind of twist is quite similar in spirit to the twisted noncommutative field theories. 
The formalism developed here will closely resemble (with generalizations and modifications which we will elaborate on) to the formalism of twisted noncommutative theories
\cite{bal-pinzul}. We then go on to construct more general twisted statistics which can be viewed as internal symmetry analogue of dipole theories \cite{dipole-keshav}.
 We also discuss the construction of interaction terms and scattering formalism
for both types of twists. We end the chapter with discussion of causality of such twisted field theories.

This chapter is based on the work published in \cite{insym}.


 \section{Brief Review of Global Symmetries In Untwisted Case}


In this section we briefly review the standard treatment of global symmetries in the usual untwisted case \cite{weinberg, greiner, greiner-sym, ait}. In view of the computations
 and generalizations done in later parts, 
we take the route of Hamiltonian formalism, instead of the more convenient Lagrangian formalism, to study the global symmetries. For sake of simplicity, we mostly 
restrict our discussion to the case of matter fields which transform as scalars under Poincaré group and transform as a fundamental representation of a given 
global $SU(N)$ symmetry group. The treatment here can be easily generalized to spinor fields, as well as, higher representations of $SU(N)$ group. 

Let $\phi_r (x)$, $r = 1, 2, \dots N $ be a set of complex scalar (under Lorentz transformation) quantum fields having mode expansion given by\footnote{ Throughout this work we 
will denote the usual bosonic/fermionic annihilation operators by the labels $c_r$ and $d_r$ whereas the twisted operators will be denoted by $a_r$ and $b_r$. The same notation will 
be followed for usual and twisted creation operators.}
\begin{eqnarray}
\phi_r (x) & = & \int \frac{d^3 p}{(2\pi)^3} \frac{1}{2E_{p}} \left[ c_r(p) e^{-i px} \, + \, d^{\dagger}_r(p) e^{i px} \right ], \nonumber \\
\phi^\dagger_r (x) & = & \int \frac{d^3 p}{(2\pi)^3} \frac{1}{2E_{p}} \left[ d_r(p) e^{-i px} \, + \, c^{\dagger}_r(p) e^{i px} \right ].
\label{modeexpansioncom}
\end{eqnarray}
The creation/annihilation operators satisfy the commutation relations 
\begin{eqnarray}
  c^\#_r(p_{1}) c^\#_s(p_{2}) & = & \eta \, c^\#_s(p_{2}) c^\#_r(p_{1}), \nonumber \\
(c^{\dagger}_r)^\# (p_{1}) (c^{\dagger}_s)^\# (p_{2}) & = & \eta \,(c^{\dagger}_s)^\# (p_{2})(c^{\dagger}_r)^\# (p_{1}), \nonumber \\
c_r (p_{1})  c^{\dagger}_s(p_{2}) & = & \eta \,  c^{\dagger}_s (p_{2}) c_r (p_{1}) + (2\pi)^3 \, 2 E_p \, \delta_{rs} \, \delta^{3}(p_{1} \, - \, p_{2}), \nonumber \\
d_r (p_{1}) d^{\dagger}_s (p_{2}) & = & \eta \,  d^{\dagger}_s (p_{2}) d_r (p_{1}) + (2\pi)^3 \, 2 E_p \, \delta_{rs} \, \delta^{3}(p_{1} \, - \, p_{2}), \nonumber \\
c_r (p_{1}) d^{\dagger}_s (p_{2}) & = & \eta \, d^{\dagger}_s (p_{2}) c_r (p_{1}), 
\label{cstcom}
\end{eqnarray}
where $c^\#_r ({p})$ stands for either of the operators $c_r (p), d_r (p)$ and $(c^{\dagger}_r)^\# (p) $ stands for either of the operator $c^\dagger_r (p), d^\dagger_r (p)$. 
Also, as these 
are bosonic operators, so $\eta = 1$ should be taken in (\ref{cstcom}). For the case of fermionic operators $\eta = -1$ should be taken. Since we are mostly concerned with 
internal symmetries, so we will usually suppress the momentum dependence of the operators and will not write them explicitly unless we need them.

Let $U(\sigma) = \exp(i \sigma_a \Lambda_a)$, $a = 1, 2, \dots (N^2 - 1) $ and $\sigma$ being $(N^2 - 1) $ arbitrary parameters (independent of spacetime coordinates), be the unitary 
representation of the group element ``$\sigma$'' of the $SU(N)$ group on Fock space. Then, if the fields $\phi_r (x)$ transform as fundamental representation of the $SU(N)$, we have
\begin{eqnarray}
 U(\sigma) \phi_r (x) U^\dagger(\sigma) & = &  \phi'_r (x) \, = \, \left( e^{-i \sigma_a T_a} \right)_{rs} \,\phi_s (x), \nonumber \\
  U(\sigma) \phi^\dagger_r (x) U^\dagger(\sigma) & = &  \phi'^\dagger_r (x) \, = \, \left( e^{i \sigma_a T_a} \right)_{sr} \,\phi^\dagger_s (x), 
\label{suntranformcom}
\end{eqnarray}
where $r,s = 1, \dots N$ and $ a = 1, \dots (N^2 - 1)$. Also, $T_a$ are $N \times N $  hermitian matrices and furnish the fundamental representation of the generators of the 
group satisfying the Lie algebra
\begin{eqnarray}
 [T_a, T_b] & = & i f_{abc}\, T_c
\label{sunlie}
\end{eqnarray}
where $f_{abc}$ are the structure constants. 

The infinitesimal version of (\ref{suntranformcom}) can be written as 
\begin{eqnarray}
 U(\epsilon) \phi_r (x) U^\dagger(\epsilon) & = &  \phi'_r (x) \, = \, \phi_r(x) - i \epsilon_{a} (T_{a})_{rs} \phi_s(x), \nonumber \\
U(\epsilon) \phi^\dagger_r (x) U^\dagger(\epsilon) & = &  \phi'^\dagger_r (x) \, = \, \phi^\dagger_{r}(x) + i \epsilon_{a} (T_{a})_{sr} \phi^\dagger_s(x).
\label{infitsuntrans}
\end{eqnarray}
The $\Lambda_a$ furnish a Fock space representation of the generators of the group, having the form
\begin{eqnarray}
 \Lambda_a & = &  \int \frac{d^3 p}{(2\pi)^3} \frac{1}{2E_{p}} \left[ (T_a)_{rs} c^\dagger_r c_s \, - \, (T_a)^\ast_{rs}  d^\dagger_r d_s \right],
\label{fockgenratorssun}
\end{eqnarray}
where $(-T^\ast_a) = (-T^T_a)$ are $N \times N $ matrices  \footnote{$T^T_a$ stands for transpose of $T_a$. Also this relation holds because of hermiticity of the generators.}
and furnish the anti-fundamental representation of the generators of the group satisfying the Lie algebra
\begin{eqnarray}
 [(-T^\ast_a), (-T^\ast_b)] & = & i f_{abc}\, (-T^\ast_c).
\label{antisunlie}
\end{eqnarray}
The operators $\Lambda_a$ also satisfy the same Lie algebra
\begin{eqnarray}
 [\Lambda_a, \Lambda_b] & = & i f_{abc}\, \Lambda_c.
\label{focksunlie}
\end{eqnarray}  
Using (\ref{fockgenratorssun}) one can immediately check the correctness of (\ref{suntranformcom}) and can deduce the transformation properties of the 
creation/annihilation operators which are given by
\begin{eqnarray}
 U(\sigma) c_r U^\dagger(\sigma) & = &  c'_r \, = \, \left( e^{-i \sigma_a T_a} \right)_{rs} \,c_s, \nonumber \\
 U(\sigma) c^\dagger_r U^\dagger(\sigma) & = &  c'^\dagger_r \, = \, \left( e^{i \sigma_a T_a} \right)_{sr} \,c^\dagger_s, \nonumber \\
U(\sigma) d_r U^\dagger(\sigma) & = &  d'_r \, = \, \left( e^{i \sigma_a T^\ast_a} \right)_{rs} \,d_s, \nonumber \\
 U(\sigma) d^\dagger_r U^\dagger(\sigma) & = &  d'^\dagger_r \, = \, \left( e^{-i \sigma_a T^\ast_a} \right)_{sr} \,d^\dagger_s.
\label{creanisuntranformcom}
\end{eqnarray}
Using (\ref{creanisuntranformcom}) and assuming that vacuum remains invariant under the transformations i.e. $ U(\sigma) | 0 \rangle = | 0 \rangle$, we can  
deduce the transformation property of state vectors, which for single-particle states, is given by
\begin{eqnarray}
 U(\sigma) | r \rangle & = &  U(\sigma) c^\dagger_r | 0 \rangle \, = \, U(\sigma) c^\dagger_r U^\dagger(\sigma) U(\sigma) | 0 \rangle 
\, = \, \left( e^{i \sigma_a T_a} \right)_{sr} \,c^\dagger_s | 0 \rangle \, = \, \left( e^{i \sigma_a T_a} \right)_{sr}  | s \rangle, \nonumber \\
 U(\sigma) \overline{| r \rangle} & = &  U(\sigma) d^\dagger_r | 0 \rangle \, = \, U(\sigma) d^\dagger_r U^\dagger(\sigma) U(\sigma) | 0 \rangle 
\, = \, \left( e^{-i \sigma_a T^\ast_a} \right)_{sr} \,d^\dagger_s | 0 \rangle \, = \, \left( e^{-i \sigma_a T^\ast_a} \right)_{sr}  \overline{| s \rangle}.\nonumber \\
\label{statevecsuntransform}
\end{eqnarray}
Similar transformation properties hold for multi-particle states. 

Having discussed the transformation properties of quantum fields and state vectors under the $SU(N)$ group, let us now consider a Hamiltonian density $\mathcal{H} (x)$ constructed 
out of the fields $\phi_r (x)$ and their canonical conjugates $\Pi_r (x)$. The operator $U(\sigma) = \exp(i \sigma_a \Lambda_a)$ transforms it as
\begin{eqnarray}
 U(\sigma) \mathcal{H} (x) U^\dagger(\sigma) & = & \mathcal{H}' (x).
\label{Hamiltoniansun}
\end{eqnarray}
The transformation of (\ref{Hamiltoniansun}) is said to be a symmetry transformation and the system is said to be having a $SU(N)$ global symmetry if the Hamiltonian density remains 
invariant under such a transformation\footnote{Since these are global transformations so it is enough for $\mathcal{H} (x)$ to be invariant, which will automatically imply 
that the Hamiltonian $H$ itself remains invariant.}. Therefore we have 
\begin{eqnarray}
 U(\sigma) \mathcal{H} (x) U^\dagger(\sigma) & = & \mathcal{H}' (x) \, = \, \mathcal{H} (x). 
\label{hamsun}
\end{eqnarray}
The above condition in turn implies that 
\begin{eqnarray}
 [H, \Lambda_a] & = & 0.
\label{symmofh}
\end{eqnarray}
From (\ref{symmofh}) we can infer that all $\Lambda_a$ are constants of motion and hence conserved quantities, called `` charge operators'' and their eigenvalues are termed as
 ``internal charges'' of the given eigenstate. Since $SU(N)$ is a nonabelian group satisfying the Lie algebra (\ref{focksunlie}), a state cannot simultaneously be 
an eigenstate of all the charge operators and hence only a subset of the charges can be simultaneously measured.  The maximal commuting subset of the charge operators is called 
 ``Cartans'' of the group and usually the eigenstates of the Cartans are 
taken as the basis states. For a $SU(N)$ group there are $N-1$ Cartans and we will denote them by $Q_m$; $1 \leq m \leq N-1 $. 

The condition (\ref{hamsun}) puts stringent constraints on the type of Hamiltonian densities allowed by the symmetry. For example, a free theory Hamiltonian density 
will satisfy (\ref{hamsun}) if and only if the masses $m_r$ of all particles are same i.e. $m_r = m_s = \cdots  = m$ and is given by
 \begin{eqnarray}
  \mathcal{H}_0 & = &  \Pi^{\dagger}_r \Pi_r  \, + \, (\partial_i \phi^{\dagger}_r) (\partial^i \phi_r) \, + \, m^2  \, \phi^{\dagger}_r \phi_r.
\label{freeham}
 \end{eqnarray}
where $\Pi_{r}$ is the canonical conjugate momentum.
The only renormalizable interaction Hamiltonian density compatible with (\ref{hamsun}) is given by
\begin{eqnarray}
 \mathcal{H}_{\rm{int}} & = &  \frac{\gamma}{4} \, \phi^{\dagger}_r \phi^{\dagger}_s \phi_r \phi_s,
\label{gaugen4susy}
\end{eqnarray}
where $r,s = 1, 2, \cdots, N$.


\subsection*{Weight Basis}


The discussion till now holds for generators written in any basis. For purpose of our work, it is convenient to write them in ``weight basis''. From now onwards we will write the 
generators in weight basis only. The obvious advantage of working in weight basis being that the Cartans are all diagonal matrices and easy to deal with.  
In this basis, we denote Cartans by $Q_m$; $1 \leq m \leq N-1 $. The other generators, denoted by $E_n$; $1 \leq n \leq N(N-1) $, are the so called
``raising/lowering'' generators. Their Fock space representation is given by  
\begin{eqnarray}
 Q_m & = &  \int \frac{d^3 p}{(2\pi)^3} \frac{1}{2E_{p}} \left[ (q_m)_{rs} c^\dagger_r c_s \, - \, (q_m)^\ast_{rs}  d^\dagger_r d_s \right], \nonumber \\
 E_n & = &  \int \frac{d^3 p}{(2\pi)^3} \frac{1}{2E_{p}} \left[ (e_n)_{rs} c^\dagger_r c_s \, - \, (e_n)^\ast_{rs}  d^\dagger_r d_s \right],
\label{fockcargensun}
\end{eqnarray}
where  creation/annihilation operators in (\ref{fockcargensun}) are labeled using weights. Also, $q_m$ and $e_n$ are $N \times N $ matrices and they together satisfy the lie 
algebra (\ref{sunlie}) of the group, furnishing the fundamental representation of the generators. 
The $N \times N $ matrices $(-q^\ast_m)$ and $(-e^\ast_n)$ furnish the anti-fundamental representation of the generators. Moreover since $q_m$ and $q^\ast_m$ are Cartans of the group 
and are written in the weight basis so
\begin{eqnarray}
[q_m , q_{m'}] & = & 0,
\label{comCartan}
\end{eqnarray}
and
\begin{eqnarray}
q_m  \, = \,q^\ast_m & = &  
  \begin{pmatrix}
  \lambda_m^1 & 0 & \cdots & 0 \\
  0 & \lambda_m^2 & \cdots & 0 \\
  \vdots & \vdots   & \ddots & \vdots \\
  0 & 0 & \cdots & \lambda_m^{N}
 \end{pmatrix}.
\label{digCartan}
\end{eqnarray}
In the weight basis we have 
\begin{eqnarray}
 \left[Q_m , c^{\dagger}_u \right] & = &  \lambda_{m}^{(u)} c^{\dagger}_u,
\label{weightcre}
\end{eqnarray}
where $\lambda_{m}^{(u)}$ is the mth component of the weight vector corresponding to the Cartan $Q_m$. 

Similarly we have
\begin{eqnarray}
 \left[Q_m , c_u \right] & = & - \lambda_{m}^{(u)} c_u, \nonumber \\
\left[Q_m , d_u \right] & = &   \lambda_{m}^{(u)} d_u, \nonumber \\
\left[Q_m , d^{\dagger}_u \right] & = &  - \lambda_{m}^{(u)} d^\dagger_u.
\label{weightani}
\end{eqnarray}
Using (\ref{weightani}) and (\ref{weightcre}) we have 
\begin{eqnarray}
 \left[Q_m , \phi_u \right] & = & - \lambda_{m}^{(u)} \phi_u, \nonumber \\
\left[Q_m , \phi^{\dagger}_u \right] & = &   \lambda_{m}^{(u)} \phi^\dagger_u.
\label{chargefields}
\end{eqnarray}
The above was a very brief review of the standard discussion of global symmetries in quantum field theories. Apart from completeness of the work, the main purpose of this section 
was to setup the notations and conventions that we use in the rest of the chapter. Keeping that in mind we restricted ourself to the discussion of only $SU(N)$ group symmetries
and to only scalar fields transforming as a fundamental representation. Other type of groups like $SO(N)$ can be discussed in a similar way. Also, within $SU(N)$
group symmetries, generalization to higher representations and discussion of transformation properties of spinor (under Lorentz transformation) fields 
can be done in a similar and straightforward way.


\section{Antisymmetric Twists}


Our main interest is to write field theories where the particles satisfy twisted commutation relations, which in general can be 
\begin{eqnarray}
a_r ({p_{1}}) a_s ({p_{2}}) & = & \zeta_1  \,a_s ({p_{2}}) a_r ({p_{1}}), \nonumber \\
a^{\dagger}_r ({p_{1}}) a^{\dagger}_s ({p_{2}}) & = & \zeta_2 \, a^{\dagger}_s ({p_{2}}) a^{\dagger}_r ({p_{1}}), \nonumber \\
\vdots \nonumber \\
a_r ({p_{1}})  a^{\dagger}_s ({p_{2}})  & = & \zeta_{n-2} \,  a^{\dagger}_s ({p_{2}}) a_r ({p_{1}}) + (2\pi)^3 \, 2 E_p \, \delta_{rs} \, \delta^{3}(p_{1} \, - \, p_{2}), \nonumber \\
b_r ({p_{1}}) b^{\dagger}_s ({p_{2}}) & = & \zeta_{n-1} \, b^{\dagger}_s ({p_{2}}) b_r ({p_{1}}) + (2\pi)^3 \, 2 E_p \, \delta_{rs} \, \delta^{3}(p_{1} \, - \, p_{2}), \nonumber \\
a_r ({p_{1}}) b^{\dagger}_s ({p_{2}}) & = & \zeta_n \, b^{\dagger}_s ({p_{2}}) a_r ({p_{1}}),  
\label{gtwistedcomrel}
\end{eqnarray}
where we are denoting the twisted creation and annihilation operators for particles and anti-particles by $a_r, b_r$ and $a^\dagger_r, b^\dagger_r$ respectively. 
Also, $\zeta_i$; $i = 1,2, \cdots n$ are some arbitrary c-numbers. Moreover, all $\zeta_i$ are not necessarily independent of each other.

In this section, we restrict ourself only to the discussion of  a specific type of twist which we call ``antisymmetric twist''. This kind of twist is quite similar in spirit to the 
twisted statistics of noncommutative field theories e.g Groenewold-Moyal (GM) plane. The formalism developed here will closely resemble (with appropriate generalizations 
and modifications) the formalism of twisted noncommutative theories \cite{bal-pinzul}. Also, the formalism developed here is true for any $SU(N)$ group with N $\geq$ 3. 
The antisymmetric twisted commutation relations for $a_r, b_r$ and $a^\dagger_r, b^\dagger_r$ are
\begin{eqnarray}
 a^\# _r ({p_{1}}) a^\#_s ({p_{2}}) & = & \eta \, e^{ i\tilde{\lambda}^{(r)} \wedge \tilde{\lambda}^{(s)}}  \, a^\#_s ({p_{2}}) a^\#_r ({p_{1}}), \nonumber \\
(a^{\dagger}_r)^\# ({p_{1}}) (a^{\dagger}_s)^\# ({p_{2}}) & = & \eta \,e^{ i\tilde{\lambda}^{(r)} \wedge \tilde{\lambda}^{(s)}} \, (a^{\dagger}_s)^\# ({p_{2}}) 
(a^{\dagger}_r)^\# ({p_{1}}), \nonumber \\
a_r ({p_{1}})  a^{\dagger}_s ({p_{2}})  & = & \eta \,e^{ - i\tilde{\lambda}^{(r)} \wedge \tilde{\lambda}^{(s)}}  \, a^{\dagger}_s ({p_{2}}) a_r ({p_{1}})
 + (2\pi)^3 \, 2 E_p \, \delta_{rs} \, \delta^{3}(p_{1} \, - \, p_{2}), \nonumber \\
b_r ({p_{1}}) b^{\dagger}_s ({p_{2}}) & = & \eta \,e^{ -i\tilde{\lambda}^{(r)} \wedge \tilde{\lambda}^{(s)}} \, b^{\dagger}_s ({p_{2}}) b_r ({p_{1}}) 
+ (2\pi)^3 \, 2 E_p \, \delta_{rs} \, \delta^{3}(p_{1} \, - \, p_{2}), \nonumber \\
a_r ({p_{1}}) b^{\dagger}_s ({p_{2}}) & = & \eta \, e^{ i\tilde{\lambda}^{(r)} \wedge \tilde{\lambda}^{(s)}} \, b^{\dagger}_s ({p_{2}}) a_r ({p_{1}}),  
\label{atwistedcomrel}
\end{eqnarray}
where  $\tilde{\lambda}^{(r)} \wedge \tilde{\lambda}^{(s)} = \tilde{\lambda}^{(r)}_{l} \tilde{\theta}_{lm} \tilde{\lambda}^{(s)}_{m} $; $l,m = 1, 2, \cdots, (N-1)$.
Right now $\tilde{\lambda}^{(r)}_{l}, \tilde{\lambda}^{(s)}_{m} $ are some arbitrary parameters whose meaning will be clarified soon. 
Also, $\tilde{\theta}_{lm} = - \tilde{\theta}_{ml}$  is a arbitrary 
real antisymmetric matrix. Moreover, $a^\#_r $ and $(a^{\dagger}_r)^\# $ stands for either of the operators $a_r, b_r$ and $a^\dagger_r, b^\dagger_r$ respectively. 
Again, for ``twisted bosons'' $\eta = 1$ and for ``twisted fermions'' $\eta = -1$ should be taken in (\ref{atwistedcomrel}). 

Using the above creation/annihilation operators we can write down complex scalar (under Lorentz transformations) quantum fields. Let $\phi_{\tilde{\theta}, r}(x)$, 
$r = 1, 2, \dots N $ be such a set of complex scalar quantum fields having mode expansion given by
\begin{eqnarray}
  \phi_{\tilde{\theta}, r} (x) & = & \int \frac{d^3 p}{(2\pi)^3} \frac{1}{2E_{p}} \left[ a_r(p) e^{-i px} \, + \, b^{\dagger}_r(p) e^{i px} \right ], \nonumber \\
\phi^\dagger_{\tilde{\theta}, r} (x) & = & \int \frac{d^3 p}{(2\pi)^3} \frac{1}{2E_{p}} \left[ b_r(p) e^{-i px} \, + \, a^{\dagger}_r(p) e^{i px} \right ].
\label{modeexpansionatwist}
\end{eqnarray}
The Fock space states can be similarly constructed using these twisted operators. To start with, we assume that the vacuum of the twisted theory is same as that of the untwisted theory. 
The reason for the above assumption will be clarified soon. The multi-particle states can be obtained by acting the twisted creation operators on the vacuum state. 
Because of the twisted statistics (\ref{atwistedcomrel}), there is an ambiguity in defining the action of the twisted creation and annihilation operators on Fock space states.
We choose to define $a^\dagger_r (p)$, $p$ being the momentum label, to be an operator which adds a particle to the right of the particle list i.e.
\begin{eqnarray}
a^\dagger_r (p) | p_1,r_1; \, p_2,r_2;\, \dots \, p_n,r_n \rangle_{\tilde{\theta}} & = & | p_1,r_1;\, p_2,r_2;\, \dots \, p_n,r_n;\, p,r \rangle_{\tilde{\theta}}.
\label{actiontanticre}
\end{eqnarray}
With this convention, the single-particle Fock space states for this twisted theory are given by 
\begin{eqnarray}
\overline{|p, r \rangle}_{\tilde{\theta}} & = &   b^\dagger_r (p) | 0 \rangle,  \nonumber \\
|p, r \rangle_{\tilde{\theta}} & = &  a^\dagger_r (p) | 0 \rangle.
\label{sat}
\end{eqnarray}
The multi-particle states are given by
\begin{eqnarray}
\overline{| p_1, r_1; \, p_2,r_2; \, \dots \, p_n,r_n \rangle}_{\tilde{\theta}} & = &  b^\dagger_{r_n} (p_n) \dots b^\dagger_{r_2} (p_2) b^\dagger_{r_1} (p_1) | 0 \rangle, \nonumber \\
|p_1, r_1; \, p_2,r_2; \, \dots \, p_n,r_n \rangle_{\tilde{\theta}} & = &  a^\dagger_{r_n} (p_n) \dots a^\dagger_{r_2} (p_2) a^\dagger_{r_1} (p_1) | 0 \rangle .
\label{mat}
\end{eqnarray}
Owing to the twisted commutation relations of (\ref{atwistedcomrel}), the state vectors also satisfy a similar twisted relation e.g. for two-particle states we have 
\begin{eqnarray}
 \overline{| p_2, r_2; \, p_1,r_1 \rangle}_{\tilde{\theta}} & = & e^{ i\tilde{\lambda}^{(r_1)} \wedge \tilde{\lambda}^{(r_2)}} 
\, \overline{| p_1, r_1; \, p_2,r_2 \rangle}_{\tilde{\theta}}, \nonumber \\
 |p_2, r_2; \, p_1,r_1 \rangle_{\tilde{\theta}} & = &  e^{ i\tilde{\lambda}^{(r_1)} \wedge \tilde{\lambda}^{(r_2)}} \, |p_1, r_1; \, p_2,r_2 \rangle_{\tilde{\theta}}.  
\label{2statecom}
\end{eqnarray}


\subsection*{Dressing Transformations}


Before going further and discussing the transformation properties of these twisted fields under $SU(N)$ group and construction of various Hamiltonian densities, we would 
like to discuss a very convenient map between the twisted creation/annihilation operators and their untwisted counterparts.  Such a map not only 
enables us to do various cumbersome manipulations on twisted operators in a convenient way but will also enable us to compare and contrast the twisted theories with their untwisted
counterparts. 

We start with noting the fact that, if we define certain composite operators as
\begin{eqnarray}
 \tilde{a}_r & = & c_r \, e^{-\frac{i}{2} \lambda^{(r)} \wedge Q}, \nonumber \\
\tilde{a}^\dagger_r & = &  e^{\frac{i}{2} \lambda^{(r)} \wedge Q} \, c^\dagger_r, \nonumber \\
 \tilde{b}_r & = & d_r \, e^{\frac{i}{2} \lambda^{(r)} \wedge Q}, \nonumber \\
\tilde{b}^\dagger_r & = &  e^{-\frac{i}{2} \lambda^{(r)} \wedge Q} \, d^\dagger_r, 
\label{dresstranformagen}
\end{eqnarray}
where $Q_m$; $m = 1, 2, \cdots, (N-1)$ are the Cartans of the $SU(N)$ group, given by (\ref{fockcargensun}) and the $\lambda^{(r)}_m$ are defined by 
(\ref{weightcre}), (\ref{weightani}). 
Also, $\lambda^{(r)} \wedge Q = \lambda^{(r)}_l \theta_{lm} Q_m $; $l, m = 1, 2, \cdots (N - 1)$, $\theta_{lm} = - \theta_{ml}$ being an arbitrary real anti-symmetric matrix.

One can check that operators in (\ref{dresstranformagen}) satisfy the same twisted commutation relations as (\ref{atwistedcomrel}) if we identify 
$\lambda^{(r)}_l \theta_{lm} \lambda^{(s)}_m = \tilde{\lambda}^{(r)}_{l} \tilde{\theta}_{lm} \tilde{\lambda}^{(s)}_{m}$. But as both $\tilde{\theta}_{lm}$ and $\theta_{lm}$ are
arbitrary matrices, the above demand is always satisfied. 

Hence we have a map between creation/annihilation operators of the twisted theory with those of the untwisted theory, which we call as ``dressing transformations'' and is given by
\begin{eqnarray}
 a_r \, \, = &   c_r \, e^{-\frac{i}{2} \lambda^{(r)} \wedge Q}   & = \,  \, e^{-\frac{i}{2} \lambda^{(r)} \wedge Q} \, c_r,  \nonumber \\
a^\dagger_r \, \, = &   c^\dagger_r \, e^{\frac{i}{2} \lambda^{(r)} \wedge Q}  & = \, \, e^{\frac{i}{2} \lambda^{(r)} \wedge Q} \, c^\dagger_r ,  \nonumber \\
 b_r  \,  \,= &  d_r \, e^{\frac{i}{2} \lambda^{(r)} \wedge Q}  & =  \, \, e^{\frac{i}{2} \lambda^{(r)} \wedge Q}  \, d_r, \nonumber \\
b^\dagger_r  \, \, = & d^\dagger_r  \,  e^{-\frac{i}{2} \lambda^{(r)} \wedge Q}  & = \, \, e^{-\frac{i}{2} \lambda^{(r)} \wedge Q} \, d^\dagger_r.
\label{dresstranformanti}
\end{eqnarray}
This dressing map extends to all operators and state vectors in the two theories and provides us with a convenient way to discuss the twisted field theories.  For twisted fields of 
(\ref{modeexpansionatwist}), we have 
\begin{eqnarray}
 \phi_{\theta, r} (x) \, \, = &  \phi_{0, r} (x) \, e^{-\frac{i}{2} \lambda^{(r)} \wedge Q} & = \, \, e^{-\frac{i}{2} \lambda^{(r)} \wedge Q} \,  \phi_{0, r} (x),\nonumber \\
 \phi^\dagger_{\theta, r} (x) \, \, = &  \phi^\dagger_{0, r} (x) \, e^{\frac{i}{2} \lambda^{(r)} \wedge Q} & = \, \, e^{\frac{i}{2} \lambda^{(r)} \wedge Q} \, \phi^\dagger_{0, r} (x),
\label{fieldsantidress}
\end{eqnarray}
where $\phi_{0, r} (x)$, $\phi^\dagger_{0, r} (x)$ are the untwisted fields given by (\ref{modeexpansioncom}) and we have put a subscript ``0'' to distinguish them from twisted fields.
Also we note that $\lambda^{(r)} \wedge \lambda^{(r)} = 0$, owing to the antisymmetry of $\theta$. Hence one can freely move the exponential terms in (\ref{dresstranformanti}) and
(\ref{fieldsantidress}) from left to right and vice versa.  The antisymmetry of the $\theta$ matrix also means that it is not possible to get twisted statistics for any 
internal symmetry group which is of rank less than 2. Thus $SU(3)$ is the smallest $SU(N)$ group for which we can have a twisted statistics of the above type.

The twisted fields satisfy the commutation relations
\begin{eqnarray}
 \phi_{\theta,r}(x) \phi_{\theta,s}(x) & = & e^{ i \lambda^{(r)} \wedge \lambda^{(s)} } \, \phi_{\theta,s}(x) \phi_{\theta,r}(x), \nonumber \\
\phi^{\dagger}_{\theta,r}(x) \phi^{\dagger}_{\theta,s}(x) & = & e^{ i \lambda^{(r)} \wedge \lambda^{(s)} } \, \phi^{\dagger}_{\theta,s}(x) \phi^{\dagger}_{\theta,r}(x),
\label{antitwistfield}
\end{eqnarray}
which can be easily checked by using (\ref{atwistedcomrel}) or alternatively by using (\ref{fieldsantidress}). 

We also note that, the number operator $N$ remains unchanged i.e.
\begin{eqnarray}
N_\theta & = &  \int \frac{d^3 p}{(2\pi)^3} \frac{1}{2E_{p}} \left[ a^\dagger_r a_r \, + \, b^\dagger_r b_r \right] \nonumber \\
& = &  \int \frac{d^3 p}{(2\pi)^3} \frac{1}{2E_{p}} \left[ c^\dagger_r c_r \, + \, d^\dagger_r d_r \right] \, = \, N_0.
\end{eqnarray}
Also we have 
\begin{eqnarray}
 \left[ Q_m , \phi_{\theta, r} (x) \right ] & = & -\lambda^{(r)}_m \phi_{\theta, r} (x),  \nonumber \\
 \left[ Q_m , \phi^{\dagger}_{\theta, r} (x) \right ] & = &  \lambda^{(r)}_m \phi^{\dagger}_{\theta, r} (x),
\label{relationcomso6}
\end{eqnarray}
which is same as the relation (\ref{chargefields}) satisfied by the untwisted fields. 

Also, if $| 0 \rangle_0$ and $| 0 \rangle_\theta$  are the vacua of untwisted and twisted theories then
\begin{eqnarray}
c_r  | 0 \rangle_0 & = &  d_r  | 0 \rangle_0 \, = \, 0,  \nonumber \\
a_r  | 0 \rangle_\theta & = &  b_r  | 0 \rangle_\theta \, = \, 0 .
\label{vaccuas}
\end{eqnarray}
But because of the dressing transformations (\ref{dresstranformanti}) we have
\begin{eqnarray}
a_r  | 0 \rangle_0 & = &  c_r \, e^{-\frac{i}{2} \lambda^{(r)} \wedge Q} | 0 \rangle_0 ,\nonumber \\
b_r | 0 \rangle_0 & = &  d_r \, e^{\frac{i}{2} \lambda^{(r)} \wedge Q} | 0 \rangle_0.   
\label{vaccuasdress}
\end{eqnarray}
If the untwisted vacuum is invariant under the $SU(N)$ group transformations i.e. if $Q_m | 0 \rangle_0 = E_n | 0 \rangle_0 = 0$ then
\begin{eqnarray}
a_r  | 0 \rangle_0 & = &  c_r  | 0 \rangle_0 \, = \, 0, \nonumber \\
b_r | 0 \rangle_0 & = &  d_r | 0 \rangle_0   \, = \, 0.
\label{vaccuasinv}
\end{eqnarray}
Hence, the vacuum of the two theories is one and same. Similarly we find that, provided the untwisted vacuum is invariant under the $SU(N)$ group transformations, 
the single-particle states in the two theories are also same i.e.
\begin{eqnarray}
\overline{| r \rangle}_\theta & = &  \overline{|r \rangle}_0 , \nonumber \\
| r \rangle_\theta & = &  |r \rangle_0 .  
\label{singlrpart}
\end{eqnarray}
In this work we will restrict ourself only to the case of the untwisted vacuum being invariant under the $SU(N)$ group transformations. 
The case of it not being invariant under $SU(N)$ transformations is also a exciting scenario, as it will lift the degeneracy of the vacua of the two theories and 
we expect it to result into new features in the twisted theory. We plan to discuss 
it in more details in a separate work. 

Now we can discuss the transformation property of the fields $\phi_{\theta, r}$ under $SU(N)$, which is given by
\begin{eqnarray}
 U(\sigma) \phi_{\theta, r} (x) U^\dagger(\sigma) & = &  \phi'_{\theta, r} (x) \, = \,\, U(\sigma) e^{-\frac{i}{2} \lambda^{(r)} \wedge Q} U^\dagger(\sigma) \,
 U(\sigma) \phi_{0, r} (x) U^\dagger(\sigma) \nonumber \\ 
& = & U(\sigma) e^{-\frac{i}{2} \lambda^{(r)} \wedge Q} U^\dagger(\sigma) \, \left( e^{-i \sigma_a T_a} \right)_{rs} \,\phi_s (x) \, 
\, = \, \xi_{(r)}(\sigma) \, \left( e^{-i \sigma_a T_a} \right)_{rs} \,\phi_s (x),  \nonumber \\
 U(\sigma) \phi^\dagger_{\theta, r} (x) U^\dagger(\sigma) & = &  \phi'^\dagger_{\theta, r} (x) \, = \, U(\sigma) \phi^\dagger_{0, r} (x) U^\dagger(\sigma) 
\, U(\sigma) e^{\frac{i}{2} \lambda^{(r)} \wedge Q} U^\dagger(\sigma) \nonumber \\ 
& = & \left( e^{i \sigma_a T_a} \right)_{sr} \phi^\dagger_s (x)  U(\sigma) e^{\frac{i}{2} \lambda^{(r)} \wedge Q} U^\dagger(\sigma) 
\, = \,\left( e^{i \sigma_a T_a} \right)_{sr} \phi^\dagger_s (x)  \xi^\dagger_{(r)}(\sigma), 
\label{suntranformanti}
\end{eqnarray}
where $\xi_{(r)}(\sigma) = U(\sigma) e^{-\frac{i}{2} \lambda^{(r)} \wedge Q} U^\dagger(\sigma)$ is a unitary operator satisfying 
$\xi_{(r)}(\sigma)\xi^\dagger_{(r)}(\sigma) =  \xi^\dagger_{(r)}(\sigma) \xi_{(r)}(\sigma) = \textbf{I}$. 
Also we have 
\begin{eqnarray}
\xi_{(r)}(\sigma) \, \left( e^{-i \sigma_a T_a} \right)_{rs} \,\phi_s (x) & = & U(\sigma) e^{-\frac{i}{2} \lambda^{(r)} \wedge Q} U^\dagger(\sigma) \,
U(\sigma) \phi_{0, r} (x) U^\dagger(\sigma) \nonumber \\ 
&  = & U(\sigma) e^{-\frac{i}{2} \lambda^{(r)} \wedge Q} \phi_{0, r} (x) U^\dagger(\sigma) \nonumber \\
&  = &  U(\sigma) \phi_{0, r} (x) e^{-\frac{i}{2} \lambda^{(r)} \wedge Q} U^\dagger(\sigma)
\nonumber \\
& = & U(\sigma) \phi_{0, r} (x) U^\dagger(\sigma) \, U(\sigma) e^{-\frac{i}{2} \lambda^{(r)} \wedge Q} U^\dagger(\sigma) \nonumber \\
&  = & \left( e^{-i \sigma_a T_a} \right)_{rs} \,\phi_s (x) \, \xi_{(r)}(\sigma).
\label{antixifree}
\end{eqnarray}

The transformation properties of the state vectors can be similarly discussed. For example, assuming that the vacuum remains invariant under the $SU(N)$ transformations 
i.e. $ U(\sigma) | 0 \rangle = | 0 \rangle$, the single-particle states transform as
\vspace{-.3cm}
\begin{eqnarray}
 U(\sigma) | r \rangle_\theta & = &  U(\sigma) a^\dagger_r | 0 \rangle \, = \, U(\sigma) a^\dagger_r U^\dagger(\sigma) U(\sigma) | 0 \rangle 
 \, = \, U(\sigma) c^\dagger_r \, e^{\frac{i}{2} \lambda^{(r)} \wedge Q} U^\dagger(\sigma) U(\sigma) | 0 \rangle \nonumber \\ 
& = &  U(\sigma) c^\dagger_r U^\dagger(\sigma)| 0 \rangle \, = \, \left( e^{i \sigma_a T_a} \right)_{sr} \,c^\dagger_s | 0 \rangle 
\, = \, \left( e^{i \sigma_a T_a} \right)_{sr} \,a^\dagger_s | 0 \rangle \, = \, \left( e^{i \sigma_a T_a} \right)_{sr}  | s \rangle_\theta, \nonumber \\
 U(\sigma) \overline{| r \rangle}_\theta & = &  U(\sigma) b^\dagger_r | 0 \rangle \, = \, U(\sigma) b^\dagger_r U^\dagger(\sigma) U(\sigma) | 0 \rangle 
 \, = \, U(\sigma) d^\dagger_r \, e^{-\frac{i}{2} \lambda^{(r)} \wedge Q} U^\dagger(\sigma) U(\sigma) | 0 \rangle \nonumber \\
& = & U(\sigma) d^\dagger_r U^\dagger(\sigma)| 0 \rangle  =  \left( e^{-i \sigma_a T^\ast_a} \right)_{sr} \,d^\dagger_s | 0 \rangle
 =  \left( e^{-i \sigma_a T^\ast_a} \right)_{sr} \,b^\dagger_s | 0 \rangle   =  \left( e^{-i \sigma_a T^\ast_a} \right)_{sr}  \overline{| s \rangle}_\theta. \nonumber \\
\label{1tstatevecsuntransform}
\end{eqnarray}

Hence, under $SU(N)$ transformations, the twisted single-particle states transform in same way as the untwisted single-particle states. Due to (\ref{singlrpart}), the transformation 
of twisted single-particle states was expected to be same as that of the untwisted state. However, twisted multi-particle states will have a different transformation 
e.g. for two-particle state we have 
\begin{eqnarray}
 U(\sigma) | r , s \rangle_\theta & = &  U(\sigma) a^\dagger_s a^\dagger_r | 0 \rangle 
\, = \,  U(\sigma) a^\dagger_s U^\dagger(\sigma) \, U(\sigma) a^\dagger_r U^\dagger(\sigma) \, U(\sigma) | 0 \rangle \nonumber \\ 
& = & U(\sigma) c^\dagger_s \, e^{\frac{i}{2} \lambda^{(s)} \wedge Q} U^\dagger(\sigma) \, 
U(\sigma) c^\dagger_r \, e^{\frac{i}{2} \lambda^{(r)} \wedge Q} U^\dagger(\sigma) \, U(\sigma) | 0 \rangle \nonumber \\ 
& = &   e^{\frac{i}{2} \lambda^{(s)} \wedge \lambda^{(r)}} \, U(\sigma) c^\dagger_s U^\dagger(\sigma) \, U(\sigma) c^\dagger_r U^\dagger(\sigma)| 0 \rangle \nonumber \\ 
& = &  e^{\frac{i}{2} \lambda^{(s)} \wedge \lambda^{(r)}} \, \left( e^{i \sigma_a T_a} \right)_{ts} \left( e^{i \sigma_a T_a} \right)_{ur} \, c^\dagger_t c^\dagger_u | 0 \rangle 
 \nonumber \\ 
& = & e^{\frac{i}{2} \lambda^{(s)} \wedge \lambda^{(r)}} \, e^{-\frac{i}{2} \lambda^{(t)} \wedge \lambda^{(u)}} \, \left( e^{i \sigma_a T_a} \right)_{ts}  
\left( e^{i \sigma_a T_a} \right)_{ur} \, a^\dagger_t a^\dagger_u| 0 \rangle \nonumber \\ 
& = &  e^{\frac{i}{2} \lambda^{(s)} \wedge \lambda^{(r)}} \, e^{-\frac{i}{2} \lambda^{(t)} \wedge \lambda^{(u)}} \, \left( e^{i \sigma_a T_a} \right)_{ts}  
\left( e^{i \sigma_a T_a} \right)_{ur} \, | u ,t \rangle_\theta ,
\label{2atstatevecsuntransform}
\end{eqnarray}
and 
\begin{eqnarray}
 U(\sigma) \overline{| r, s \rangle}_\theta & = & U(\sigma) b^\dagger_s b^\dagger_r | 0 \rangle 
\, = \,  U(\sigma) b^\dagger_s U^\dagger(\sigma) \, U(\sigma) b^\dagger_r U^\dagger(\sigma) \, U(\sigma) | 0 \rangle \nonumber \\ 
& = & U(\sigma) d^\dagger_s \, e^{-\frac{i}{2} \lambda^{(s)} \wedge Q} U^\dagger(\sigma) \, 
U(\sigma) d^\dagger_r \, e^{-\frac{i}{2} \lambda^{(r)} \wedge Q} U^\dagger(\sigma) \, U(\sigma) | 0 \rangle\nonumber \\
& = &  e^{\frac{i}{2} \lambda^{(s)} \wedge \lambda^{(r)}} \, U(\sigma) d^\dagger_s U^\dagger(\sigma) \, U(\sigma) d^\dagger_r U^\dagger(\sigma)| 0 \rangle \nonumber \\ 
& = & e^{\frac{i}{2} \lambda^{(s)} \wedge \lambda^{(r)}}\, \left( e^{-i \sigma_a T^\ast_a} \right)_{ts}\, \left( e^{-i \sigma_a T^\ast_a} \right)_{ur}\,d^\dagger_t d^\dagger_u| 0 \rangle
\nonumber \\ 
& = & e^{\frac{i}{2} \lambda^{(s)} \wedge \lambda^{(r)}}\, e^{-\frac{i}{2} \lambda^{(t)} \wedge \lambda^{(u)}}\, \left( e^{-i \sigma_a T^\ast_a} \right)_{ts}\, 
\left( e^{-i \sigma_a T^\ast_a} \right)_{ur}\, b^\dagger_t b^\dagger_u| 0 \rangle \nonumber \\ 
& = &  e^{\frac{i}{2} \lambda^{(s)} \wedge \lambda^{(r)}}\, e^{-\frac{i}{2} \lambda^{(t)} \wedge \lambda^{(u)}}\, \left( e^{-i \sigma_a T^\ast_a} \right)_{ts}\, 
\left( e^{-i \sigma_a T^\ast_a} \right)_{ur}\, \overline{| u, t \rangle}_\theta.
\label{2btstatevecsuntransform}
\end{eqnarray}
Similarly all other multi-particle states follow twisted transformation rules. 

Before we write down field theories using $\phi_{\theta, r}$ fields, we define a new multiplication rule for the product of two fields. We define the `` antisymmetric star-product '' 
$\star$ as follows
\begin{eqnarray}\hspace{-.6cm}
 \phi^\#_{\theta, r} (x) \star \phi^\#_{\theta, s}(y) & = &  \phi^\#_{\theta, r} (x) \, e^{-\frac{i}{2} \overleftarrow{Q} \wedge \overrightarrow{Q} } \phi^\#_{\theta, s} (y)\nonumber \\
& = &  \phi^\#_{\theta, r} (x) \phi^\#_{\theta, s} (y) \, - \, \frac{i}{2} \theta_{lm} \left[ Q_l, \phi^\#_{\theta, r} (x)\right] \left[ Q_m, \phi^\#_{\theta, s} (y) \right]\nonumber \\
& + & \frac{1}{2!} \frac{i}{2} \theta_{lm} \, \frac{i}{2} \theta_{np} \left[ Q_l, \left[ Q_n,\phi^\#_{\theta, r} (x)\right]\right] 
\left[ Q_m, \left[ Q_p,\phi^\#_{\theta, s} (y) \right] \right]
\, + \, \cdots,
\label{starproductanti}
\end{eqnarray}
where $\phi^\#_{\theta, r}$ stands for either of $\phi_{\theta, r}$ or $\phi^\dagger_{\theta, r}$. It should be remarked that the above defined $\star$-product is infact the 
internal symmetry analogue of the widely studied Moyal-product \cite{bal-pinzul}. 

Note that, owing to the relation (\ref{relationcomso6}), we have 
\begin{eqnarray}
 \phi^\#_{\theta, r} (x) \star \phi^\#_{\theta, s}(y) & = & e^{-\frac{i}{2} (\pm \lambda^{(r)}) \wedge (\pm \lambda^{(s)}) } \phi^\#_{\theta, r} (x) 
\, \cdot \, \phi^\#_{\theta, s}(y),
\label{starproductantirel}
\end{eqnarray}
where $\lambda^{(r)}$ should be taken if $\phi^\#_{\theta, r} = \phi^\dagger_{\theta, r}$ and $-\lambda^{(r)}$ should be taken if $\phi^\#_{\theta, r} = \phi_{\theta, r}$.
Hence, multiplying two fields with a star-product is nothing but multiplying a certain phase factor to ordinary product of fields. Nonetheless, using the star-product 
greatly simplifies things and it should be regarded just as a shorthand notation for the phases that are present in a particular term in the Hamiltonian density. These 
simplifications will be further elaborated when we construct the interaction terms for twisted fields.

The star-product introduced here has the property that
\begin{eqnarray}
 \phi^\#_{\theta,r} \, \star \, \phi^\#_{\theta, s} & = &  \phi^\#_{\theta, s} \, \star \, \phi^\#_{\theta,r}, \nonumber \\
 \phi^\#_{\theta,r} \, \star \, \left( \phi^\#_{\theta, s} \, \star \, \phi^\#_{\theta, t} \right) 
& = & \left( \phi^\#_{\theta,r} \, \star \, \phi^\#_{\theta, s} \right)\, \star \, \phi^\#_{\theta, t}, 
\label{idanti1}
\end{eqnarray}
and due to the antisymmetry we have 
\begin{eqnarray}
 \phi^\#_{\theta,r} \, \star \, \phi^\#_{\theta, r} & = &  \phi^\#_{\theta, r} \cdot \phi^\#_{\theta,r}. 
\label{idanti2}
\end{eqnarray}
Also, because of the dressing transformations (\ref{fieldsantidress}) we have
\begin{eqnarray}
 \phi^\#_{\theta,r_1} \, \star \, \phi^\#_{\theta, r_2} \, \star \, \cdots \, \star \, \phi^\#_{\theta, r_n} & = & \phi^\#_{0,r_1} \, \phi^\#_{0, r_2} \, \cdots \, \phi^\#_{0, r_n} 
\, e^{\frac{i}{2} \left( \pm \lambda^{(r_1)} \, \pm \,  \lambda^{(r_2)} \, \pm \, \cdots \, \pm \,  \lambda^{(r_n)} \right) \wedge Q },
\label{idanti3}
\end{eqnarray}
where $ + \lambda^{(r)}$ is to be taken if $ \phi^\#_{\theta,r} = \phi^\dagger_{\theta,r}$ and $ - \lambda^{(r)}$ if $ \phi^\#_{\theta,r} = \phi_{\theta,r}$.

Having introduced the $\star$-product we can now discuss how to write Hamiltonian densities using twisted fields. For any given untwisted Hamiltonian its twisted counterpart 
should be written by replacing the untwisted fields $\phi_{0,r}$ by the twisted fields $\phi_{\theta,r}$ and the ordinary product between fields by the $\star$-product. 
Hence the free theory Hamiltonian density $\mathcal{H}_{\theta, F}$ in terms of twisted fields becomes
 \begin{eqnarray}
  \mathcal{H}_{\theta, F} & = &  \Pi^{\dagger}_{\theta,r} \star \Pi_{\theta,r}  \, + \, (\partial_i \phi^{\dagger}_{\theta,r}) \star(\partial^i \phi_{\theta,r})
 \, + \, m^2  \, \phi^{\dagger}_{\theta,r} \star \phi_{\theta,r} \nonumber \\
& = &  \Pi^{\dagger}_{\theta,r} \Pi_{\theta,r}  \, + \, (\partial_i \phi^{\dagger}_{\theta,r}) (\partial^i \phi_{\theta,r})  \, + \, m^2  \, \phi^{\dagger}_{\theta,r} \phi_{\theta,r},
\label{freeantiham}
 \end{eqnarray}
where $\Pi_{\theta,r}$ is the canonical conjugate of $\phi_{\theta,r}$; $r = 1, 2, \cdots, N$ and the last line in (\ref{freeantiham}) is obtained using (\ref{idanti2}). 
This Hamiltonian density is invariant under the $SU(N)$ global transformations which can be explicitly checked by using (\ref{suntranformanti}). 

The renormalizable $SU(N)$ invariant interaction Hamiltonian density compatible with (\ref{hamsun}) is given by
\begin{eqnarray}
 \mathcal{H}_{\theta,\rm{Int}} & = &  \frac{\gamma}{4} \, \phi^{\dagger}_{\theta,r}  \star  \phi^{\dagger}_{\theta,s}  \star \phi_{\theta,r}  \star \phi_{\theta,s} \nonumber \\
& = & \frac{\gamma}{4}\, e^{- i \lambda^{(r)} \wedge \lambda^{(s)}} \, \phi^{\dagger}_{\theta,r} \phi^{\dagger}_{\theta,s} \phi_{\theta,r} \phi_{\theta,s},
\label{intantiham}
\end{eqnarray}
where $r,s = 1, 2, \cdots, N$. 

Using (\ref{suntranformanti}) one can check that (\ref{intantiham}) is indeed invariant under $SU(N)$ group. Alternatively, one can also use the dressing transformation 
(\ref{fieldsantidress}) and the identity (\ref{idanti2}) to write everything in terms of the untwisted fields and then apply the $SU(N)$ transformations given by 
(\ref{suntranformcom}) to check for the invariance of the Hamiltonian density. 

From (\ref{intantiham}) it is clear that unlike the untwisted case where $SU(N)$ invariance forces all the interaction terms in (\ref{gaugen4susy}) to have the same coupling $\gamma$, 
here the demand of $SU(N)$ invariance forces the various terms to have different couplings related with each other by phases of the type $e^{\pm i \lambda^{(r)} \wedge \lambda^{(s)}}$.

Although in this section we restricted our discussion only to scalar fields and twisted bosons but it is easy to generalize the discussion to include twisted fermions and 
spinor fields. For the discussion of twisted fermions we have to 
consider anticommuting creation/annihilation operators. The twisted fermions will again satisfy (\ref{atwistedcomrel}) but with $\eta = -1$. One can again write down $SU(N)$
invariant field theories involving such twisted fermions in a way very similar to the one we discussed.  Also, the above discussion can be easily 
generalized to higher dimensional representations of $SU(N)$ group as well as to other symmetry groups like $SO(N)$.


\subsection{$S$-matrix Elements}


In the previous section we set up the formalism for writing down field theories with a special type of twisted statistics which we called `` antisymmetric twisted statistics''.
We showed that using such twisted fields we can write down $SU(N)$ invariant interactions. We now want to discuss the possibility of experimental signatures of the twisted 
field theories. We start our discussion from scattering processes. 

Let us first start with the $SU(N)$ invariant free Hamiltonian density $\mathcal{H}_{\theta, F}$. Using the dressing transformation (\ref{fieldsantidress}) and the 
identity (\ref{idanti3}) we have
 \begin{eqnarray}
  \mathcal{H}_{\theta, F} & = &  \Pi^{\dagger}_{\theta,r} \star \Pi_{\theta,r}  \, + \, (\partial_i \phi^{\dagger}_{\theta,r}) \star(\partial^i \phi_{\theta,r})
 \, + \, m^2  \, \phi^{\dagger}_{\theta,r} \star \phi_{\theta,r} \nonumber \\
& = &  \Pi^{\dagger}_{0,r} \Pi_{0,r}  \, + \, (\partial_i \phi^{\dagger}_{0,r}) (\partial^i \phi_{0,r}) \, + \, m^2  \, \phi^{\dagger}_{0,r} \phi_{0,r}
\, = \,  \mathcal{H}_{0,F} .
\label{freeantihamrel}
 \end{eqnarray}
Hence the Hamiltonian for twisted free theory is same as its untwisted counterpart.
Next we look at the renormalizable $SU(N)$ invariant interaction Hamiltonian density which is 
 \begin{eqnarray}
\hspace{-.5cm}  
\mathcal{H}_{\theta, \rm{Int}} & = &  \Pi^{\dagger}_{\theta,r} \star \Pi_{\theta,r}  \, + \, (\partial_i \phi^{\dagger}_{\theta,r}) \star(\partial^i \phi_{\theta,r})
 \, + \, m^2  \, \phi^{\dagger}_{\theta,r} \star \phi_{\theta,r} 
\, + \,  \frac{\gamma}{4} \, \phi^{\dagger}_{\theta,r}  \star  \phi^{\dagger}_{\theta,s}  \star \phi_{\theta,r}  \star \phi_{\theta,s} \nonumber \\
& = &  \Pi^{\dagger}_{0,r} \Pi_{0,r}  +  (\partial_i \phi^{\dagger}_{0,r}) (\partial^i \phi_{0,r})  +  m^2   \phi^{\dagger}_{0,r} \phi_{0,r}
 +   \frac{\gamma}{4}  \phi^{\dagger}_{0,r}  \phi^{\dagger}_{0,s}  \phi_{0,r}  \phi_{0,s} 
 \, = \, \mathcal{H}_{0, \rm{Int}} ,
\label{intantihamrel}
 \end{eqnarray}
where to obtain the last line we have again used the dressing transformation (\ref{fieldsantidress}) and the identity (\ref{idanti3}). So even the $SU(N)$ invariant interaction 
Hamiltonian density for the two theories turns out to be same. But the in/out states for the twisted theory also contain information about twisted statistics. 
So we should look at $S$-matrix elements which can still provide information about twisted statistics. Let us take a typical $S$-matrix element, say for the scattering process of 
$\phi_{\theta, r} \phi_{\theta, s} \rightarrow \phi_{\theta, r} \phi_{\theta, s}$. Then we have 
 \begin{eqnarray}
  S\left[ \phi_{\theta, r} \phi_{\theta, s} \rightarrow \phi_{\theta, r} \phi_{\theta, s}  \right] 
& = & \leftidx{_{out,\theta}}{\left \langle r s | r s  \right \rangle }{_{ \theta,in}} 
\, = \, \leftidx{_{\theta}}{\left \langle r s | S_\theta | r s \right \rangle }{_{ \theta}},
\label{smatanti}
 \end{eqnarray}
where $S_\theta = \mathcal{T}  \exp \left[-i\int^{\infty}_{-\infty} d^{4}z \mathcal{H}_{\theta, \rm{ Int}} (z) \right] $ is the $S$-operator and we have denoted the two-particle 
in and out states by $\left| r s \right \rangle_\theta = a^\dagger_s a^\dagger_r | 0 \rangle$. Because of (\ref{intantihamrel}) we have 
\begin{eqnarray}
 S_\theta & = & \mathcal{T}  \exp \left[-i\int^{\infty}_{-\infty} d^{4}z \mathcal{H}_{\theta, \rm{Int}} (z) \right]  
\, = \, \mathcal{T}  \exp \left[-i\int^{\infty}_{-\infty} d^{4}z \mathcal{H}_{0, \rm{Int}} (z) \right]  \, = \, S_0.
\label{sopanti}
\end{eqnarray}
Also we have 
\begin{eqnarray}
 \left| r s \right \rangle_\theta & = & a^\dagger_s a^\dagger_r  \left| 0 \right\rangle 
\, = \, c^\dagger_s \, e^{\frac{i}{2} \lambda^{(s)} \wedge Q} \, c^\dagger_r \, e^{\frac{i}{2} \lambda^{(r)} \wedge Q} \left| 0 \right\rangle 
\, = \, e^{\frac{i}{2}\lambda^{(s)} \wedge \lambda^{(r)} } \,  c^\dagger_s \, c^\dagger_r \, e^{\frac{i}{2} ( \lambda^{(r)} + \lambda^{(s)}) \wedge Q} \left| 0 \right\rangle 
 \nonumber \\
& = & e^{\frac{i}{2}\lambda^{(s)} \wedge \lambda^{(r)} } \,  c^\dagger_s \, c^\dagger_r \,\left| 0 \right\rangle  
\, = \, e^{\frac{i}{2}\lambda^{(s)} \wedge \lambda^{(r)} } \, \left| r s \right \rangle_0.
\label{stateanti}
\end{eqnarray}
Using (\ref{stateanti}) and (\ref{sopanti}) we get
 \begin{eqnarray}
  S\left[ \phi_{\theta, r} \phi_{\theta, s} \rightarrow \phi_{\theta, r} \phi_{\theta, s}  \right] & = & 
\leftidx{_{0}}{\left \langle r s | \, e^{\frac{-i}{2}\lambda^{(s)} \wedge \lambda^{(r)} } \, S_0 \, e^{\frac{i}{2}\lambda^{(s)} \wedge \lambda^{(r)} } \, | r s \right \rangle }{_{ 0}}
\nonumber \\
& = & \leftidx{_{0}}{\left \langle r s | \, S_0 \, | r s \right \rangle }{_{ 0}} \nonumber \\
& = &  S\left[ \phi_{0, r} \phi_{0, s} \rightarrow \phi_{0, r} \phi_{0, s}  \right].
\label{smatantirel}
\end{eqnarray}
So it turns out that, even the $S$-matrix elements for the two theories are same. It seems like the twisted $SU(N)$ invariant theory is indistinguishable from an untwisted 
$SU(N)$ invariant
 theory but we should take note of the fact that the particles in the two theory follow different type of statistics. Perhaps the best place to look for potential signatures
of such particles is to look at the statistical properties and to construct observables which depend crucially on the statistics followed by these particles.
 
One should also note that this indistinguishability arised because we demanded that our Hamiltonian density remains invariant under $SU(N)$ transformations and that 
the twisted vacuum is not only same 
as untwisted vacuum but also annihilates all the charge operators. Dropping either of these two demands makes the two theories distinct. We now briefly discuss the 
first scenario i.e. the case where the Hamiltonian density does not remain invariant under $SU(N)$ transformations. We plan to consider the second scenario in more 
details in a separate work.   

Let us take the Hamiltonian density (\ref{intantihamrel}) but with fields multiplied not by $\star$-products but by ordinary products i.e.
\begin{eqnarray}
  \mathcal{H}_{\theta} & = &  \Pi^{\dagger}_{\theta,r} \Pi_{\theta,r}  \, + \, (\partial_i \phi^{\dagger}_{\theta,r}) (\partial^i \phi_{\theta,r})
 \, + \, m^2  \, \phi^{\dagger}_{\theta,r} \phi_{\theta,r} 
\, + \,  \frac{\gamma}{4} \, \phi^{\dagger}_{\theta,r}  \phi^{\dagger}_{\theta,s} \phi_{\theta,r}  \phi_{\theta,s} \, + \, \rm{h. c}.
\label{hamwithoutstar}
\end{eqnarray}
Since $\phi_{\theta, r}$ are noncommuting fields so there is an ambiguity in ordering of operators in the interaction term and there are other inequivalent terms 
which one can write. The full Hamiltonian density with all $\phi^4$ type terms can be written as 
 \begin{eqnarray}
  \mathcal{H}_{\theta} & = &  \Pi^{\dagger}_{\theta,r} \Pi_{\theta,r}  \, + \, (\partial_i \phi^{\dagger}_{\theta,r}) (\partial^i \phi_{\theta,r})
 \, + \, m^2  \, \phi^{\dagger}_{\theta,r} \phi_{\theta,r} 
\, + \,  \frac{\gamma}{4} \, \left[ \phi^{\dagger}_{\theta,r}  \phi^{\dagger}_{\theta,s} \phi_{\theta,r}  \phi_{\theta,s} 
\, + \, \phi^{\dagger}_{\theta,s} \phi^{\dagger}_{\theta,r} \phi_{\theta,r}  \phi_{\theta,s}  \right . \nonumber \\  
& + & \left. \phi^{\dagger}_{\theta,s} \phi^{\dagger}_{\theta,r}   \phi_{\theta,s} \phi_{\theta,r}
\, + \, \phi^{\dagger}_{\theta,r}  \phi^{\dagger}_{\theta,s}   \phi_{\theta,s} \phi_{\theta,r}  \, + \, \cdots  \right ] \, + \, \rm{h. c}.
\label{genhamwithoutstar}
\end{eqnarray}
One can in principle write 24 such terms and $\cdots$ represents the other terms which we have not written. Some of these 24 terms will be equivalent to other terms but 
unlike the untwisted case not all of them are equal to each other. Moreover, this Hamiltonian density has no $SU(N)$ symmetry. The easiest way to see that is by using the 
dressing transformation (\ref{fieldsantidress}) and writing it in terms of untwisted fields 
 \begin{eqnarray}
\mathcal{H}_{0} & = &  \Pi^{\dagger}_{0,r} \Pi_{0,r}  \, + \, (\partial_i \phi^{\dagger}_{0,r}) (\partial^i \phi_{0,r})
 \, + \, m^2  \, \phi^{\dagger}_{0,r} \phi_{0,r}
\, + \,  \frac{\gamma}{4} \,e^{\frac{i}{2}\lambda^{(r)} \wedge \lambda^{(s)} } \,  \phi^{\dagger}_{0,r}  \phi^{\dagger}_{0,s} \phi_{0,r}  \phi_{0,s}  \nonumber \\  
& + &  \frac{\gamma}{4} \, \phi^{\dagger}_{0,s} \phi^{\dagger}_{0,r} \phi_{0,r}  \phi_{0,s}
\, + \, \frac{\gamma}{4} \,e^{-\frac{i}{2}\lambda^{(r)} \wedge \lambda^{(s)} } \,  \phi^{\dagger}_{0,s} \phi^{\dagger}_{0,r}   \phi_{0,s} \phi_{0,r}  
\, + \, \frac{\gamma}{4} \, \phi^{\dagger}_{0,r}  \phi^{\dagger}_{0,s}   \phi_{0,s} \phi_{0,r}  \, + \, \cdots   \, + \, \rm{h. c}. \nonumber \\ 
\label{genhamrelwithoutstar}
\end{eqnarray}
Because of the presence of $e^{\pm\frac{i}{2}\lambda^{(r)} \wedge \lambda^{(s)} }$  type phases, clearly this Hamiltonian density has no symmetry. Infact, (\ref{genhamwithoutstar}) is equivalent 
to a marginally deformed $SU(N)$ Hamiltonian density. We will discuss more about such Hamiltonian densities in the next section.

Similarly, one can consider many more Hamiltonians which explicitly break $SU(N)$ invariance e.g. let us consider the interaction Hamiltonian density given by
\begin{eqnarray}
  \mathcal{H}_{\theta,\rm{Int}} & = &  \frac{\gamma}{4} \, \phi^{\dagger}_{\theta,r}  \star  \phi^{\dagger}_{\theta,r}  \star \phi^{\dagger}_{\theta,s}  \star \phi_{\theta,s} 
\, + \, \rm{h.c}.
\label{hamsunvioanti}
\end{eqnarray}
Unlike (\ref{genhamwithoutstar}) whose untwisted counterpart was $SU(N)$ invariant, even the untwisted counterpart of this Hamiltonian density
\begin{eqnarray}
 \mathcal{H}_{0,\rm{Int}} & = &  \frac{\gamma}{4} \, \phi^{\dagger}_{0,r}    \phi^{\dagger}_{0,r}  \phi^{\dagger}_{0,s}   \phi_{0,s} 
\, + \, \rm{h.c},
\label{hamsunviocom}
\end{eqnarray}
is not $SU(N)$ invariant. 

But now (\ref{hamsunvioanti}) is not even equivalent to any local untwisted Hamiltonian density, as after using dressing transformation we have
\begin{eqnarray}
 \mathcal{H}_{\theta,\rm{Int}} & = &  \frac{\gamma}{4} \, \phi^{\dagger}_{0,r} \phi^{\dagger}_{0,r} \phi^{\dagger}_{0,s} \phi_{0,s} \, e^{ i \lambda^{(r)} \wedge Q }
\, + \, \rm{h.c},
\label{hamsunvioantirel}
\end{eqnarray}
which is nonlocal because of the presence of nonlocal operators $Q_m$. 

So we find that, if we demand the twisted Hamiltonians to be $SU(N)$ invariant then they turn out to indistinguishable from untwisted Hamiltonians. But if we relax the demand of $SU(N)$
invariance then the two Hamiltonians are not exactly same. Infact many of such $SU(N)$ breaking Hamiltonians like (\ref{hamsunvioanti}) turn out to be nonlocal and can't be mapped to any
local untwisted Hamiltonian. Even in the case of $SU(N)$ invariant Hamiltonians, one can possibly construct observables which show signatures of the underlying twisted statistics.


\subsection{The $\mathcal{N} = 4$ SUSY Hamiltonian and its Marginal ($\beta$-) Deformations}


As a specific example of equivalence between twisted interaction Hamiltonians of the type (\ref{genhamwithoutstar}) and untwisted marginally deformed $SU(N)$ Hamiltonians, 
let us look at the scalar matter sector of $\mathcal{N}  =  4$ supersymmetric (SUSY) Yang-Mills theory in four dimensions and its marginal deformations \cite{Leigh:1995ep}. 
Although it is not difficult to generalize the discussion to include the fermionic sector also but to illustrate our point it is sufficient to show the equivalence only for
the scalar sector. 
The scalar sector of the SUSY theory consists of six real scalars $\phi_{0,r}$; $r = 1, 2, \cdots 6$ having a $SO(6)$ global symmetry. The scalars transform as fundamental 
representation of $SO(6)$ group or equivalently as the 6-dimensional representation of the $SU(4)$ group \cite{Frolov:2005iq, Frolov:2005dj}. 
These six real scalars can also be combined to form 3 complex scalars $\Phi_{0,r}$; $r = 1,2,3$. When written in terms of the complex fields only the $SU(3)$ subgroup of 
the full symmetry group is apparent and the three complex fields transform as the fundamental representation of $SU(3)$ group. The interaction term is given by
\begin{eqnarray}
 \mathcal{H}_{0,\rm{int}} & = &  \frac{g^2}{2} f_{ijk} f^{i}_{lm}\,  \epsilon_{rst}\epsilon^{uvt} \, \Phi^{\dagger}_{0,r} \Phi^{\dagger}_{0,s} \Phi_{0,u} \Phi_{0,v}.
\label{gaugeneral4susy}
\end{eqnarray}
Since we are not concerned with the details of the gauge theory, we will ``switch off'' the gauge interaction. So without gauge interactions we have (denoting coupling constant by 
$\frac{\tilde{\gamma}'}{4}$)
\begin{eqnarray}
 \mathcal{H}_{0,\rm{int}} & = &  \frac{\tilde{\gamma}'}{4} \,  \epsilon_{rst}\epsilon^{uvt} \, \Phi^{\dagger}_{0,r} \Phi^{\dagger}_{0,s} \Phi_{0,u} \Phi_{0,v} \nonumber \\
& = & \frac{\tilde{\gamma}'}{4} \, ( \delta^u_r \delta^v_s \, - \, \delta^v_r \delta^u_s )\, \Phi^{\dagger}_{0,r} \Phi^{\dagger}_{0,s} \Phi_{0,u} \Phi_{0,v} \nonumber \\
& = &\frac{\tilde{\gamma}'}{4} \, \Phi^{\dagger}_{0,r} \Phi^{\dagger}_{0,s} \Phi_{0,r} \Phi_{0,s} 
\; - \; \frac{\tilde{\gamma}'}{4} \, \Phi^{\dagger}_{0,r} \Phi^{\dagger}_{0,s} \Phi_{0,s} \Phi_{0,r}.
\label{gauge4susy}
\end{eqnarray}
Without the gauge interactions, the fields commute i.e. $ [\Phi_{0,r}\, , \, \Phi_{0,s}] \; = \; [\Phi^{\dagger}_{0,r}\, , \, \Phi^{\dagger}_{0,s}] \; = \; 0 $. 
Hence, in the untwisted case the interaction Hamiltonian vanishes \cite{Sohnius:1985qm}. Let us see what happens if we replace the untwisted fields by twisted fields in 
(\ref{gauge4susy}). In that case we have
\begin{eqnarray}
 \mathcal{H}_{\theta,\rm{int}} & = & \frac{\tilde{\gamma}'}{4} \, \Phi^{\dagger}_{\theta,r} \Phi^{\dagger}_{\theta,s} \Phi_{\theta,r} \Phi_{\theta,s} 
\; - \; \frac{\tilde{\gamma}'}{4}\, \Phi^{\dagger}_{\theta,r} \Phi^{\dagger}_{\theta,s} \Phi_{\theta,s} \Phi_{\theta,r}.
\label{tgaugen4susy}
\end{eqnarray}
Using the relations
 \begin{eqnarray}
 \Phi_{\theta,r}(x) \Phi_{\theta,s}(x) & = & e^{ i \lambda^{(r)} \wedge \lambda^{(s)} } \, \Phi_{\theta,s}(x) \Phi_{\theta,r}(x), \nonumber \\
\Phi^{\dagger}_{\theta,r}(x) \Phi^{\dagger}_{\theta,s}(x) & = & e^{ i \lambda^{(r)} \wedge \lambda^{(s)} } \, \Phi^{\dagger}_{\theta,s}(x) \Phi^{\dagger}_{\theta,r}(x), 
\label{so6flipfield}
\end{eqnarray}
we have 
\begin{eqnarray}
 \mathcal{H}_{\theta,\rm{int}} & = & \frac{\tilde{\gamma}'}{4} \, \Phi^{\dagger}_{\theta,r} \Phi^{\dagger}_{\theta,s} \Phi_{\theta,r} \Phi_{\theta,s}
 \; - \; \frac{\tilde{\gamma}'}{4} \,e^{-i \lambda^{(r)}\wedge \lambda^{(s)}} \, \Phi^{\dagger}_{\theta,r} \Phi^{\dagger}_{\theta,s} \Phi_{\theta,r} \Phi_{\theta,s} \nonumber \\
& = & \frac{\tilde{\gamma}'}{4} \,\left( 1\; - \; e^{-i \lambda^{(r)}\wedge \lambda^{(s)}} \right)\, \Phi^{\dagger}_{\theta,r} \Phi^{\dagger}_{\theta,s} \Phi_{\theta,r} 
\Phi_{\theta,s} .
\label{tgauge4susy}
\end{eqnarray}
So the twisted interaction Hamiltonian density does not vanish. We can use the dressing transformations between twisted and untwisted fields 
to write it in terms of untwisted fields only. Then we have   
\begin{eqnarray}
 \mathcal{H}_{\theta,\rm{int}} & = & \frac{\tilde{\gamma}'}{4} \, \left( 1\; - \; e^{-i \lambda^{(r)}\wedge \lambda^{(s)}} \right)\, \Phi^{\dagger}_{0,r} \, 
e^{\frac{i}{2}\lambda^{(r)}\wedge Q}\, 
\Phi^{\dagger}_{0,s} \, e^{\frac{i}{2}\lambda^{(s)}\wedge Q}\, \Phi_{0,r} \, e^{\frac{-i}{2}\lambda^{(r)}\wedge Q}\, \Phi_{0,s} \, e^{\frac{-i}{2}\lambda^{(s)}\wedge Q}\, \nonumber \\
& = & \frac{\tilde{\gamma}'}{4} \, \left( 1\; - \; e^{-i \lambda^{(r)}\wedge \lambda^{(s)}} \right)\,e^{i \lambda^{(r)}\wedge \lambda^{(s)}}\, \Phi^{\dagger}_{0,r} 
\Phi^{\dagger}_{0,s} \Phi_{0,r} \Phi_{0,s} \nonumber \\
& = & \frac{\tilde{\gamma}'}{4} \, \left(e^{i \lambda^{(r)}\wedge \lambda^{(s)}} \; - \; 1 \right)\,\Phi^{\dagger}_{0,r} \Phi^{\dagger}_{0,s} \Phi_{0,r} \Phi_{0,s}.
\label{tcgauge4susy}
\end{eqnarray}
where r,s = 1,2,3. Expanding in terms of component fields we get
\begin{eqnarray}
 \mathcal{H}_{\theta,\rm{int}} & = & \frac{\tilde{\gamma}'}{4} \, \left[ \left(e^{i \lambda^{(1)}\wedge \lambda^{(1)}} \; - \; 1 \right)
\,\Phi^{\dagger}_{0,1} \Phi^{\dagger}_{0,1} \Phi_{0,1} \Phi_{0,1} 
\, + \, \left(e^{i \lambda^{(2)}\wedge \lambda^{(2)}} \; - \; 1 \right)\,\Phi^{\dagger}_{0,2} \Phi^{\dagger}_{0,2} \Phi_{0,2} \Phi_{0,2} 
\right. \nonumber \\
& + &  \left(e^{i \lambda^{(3)}\wedge \lambda^{(3)}} \; - \; 1 \right)\,\Phi^{\dagger}_{0,3} \Phi^{\dagger}_{0,3} \Phi_{0,3} \Phi_{0,3} 
\, + \,\left(e^{i \lambda^{(1)}\wedge \lambda^{(2)}} \; - \; 1 \right)\,\Phi^{\dagger}_{0,1} \Phi^{\dagger}_{0,2} \Phi_{0,1} \Phi_{0,2} \nonumber \\
& + &  \left(e^{i \lambda^{(1)}\wedge \lambda^{(3)}} \; - \; 1 \right)\,\Phi^{\dagger}_{0,1} \Phi^{\dagger}_{0,3} \Phi_{0,1} \Phi_{0,3}  
\, + \, \left(e^{i \lambda^{(2)}\wedge \lambda^{(3)}} \; - \; 1 \right)\,\Phi^{\dagger}_{0,2} \Phi^{\dagger}_{0,3} \Phi_{0,2} \Phi_{0,3} \nonumber \\
& + &   \left(e^{i \lambda^{(2)}\wedge \lambda^{(1)}} \; - \; 1 \right)\,\Phi^{\dagger}_{0,2} \Phi^{\dagger}_{0,1} \Phi_{0,2} \Phi_{0,1} 
\, + \,\left(e^{i \lambda^{(3)}\wedge \lambda^{(1)}} \; - \; 1 \right)\,\Phi^{\dagger}_{0,3} \Phi^{\dagger}_{0,1} \Phi_{0,3} \Phi_{0,1} \nonumber \\
& + & \left. \left(e^{i \lambda^{(3)}\wedge \lambda^{(2)}} \; - \; 1 \right)\,\Phi^{\dagger}_{0,3} \Phi^{\dagger}_{0,2} \Phi_{0,3} \Phi_{0,2}  \right].
\label{tcexpandgauge4susy}
\end{eqnarray}
Noting the fact that $e^{i \lambda^{(r)}\wedge \lambda^{(r)}} \, = \, 1$ and  $e^{i \lambda^{(s)}\wedge \lambda^{(r)}} \, = \, e^{-i \lambda^{(r)}\wedge \lambda^{(s)}} $, we 
can simplify (\ref{tcexpandgauge4susy}) and get 
\begin{eqnarray}
 \mathcal{H}_{\theta,\rm{int}} & = & \frac{\tilde{\gamma}'}{4} \, \left[ \left(e^{i \lambda^{(1)}\wedge \lambda^{(2)}} 
\; + \; e^{-i \lambda^{(1)}\wedge \lambda^{(2)}} \; - \;2 \right)\,\Phi^{\dagger}_{0,1} \Phi^{\dagger}_{0,2} \Phi_{0,1} \Phi_{0,2} \right. \nonumber \\
& + & \left(e^{i \lambda^{(2)}\wedge \lambda^{(3)}} \; + \; e^{-i \lambda^{(2)}\wedge \lambda^{(3)}} \; - \;2 \right)
 \Phi^{\dagger}_{0,2} \Phi^{\dagger}_{0,3} \Phi_{0,2} \Phi_{0,3} \nonumber \\
& + & \left. \left( e^{i \lambda^{(3)}\wedge \lambda^{(1)}} \; + \; e^{-i \lambda^{(3)}\wedge \lambda^{(1)}} \; - \;2 \right) \,
\Phi^{\dagger}_{0,3} \Phi^{\dagger}_{0,1} \Phi_{0,3} \Phi_{0,1} \right] \nonumber \\
& = & -\frac{\tilde{\gamma}'}{4} \, \left[ 4 \sin^2 \left\{ \frac{\lambda^{(1)}\wedge \lambda^{(2)}}{2} \right\}\, \Phi^{\dagger}_{0,1} \Phi^{\dagger}_{0,2} 
\Phi_{0,1} \Phi_{0,2} \, + \,  4 \sin^2 \left\{ \frac{\lambda^{(2)}\wedge \lambda^{(3)}}{2} \right\}  \right. \nonumber \\
&  & \left.\, \Phi^{\dagger}_{0,2} \Phi^{\dagger}_{0,3} \Phi_{0,2} \Phi_{0,3}  
\, + \, 4 \sin^2 \left\{ \frac{\lambda^{(3)}\wedge \lambda^{(1)}}{2} \right\}\,\Phi^{\dagger}_{0,3} \Phi^{\dagger}_{0,1} \Phi_{0,3} \Phi_{0,1} \right] 
\nonumber \\
& = &  \frac{\gamma_{12}}{2} \Phi^{\dagger}_{0,1} \Phi^{\dagger}_{0,2} \Phi_{0,1} \Phi_{0,2}  
 +  \frac{\gamma_{23}}{2}  \Phi^{\dagger}_{0,2} \Phi^{\dagger}_{0,3} \Phi_{0,2} \Phi_{0,3}  
 +  \frac{\gamma_{31}}{2}  \Phi^{\dagger}_{0,3} \Phi^{\dagger}_{0,1} \Phi_{0,3} \Phi_{0,1}.
\label{tcepgauge4susy}
\end{eqnarray}
We now show that (\ref{tcepgauge4susy}) is equivalent to marginal deformations of the scalar part of $\mathcal{N} = 4$ SUSY theory with gauge interactions switched off. 
The ``Marginally Deformed'' $\mathcal{N} = 4$ SUSY Hamiltonian density is given by \cite{Frolov:2005iq, Frolov:2005dj}
\begin{eqnarray}
\mathcal{H}_{0, \rm{int}} & = & \frac{\gamma}{4} \; Tr \left [ \left | \Phi_{0,1} \Phi_{0,2} - e^{-2i \pi \beta_{12}} \Phi_{0,2} \Phi_{0,1} \right |^2  
\; + \; \left | \Phi_{0,2} \Phi_{0,3} - e^{-2i \pi \beta_{23}}\Phi_{0,3} \Phi_{0,2} \right |^2  \right. \nonumber \\
& + & \left. \left| \Phi_{0,3} \Phi_{0,1} - e^{-2i \pi \beta_{31}} \Phi_{0,1} \Phi_{0,3} \right|^2 \right] 
 +  \frac{\tilde{\gamma}}{4}  Tr \left [ \left \{ [\Phi_{0,1} , \Phi^{\dagger}_{0,1}] + [\Phi_{0,2} , \Phi^{\dagger}_{0,2}] 
+ [\Phi_{0,3} , \Phi^{\dagger}_{0,3}] \right\}^2 \right ],\nonumber \\
\label{mdgaugen4susy}
\end{eqnarray}
where the trace is over gauge index of the gauge group $SU(N)$. Since we are not interested in gauge fields, so we switch off the gauge interactions. 
The $ \mathcal{H}_{0, \rm{int}}$ then takes the form
\begin{eqnarray}
 \mathcal{H}_{0, \rm{int}} & = &  \frac{\gamma}{4} \left| \Phi_{0,1} \Phi_{0,2} - e^{-2i \pi \beta_{12}} \Phi_{0,2} \Phi_{0,1} \right|^2  
\; + \; \frac{\gamma}{4} \left| \Phi_{0,2} \Phi_{0,3} - e^{-2i \pi \beta_{23}} \Phi_{0,3} \Phi_{0,2} \right |^2    \nonumber \\
& + & \frac{\gamma}{4} \left | \Phi_{0,3} \Phi_{0,1} - e^{-2i \pi \beta_{31}} \Phi_{0,1} \Phi_{0,3} \right |^2 
\; + \;  \frac{\tilde{\gamma}}{4} \left \{ [\Phi_{0,1} , \Phi^{\dagger}_{0,1}] + [\Phi_{0,2} , \Phi^{\dagger}_{0,2}] + [\Phi_{0,3} , \Phi^{\dagger}_{0,3}] \right\}^2  \nonumber \\
& = &  \frac{\gamma}{4} \left | \Phi_{0,1} \Phi_{0,2} - e^{-2i \pi \beta_{12}} \Phi_{0,2} \Phi_{0,1} \right |^2  
\; + \; \frac{\gamma}{4} \left | \Phi_{0,2} \Phi_{0,3} - e^{-2i \pi \beta_{23}} \Phi_{0,3} \Phi_{0,2} \right |^2  \nonumber \\
& + & \frac{\gamma}{4} \left | \Phi_{0,3} \Phi_{0,1} - e^{-2i \pi \beta_{31}} \Phi_{0,1} \Phi_{0,3} \right |^2 ,
\label{mdnfoursusy}
\end{eqnarray}
where to obtain the last line in (\ref{mdnfoursusy}) we have used the fact that for untwisted fields  
\begin{eqnarray}
 [\Phi_{0,1} (x) , \Phi^{\dagger}_{0,1} (x)] & = & [\Phi_{0,2} (x) , \Phi^{\dagger}_{0,2} (x)] \quad = \quad [\Phi_{0,3} (x) , \Phi^{\dagger}_{0,3} (x)] \quad = \quad 0.
\label{sterm}
\end{eqnarray}
We can further simplify (\ref{mdnfoursusy}) and get 
\begin{eqnarray}
 \mathcal{H}_{0, \rm{int}} & = &   \frac{\gamma}{2}(1 - \cos 2\sigma_{12})\,\Phi^{\dagger}_{0,1} \Phi^{\dagger}_{0,2} \Phi_{0,1} \Phi_{0,2}  
\; + \;  \frac{\gamma}{2}(1 - \cos 2\sigma_{23})\,\Phi^{\dagger}_{0,2} \Phi^{\dagger}_{0,3}  \Phi_{0,2} \Phi_{0,3}  \nonumber \\
& + &  \frac{\gamma}{2} (1 - \cos 2\sigma_{31}) \, \Phi^{\dagger}_{0,3} \Phi^{\dagger}_{0,1} \Phi_{0,3} \Phi_{0,1} \nonumber \\
& = &  -\tilde{\gamma}' \sin^2 \sigma_{12} \: \Phi^{\dagger}_{0,1} \Phi^{\dagger}_{0,2} \Phi_{0,1} \Phi_{0,2}  
\; - \; \tilde{\gamma}' \sin^2 \sigma_{23} \: \Phi^{\dagger}_{0,2} \Phi^{\dagger}_{0,3} \Phi_{0,2} \Phi_{0,3} \nonumber \\ 
& - & \tilde{\gamma}' \sin^2 \sigma_{31} \: \Phi^{\dagger}_{0,3} \Phi^{\dagger}_{0,1} \Phi_{0,3} \Phi_{0,1} \nonumber \\
& = &  \frac{\gamma_{12}}{2}\: \Phi^{\dagger}_{0,1} \Phi^{\dagger}_{0,2} \Phi_{0,1} \Phi_{0,2}  \; + \; \frac{\gamma_{23}}{2} \: \Phi^{\dagger}_{0,2} \Phi^{\dagger}_{0,3} \Phi_{0,2} \Phi_{0,3} 
 \; + \; \frac{\gamma_{31}}{2} \: \Phi^{\dagger}_{0,3} \Phi^{\dagger}_{0,1} \Phi_{0,3} \Phi_{0,1} \nonumber \\
& = &  \mathcal{H}_{\theta, \rm{int}},
\label{mdn4susy}
\end{eqnarray}
where we have identified $2\sigma_{rs} =  - 2\pi\beta_{rs} = \lambda^{(r)}\wedge \lambda^{(s)}$ and $\tilde{\gamma}' = -\gamma$. Since $\sigma_{rs}$ 
and $\lambda^{(r)}\wedge \lambda^{(s)}$ are arbitrary parameters (due to arbitrariness of the components of $\theta$ matrix and of $\beta_{rs}$) so the above demand 
can be always satisfied. Hence, we infer that twisted scalar interaction Hamiltonian density (\ref{tgaugen4susy}) is equivalent to untwisted marginally deformed 
scalar interaction Hamiltonian density of (\ref{mdnfoursusy}). Interested readers can also look at \cite{Beisert:2005if} for a similar work.


\section{Generic Twists}


So far we have restricted our discussion to only a very particular type of deformed statistics, which we called ``antisymmetric twisted statistics''. 
Such a twist is characterized by the the commutation relations (\ref{atwistedcomrel}) and (\ref{antitwistfield}). We called it an antisymmetric twist because the 
$\theta$ matrix characterizing it was a antisymmetric matrix. Now we want to discuss twists which are more general in nature. 

Let us start with considering a more general dressing transformation, which is  
\begin{eqnarray}
 a^R_r & = &   c_r \, e^{-\frac{i}{2} \lambda^{(r)}_l \tilde{\theta}_{lm} Q_m},   \nonumber \\
(a_r^R)^\dagger & = &   e^{\frac{i}{2} \lambda^{(r)}_l \tilde{\theta}^\ast_{lm} Q_m} \, c^\dagger_r , \nonumber \\
 b^R_r  & = &  d_r \,  e^{\frac{i}{2} \lambda^{(r)}_l \tilde{\theta}^\ast_{lm} Q_m} , \nonumber \\
(b_r^R)^\dagger  & = &  e^{-\frac{i}{2} \lambda^{(r)}_l \tilde{\theta}_{lm} Q_m} \, d^\dagger_r , 
\label{dresstranformrmg}
\end{eqnarray}
where $\tilde{\theta}_{lm}$ is some arbitrary complex matrix and we have put a subscript $R$ to distinguish these transformations from the other possible transformations which are
\begin{eqnarray}
 a^L_r & = &    e^{-\frac{i}{2} \lambda^{(r)}_l \tilde{\theta}_{lm} Q_m}  \, c_r , \nonumber \\
(a_r^L)^\dagger & = &  c^\dagger_r \,  e^{\frac{i}{2} \lambda^{(r)}_l \tilde{\theta}^\ast_{lm} Q_m},   \nonumber \\
 b^L_r  & = &    e^{\frac{i}{2} \lambda^{(r)}_l \tilde{\theta}^\ast_{lm} Q_m} \, d_r , \nonumber \\
(b_r^L)^\dagger  & = &  d^\dagger_r \, e^{-\frac{i}{2} \lambda^{(r)}_l \tilde{\theta}_{lm} Q_m}.  
\label{dresstranformlmg}
\end{eqnarray}  
Unlike the antisymmetric twist case, these two transformations are not equivalent but are related to each other as 
 \begin{eqnarray}
 a^R_r & = &   c_r \, e^{-\frac{i}{2} \lambda^{(r)}_l \tilde{\theta}_{lm} Q_m}   
\, = \,  e^{-\frac{i}{2} \lambda^{(r)}_l \tilde{\theta}_{lm} \lambda^{(r)}_m} \,  e^{-\frac{i}{2} \lambda^{(r)}_l \tilde{\theta}_{lm} Q_m}  \, c_r  
\, = \,e^{-\frac{i}{2} \lambda^{(r)}_l \tilde{\theta}_{lm} \lambda^{(r)}_m} \,  a^L_r .
\label{relarnal}
\end{eqnarray}
Similar relations hold for all other operators. 

Now, consider the operator $N^R_r = (a_r^R)^\dagger a^R_r $. We have
\begin{eqnarray}
N^R_r & = & (a_r^R)^\dagger a^R_r \, = \, e^{\frac{i}{2} \lambda^{(r)}_l \tilde{\theta}^\ast_{lm} Q_m}\, c^\dagger_r\, c_r \, e^{-\frac{i}{2} \lambda^{(r)}_l \tilde{\theta}_{lm} Q_m}
\, = \,  c^\dagger_r\, c_r \, e^{\frac{i}{2}  \lambda^{(r)}_l ( \tilde{\theta}^\ast_{lm} \, - \, \tilde{\theta}_{lm} ) Q_m}.
\label{genrnumberop}
\end{eqnarray}
In this work we will restrict to the case of $N^R_r = c^\dagger_r\, c_r \, = \, N_{0,r}$, so that atleast the twisted free Hamiltonian is equivalent to its untwisted counterpart. 
The above condition implies that $\tilde{\theta}^\ast_{lm} \, = \, \tilde{\theta}_{lm} $ i.e. all elements of  $\tilde{\theta}$ are real. Hence forth we will restrict ourself to 
only real $\tilde{\theta}$ and denote it simply by $\theta$. Similar conditions will hold for left twists. 

Also we introduce the compact notation 
\begin{eqnarray}
\lambda^{(r)}_l \theta_{lm} Q_m & = & \lambda^{(r)} \vee Q_m  \quad \text{where} \, \theta \, \text{is a real arbitrary matrix }, \nonumber \\
\lambda^{(r)}_l \theta'_{lm} Q_m & = & \lambda^{(r)} \wedge Q_m  \quad \text{where}  \, \theta'_{lm}  =  - \theta'_{ml}  \, \text{is a real antisymmetric matrix}.
\label{veeproduct}
\end{eqnarray}
So the dressing transformations of (\ref{dresstranformlmg}) and (\ref{dresstranformrmg}) in this notation becomes
\begin{eqnarray}
 a^R_r & = &   c_r \, e^{-\frac{i}{2} \lambda^{(r)} \vee Q} ,  \nonumber \\
(a_r^R)^\dagger & = &   e^{\frac{i}{2} \lambda^{(r)}\vee Q} \, c^\dagger_r , \nonumber \\
 b^R_r  & = &  d_r \,  e^{\frac{i}{2} \lambda^{(r)} \vee Q} , \nonumber \\
(b_r^R)^\dagger  & = &  e^{-\frac{i}{2} \lambda^{(r)} \vee Q} \, d^\dagger_r , \nonumber \\
 a^L_r & = &    e^{-\frac{i}{2} \lambda^{(r)} \vee Q}  \, c_r , \nonumber \\
(a_r^L)^\dagger & = &  c^\dagger_r \,  e^{\frac{i}{2} \lambda^{(r)} \vee Q} ,  \nonumber \\
 b^L_r  & = &    e^{\frac{i}{2} \lambda^{(r)} \vee Q} \, d_r , \nonumber \\
(b_r^L)^\dagger  & = &  d^\dagger_r \, e^{-\frac{i}{2} \lambda^{(r)} \vee Q}.  
\label{dresstranformlrvee}
\end{eqnarray}  
The creation/annihilation operators defined in (\ref{dresstranformlrvee}) satisfy twisted statistics of the form
\begin{eqnarray}
 a^R_r \, a^R_s & = &   c_r \, e^{-\frac{i}{2} \lambda^{(r)} \vee Q} \,  c_s \, e^{-\frac{i}{2} \lambda^{(s)} \vee Q}  
\, = \, e^{\frac{i}{2} \lambda^{(r)} \vee \lambda^{(s)}} \,c_r \,c_s \, e^{-\frac{i}{2} \lambda^{(r)} \vee Q} \, e^{-\frac{i}{2} \lambda^{(s)} \vee Q}  \nonumber \\
& = & \eta \, e^{\frac{i}{2} \lambda^{(r)} \vee \lambda^{(s)}} \,c_s \,c_r \, e^{-\frac{i}{2} \lambda^{(r)} \vee Q} \, e^{-\frac{i}{2} \lambda^{(s)} \vee Q} \nonumber \\
& = & \eta \, e^{\frac{i}{2} \lambda^{(r)} \vee \lambda^{(s)}} \, e^{-\frac{i}{2} \lambda^{(s)} \vee \lambda^{(r)}} \,c_s \, e^{-\frac{i}{2} \lambda^{(s)} \vee Q} \,c_r \,
 e^{-\frac{i}{2} \lambda^{(r)} \vee Q} \nonumber \\
& = & \eta \, e^{\frac{i}{2} \lambda^{(r)}_l \theta_{lm} \lambda^{(s)}_m} \, e^{-\frac{i}{2} \lambda^{(s)}_l \theta_{lm} \lambda^{(r)}_m} \,c_s \, e^{-\frac{i}{2} \lambda^{(s)} \vee Q} 
\,c_r \,  e^{-\frac{i}{2} \lambda^{(r)} \vee Q} \nonumber \\
& = & \eta \, e^{\frac{i}{2} \lambda^{(r)}_l \theta_{lm} \lambda^{(s)}_m} \, e^{-\frac{i}{2} \lambda^{(r)}_l \theta_{ml} \lambda^{(s)}_m} \,c_s \, e^{-\frac{i}{2} \lambda^{(s)} \vee Q} 
\,c_r \,  e^{-\frac{i}{2} \lambda^{(r)} \vee Q} \nonumber \\
& = & \eta \, e^{\frac{i}{2} \lambda^{(r)}_l (\theta_{lm} - \theta_{ml})\lambda^{(s)}_m} \, c_s \, e^{-\frac{i}{2} \lambda^{(s)} \vee Q} 
\,c_r \,  e^{-\frac{i}{2} \lambda^{(r)} \vee Q} \nonumber \\
& = & \eta \, e^{i \lambda^{(r)} \wedge \lambda^{(s)}} \,a^R_s \, a^R_r .
\label{stwiststat}
\end{eqnarray}
where we have denoted $i \lambda^{(r)} \wedge \lambda^{(s)} = i \lambda^{(r)}_l \theta'_{lm}\lambda^{(s)}_m = \frac{i}{2} \lambda^{(r)}_l (\theta_{lm} - \theta_{ml})\lambda^{(s)}_m $,
$\theta'_{lm}$ being an antisymmetric matrix given by $2 \theta'_{lm} = \theta_{lm} - \theta_{ml}$. 

Similarly we find that 
\begin{eqnarray}
(a^R_r)^{\dagger} \, (a^R_s)^{\dagger} & = & \eta \,e^{ i \lambda^{(r)} \wedge \lambda^{(s)}} \, (a^R_s)^{\dagger} \, (a^R_r)^{\dagger},  \nonumber \\
 a^R_r \,  (a^R_s)^{\dagger} & = & \eta \,e^{ - i\lambda^{(r)} \wedge \lambda^{(s)}}  \, (a^R_s)^{\dagger} \,  a^R_r  
\,  + \, (2\pi)^3 \, 2 E_p \, \delta_{rs} \, \delta^{3}(p_{1} \, - \, p_{2}) ,\nonumber \\
b^R_r \, b^R_s & = & \eta \,e^{ i \lambda^{(r)} \wedge \lambda^{(s)}} \, b^R_s \, b^R_r , \nonumber \\
(b^R_r)^{\dagger} \, (b^R_s)^{\dagger} & = & \eta \,e^{ i \lambda^{(r)} \wedge \lambda^{(s)}} \, (b^R_s)^{\dagger} \, (b^R_r)^{\dagger},  \nonumber \\
b^R_r \, (b^R_s)^{\dagger} & = & \eta \,e^{ -i \lambda^{(r)} \wedge \lambda^{(s)}} \,  (b^R_s)^{\dagger} \, b^R_r
+ (2\pi)^3 \, 2 E_p \, \delta_{rs} \, \delta^{3}(p_{1} \, - \, p_{2}),
\label{stwistedcomrel}
\end{eqnarray}
which are all same as in (\ref{atwistedcomrel}). So making $\theta$ matrix arbitrary but same for every particle species does not result in a different twisted statistics.

To get possibly other types of twists, we have to take more general dressing transformations of the type 
\begin{eqnarray}
 a^R_r & = &   c_r \, e^{-\frac{i}{2} \lambda^{(r)}_l \tilde{\theta}^{(r)}_{lm} Q_m} ,  \nonumber \\
(a_r^R)^\dagger & = &   e^{\frac{i}{2} \lambda^{(r)}_l (\tilde{\theta}^{(r)})^\ast_{lm} Q_m} \, c^\dagger_r , \nonumber \\
 b^R_r  & = &  d_r \,  e^{\frac{i}{2} \lambda^{(r)}_l (\tilde{\theta}^{(r)})^\ast_{lm} Q_m} , \nonumber \\
(b_r^R)^\dagger  & = &  e^{-\frac{i}{2} \lambda^{(r)}_l \tilde{\theta}^{(r)}_{lm} Q_m} \, d^\dagger_r . 
\label{dresstranformmrmg}
\end{eqnarray}
Here $\tilde{\theta}^{(r)}$ is an arbitrary matrix but is not same for all particle species i.e.  $\tilde{\theta}^{(r)}_{lm} \neq \tilde{\theta}^{(s)}_{lm}$. 

Also, the left twists can be defined as  
\begin{eqnarray}
 a^L_r & = &    e^{-\frac{i}{2} \lambda^{(r)}_l \tilde{\theta}^{(r)}_{lm} Q_m}  \, c_r , \nonumber \\
(a_r^L)^\dagger & = &  c^\dagger_r \,  e^{\frac{i}{2} \lambda^{(r)}_l (\tilde{\theta}^{(r)})^\ast_{lm} Q_m},   \nonumber \\
 b^L_r  & = &    e^{\frac{i}{2} \lambda^{(r)}_l (\tilde{\theta}^{(r)})^\ast_{lm} Q_m} \, d_r , \nonumber \\
(b_r^L)^\dagger  & = &  d^\dagger_r \, e^{-\frac{i}{2} \lambda^{(r)}_l \tilde{\theta}^{(r)}_{lm} Q_m}.  
\label{dresstranformmlmg}
\end{eqnarray}  
As before, the left and right twists are not equivalent but are related to each other as 
 \begin{eqnarray}
 a^R_r & = &   c_r \, e^{-\frac{i}{2} \lambda^{(r)}_l \tilde{\theta}^{(r)}_{lm} Q_m}   
\, = \,  e^{-\frac{i}{2} \lambda^{(r)}_l \tilde{\theta}^{(r)}_{lm} \lambda^{(r)}_m} \,  e^{-\frac{i}{2} \lambda^{(r)}_l \tilde{\theta}^{(r)}_{lm} Q_m}  \, c_r  
\, = \,e^{-\frac{i}{2} \lambda^{(r)}_l \tilde{\theta}^{(r)}_{lm} \lambda^{(r)}_m} \,  a^L_r .
\label{mrelarnal}
\end{eqnarray}
Similar relations hold for all other operators. 

Again we consider the operator $N^R_r = (a_r^R)^\dagger a^R_r $ 
\begin{equation}
N^R_r \, = \, (a_r^R)^\dagger a^R_r \, = \, e^{\frac{i}{2} \lambda^{(r)}_l (\tilde{\theta}^{(r)})^\ast_{lm} Q_m}\, c^\dagger_r\, c_r 
\, e^{-\frac{i}{2} \lambda^{(r)}_l \tilde{\theta}^{(r)}_{lm} Q_m}
\, = \,  c^\dagger_r\, c_r \, e^{\frac{i}{2}  \lambda^{(r)}_l ( (\tilde{\theta}^{(r)})^\ast_{lm} \, - \, \tilde{\theta}^{(r)}_{lm} ) Q_m}.
\label{genrnumberopm}
\end{equation}
We restrict to the case of $N^R_r = c^\dagger_r\, c_r \, = \, N_{0,r}$, so that the twisted free Hamiltonian is equivalent to the untwisted one. 
The above condition implies that $(\tilde{\theta}^{(r)})^\ast_{lm} \, = \, \tilde{\theta}^{(r)}_{lm} $ i.e. all elements of  $\tilde{\theta}^{(r)}$ are real. 
Similar conditions hold for left twists. Henceforth, we will restrict to only real $\tilde{\theta}^{(r)}_{lm}$ and will denote it simply by $\theta^{(r)}_{lm}$.  

Also, we introduce the compact notation 
\begin{eqnarray}
\lambda^{(r)}_l \theta^{(r)}_{lm}  & = & 2 \alpha^{(r)}_m.
\label{veeproductm}
\end{eqnarray}
So the dressing transformations of (\ref{dresstranformmrmg}) and (\ref{dresstranformmlmg}) in this notation become
\begin{eqnarray}
 a^R_r & = &   c_r \, e^{-i \alpha^{(r)}_m Q_m} \, \equiv \, c_r \, e^{-i \alpha^{(r)}  Q} ,   \nonumber \\
(a_r^R)^\dagger & = &  e^{i \alpha^{(r)}_m Q_m} \, c^\dagger_r \, \equiv \, e^{i \alpha^{(r)} Q} \, c^\dagger_r , \nonumber \\
 b^R_r  & = &  d_r \, e^{i \alpha^{(r)}_m Q_m}  \, \equiv \,  d_r \, e^{i \alpha^{(r)} Q} , \nonumber \\
(b_r^R)^\dagger  & = & e^{-i \alpha^{(r)}_m Q_m} \, d^\dagger_r  \, \equiv \, e^{-i \alpha^{(r)} Q} \, d^\dagger_r , \nonumber \\
 a^L_r & = &   e^{-i \alpha^{(r)}_m Q_m}  \, c_r  \, \equiv \, e^{-i \alpha^{(r)} Q}  \, c_r , \nonumber \\
(a_r^L)^\dagger & = &  c^\dagger_r \, e^{i \alpha^{(r)}_m Q_m}  \, \equiv \,  c^\dagger_r \, e^{i \alpha^{(r)} Q} ,  \nonumber \\
 b^L_r  & = &  e^{i \alpha^{(r)}_m Q_m} \, d_r  \, \equiv \,  e^{i \alpha^{(r)} Q} \, d_r ,  \nonumber \\
(b_r^L)^\dagger  & = &  d^\dagger_r \, e^{-i \alpha^{(r)}_m Q_m}  \, \equiv \,   d^\dagger_r \, e^{-i \alpha^{(r)} Q}. 
\label{dresstranformalpha}
\end{eqnarray}
The creation/annihilation operators defined by (\ref{dresstranformalpha}) satisfy twisted statistics of the form
\begin{eqnarray}
 a^R_r \, a^R_s & = &   c_r \, e^{-i \alpha^{(r)} Q}  \,  c_s \, e^{-i \alpha^{(s)} Q}   
\, = \, e^{ i \alpha^{(r)}\lambda^{(s)}} \,c_r \,c_s \, e^{-i \alpha^{(s)} Q} \,  e^{-i \alpha^{(r)} Q}  \nonumber \\
& = & \eta \, e^{ i \alpha^{(r)}\lambda^{(s)}} \,c_s \,c_r \, e^{-i \alpha^{(s)} Q} \,  e^{-i \alpha^{(r)} Q} \nonumber \\
& = & \eta \, e^{ i \alpha^{(r)}\lambda^{(s)}} \, e^{ - i \alpha^{(s)}\lambda^{(r)}}  \,c_s \, e^{-i \alpha^{(s)} Q} \,c_r \, e^{-i \alpha^{(r)} Q} \nonumber \\
& = & \eta \, e^{\frac{i}{2} \lambda^{(r)}_l \theta^{(r)}_{lm} \lambda^{(s)}_m} \, e^{-\frac{i}{2} \lambda^{(s)}_l \theta^{(s)}_{lm} \lambda^{(r)}_m} \, a_s \, a_r \nonumber \\
& = & \eta \, e^{\frac{i}{2} \lambda^{(r)}_l \theta^{(r)}_{lm} \lambda^{(s)}_m} \, e^{-\frac{i}{2} \lambda^{(r)}_l \theta^{(s)}_{ml} \lambda^{(s)}_m} \, a_s \, a_r \nonumber \\
& = & \eta \, e^{\frac{i}{2} \lambda^{(r)}_l \left(\theta^{(r)}_{lm} - \theta^{(s)}_{ml} \right) \lambda^{(s)}_m} \, a_s \, a_r .
\label{mstwiststat}
\end{eqnarray}
Since, $\theta^{(r)}_{lm} \neq \theta^{(s)}_{lm}$, and the $\theta$s are arbitrary matrices so $2 \theta'_{lm} = \theta^{(r)}_{lm} - \theta^{(s)}_{ml}$ also remains an arbitrary matrix
and hence (\ref{mstwiststat}) gives more general twisted statistics. Also for twisted bosons $\eta = 1$ should be taken and for twisted fermions $\eta = -1$ is to be taken. 

Similarly we find that 
\begin{eqnarray}
(a^R_r)^{\dagger} \, (a^R_s)^{\dagger} & = & \eta \, e^{ i \left( \alpha^{(r)}\lambda^{(s)} -  \alpha^{(s)}\lambda^{(r)} \right)} \, (a^R_s)^{\dagger} \, (a^R_r)^{\dagger}  
\, = \, \eta \, e^{\frac{i}{2} \lambda^{(r)}_l \left(\theta^{(r)}_{lm} - \theta^{(s)}_{ml} \right) \lambda^{(s)}_m}  \, (a^R_s)^{\dagger} \, (a^R_r)^{\dagger}, \nonumber \\
b^R_r \, b^R_s & = & \eta \, e^{ i \left( \alpha^{(r)}\lambda^{(s)} -  \alpha^{(s)}\lambda^{(r)} \right)} \, b^R_s \, b^R_r  
\, = \, \eta \, e^{\frac{i}{2} \lambda^{(r)}_l \left(\theta^{(r)}_{lm} - \theta^{(s)}_{ml} \right) \lambda^{(s)}_m}  \, b^R_s \, b^R_r  , \nonumber \\
(b^R_r)^{\dagger} \, (b^R_s)^{\dagger} & = & \eta \, e^{ i \left( \alpha^{(r)}\lambda^{(s)} -  \alpha^{(s)}\lambda^{(r)} \right)} \, (b^R_s)^{\dagger} \, (b^R_r)^{\dagger}  
\, = \, \eta \, e^{\frac{i}{2} \lambda^{(r)}_l \left(\theta^{(r)}_{lm} - \theta^{(s)}_{ml} \right) \lambda^{(s)}_m}  \, (b^R_s)^{\dagger} \, (b^R_r)^{\dagger} , \nonumber \\
 a^R_r \,  (a^R_s)^{\dagger} & = & \eta \, e^{ - i \left( \alpha^{(r)}\lambda^{(s)} -  \alpha^{(s)}\lambda^{(r)} \right)}   \, (a^R_s)^{\dagger} \,  a^R_r  
\,  + \, (2\pi)^3 \, 2 E_p \, \delta_{rs} \, \delta^{3}(p_{1} \, - \, p_{2}) \nonumber \\
& = & \, \eta \, e^{-\frac{i}{2} \lambda^{(r)}_l \left(\theta^{(r)}_{lm} - \theta^{(s)}_{ml} \right) \lambda^{(s)}_m}  \, (a^R_s)^{\dagger} \,  a^R_r  
\,  + \, (2\pi)^3 \, 2 E_p \, \delta_{rs} \, \delta^{3}(p_{1} \, - \, p_{2}) ,\nonumber \\
b^R_r \, (b^R_s)^{\dagger} & = & \eta \, e^{ -i \left( \alpha^{(r)}\lambda^{(s)} -  \alpha^{(s)}\lambda^{(r)} \right)} \, (b^R_s)^{\dagger} \, b^R_r
+ (2\pi)^3 \, 2 E_p \, \delta_{rs} \, \delta^{3}(p_{1} \, - \, p_{2}) \nonumber \\
& = & \eta \, e^{-\frac{i}{2} \lambda^{(r)}_l \left(\theta^{(r)}_{lm} - \theta^{(s)}_{ml} \right) \lambda^{(s)}_m} \,  (b^R_s)^{\dagger} \, b^R_r
+ (2\pi)^3 \, 2 E_p \, \delta_{rs} \, \delta^{3}(p_{1} \, - \, p_{2}), \nonumber \\
a^R_r \, (b^R_s)^{\dagger} & = & \eta \, e^{ i \left( \alpha^{(r)}\lambda^{(s)} -  \alpha^{(s)}\lambda^{(r)} \right)} \, (b^R_s)^{\dagger} \,  a^R_r
\, = \, \eta \, e^{\frac{i}{2} \lambda^{(r)}_l \left(\theta^{(r)}_{lm} - \theta^{(s)}_{ml} \right) \lambda^{(s)}_m}  \, (b^R_s)^{\dagger} \,  a^R_r ,\nonumber \\
a^R_r \, b^R_s & = & \eta \, e^{ - i \left( \alpha^{(r)}\lambda^{(s)} -  \alpha^{(s)}\lambda^{(r)} \right)} \, b^R_s \, a^R_r
\, = \, \eta \, e^{- \frac{i}{2} \lambda^{(r)}_l \left(\theta^{(r)}_{lm} - \theta^{(s)}_{ml} \right) \lambda^{(s)}_m}  \, b^R_s \, a^R_r.
\label{mstwistedcomrel}
\end{eqnarray}
From (\ref{mstwistedcomrel}) it is clear that the antisymmetric twist discussed in the previous section is just a special case of this generic twist. If we take 
$\theta^{(r_1)} = \theta^{(r_2)} = \cdots = \theta^{(r_N)} = \theta$ and $\theta_{lm} = - \theta_{ml}$ in (\ref{mstwistedcomrel}), we will recover back the antisymmetric twisted
statistics of the previous section. Moreover, unlike the case of antisymmetric twist, where due to antisymmetry of the $\theta$ matrix, it was not possible to get twisted 
statistics for an internal symmetry group of rank less than 2, in this case, we can have twisted statistics for SU(2) as well as U(1) group.

Using the twisted creation/annihilation operators, the left and right twisted quantum fields $\phi^{L,R}_{\theta,r}$ can be composed as \footnote{One can also compose fields with 
only left twisted or right twisted creation/annihilation 
operators but fields theories with such quantum fields are tricky to write and one has to introduce quantities like ``complex mass'' to write these theories. 
We will not discuss them in this work. }
\begin{eqnarray}
\phi^R_{\theta, r} (x) & = & \int \frac{d^3 p}{(2\pi)^3} \frac{1}{2E_{p}} \left[ a^R_r(p) e^{-i px} \, + \, (b^L_r)^{\dagger}(p) e^{i px} \right ], \nonumber \\
\phi^L_{\theta, r} (x) & = & \int \frac{d^3 p}{(2\pi)^3} \frac{1}{2E_{p}} \left[ a^L_r(p) e^{-i px} \, + \, (b^R_r)^\dagger (p) e^{i px} \right ], \nonumber \\
(\phi^R_{\theta, r})^\dagger (x) & = & \int \frac{d^3 p}{(2\pi)^3} \frac{1}{2E_{p}} \left[ b^L_r(p) e^{-i px} \, + \, (a^R_r)^{\dagger}(p) e^{i px} \right ], 
\nonumber 
\end{eqnarray}
\begin{eqnarray}
(\phi^L_{\theta, r})^\dagger (x) & = & \int \frac{d^3 p}{(2\pi)^3} \frac{1}{2E_{p}} \left[ b^R_r(p) e^{-i px} \, + \, (a^L_r)^\dagger (p) e^{i px} \right ] .
\label{modeexpansionagen}
\end{eqnarray}

Using the dressing transformations of (\ref{dresstranformalpha}), it is easy to check that the above defined fields satisfy the dressing transformations 
\begin{eqnarray}
\phi^R_{\theta, r} (x) & = &  \phi_{0, r} (x) \, e^{-i \alpha^{(r)}  Q},  \nonumber \\
\phi^L_{\theta, r} (x) & = &  e^{-i \alpha^{(r)}  Q} \, \phi_{0, r} (x), \nonumber \\
(\phi^R_{\theta, r})^\dagger (x) & = & e^{i \alpha^{(r)}  Q} \, \phi^\dagger_{0, r} (x), \nonumber \\
(\phi^L_{\theta, r})^\dagger (x) & = &  \phi^\dagger_{0, r} (x) \, e^{i \alpha^{(r)}  Q}. 
\label{dressfieldsg}
\end{eqnarray}
Also, we have 
\begin{eqnarray}
 \left[ Q_m , \phi^{L,R}_{\theta, r} (x) \right ] & = & -\lambda^{(r)}_m \phi^{L,R}_{\theta, r} (x),  \nonumber \\
 \left[ Q_m , (\phi^{L,R}_{\theta, r})^{\dagger} (x) \right ] & = &  \lambda^{(r)}_m (\phi^{L,R}_{\theta, r})^{\dagger} (x).
\label{chargegen}
\end{eqnarray}
As before, the Fock space states can be constructed using these twisted operators. We assume (with similar justification as for the case of antisymmetric twists) that the vacuum 
of the twisted theory is same as that for untwisted theory. The multi-particle states can be obtained by acting the twisted creation operators on the vacuum state. 
Because of the twisted statistics (\ref{mstwistedcomrel}), there is an ambiguity in defining the action of the twisted creation and annihilation operators on Fock space states.
Like the previous case, we choose to define $a^\dagger_r (p)$, $p$ being the momentum label and $a^\dagger_r$ standing for either of the left or right twisted creation operators, 
to be an operator which adds a particle to the right of the particle list i.e.
\begin{eqnarray}
a^\dagger_r (p) | p_1,r_1; \, p_2,r_2;\, \dots \, p_n,r_n \rangle_{\theta} & = & | p_1,r_1;\, p_2,r_2;\, \dots \, p_n,r_n;\, p,r \rangle_{\theta} .
\label{actiontcre}
\end{eqnarray}
With this convention, the single-particle Fock space states for the twisted theory are given by 
\begin{eqnarray}
\overline{|p, r \rangle}_{\theta} & = &   b^\dagger_r (p) | 0 \rangle , \nonumber \\
|p, r \rangle_{\theta} & = &  a^\dagger_r (p) | 0 \rangle.
\label{gsat}
\end{eqnarray}
The multi-particle states are given by
\begin{eqnarray}
\overline{| p_1, r_1; \, p_2,r_2; \, \dots \, p_n,r_n \rangle}_{\theta} & = &  b^\dagger_{r_n} (p_n) \dots b^\dagger_{r_2} (p_2) b^\dagger_{r_1} (p_1) | 0 \rangle ,\nonumber \\
|p_1, r_1; \, p_2,r_2; \, \dots \, p_n,r_n \rangle_{\theta} & = &  a^\dagger_{r_n} (p_n) \dots a^\dagger_{r_2} (p_2) a^\dagger_{r_1} (p_1) | 0 \rangle .
\label{gmat}
\end{eqnarray}
Owing to the twisted commutation relations of (\ref{mstwistedcomrel}), the state vectors also satisfy a similar twisted relation e.g. for two-particle states we have 
\begin{eqnarray}
\overline{| p_2, r_2; \, p_1,r_1 \rangle}_{\theta} & = & e^{ i \left( \alpha^{(r)}\lambda^{(s)} - \alpha^{(s)}\lambda^{(r)} \right)} 
\,\overline{| p_1, r_1; \, p_2,r_2 \rangle}_{\theta}, \nonumber \\
 |p_2, r_2; \, p_1,r_1 \rangle_{\theta} & = &  e^{ i \left( \alpha^{(r)}\lambda^{(s)} -  \alpha^{(s)}\lambda^{(r)} \right)}  \, |p_1, r_1; \, p_2,r_2 \rangle_{\theta} . 
\label{g2statecom}
\end{eqnarray}

The $SU(N)$ transformations of the twisted fields can be discussed in a way similar to the previous section. For example, the fields $\phi^L_{\theta, r}$ 
transform under $SU(N)$ as
\begin{eqnarray}
U(\sigma) \phi^L_{\theta, r} (x) U^\dagger(\sigma) & = &  \phi^{L'}_{\theta, r} (x) \, = \,\, U(\sigma) e^{-i \alpha^{(r)}  Q} \, U^\dagger(\sigma) \,
U(\sigma) \phi_{0, r} (x) U^\dagger(\sigma) \nonumber \\ 
& = & U(\sigma) e^{-i \alpha^{(r)}  Q} U^\dagger(\sigma) \, \left( e^{-i \sigma_a T_a} \right)_{rs} \,\phi_s (x) \, 
\, = \, \zeta_{(r)}(\sigma) \, \left( e^{-i \sigma_a T_a} \right)_{rs} \,\phi_s (x),  \nonumber \\
U(\sigma) (\phi^L_{\theta, r})^\dagger (x) U^\dagger(\sigma) & = & (\phi^{L'}_{\theta, r})^\dagger  (x) \, = \, U(\sigma) \phi^\dagger_{0, r} (x) U^\dagger(\sigma) 
\, U(\sigma) e^{i \alpha^{(r)}  Q} U^\dagger(\sigma) \nonumber \\ 
& = & \left( e^{i \sigma_a T_a} \right)_{sr} \phi^\dagger_s (x)  U(\sigma)  e^{i \alpha^{(r)}  Q} U^\dagger(\sigma) 
\, = \,\left( e^{i \sigma_a T_a} \right)_{sr} \phi^\dagger_s (x) \zeta^\dagger_{(r)}(\sigma), 
\label{suntranformgen}
\end{eqnarray}
where $\zeta_{(r)}(\sigma) = U(\sigma)  e^{-i \alpha^{(r)}  Q} U^\dagger(\sigma)$ is a unitary operator satisfying 
$\zeta_{(r)}(\sigma)\zeta^\dagger_{(r)}(\sigma) = $ \, $\zeta^\dagger_{(r)}(\sigma) \zeta_{(r)}(\sigma) = \textbf{I}$. Similar relations hold for $\phi^R_{\theta, r}$ fields also. 

The transformation properties of the state vectors can be similarly discussed. For example, assuming that vacuum remains invariant under the transformations 
i.e. $ U(\sigma) | 0 \rangle = | 0 \rangle$, the single-particle states transform as
\begin{eqnarray}
 U(\sigma) | r \rangle_\theta & = &  U(\sigma) a^\dagger_r | 0 \rangle \, = \, U(\sigma) a^\dagger_r U^\dagger(\sigma) U(\sigma) | 0 \rangle 
 \, = \, U(\sigma) c^\dagger_r \, e^{i \alpha^{(r)}  Q} \, U^\dagger(\sigma) U(\sigma) | 0 \rangle \nonumber \\ 
& = &  U(\sigma) c^\dagger_r U^\dagger(\sigma)| 0 \rangle \, = \, \left( e^{i \sigma_a T_a} \right)_{sr} \,c^\dagger_s | 0 \rangle 
\, = \, \left( e^{i \sigma_a T_a} \right)_{sr} \,a^\dagger_s | 0 \rangle \, = \, \left( e^{i \sigma_a T_a} \right)_{sr}  | s \rangle_\theta, \nonumber \\
 U(\sigma) \overline{| r \rangle}_\theta & = &  U(\sigma) b^\dagger_r | 0 \rangle \, = \, U(\sigma) b^\dagger_r U^\dagger(\sigma) U(\sigma) | 0 \rangle 
 \, = \, U(\sigma) d^\dagger_r \, e^{-i \alpha^{(r)}  Q} \, U^\dagger(\sigma) U(\sigma) | 0 \rangle \nonumber \\
& = & U(\sigma) d^\dagger_r U^\dagger(\sigma)| 0 \rangle \, = \, \left( e^{-i \sigma_a T^\ast_a} \right)_{sr} \,d^\dagger_s | 0 \rangle
\, = \, \left( e^{-i \sigma_a T^\ast_a} \right)_{sr} \,b^\dagger_s | 0 \rangle  \, = \, \left( e^{-i \sigma_a T^\ast_a} \right)_{sr}  \overline{| s \rangle}_\theta .\nonumber \\
\label{1tstesuntrans}
\end{eqnarray}
Again, the multi-particle states follow twisted transformation rules, e.g. the two-particle states transform as
\begin{eqnarray}
 U(\sigma) | r , s \rangle_\theta & = & e^{i \alpha^{(s)} \lambda^{(r)}} \, e^{-i \alpha^{(t)} \lambda^{(u)}} \, \left( e^{i \sigma_a T_a} \right)_{ts}  
\left( e^{i \sigma_a T_a} \right)_{ur} \, | u ,t \rangle_\theta, \nonumber \\
U(\sigma) \overline{| r, s \rangle}_\theta & = & e^{i \alpha^{(s)} \lambda^{(r)}} \, e^{-i \alpha^{(t)} \lambda^{(u)}}  \, \left( e^{-i \sigma_a T^\ast_a} \right)_{ts}\, 
\left( e^{-i \sigma_a T^\ast_a} \right)_{ur}\, \overline{| u, t \rangle}_\theta.
\label{2atstatvecsuntrans}
\end{eqnarray}
Since the left and right twisted fields have analogous properties, so henceforth we will consider only left twisted fields and will drop the superscript `` L '' from it. 
All the computations and conclusions applicable to left twisted fields can be equally applied to right twisted fields. 

Now we have to define the analogue of star-product of previous section. We define the ``generic star-product'' $\ast$ as 
\begin{eqnarray}
 \phi^\#_{\theta, r} (x) \ast \phi^\#_{\theta, s}(y) & = &  \phi^\#_{\theta, r} (x) \, e^{ \frac{i}{2} \left( \left( \pm \alpha^{(s)} \right) \overleftarrow{Q} 
- \left(\pm \alpha^{(r)} \right) \overrightarrow{Q} \right)} \, \phi^\#_{\theta, s} (y)\nonumber \\
& = &  \phi^\#_{\theta, r} (x) \phi^\#_{\theta, s} (y) \, + \, \frac{i}{2} \left\{ \left(\pm \alpha^{(s)}_l \right) \left[ Q_l, \phi^\#_{\theta, r} (x)\right] 
- \left(\pm \alpha^{(r)}_m \right) \left[ Q_m, \phi^\#_{\theta, s} (y) \right] \right \} \nonumber \\
& + & \frac{1}{2!} \left(\frac{i}{2}\right)^2 \, \left\{ \left(\pm \alpha^{(s)}_l \right) \left(\pm \alpha^{(s)}_n \right) \left[ Q_l, \left[ Q_n,\phi^\#_{\theta, r} (x)\right]\right] 
\right. \nonumber \\
& - & \left. \left(\pm \alpha^{(r)}_m \right) \left(\pm \alpha^{(r)}_p \right) \left[ Q_m, \left[ Q_p,\phi^\#_{\theta, s} (y) \right] \right] \right \} \, + \,  \cdots,
\label{istarproduct}
\end{eqnarray}
where $ + \alpha^{(r)}$ is to be taken if the field $\phi^\#_{\theta, r}$ stands for  $\phi^\dagger_{\theta, r}$ and $ - \alpha^{(r)}$ 
if  $\phi^\#_{\theta, r}$ stands for  $\phi_{\theta, r}$. 

Due to the relation (\ref{chargegen}), we have 
\begin{eqnarray}
 \phi^\#_{\theta, r} (x) \ast \phi^\#_{\theta, s}(y) & = &   e^{ \frac{i}{2} \left( \left( \pm \alpha^{(s)}\right) \left( \pm \lambda^{(r)}\right) 
- \left( \pm \alpha^{(r)}\right) \left( \pm \lambda^{(s)}\right) \right)}  \,\phi^\#_{\theta, r} (x) \, \cdot \, \phi^\#_{\theta, s}(y),
\label{astproductrel}
\end{eqnarray}
where $\alpha^{(r)}$, $\lambda^{(r)}$ should be taken if $\phi^\#_{\theta, r} = \phi^\dagger_{\theta, r}$ and $ - \alpha^{(r)}$, $-\lambda^{(r)}$ should be taken if 
$\phi^\#_{\theta, r} = \phi_{\theta, r}$. So again the $\ast$-product is nothing but multiplying a certain phase factor to ordinary product. Nonetheless like in the 
previous case, using $\ast$-product will simplify things and the product should be regarded as a compact notation used for convenience. 

Having defined the product rule on single fields we have to now define how it acts on a composition of fields like $\left(\phi^\#_{\theta, r} \phi^\#_{\theta, s}\right)$.
Demanding that our product remains associative, the action of the $\ast$-product on composition of fields is defined as
\begin{eqnarray}
 \phi^\#_{\theta, r} \ast \left( \phi^\#_{\theta, s} \phi^\#_{\theta, t} \right) & = &  \phi^\#_{\theta, r} \, 
e^{ \frac{i}{2} \left( \left( \pm \alpha^{(s)} \pm \alpha^{(t)} \right) \overleftarrow{Q} - \left(\pm \alpha^{(r)} \right) \overrightarrow{Q} \right)} \,
\left( \phi^\#_{\theta, s} \phi^\#_{\theta, t} \right) \nonumber \\
& = &  \phi^\#_{\theta, r} \phi^\#_{\theta, s} \phi^\#_{\theta, t} \, + \, \frac{i}{2} \left\{ \left(\pm \alpha^{(s)}_l \pm \alpha^{(t)}_l \right) 
\left[ Q_l, \phi^\#_{\theta, r} \right] - \left(\pm \alpha^{(r)}_m \right) \left[ Q_m, \left( \phi^\#_{\theta, s} \phi^\#_{\theta, t} \right) \right] \right \} \nonumber \\
& + & \frac{1}{2!} \left(\frac{i}{2}\right)^2 \, \left\{ \left(\pm \alpha^{(s)}_l \pm \alpha^{(t)}_l \right) \left(\pm \alpha^{(s)}_n \pm \alpha^{(t)}_n \right)
 \left[ Q_l, \left[ Q_n,\phi^\#_{\theta, r} \right]\right] \right. \nonumber \\
& - & \left. \left(\pm \alpha^{(r)}_m \right) \left(\pm \alpha^{(r)}_p \right) \left[ Q_m, \left[ Q_p,\left( \phi^\#_{\theta, s} \phi^\#_{\theta, t} \right) \right] \right] \right \} 
\, + \, \cdots
\label{3starproduct}
\end{eqnarray}
Similar action of $\ast$-product applies for composition of multiple fields
\begin{eqnarray}
\left( \phi^\#_{\theta, r_1} \cdots \phi^\#_{\theta, r_N} \right) \ast \left( \phi^\#_{\theta, s_1} \cdots \phi^\#_{\theta, s_M} \right) 
& = &  \left( \phi^\#_{\theta, r_1} \cdots \phi^\#_{\theta, r_N} \right) \, 
e^{\frac{i}{2} \left(\left(\pm \alpha^{(s_1)}\cdots \pm \alpha^{(s_M)} \right) \overleftarrow{Q} - \left(\pm \alpha^{(r_1)}\cdots \pm \alpha^{(r_N)} \right)\overrightarrow{Q} \right)} \, 
\nonumber \\
& & \left( \phi^\#_{\theta, s_1} \cdots \phi^\#_{\theta, s_M} \right) .
\label{mstarproduct}
\end{eqnarray}
The above defined $\ast$-product can be viewed as the internal space analogue of another widely studied spacetime product called ``Dipole Product'' \cite{dipole-keshav}.

The star-product introduced here has the property that
\begin{eqnarray}
 \phi^\#_{\theta,r} \, \ast  \, \phi^\#_{\theta, s} & = &  \phi^\#_{\theta, s} \, \ast  \, \phi^\#_{\theta,r}, \nonumber \\
 \phi^\#_{\theta,r} \, \ast  \, \left( \phi^\#_{\theta, s} \, \ast  \, \phi^\#_{\theta, t} \right) 
& = & \left( \phi^\#_{\theta,r} \, \ast  \, \phi^\#_{\theta, s} \right)\, \ast  \, \phi^\#_{\theta, t} ,
\label{astidanti1}
\end{eqnarray}
and by construction we have
\begin{eqnarray}
 \phi^\#_{\theta,r} \, \ast  \, \phi^\#_{\theta, r} & = &  \phi^\#_{\theta, r} \cdot \phi^\#_{\theta,r} .
\label{astidanti2}
\end{eqnarray}
Also, because of the dressing transformations (\ref{fieldsantidress}) we have
\begin{eqnarray} \hspace{-.3cm}
 \phi^\#_{\theta,r_1} \, \ast  \, \phi^\#_{\theta, r_2} \, \ast  \, \cdots \, \ast  \, \phi^\#_{\theta, r_n} & = & \kappa \, \phi^\#_{0,r_1} \, \phi^\#_{0, r_2} \, \cdots \, 
\phi^\#_{0, r_n} \, e^{i \left( \pm \alpha^{(r_1)} \, \pm \,  \alpha^{(r_2)} \, \pm \, \cdots \, \pm \,  \alpha^{(r_n)} \right) Q },
\label{astidanti3}
\end{eqnarray}
where $\kappa $ is a phase factor whose explicit form depends on whether  $ \phi^\#_{\theta,r}$ stands for $\phi^\dagger_{\theta,r}$ or $ \phi_{\theta,r}$ field.
Also, $ + \alpha^{(r)}$ is to be taken if $ \phi^\#_{\theta,r} = \phi^\dagger_{\theta,r}$ and $ -\alpha^{(r)}$ if $ \phi^\#_{\theta,r} = \phi_{\theta,r}$. 

Having defined the $\ast$-product, the field theories can be conveniently written by following the rule that, for any given untwisted Hamiltonian its twisted counterpart 
should be written by replacing the untwisted fields $\phi_{0,r}$ by the twisted fields $\phi_{\theta,r}$ and the ordinary product between fields by the $\ast$-product. 

Using the above rule, the free theory Hamiltonian density $\mathcal{H}_{\theta, F}$ in terms of twisted fields can be written as
 \begin{eqnarray}
  \mathcal{H}_{\theta, F} & = &  \Pi^{\dagger}_{\theta,r} \ast  \Pi_{\theta,r}  \, + \, (\partial_i \phi^{\dagger}_{\theta,r}) \ast (\partial^i \phi_{\theta,r})
 \, + \, m^2  \, \phi^{\dagger}_{\theta,r} \ast  \phi_{\theta,r} \nonumber \\
& = &  \Pi^{\dagger}_{\theta,r} \Pi_{\theta,r}  \, + \, (\partial_i \phi^{\dagger}_{\theta,r}) (\partial^i \phi_{\theta,r})  \, + \, m^2  \, \phi^{\dagger}_{\theta,r} \phi_{\theta,r},
\label{freetham}
 \end{eqnarray}
where $\Pi_{\theta,r}$ is the canonical conjugate of $\phi_{\theta,r}$; $r = 1, 2, \cdots, N$ and to obtain the last line in (\ref{freetham}) we have used (\ref{astidanti2}). 
The Hamiltonian density in (\ref{freetham}) is invariant under the $SU(N)$ global transformations which can be explicitly checked by using (\ref{suntranformgen}). 

The renormalizable $SU(N)$ invariant interaction Hamiltonian density is given by
\begin{eqnarray}
\mathcal{H}_{\theta,\rm{Int}} & = &  \frac{\gamma}{4} \, \phi^{\dagger}_{\theta,r}  \ast   \phi^{\dagger}_{\theta,s}  \ast  \phi_{\theta,r}  \ast  \phi_{\theta,s} \nonumber \\
& = & \frac{\gamma}{4}\, e^{ i \left( \alpha^{(s)}\lambda^{(r)} -  \alpha^{(r)}\lambda^{(s)} \right)} \, \phi^{\dagger}_{\theta,r} \phi^{\dagger}_{\theta,s} 
\phi_{\theta,r} \phi_{\theta,s},
\label{intgenham}
\end{eqnarray}
where $r,s = 1, 2, \cdots, N$. Using (\ref{suntranformgen}) one can check that (\ref{intgenham}) is indeed invariant under $SU(N)$ group. 
The presence of phases of the type $e^{ i \left( \alpha^{(s)}\lambda^{(r)} -  \alpha^{(r)}\lambda^{(s)} \right)} $ in (\ref{intgenham}) means that unlike the untwisted case,
the demand of $SU(N)$ invariance forces the various terms to have different couplings related with each other in a specific way. 

Again like the previous section, the discussion in this section can be generalized in a straightforward manner to include spinor fields and twisted fermions. For that we have to take
anticommuting creation/annihilation operators and the twisted fermions will again satisfy (\ref{mstwistedcomrel}) but with $\eta = -1$. Also, the above discussion can be easily 
generalized to higher dimensional representations of $SU(N)$ group as well as to other symmetry groups like $SO(N)$.


\subsection{$S$-matrix Elements}


Like the previous case of antisymmetric twists, here also by construction we have ensured that the twisted $SU(N)$ invariant free Hamiltonian density $\mathcal{H}_{\theta, F}$ 
remains same as the untwisted one i.e.
 \begin{eqnarray}
  \mathcal{H}_{\theta, F} & = &  \Pi^{\dagger}_{\theta,r} \ast \Pi_{\theta,r}  \, + \, (\partial_i \phi^{\dagger}_{\theta,r}) \ast(\partial^i \phi_{\theta,r})
 \, + \, m^2  \, \phi^{\dagger}_{\theta,r} \ast \phi_{\theta,r} \nonumber \\
& = &  \Pi^{\dagger}_{0,r} \Pi_{0,r}  \, + \, (\partial_i \phi^{\dagger}_{0,r}) (\partial^i \phi_{0,r}) \, + \, m^2  \, \phi^{\dagger}_{0,r} \phi_{0,r}
\, = \,  \mathcal{H}_{0,F}, 
\label{freegenhamrel}
 \end{eqnarray}
where to obtain the last line we have used the dressing transformation (\ref{dressfieldsg}) and the relation (\ref{astidanti3}). Similarly, the twisted $SU(N)$ invariant interaction
term can be shown to be same as the untwisted one
 \begin{eqnarray} \hspace{-.6cm}
  \mathcal{H}_{\theta, \rm{Int}} & = &  \Pi^{\dagger}_{\theta,r} \ast \Pi_{\theta,r}  \, + \, (\partial_i \phi^{\dagger}_{\theta,r}) \ast(\partial^i \phi_{\theta,r})
 \, + \, m^2  \, \phi^{\dagger}_{\theta,r} \ast \phi_{\theta,r} 
\, + \,  \frac{\gamma}{4} \, \phi^{\dagger}_{\theta,r}  \ast  \phi^{\dagger}_{\theta,s}  \ast \phi_{\theta,r}  \ast \phi_{\theta,s} \nonumber \\
& = &  \Pi^{\dagger}_{0,r} \Pi_{0,r}   +  (\partial_i \phi^{\dagger}_{0,r}) (\partial^i \phi_{0,r})  +  m^2   \phi^{\dagger}_{0,r} \phi_{0,r}
 +   \frac{\gamma}{4}  \phi^{\dagger}_{0,r}  \phi^{\dagger}_{0,s}  \phi_{0,r}  \phi_{0,s} 
 \, = \, \mathcal{H}_{0, \rm{Int}} .
\label{intgenhamrel}
 \end{eqnarray}
Since the twisted in/out states also contain information about twisted statistics so we should again look at the $S$-matrix elements. For a typical $S$-matrix element, like for the 
scattering process of $\phi_{\theta, r} \phi_{\theta, s} \rightarrow \phi_{\theta, r} \phi_{\theta, s}$ we have
 \begin{eqnarray}
  S\left[ \phi_{\theta, r} \phi_{\theta, s} \rightarrow \phi_{\theta, r} \phi_{\theta, s}  \right] 
& = & \leftidx{_{out,\theta}}{\left \langle r s | r s  \right \rangle }{_{ \theta,in}} 
\, = \, \leftidx{_{\theta}}{\left \langle r s | S_\theta | r s \right \rangle }{_{ \theta}},
\label{smatgen}
 \end{eqnarray}
where $S_\theta = \mathcal{T}  \exp \left[-i\int^{\infty}_{-\infty} d^{4}z \mathcal{H}_{\theta, \rm{ Int}} (z) \right] $ is the $S$-operator and we have denoted the two-particle in and 
out states by $\left| r s \right \rangle_\theta = a^\dagger_s a^\dagger_r | 0 \rangle$. Because of (\ref{intgenhamrel}) we have 
\begin{eqnarray}
 S_\theta & = & \mathcal{T}  \exp \left[-i\int^{\infty}_{-\infty} d^{4}z \mathcal{H}_{\theta, \rm{Int}} (z) \right]  
\, = \, \mathcal{T}  \exp \left[-i\int^{\infty}_{-\infty} d^{4}z \mathcal{H}_{0, \rm{Int}} (z) \right]  \, = \, S_0.
\label{sopgen}
\end{eqnarray}
Also we have 
\begin{eqnarray}
 \left| r s \right \rangle_\theta & = & a^\dagger_s a^\dagger_r  \left| 0 \right\rangle 
\, = \, e^{i \alpha^{(s)} \lambda^{(r)} } \,  c^\dagger_s \, c^\dagger_r \,\left| 0 \right\rangle  
\, = \, e^{i \alpha^{(s)} \lambda^{(r)} } \, \left| r s \right \rangle_0.
\label{stategen}
\end{eqnarray}
Using (\ref{stategen}) and (\ref{sopgen}) we have
 \begin{eqnarray}
  S\left[ \phi_{\theta, r} \phi_{\theta, s} \rightarrow \phi_{\theta, r} \phi_{\theta, s}  \right] & = & 
\leftidx{_{0}}{\left \langle r s \left| \, e^{-i \alpha^{(s)} \lambda^{(r)} } \, S_0 \, e^{i \alpha^{(s)} \lambda^{(r)} } \, \right| r s \right \rangle }{_{ 0}}
\nonumber \\
& = & \leftidx{_{0}}{\left \langle r s | \, S_0 \, | r s \right \rangle }{_{ 0}} \nonumber \\
& = &  S\left[ \phi_{0, r} \phi_{0, s} \rightarrow \phi_{0, r} \phi_{0, s}  \right] .
\label{smatgenrel}
\end{eqnarray}
So like the antisymmetric twist case, the $S$-matrix elements of twisted $SU(N)$ invariant theory are same as that of the untwisted $SU(N)$ invariant theory and it is difficult to 
distinguish between the two theories. One can equally regard a $SU(N)$ invariant $S$-matrix element as due to untwisted fields with local interaction terms or due to nonlocal
twisted fields. 

Again dropping either the demand of invariance of vacuum or invariance of the interaction term under $SU(N)$ transformations, will result into twisted theories 
being different from the untwisted
ones. For example if we do not multiply fields with $\ast$-product then the twisted Hamiltonian density with quartic interactions can be written as    
 \begin{eqnarray}
  \mathcal{H}_{\theta} & = &  \Pi^{\dagger}_{\theta,r} \Pi_{\theta,r}  \, + \, (\partial_i \phi^{\dagger}_{\theta,r}) (\partial^i \phi_{\theta,r})
 \, + \, m^2  \, \phi^{\dagger}_{\theta,r} \phi_{\theta,r} 
\, + \,  \frac{\gamma}{4} \, \left[ \phi^{\dagger}_{\theta,r}  \phi^{\dagger}_{\theta,s} \phi_{\theta,r}  \phi_{\theta,s} 
\, + \, \phi^{\dagger}_{\theta,s} \phi^{\dagger}_{\theta,r} \phi_{\theta,r}  \phi_{\theta,s} \right . \nonumber \\  
& + & \left.  \phi^{\dagger}_{\theta,s} \phi^{\dagger}_{\theta,r}   \phi_{\theta,s} \phi_{\theta,r} 
\, + \, \phi^{\dagger}_{\theta,r}  \phi^{\dagger}_{\theta,s}   \phi_{\theta,s} \phi_{\theta,r}  \, + \, \cdots  \right ] \, + \, \rm{h. c}.
\label{genhamwithoutast}
\end{eqnarray}
where $\cdots$ represents the other 24 possible terms which we can write. Some of these 24 terms are equivalent to other terms but 
unlike the untwisted case not all of them are equal to each other. Moreover, this Hamiltonian has no $SU(N)$ symmetry and it maps to the marginally deformed Hamiltonian of the 
untwisted theory. 

The interaction Hamiltonian given by
\begin{eqnarray}
  \mathcal{H}_{\theta,\rm{Int}} & = &  \frac{\gamma}{4} \, \phi^{\dagger}_{\theta,r}  \ast  \phi^{\dagger}_{\theta,r}  \ast \phi^{\dagger}_{\theta,s}  \ast \phi_{\theta,s} 
\, + \, \rm{h.c},
\label{hamsunviogen}
\end{eqnarray}
is not even equivalent to any local untwisted Hamiltonian, and maps to 
\begin{eqnarray}
 \mathcal{H}_{0,\rm{Int}} & = &  \frac{\gamma}{4} \, \phi^{\dagger}_{0,r} \phi^{\dagger}_{0,r} \phi^{\dagger}_{0,s} \phi_{0,s} \, e^{ i \lambda^{(r)} \wedge Q }
\, + \, \rm{h.c},
\label{hamsunviogenrel}
\end{eqnarray}
which is nonlocal because of the presence of nonlocal operators $Q_m$. 

In this section and the preceding one we constructed field theories involving nonlocal fields having twisted statistics. We only restricted to a small subset of all such possible 
twisted theories. One can consider various generalizations of such twisted theories and we plan to discuss more of them in later works.


\section{Causality of Twisted Field Theories}


In this section we briefly address the issue of causality of the twisted quantum field theories. As it turns out, the twisted fields and hence the twisted field theories 
constructed out of them are in general non-causal but inspite of that one can construct Hamiltonian densities like the $SU(N)$ invariant ones, which are causal and 
also satisfy cluster decomposition principle. 


\subsection{Commutative Case}


For sake of completeness, we start with reviewing the discussion of causality in the untwisted case. Again we limit our discussion only to scalar fields but similar arguments
(with appropriate modifications) also hold for spinor fields and anti commuting operators. In the untwisted case, for complex scalar fields $\phi_{0,r}(x)$; $r = 1,2, \cdots N$
we have 
\begin{eqnarray}
i \Delta^{0}_{rs}(x-y) & = & \left< 0 \left|\left[\phi_{0, r}(x), \phi^{\dagger}_{0, s}(y) \right] \right| 0 \right> \nonumber \\
& = & - \delta_{rs}\int \frac{d^3 p}{(2\pi)^3} \frac{\sin\,p(x-y)}{E_{p}}.
\label{comspacecommutator}
\end{eqnarray}
It can be checked that for space like separations i.e. $(x - y)^2 < 0$ we have 
 \begin{eqnarray}
  i \Delta^{0}_{rs}(x-y) & = & \left< 0 \left|\left[\phi_{0, r}(x), \phi^{\dagger}_{0, s}(y) \right] \right| 0 \right> \; = \; 0 \qquad \text{for} \; (x-y)^2 \, < \, 0.
\label{comcausalitycon}
 \end{eqnarray}
Also, we have 
\begin{eqnarray}
 \left < 0 |[\phi_{0, r}(x), \phi_{0, s}(y) ]| 0 \right > \; = \; \left< 0 \left| [\phi^{\dagger}_{0, r}(x), \phi^{\dagger}_{0, s}(y)] \right|0 \right > \; = \; 0.
\label{comothercom}
\end{eqnarray}

Using (\ref{comcausalitycon}) and (\ref{comothercom}) it can be easily shown that any local operator which is a functional of $\phi_{0, r}$, $\phi^{\dagger}_{0, s}$ and 
their derivatives will also follow a similar relation e.g consider a generic local bilinear operator 
\begin{eqnarray}
\Xi^0(x) = \phi^{\dagger}_{0, r}(x) \xi_{rs}(x)\phi_{0, s}(x), 
\label{combilinearop}
\end{eqnarray}
where $\xi_{rs}(x)$ can be either a c-number valued function or a differential operator. Now, we have
\begin{eqnarray}
& & [\Xi^0(x) \,, \, \Xi^0(y)] \, = \, [\phi^{\dagger}_{0, r}(x) \xi_{rs}(x)\phi_{0, s}(x) \,,\, \phi^{\dagger}_{0,u} (y) \xi_{uv}(y)\phi_{0, v} (y) ] \nonumber \\
& &  =  \xi_{rs}(x) \xi_{uv}(y)\: [\phi^{\dagger}_{0, r}(x)\phi_{0, s}(x) \,,\,\phi^{\dagger}_{0, u} (y)\phi_{0, v} (y) ] \nonumber \\
 & & =  \xi_{rs}(x) \xi_{uv}(y)\: \left\{\phi^{\dagger}_{0, r}(x)\phi^{\dagger}_{0,u} (y)\: [\phi_{0, s}(x) \,,\, \phi_{0, v}(y)] 
\; +\; \phi^{\dagger}_{0, r}(x) \: [\phi_{0, s}(x) \,,\, \phi^{\dagger}_{0, u} (y)]\: \phi_{0,v} (y)
\right. \nonumber \\
& & +  \left. \phi^{\dagger}_{0, u} (y)\:[\phi^{\dagger}_{0, r}(x) \,,\, \phi_{0, v} (y)]\:\phi_{0, s}(x) 
\; +\; [\phi^{\dagger}_{0, r}(x) \,,\, \phi^{\dagger}_{0, u} (y)] \: \phi_{0, s}(x) \phi_{0, v}(y) \right\} \nonumber \\
& &  =  \xi_{rs}(x) \xi_{uv}(y)\: \left\{ \phi^{\dagger}_{0, r}(x) \: i \delta_{su}\Delta_{su}(x-y)\:\phi_{0, v} (y) 
\; +\;  \phi^{\dagger}_{0, u} (y)\: (-i) \delta_{rv}\Delta_{rv}(y-x)\:\phi_{0, s}(x) \right\}
\nonumber \\
& & =  0 \qquad \text{for} \; (x-y)^2 \, < \, 0.
\label{comopcom}
\end{eqnarray}

Similarly it can be shown that self commutator of other local operators (at two different spacetime labels) which are functional of $\phi_{0, r}$, $\phi^{\dagger}_{0, r}$ and 
their derivatives will
always vanish for $(x-y)^2 \, < \, 0 $. In particular it is straight forward to see that the self commutator of a local Hamiltonian density at two different spacetime labels always 
vanish for $(x-y)^2 \, < \, 0 $ i.e.
\begin{eqnarray}
 [\mathcal{H}(x), \mathcal{H} (y)] & = & 0 \qquad \text{for} \; (x-y)^2 \, < \, 0.
\label{comhamcon}
\end{eqnarray}
 Hence, in untwisted theory, (\ref{comcausalitycon}) is a sufficient although not necessary condition for the theory to be causal.


\subsection{Twisted Case}


In the twisted case, the relation analogous to (\ref{comcausalitycon}) does not hold. So for twisted case we have
\begin{eqnarray}
i \Delta^{\theta}_{rs}(x-y) & = & \left< 0 \left|\left[\phi_{\theta, r}(x), \phi^{\dagger}_{\theta, s}(y) \right] \right| 0 \right> \nonumber \\
& = &  \left< 0 \left|\left[ e^{-i \alpha^{(r)}  Q} \, \phi_{0, r}(x) \,,\, \phi^{\dagger}_{0, s}(y) e^{i \alpha^{(s)}  Q}\,  \right] \right| 0 \right> 
\nonumber \\
& = &  \left< 0 \left| \left \{ e^{-i \alpha^{(r)}  Q} \, \left[ \phi_{0, r}(x)\,,\,\phi^{\dagger}_{0, s}(y) \right]  \,  e^{i \alpha^{(s)} Q}
\; + \;  \left[ e^{-i \alpha^{(r)}  Q} \,,\, \phi^{\dagger}_{0, s}(y)\right] \, \phi_{0, r}(x)\, e^{i \alpha^{(s)} Q}
\right. \right. \right. \nonumber \\
& + & \left. \left.\left.  \phi^{\dagger}_{0, s}(y) \, e^{-i \alpha^{(r)}  Q} \, \left[\phi_{0, r}(x) \,,\,  e^{i \alpha^{(s)} Q} \right]
\; + \; \phi^{\dagger}_{0, s}(y) \,\left[ e^{-i \alpha^{(r)}  Q} \,,\, e^{i \alpha^{(s)} Q} \right] \, \phi_{0, r}(x) \right\}\right| 0 \right>
\nonumber \\
& = &  \left< 0 \left|\left[ \phi_{0, r}(x)\,,\,\phi^{\dagger}_{0, s}(y) \right]\right| 0 \right> 
\; + \; \left< 0 \left| e^{-i \alpha^{(r)} \lambda^{(s)}} \, \phi^{\dagger}_{0, s}(y) \, \phi_{0, r}(x)\, \right| 0 \right> \nonumber \\
& + & \left< 0 \left| e^{i \alpha^{(s)}  \lambda^{(r)}} \, \phi^{\dagger}_{0, s}(y) \, e^{-i \alpha^{(r)}  Q} \, \phi_{0, r}(x) \, \right| 0 \right> 
\nonumber \\
& = & i \delta_{rs} \Delta^{0}_{rs}(x - y)  +   \left\{ e^{-i \alpha^{(r)} \lambda^{(s)}}  +  e^{i (\alpha^{(s)} + \alpha^{(r)} )\lambda^{(r)}} \right\}
 \left< 0 \left|\phi^{\dagger}_{0, s}(y) \phi_{0, r}(x)\right| 0 \right> ,
\label{twistedcomrel}
\end{eqnarray}
where in last two steps we have used the fact that $Q |0> = |0>$. Let us denote $ \mathbf{A} = \left< 0 \left|\phi^{\dagger}_{0, s}(y) \phi_{0, r}(x)\right| 0 \right> $. 
It can be easily seen that although $\Delta^{0}_{rs}(x - y) $ vanishes for space-like separation but $\mathbf{A}$ does not vanish. For example, let us take the special 
case of $(x^0 - y^0) = 0$ and $(\vec{x} - \vec{y}) = \vec{z}$. This is a special case of space like separation i.e in this case $(x -y)^2 < 0$. Therefore, we have
\begin{eqnarray}
 \mathbf{A} & = & \left< 0 \left|\phi^{\dagger}_{0, s}(y) \phi_{0, r}(x)\right| 0 \right> \nonumber \\
& = & \int \frac{d^3 p}{(2\pi)^3} \frac{1}{2E_p} \int \frac{d^3 q}{(2\pi)^3} \frac{1}{2E_q}
\left< 0 \left|\left( c^{\dagger}_s(q)e^{iqy} \, + \, d_s(q)e^{-iqy} \right)\left( c_r(p)e^{-ipx} \, + \, d^{\dagger}_r(p)e^{ipx} \right)\right| 0 \right> \nonumber \\
& = & \int \frac{d^3 p}{(2\pi)^3} \frac{1}{2E_p} \int \frac{d^3 q}{(2\pi)^3} \frac{1}{2E_q} \left<0\left|d_s(q)d^{\dagger}_r(p)\right|0\right> e^{ipx} e^{-iqy}
\nonumber \\
& = & \int \frac{d^3 p}{(2\pi)^3} \frac{1}{2E_p} \int \frac{d^3 q}{(2\pi)^3} \frac{1}{2E_q} (2\pi)^3 2E_p \delta_{rs} \delta^3(p - q) e^{ipx} e^{-iqy} \nonumber \\
& = & \delta_{rs} \int \frac{d^3 p}{(2\pi)^3} \frac{1}{2E_p} e^{-i\vec{p}\vec{z}} \qquad \qquad \text{for} \; (x^0 - y^0) = 0 \; \text{and} \; (\vec{x} - \vec{y}) = \vec{z}.
\label{twistA}
\end{eqnarray}
Going to polar coordinates we have 
\begin{eqnarray}
  \mathbf{A} & = & 2\pi \delta_{rs} \int^{\infty}_{0} \frac{d p}{(2\pi)^3} \frac{|\vec{p}|}{\sqrt{|\vec{p}|^2 + m^2}} \frac{\sin |\vec{p}||\vec{z}|}{|\vec{z}|} \nonumber \\
& = & \frac{m\delta_{rs}}{4\pi^2 |\vec{z}|}K_1(m|\vec{z}|) ,
\label{twistedA}
\end{eqnarray}
where $K_1$ is the Hankel function. Clearly $\mathbf{A}$ does not vanish for all space-like separations. So using (\ref{twistedA}) in (\ref{twistedcomrel}) we have
\begin{eqnarray}
i \Delta^{\theta}_{rs}(x-y) & = &  i \delta_{rs} \Delta^{0}_{rs}(x - y) \; + \; \left( e^{-i \alpha^{(r)} \lambda^{(s)}} \, + \, e^{i (\alpha^{(s)} + \alpha^{(r)} )\lambda^{(r)}} \right)
\mathbf{A},
\label{twistedcausal}
\end{eqnarray}
which for a particular special case of space-like separations i.e. $(x-y)^2 \, = \, z^2 < 0 $  and $(x^0 - y^0) = 0$ is 
\begin{eqnarray}
 i \Delta^{\theta}_{rs}(x-y) & = &  \left( e^{-i \alpha^{(r)} \lambda^{(s)}} \, + \, e^{i (\alpha^{(s)} + \alpha^{(r)} )\lambda^{(r)}} \right) 
\frac{m\delta_{rs}}{4\pi^2 |\vec{z}|}K_1(m|\vec{z}|) \neq 0 .
\label{twistedcausality}
\end{eqnarray}

Hence, unlike the untwisted case, the twisted fields don't commute for all space-like separations. An immediate consequence of (\ref{twistedcausality}) is that it
can no longer be guaranteed that, the self-commutator at different spacetime labels, of all operators which are functional of the twisted fields (or their derivatives) will vanish
for space-like separations. In particular, following computations similar to (\ref{comopcom}), it can be shown that the self-commutator of generic twisted bilinear operators 
$\Xi^\theta(x) = \hat{\phi}^{\dagger}_{\theta, r}(x) \xi_{rs}(x)\hat{\phi}_{\theta, s}(x)$ does not vanish for all space-like separations i.e.
\begin{eqnarray}
  [\Xi^\theta(x)\, , \, \Xi^\theta(y)] & \neq & 0 \qquad \text{for all} \; (x-y)^2 \; < \: 0 .
\label{twistedbilinear}
 \end{eqnarray}

A similar result will follow for any generic operator formed from these twisted fields. But as remarked earlier, the condition (\ref{comcausalitycon}) (or its twisted analogue) is
just a sufficient condition and by no means it is a necessary condition. Infact even in untwisted case, (\ref{comcausalitycon}) is not satisfied by fermionic fields. Therefore
inspite of (\ref{twistedcausality}) it is still possible to construct twisted Hamiltonian densities which satisfy causality constraints 
\begin{eqnarray}
[\hat{\mathcal{H}}(x)\, , \, \hat{\mathcal{H}}(y)] & = & 0 \qquad \text{for} \; (x-y)^2 \; < \: 0 .
\label{twistedcausalham}
\end{eqnarray} 
It is easy to see that the twisted $SU(N)$ invariant Hamiltonian densities of (\ref{intantihamrel}) and (\ref{intgenhamrel}) satisfy the causality condition. 
But a generic Hamiltonian density constructed out of twisted fields will not necessarily satisfy (\ref{twistedcausalham}). For example, the nonlocal 
Hamiltonian densities of (\ref{hamsunvioanti}) and (\ref{hamsunviogen}) are noncausal.


\section{Conclusions}


In this chapter we discussed the possibility of having twisted statistics by deformation of internal symmetries. We constructed two such deformed statistics and discussed field 
theories for such deformed fields. We showed that both type of twisted quantum fields discussed in this chapter, satisfy commutation relations different from the usual bosonic/fermionic 
commutation relations. Such twisted fields by construction (and in view of CPT theorem) are nonlocal in nature. We showed that inspite of the basic ingredient fields being nonlocal,  
it is possible to construct local interaction Hamiltonians which satisfy cluster decomposition principle and are Lorentz invariant. 

We first discussed a specific type of twist called ``antisymmetric twist''. This kind of twist is quite similar in spirit to the twisted noncommutative field theories. 
The formalism developed for antisymmetric twists was analogous (with appropriate generalizations and modifications) to the formalism of twisted noncommutative theories. 
We then constructed interaction terms using such twisted fields and discussed the scattering problem for such theories. We found that the twisted $SU(N)$ invariant interaction
Hamiltonian as well as $S$-matrix elements are identical to their untwisted analogues and hence by doing a scattering experiment it is rather difficult to distinguish between 
a twisted and a untwisted theory. We further showed that relaxing the demand of $SU(N)$ invariance leads to differences between the two theories and for certain interaction 
terms the twisted theory is nonlocal although its analogous untwisted theory is local. As an interesting application of these ideas we showed that the marginal ($\beta$-)
deformations of the scalar matter sector of $N=4$ SUSY Hamiltonian density can be described in terms of a twisted interaction Hamiltonian density and hence the twisted internal
symmetries can be used to significantly simplify the discussion of such theories.  
 
We then constructed more general twisted statistics which can be viewed as internal symmetry analogue of dipole theories. We also discussed the construction of interaction 
terms and scattering formalism for it. The main results for general twists are same as those for the antisymmetric twist. 

We ended the chapter with discussion of causality of the twisted field theories. We showed that the twisted fields are noncausal and hence a generic observable constructed out 
of them is also noncausal. Inspite of this it is possible to construct certain interaction Hamiltonians, e.g. the $SU(N)$ invariant interaction Hamiltonian, which are causal and 
satisfy cluster decomposition principle. In view of the nonlocal nature of the twisted fields, these field theories (with appropriate generalizations) have the potential 
to circumvent the Coleman-Mandula no-go theorem \cite{coleman}. We plan to discuss such theories in future works.
Also, in this work we did not discuss the possibility of spontaneous symmetry breaking. Such a scenario is quite interesting but it requires a separate discussion. 
We plan to discuss it also in a future work.


\chapter{Thermal Correlation Functions}
\label{chap:chap4}
As effects of noncommutativity become important at short distances, we expect that 
important implications occur in early cosmology with its attendant high temperatures. 
Moreover, the noncommutative scale need not always be as small as Planck scale. 
For example, in presence of large extra dimensions, the ``effective Planck scale'' can be at much lower 
energies (usually taken between TeV scale and GUT scale, depending on details of specific models). This results in appearance of noncommutativity 
at scales much larger than Planck scale. 
Such large scale noncommutativity, if present, is of particular interest as its effect can be detected in present day or near future experiments.
To this end, it is 
important to formulate the thermodynamics of quantum field theories on such noncommutative spacetimes. The noncommutativity 
contributes an additional subtlety to this issue, in that the usual facility of working with a finite volume $V$ 
and then taking $V\rightarrow \infty$ is not available to us. Thus the appropriate starting point for 
any discussion of quantum thermodynamics is the KMS condition (see for instance, \cite{kubo,martin,haag}). 
Given an operator $A$ (which may for instance 
be constructed from products of quantum fields, or from products of creation or annihilation operators) in 
the Heisenberg representation, its time evolution is given by
$A(\tau) = e^{i \mathcal{H} \tau} A  e^{-i \mathcal{H} \tau}, \quad \mathcal{H} = H -\mu N$
where $\mathcal{H}$ is the grand canonical Hamiltonian. It is important 
to emphasize that the 
$\tau$ appearing in the above equation is not the coordinate time $x^{0}$, but the parameter of 
time evolution.

For any two operators $A$ and $B$, we define the retarded function as
\begin{equation} 
G_{AB}(\tau-\tau') \equiv -i\theta(\tau-\tau')\langle\langle  [A(\tau), B(\tau')]\rangle\rangle  
= -i\theta (\tau-\tau')[\langle A(\tau)B(\tau')\rangle -\eta' \langle B(\tau')A(\tau)\rangle ]
\label{green}
\end{equation}
where $\theta(x)$ is the Heavyside step-function, 
 $\langle  X  \rangle = \frac{{\rm Tr}\, [e^{-\beta\mathcal{H}}X]}{Z} 
 \equiv {\rm Tr}\, [\rho  X]$, $\rho = \frac{e^{-\beta\mathcal{H}}}{Z}$ and $Z = {\rm Tr}\,[e^{-\beta\mathcal{H}}].$
The $\tau$-independent function $\eta'$ can be chosen so that $G_{AB}$ satisfies a conveniently 
simple differential equation, as we shall show. Advanced and causal functions can be defined similarly, 
though we will not need them here. 

We will follow the strategy outlined by \cite{zubarev} for evaluating correlators of interest, making use of 
the relation between 
$G_{AB}(\tau-\tau')$, the thermal correlation functions $\mathcal{F}_{AB}(\tau-\tau')=
\langle A(\tau)B(\tau')\rangle$ and $\mathcal{F}_{BA}(\tau-\tau')=\langle B(\tau')A(\tau)\rangle$, and the spectral density $J_{BA}(\omega)$ defined by 
\begin{equation}
\mathcal{F}_{BA}(\tau-\tau') = \int^\infty _{-\infty} J_{BA}(\omega)e^{-i\omega(\tau-\tau')} 
 d\omega
 \label{spec1}
\end{equation}
Thermodynamic equilibrium (i.e. $G_{AB}, \mathcal{F}_{AB}$ etc are functions of 
$(\tau-\tau')$ only) and cyclicity of trace imply that
\begin{equation}
\mathcal{F}_{AB}(\tau-\tau') = \int^\infty _{-\infty} J_{BA}(\omega)e^{\beta\omega}
e^{-i\omega(\tau-\tau')}d\omega, 
\label{spec2}
\end{equation}
i.e. $\mathcal{F}_{AB}$ and $\mathcal{F}_{BA}$ satisfy the KMS condition \cite{haag}.

Heisenberg equations of motion for $A(\tau)$ and $B(\tau)$ imply that $G_{AB}$ satisfies
\begin{equation}
i\frac{dG_{AB}}{d\tau} = \delta(\tau-\tau') \langle A(\tau) B (\tau) - \eta' B(\tau)  A(\tau) ]\rangle
 + \langle\langle \{ A(\tau) \mathcal{H} - \mathcal{H} A(\tau); B(\tau')\} \rangle\rangle
\label{motion}
\end{equation} 
Using (\ref{spec2})  and the integral representation $\theta(\tau-\tau') = 
\frac{i}{2\pi}\int ^{\infty}_{-\infty} dx \frac{e^{-ix(\tau-\tau')}}{x+i\epsilon}$, the Fourier transform  
$G_{AB}(E) \equiv \frac{1}{2\pi} \int ^{\infty}_{-\infty}G_{AB}(\tau)e^{iE\tau}d\tau$ can be written as
\begin{equation}
G_{AB}(E) = \frac{1}{2\pi} \int ^{\infty}_{-\infty} J_{BA}(\omega)(e^{\beta\omega} - \eta') 
\frac{d\omega}{E-\omega +i\epsilon}
\label{fougre}
\end{equation}
Using (\ref{fougre}) and the delta function representation $\delta(x) = \frac{1}{2\pi i} \left \{ \frac{1}{x-i\epsilon} -  \frac{1}{x+i\epsilon} \right\} $ we get
\begin{equation}
 G_{AB}(\omega + i\epsilon) - G_{AB}(\omega - i\epsilon) = -i J_{BA}(\omega)(e^{\beta\omega-\eta'})
\label{specngre} 
\end{equation}
Using (\ref{specngre}) in (\ref{spec1},\ref{spec2}) we get
\begin{eqnarray}
\mathcal{F}_{BA}(\tau-\tau') &=& \int ^{\infty}_{-\infty} \frac{G_{AB}(E + i\epsilon)- G_{AB}(E - i\epsilon)}{e^{\beta E-\eta ^{'}}} e^{-iE(\tau-\tau')}dE, \label{corrba}\\
\mathcal{F}_{AB}(\tau' - \tau) &=& \int ^{\infty}_{-\infty} \frac{G_{AB}(E + i\epsilon)- G_{AB}(E - i\epsilon)}{e^{\beta E-\eta ^{'}}} e^{\beta E} e^{-i E(\tau-\tau')}dE
\label{corrab}
\end{eqnarray}
For a perfect quantum gas, the (grand canonical) hamiltonian  is
\begin{equation}
\mathcal{H}= H - \mu N = \int \frac{d^{3}\bold{k}}{(2\pi)^{3}2\omega _{\bold{k}}}(\omega _{\bold{k}} - \mu)a^{\dagger}_{\bold{k}}a_{\bold{k}} 
\label{haml} 
\end{equation}
where the $a^{\dagger}_{\bold{k}} $  and $a_{\bold{k}} $ satisfy (\ref{tcom}).
Substituting 
$A(\tau)=a_{\bold{p_{1}}}(\tau), B(\tau')=a^{\dagger}_{\bold{p_{2}}}(\tau')$ in (\ref{green}), we 
find that
\begin{equation}
G_{\bold{p_{1}}\bold{p_{2}}} \equiv  - i \theta (\tau -\tau')[\langle a_{\bold{p_{1}}}(\tau)
a^{\dagger}_{\bold{p_{2}}}(\tau')\rangle - \eta' \langle a^{\dagger}_{\bold{p_{2}}}(\tau')
a_{\bold{p_{1}}}(\tau)\rangle]
\label{twogreen}
\end{equation}
satisfies
\begin{equation}
i\frac{dG_{\bold{p_{1}}\bold{p_{2}}}}{d\tau} = (2\pi)^{3}2(p_{10})\delta(\tau -\tau')
\delta^{3}(\bold{p_{1}}-\bold{p_{2}}) + \left(\omega _{\bold{p_{1}}} - \mu\right) 
G_{\bold{p_{1}}\bold{p_{2}}}(\tau-\tau')
\label{twomotion}
\end{equation}
if we choose $\eta' = \eta  e^{-ip_{1}\wedge p_{2}} $.

The Fourier transform $ G_{\bold{p_{1}}\bold{p_{2}}}(E) $  
of $G_{\bold{p_{1}}\bold{p_{2}}} (\tau-\tau')$ is easily obtained:
\begin{equation}
G_{\bold{p_{1}}\bold{p_{2}}}(E) = \frac{1}{2\pi}\frac{(2\pi)^{3} 2(p_{10}) \delta^{3}(\bold{p_{1}} - 
\bold{p_{2}})}{E-\left(\omega _{\bold{p_{1}}}-\mu\right)}
\label{gfinal}
\end{equation}

Using (\ref{corrba}) and putting $ \tau = \tau'$, we get 
\begin{equation}
\langle a^{\dagger}_{\bold{p_{2}}}a_{\bold{p_{1}}}\rangle = \frac{(2\pi)^{3} 2(p_{10}) 
\delta^{3}(\bold{p_{1}}-\bold{p_{2}})}{e^{\beta \left(\omega _{\bold{p_{1}}}-\mu\right)}-\eta e^{-ip_{1}\wedge p_{2}}}
\end{equation}
Since $ p_{1}\wedge p_{2} = 0 $ if $ \bold{p_{1}} =  \bold{p_{2}} $, we have
\begin{equation}
\langle a^{\dagger}_{\bold{p_{2}}}a_{\bold{p_{1}}}\rangle = \frac{(2\pi)^{3} 2(p_{10}) 
\delta^{3}(\bold{p_{1}}-\bold{p_{2}})}{e^{\beta \left(\omega _{\bold{p_{1}}}-\mu\right)}-\eta}
\label{twopoint}
\end{equation}
which is same as the commutative correlation function. 

This result is not unexpected: translational invariance (or equivalently, energy-momentum conservation)
forces this upon us. Higher correlators however will not be so severely restricted by 
translational invariance. For instance, to calculate $\langle a^{\dagger}_{\bold{p_{1}}} 
a^{\dagger}_{\bold{p_{2}}} a_{\bold{p_{3}}} a_{\bold{p_{4}}}\rangle$, we substitute 
$A(\tau)= a_{\bold{p_{4}}}(\tau)$ and $B(\tau')= a^{\dagger}_{\bold{p_{1}}}(\tau') 
a^{\dagger}_{\bold{p_{2}}}(\tau')a_{\bold{p_{3}}}(\tau') $ in (\ref{green}):
\begin{equation}
G_{\bold{p_{4}}\bold{p_{1}}\bold{p_{2}}\bold{p_{3}}} = -i \theta(\tau - \tau')\left[\langle 
a_{\bold{p_{4}}}(\tau)a^{\dagger}_{\bold{p_{1}}}(\tau')a^{\dagger}_{\bold{p_{2}}}(\tau') 
a_{\bold{p_{3}}}(\tau')\rangle  - \eta' \langle a^{\dagger}_{\bold{p_{1}}}(\tau') 
a^{\dagger}_{\bold{p_{2}}}(\tau')a_{\bold{p_{3}}}(\tau')a_{\bold{p_{4}}}(\tau) \rangle\right]
\label{4green}
\end{equation}
This satisfies
\begin{eqnarray} 
& & i \frac{dG_{\bold{p_{4}}\bold{p_{1}}\bold{p_{2}}\bold{p_{3}}}}{d\tau}  =  \delta (\tau -\tau') 
(2\pi)^{3}\left[2(p_{10})\delta ^{3}(\bold{p_{4}}-\bold{p_{1}}) \langle a^{\dagger}_{\bold{p_{2}}}(\tau) 
a_{\bold{p_{3}}}(\tau) \rangle \right. \nonumber \\
& & +  \left.   2 \eta (p_{20}) \delta ^{3}(\bold{p_{4}}-\bold{p_{2}}) 
e^{-ip_{4}\wedge p_{1}} \langle a^{\dagger}_{\bold{p_{1}}}(\tau)a_{\bold{p_{3}}}(\tau)\rangle \right] 
+ \left(\omega _{\bold{p_{4}}}-\mu\right)G_{\bold{p_{4}}\bold{p_{1}}\bold{p_{2}}\bold{p_{3}}}
\label{fourgreen}
\end{eqnarray}
for the choice $ \eta'= \eta e^{-ip_{4}\wedge (p_{1}+ p_{2} - p_{3})}$. \\
The Fourier transform $ G_{\bold{p_{4}}\bold{p_{1}}\bold{p_{2}}\bold{p_{3}}}$ is
\begin{eqnarray} 
G_{\bold{p_{4}}\bold{p_{1}}\bold{p_{2}}\bold{p_{3}}}(E) & = & \frac{1}{2\pi} \frac{(2\pi)^{3}}{E - \omega _{\bold{p_{4}}}}[2(p_{1})_{0}\delta ^{3}(\bold{p_{1}}-\bold{p_{4}})
\langle a^{\dagger}_{\bold{p_{2}}}(\tau)a_{\bold{p_{3}}}(\tau) \rangle  \nonumber \\ 
& + & 2 \eta (p_{2})_{0} e^{i p_{1}\wedge p_{4}}\delta ^{3}(\bold{p_{2}}-\bold{p_{4}})\langle a^{\dagger}_{\bold{p_{1}}}(\tau)a_{\bold{p_{3}}}(\tau)\rangle]
\label{fougreen}
\end{eqnarray}
Using (\ref{corrba}) and putting $ \tau = \tau'$ we get 
\begin{eqnarray}
\langle a^{\dagger}_{\bold{p_{1}}}a^{\dagger}_{\bold{p_{2}}}a_{\bold{p_{3}}}a_{\bold{p_{4}}} \rangle 
& = & \frac{(2\pi)^{3}(2p_{10})}{e^{\beta \left(\omega _{\bold{p_{1}}}-\mu\right)} - \eta} 
\frac{(2\pi)^{3}(2p_{20})}{e^{\beta \left(\omega _{\bold{p_{2}}}-\mu\right)} - \eta} \left[\delta ^{3}
(\bold{p_{1}}-\bold{p_{4}})\delta ^{3}(\bold{p_{2}}-\bold{p_{3}}) \right.\nonumber \\ 
& + & \left. \eta e^{i p_{1}\wedge p_{2}}\delta ^{3}(\bold{p_{1}}-\bold{p_{3}})\delta ^{3}(\bold{p_{2}} - 
\bold{p_{4}}) \right]
\label{fin4cor}
\end{eqnarray}
This four-point correlator
differs from its commutative counterpart by appearance of the $\theta$-dependent phase 
$e^{i p_{1}\wedge p_{2}}$ in the second term.
$N$-point correlators can also be determined, by first defining the {\it twisted 
commutator}. Consider the operators $a^{\#}_{\bold{p}_{i}} $ and 
$a^{\#}_{\bold{p}_{j}}$, which stand for either creation or annihilation operators corresponding to 
the momentum state $\bold{p}_{i} $ and $ \bold{p}_{j}$ respectively. We define 
\begin{equation}
[a^{\#}_{\bold{p}_{i}},a^{\#}_{\bold{p}_{j}}]_{\theta} \equiv a^{\#}_{\bold{p}_{i}} a^{\#}_{\bold{p}_{j}} - 
\eta e^{i(\alpha_{\bold{p}_{i}\bold{p}_{j}}) p_{i}\wedge p_{j}} a^{\#}_{\bold{p}_{j}} a^{\#}_{\bold{p}_{i}}  
\label{tscm}
\end{equation}
where $\alpha_{\bold{p}_{i}\bold{p}_{j}}$ is 1 if $a^{\#}_{\bold{p}_{i} } $ and $a^{\#}_{\bold{p}_{j}} $ are 
of same type ({\it i.e.} are both creation or both annihilation operators), else is equal to $-1$. 
The commutation relations (\ref{tcom}) imply that
\begin{eqnarray}
[a^{\#}_{\bold{p}_{i}},a^{\#}_{\bold{p}_{j}}]_{\theta} & = & 0  \quad \quad
\mbox{if $ a^{\#}_{\bold{p}_{i}} $ and $a^{\#}_{\bold{p}_{j}} $ are of same type} \nonumber \\  
 & = & - \eta (2\pi)^{3}2(p_{10}) \delta^{3}({\bold p_i} -{\bold p_j}) \quad  \mbox{if $a^{\#}_{\bold{p}_{i}}=a^{\dagger}_{\bold{p}_{i}}$ 
and $a^{\#}_{\bold{p}_{j}}=a_{\bold{p}_{j}}$} \nonumber\\
&=& (2\pi)^{3}2(p_{10})\delta^{3}({\bold p_i} -{\bold p_j})  \quad \mbox{if $a^{\#}_{\bold{p}_{i}}=a_{\bold{p}_{i}}$ 
and $a^{\#}_{\bold{p}_{j}}=a^{\dagger}_{\bold{p}_{j}}$} \nonumber
\end{eqnarray}
Using (\ref{twopoint}), we see that
\begin{equation}
\langle a^{\#}_{\bold{p}_{i}} a^{\#}_{\bold{p}_{j}} \rangle=\frac{[a^{\#}_{\bold{p}_{i}},a^{\#}_{\bold{p}_{j}}]_{\theta} }
{1-\eta e^{\alpha\beta \left(\omega _{\bold{p_{i}}}-\mu\right)}}
\label{EQ36}
\end{equation}
where
\begin{eqnarray}
\alpha&=&1 \quad \mbox{if $a^{\#}_{\bold{p}_{i}} $ is a creation operator} \nonumber\\
&=&-1 \quad \mbox{if $a^{\#}_{\bold{p}_{i}}$ is an annihilation operator}    
\end{eqnarray}
 
Let us consider $\langle a^{\#}_{\bold{p}_{1}} a^{\#}_{\bold{p}_{2}}... a^{\#}_{\bold{p}_{N}} \rangle$.
Using (\ref{tscm}) repeatedly we bring $a^{\#}_{\bold{p}_{1}}$ on the right side of the sequence and then using 
the cyclic property of trace we get 
\begin{eqnarray}
\langle a^{\#}_{\bold{p}_{1}} a^{\#}_{\bold{p}_{2}}... a^{\#}_{\bold{p}_{N}} \rangle & = & 
\sum_{j=1}^{N-1}\eta^{j-1}e^{i\phi_j} [a^{\#}_{\bold{p}_{1}},a^{\#}_{\bold{p}_{j+1}}]_{\theta}
\langle \widehat{a^{\#}_{\bold{p}_{1}}} a^{\#}_{\bold{p}_{2}} ...\widehat {a^{\#}_{\bold{p}_{j+1}}} \cdots 
a^{\#}_{\bold{p}_{N}}\rangle \nonumber\\
&+&\eta^{N-1}e^{i\phi_{N}} \langle a^{\#}_{\bold{p}_{1}} \rho a^{\#}_{\bold{p}_{2}}  a^{\#}_{\bold{p}_{3}} 
\cdots  a^{\#}_{\bold{p}_{N}}\rangle 
\end{eqnarray}
where \hspace{.5mm} $\widehat{}$ \hspace{.5mm} on an operator denotes the absence of this operator from the sequence. The phase 
$\phi_j$ is given by
\begin{equation}
\phi_j=\sum_{i=1}^{j}\alpha_{1i} \bold{p}_{1} \wedge \bold{p}_{i}
\end{equation}
For the Hamiltonian (\ref{haml}),
\begin{equation}
\rho \,a^{\#}_{\bold p_i}=e^{-\alpha\beta \left(\omega _{\bold{p_{i}}}-\mu\right)} a^{\#}_{\bold p_i} \,
\rho
\label{EQ41}
\end{equation} 
Using (\ref{EQ41}) and (\ref{EQ36}) we can finally write
\begin{equation}
\langle  a^{\#}_{\bold{p}_{1}} a^{\#}_{\bold{p}_{2}}... a^{\#}_{\bold{p}_{N}}  \rangle= \Bigl(\sum_{j=1}^{N-1}\eta^{j-1}e^{i\phi_j}
 \langle a^{\#}_{\bold{p}_{1}}a^{\#}_{\bold{p}_{j+1}} \rangle
\langle \widehat{a^{\#}_{\bold{p}_{1}}} a^{\#}_{\bold{p}_{2}} ...\widehat {a^{\#}_{\bold{p}_{j+1}}}...a^{\#}_{\bold{p}_{N}}\rangle \Bigr)\xi(\beta, N,\omega_{\bold p_1})
\end{equation}
where $\xi(\beta, N,\omega_{\bold p_i})$ is given by
\begin{equation}
\xi(\beta, N,\omega_{\bold p_i})=\frac{1-\eta e^{\alpha\beta \left(\omega _{\bold{p_{i}}}-\mu\right)}}{1-\eta^{N-1}e^{i\phi_N}
e^{\alpha\beta \left(\omega _{\bold{p_{i}}}-\mu\right)}}
\end{equation}
This is the thermal version of Wick's theorem adapted to twisted quantum fields: the $N$-point correlator
is expressed in terms of the $(N-2)$-point correlators. 

This chapter is based on the work published in \cite{thcor}.


\chapter{Intensity Correlations and HBT Effect}
\label{chap:chap5}
Cosmic Rays with energies around $10^{18}$ eV and higher are called as Ultra High Energy Cosmic Rays (UHECRs) \cite{book,nag,bhat}. They are the highest energy particles known to us 
and can have energies $10^7$ times more than that produced by LHC. Inspite of recent advancements (both theoretical and experimental), UHECRs pose a considerable theoretical 
challenge: the source and mechanism of origin of such high energy particles \cite{pierre4,pierre5} and their composition are areas of active research \cite{pierre2,pierre3,wilk}. 

Due to their extremely high energies, UHECRs are not only an excellent arena for testing the validity of known laws of physics \cite{coleman,grillo} but are also some 
of the best places to look for signatures, if any, of new physics e.g. theories with Lorentz violation and/or deformed dispersion relations \cite{wolfgang,sigl,luca,stecker,stecker1}.
In this chapter we aim to show that UHECRs can be used to look for signatures of a particular model of nonlocalities coming from the underlying noncommutative nature of spacetime at
short distances. 

As argued in first chapter, simple intuitive arguments involving standard quantum mechanics uncertainty relations suggest that at length scales close to Planck length, 
strong gravity effects will limit the spatial as well  as temporal resolution beyond some fundamental length scale ($ l_p \approx$ Planck Length), leading to space - space as 
well as space - time uncertainties \cite{dop}. One cannot probe spacetime with a resolution below this scale i.e. spacetime becomes “fuzzy” below this scale, resulting into 
noncommutative spacetime. 
This noncommutative scale need not always be as small as Planck scale. For instance if there are large extra dimensions, then the ``effective Planck scale'' can be at much lower 
energies (usually taken between TeV scale and GUT scale, depending on specific models), resulting in appearance of noncommutativity at scales much larger than Planck scale. 
Such large scale noncommutativity is of particular interest as its effect can be detected in present day or near future experiments. In this chapter we look 
for signatures of such large scale noncommutativity in UHECRs.  

Because of (\ref{tcom}) the quantum fields written on G.M plane, unlike ordinary quantum fields, follow a unusual statistics which we call as twisted statistics. Twisted statistics 
are a unique feature of fields on G.M plane and can be used to search for signals of noncommutativity: because of the twisted commutation relations, interesting new effects like 
Pauli forbidden transitions \cite{bal-pauli,pramod-pauli} can arise. The effect of twisted statistics also manifests itself in certain thermodynamic quantities \cite{basu-th}, 
\cite{basu1-th}.  
The two-point distribution functions remain unchanged
\begin{eqnarray} 
 \left\langle a^{\dagger}_{\bold{p_{1}}}a_{\bold{p_{2}}}\right\rangle & = & 2\left(p_{1}\right)_{0} \, N^{(T)}_{\bold{p_{1}}} \, \delta^{3}(\bold{p_{1}}-\bold{p_{2}}) 
\label{2cor}
\end{eqnarray}
where $ N^{(T)}_{\bold{p}} =  \frac{1}{e^{\beta E_{\bold{p}}}-\eta}$ is the thermal distribution, but for example the quantity 
$ \left\langle a^{\dagger}_{\bold{p_{1}}} a^{\dagger}_{\bold{p_{2}}} a_{\bold{p_{3}}}a_{\bold{p_{4}}}\right\rangle$ gets changed \cite{basu-th}.
\begin{eqnarray} 
\left\langle a^{\dagger}_{\bold{p_{1}}} a^{\dagger}_{\bold{p_{2}}} a_{\bold{p_{3}}}a_{\bold{p_{4}}}\right\rangle 
& = & 2\left(p_{1}\right)_{0} 2\left(p_{2}\right)_{0} N^{(T)}_{\bold{p_{1}}} N^{(T)}_{\bold{p_{2}}} \left[ \delta^{3}(\bold{p_{1}}-\bold{p_{4}})\delta^{3}(\bold{p_{2}}-\bold{p_{3}}) \right.
\nonumber \\
& + & \left. \eta e^{i p_{1} \wedge p_{2}} \delta^{3}(\bold{p_{1}}-\bold{p_{3}})\delta^{3}(\bold{p_{2}}-\bold{p_{4}}) \right]
\label{4cor}
\end{eqnarray} 
The above differs from the commutative expression by the appearance of the factor $ e^{i p_{1} \wedge p_{2}} $ in the second term.

One can easily check, following a analysis similar to that done in \cite{basu-th}, that (\ref{2cor}) and (\ref{4cor}) are true not only for thermal distribution  $ N^{(T)}_{\bold{p}}$ 
but for any arbitrary wavepacket $ f(\bold{p})$ i.e. 
\begin{eqnarray} 
  \left\langle a^{\dagger}_{\bold{p_{1}}}a_{\bold{p_{2}}}\right\rangle & = &  2\left(p_{1}\right)_{0} \, f(\bold{p_{1}}) \, \delta^{3}(\bold{p_{1}}-\bold{p_{2}}) 
\label{g2cor}
\end{eqnarray} 
and
\begin{eqnarray} 
 \left\langle a^{\dagger}_{\bold{p_{1}}} a^{\dagger}_{\bold{p_{2}}} a_{\bold{p_{3}}}a_{\bold{p_{4}}}\right\rangle 
& = & 2\left(p_{1}\right)_{0} 2\left(p_{2}\right)_{0} f(\bold{p_{1}}) f(\bold{p_{2}}) \left[ \delta^{3}(\bold{p_{1}}-\bold{p_{4}})\delta^{3}(\bold{p_{2}}-\bold{p_{3}}) \right.
\nonumber \\
& + & \left. \eta e^{i p_{1} \wedge p_{2}} \delta^{3}(\bold{p_{1}}-\bold{p_{3}})\delta^{3}(\bold{p_{2}}-\bold{p_{4}})\right] .
\label{g4cor}
\end{eqnarray}

Here we discuss the consequences of (\ref{g4cor}) to the HBT correlation functions.

Hanbury-Brown Twiss (HBT) effect \cite{han} is the interference effect between intensities measured by two detectors when a beam of identical particles is projected on them, 
with the intensities recorded by the two detectors operating simultaneously. This intensity increases for bosons (and decreases for fermions) when compared with the intensities recorded 
by the same two detectors, if only one is operated at a time.  The correlation function for HBT effect is defined as
\begin{eqnarray} 
C = \frac{\langle I_{1}\cdot I_{2}\rangle}{\langle I^{'}_{1}\rangle \langle I^{'}_{2}\rangle} \; ; \qquad \qquad \text{where}
\end{eqnarray} 
$ I_{1} $, $ I_{2} $  $=$ intensities recorded by the two detectors respectively, when both are operated simultaneously.\\
$ I^{'}_{1} $ $=$  intensity recorded by first detector when the second detector is not operating. \\
$ I^{'}_{2} $ $=$  intensity recorded by second detector when the first detector is not operating. \\

The HBT correlation function $C$ obeys
\begin{eqnarray} 
 & & C = 1 \qquad \text{for distinguishable particles} \nonumber\\
& & C > 1 \qquad \text{for bosons (bunching effect)} \nonumber\\
& & C < 1 \qquad \text{for fermions (anti-bunching effect)}
\end{eqnarray} 

In the commutative case, for a beam of identical (massless, scalar \footnote{ Scalar bosons are taken for sake of simplicity but the analysis presented here can be easily generalized 
to higher spin bosons.}) bosons, the HBT correlation function can be written as \cite{glauber,glauber1}
\begin{eqnarray} 
C^{(B)}_{0} =  \frac{\left\langle \phi^{(-)}_{0}(\bold{y_{1}})\phi^{(-)}_{0}(\bold{y_{2}})\phi^{(+)}_{0}(\bold{y_{2}})\phi^{(+)}_{0}(\bold{y_{1}}) \right\rangle}
{\left\langle \phi^{(-)}_{0}(\bold{y_{1}})\phi^{(+)}_{0}(\bold{y_{1}}) \right\rangle \; 
\left\langle \phi^{(-)}_{0}(\bold{y_{2}})\phi^{(+)}_{0}(\bold{y_{2}}) \right\rangle}
\label{ccorb}
\end{eqnarray} 
where $\bold{y_{1}}$ and $\bold{y_{2}}$ are the position of the two detectors, $\phi^{(+)}_{0}(\bold{y}) = \int \frac{d^{3}\bold{p}}{(2\pi)^{3}} \, \frac{1}{2E_{\bold{p}}} \,
c_{p} \, e^{i\bold{p}\cdot\bold{y}} $ is the positive frequency part and
 $\phi^{(-)}_{0}(\bold{y}) = \int \frac{d^{3}\bold{p}}{(2\pi)^{3}} \, \frac{1}{2E_{\bold{p}}} \, c^{\dagger}_{p} \, e^{-i\bold{p}\cdot\bold{y}} $ is the negative frequency part of 
the bosonic quantum field $\phi_{0}$ \footnote{ We denote the usual bosonic (fermionic) fields by $\phi_{0}$ ($\psi_{0}$) and their 
twisted counterparts are denoted by $\phi_{\theta}$ ($\psi_{\theta}$). Also, the usual creation/annihilation operators are denoted by $c^{\dagger}_p$ 
and $c_p$ whereas the twisted ones are denoted by $a^{\dagger}_p$ and $a_p$ respectively. }. Notice that we have deliberately taken massless fields because we are interested
in looking for HBT correlation functions of ultra-relativistic particles. 

To account for the uncertainties in the energy measurements, which at present are quite significant for UHECRs, the incoming beam should be taken as a wavepacket, 
instead of plane waves. The choice of an appropriate wavepacket is potentially the only place where the information about the details of production mechanism or source of origin of 
UHECRs can reside. Central Limit Theorem tells us that the mean of a sufficiently large number of independent random variables, each with finite mean and 
variance, will be approximately distributed like a Gaussian, and hence it is a good first approximation to take the wavepacket as a Gaussian wavepacket 
$ f(\bold{p}) = N e^{- \alpha(\bold{p}- \bold{p_{0}})^{2}}$ centered around some mean momentum $\bold{p_{0}}$. 

Taking  the wavepacket to be this Gaussian, restricting ourself to only coincidence measurements and using the standard integrals \cite{grad}, the correlation function turns out to be 
\begin{eqnarray}
C^{(B)}_{0} = 1 + e^{- \frac{ \bold{y}^{2}}{2\alpha}}
\label{corgaus}
\end{eqnarray}
As clear from (\ref{corgaus}), $C^{(B)}_{0}$ depends only on  y $ = |\bold{y_{1}} - \bold{y_{2}}|$ the separation between detectors and on $\alpha $. There is no dependence on the mean
 momentum $\bold{p_{0}}$ \cite{baym,baymbook}. 

Similarly, for a beam of identical (chiral) fermions, the HBT correlation function can be written as 
\begin{eqnarray} 
C^{(F)}_{0} = \frac{\left\langle \overline{\psi}^{(-)}_{0}(\bold{y_{1}})\gamma_{0}\overline{\psi}^{(-)}_{0}(\bold{y_{2}})\gamma_{0}\frac{1}{2}(1\pm \gamma_{5})\psi^{(+)}_{0}(\bold{y_{2}})
\frac{1}{2}(1\pm \gamma_{5})\psi^{(+)}_{0}(\bold{y_{1}}) \right\rangle}{\left\langle\overline{\psi}^{(-)}_{0}(\bold{y_{1}})
\gamma_{0}\frac{1}{2}(1\pm \gamma_{5})\psi^{(+)}_{0}(\bold{y_{1}}) \right\rangle \left\langle \overline{\psi}^{(-)}_{0}(\bold{y_{2}})\gamma_{0}
\frac{1}{2}(1\pm \gamma_{5})\psi^{(+)}_{0}(\bold{y_{2}}) \right\rangle}
\label{ccorf}
\end{eqnarray} 
where, as before, $\bold{y_{1}}$ and $\bold{y_{2}}$ are the position of the two detectors, $\psi^{(+)}_{0}(\bold{y}) =  \int \frac{d^{3}\bold{p}}{(2\pi)^{3}} $ \\ 
$ \frac{1}{2E_{\bold{p}}} \,\sum_{s} \, u_{s, \bold{p}} \,c_{s,p} \, e^{i\bold{p}\cdot\bold{y}} $ is the positive frequency part and 
$\overline{\psi}^{(-)}_{0}(\bold{y}) = \int \frac{d^{3}\bold{p}}{(2\pi)^{3}}
 \, \frac{1}{2E_{\bold{p}}} \, \sum_{s} \, \overline{u}_{s, \bold{p}}$ \\ $ c^{\dagger}_{s,p} \, e^{-i\bold{p}\cdot\bold{y}} $  is the negative frequency part of the fermionic quantum 
field $\psi_{0}$. 

Since the whole analysis is done keeping ultra-relativistic particles in back of our mind, here we look for correlation between only chiral fermions (left-left or right-right): 
at such high energies, the particles are effectively massless and hence chiral fermions are the more appropriate ones to deal with \footnote{ $C_0^{(F)} = 1$ between particles 
with opposite helicities (i.e. between left-right or right-left), as at ultra-relativistic energies, they behave like distinguishable particles}.

Taking the incoming beam as a Gaussian wavepacket $ f(\bold{p}) = N e^{- \alpha(\bold{p}- \bold{p_{0}})^{2}}$ centered around some mean momentum $\bold{p_{0}}$ and restricting ourself
 to only coincidence measurements, the correlation function turns out to be 
\begin{eqnarray} 
 C^{(F)}_{0} & = & 1  - \frac{e^{-\frac{\bold{y}^{2}}{2\alpha}}}{2}  - \frac{2}{9 \pi \alpha} e^{-2\alpha \bold{p_{0}}^{2}}\left(\bold{y}^{2} + 4\alpha^{2}\bold{p_{0}^{2}}\right)    
 \leftidx{_1}{F}{_1} \left[2;\frac{5}{2};\frac{-1}{4\alpha}\left(\bold{y}^{2} - 4\alpha^{2}\bold{p_{0}}^{2} - 4i\alpha\bold{y}\cdot\bold{p_{0}} \right)\right] \nonumber \\
& & \leftidx{_1}{F}{_1} \left[2;\frac{5}{2};\frac{-1}{4\alpha}\left(\bold{y}^{2} - 4\alpha^{2}\bold{p_{0}}^{2} + 4i\alpha\bold{y}\cdot\bold{p_{0}} \right)\right]
\label{comcor}
\end{eqnarray} 
where $\bold{y} = \bold{y_{2}} -\bold{y_{1}} $ is the separation between the two detectors and $ \leftidx{_1}{F}{_1}\left(\alpha;\gamma;z\right)$ is the degenerate hypergeometric function.

We observe that in this case, the correlation function depends on the separation between detectors $\bold{y} $, on the mean momentum $ \bold{p_{0}}$ of the wavepacket and the angles
between $\bold{y} $ and $ \bold{p_{0}}$.

In the noncommutative case, there are two important differences. Firstly, the ($\cdot$) product between fields evaluated at the same point has to be replaced by ($\ast$) 
product and secondly the expectation value is changed in accordance with (\ref{4cor}), as the quantum fields are now composed of twisted creation/annihilation operators. 
Hence in noncommutative case for twisted (massless, scalar) bosonic particles, we have 
\begin{eqnarray}
C^{(B)}_{\theta} = \frac{\left\langle \phi^{(-)}(\bold{y_{1}})\left \{ \phi^{(-)}(\bold{y_{2}})\ast_{\bold{y_{2}}}\phi^{(+)}(\bold{y_{2}}) \right \}
\ast_{\bold{y_{1}}}\phi^{(+)}(\bold{y_{1}}) \right\rangle}
{\left\langle \phi^{(-)}(\bold{y_{1}})\ast_{\bold{y_{1}}}\phi^{(+)}(\bold{y_{1}}) \right\rangle 
 \left\langle \phi^{(-)}(\bold{y_{2}})\ast_{\bold{y_{2}}}\phi^{(+)}(\bold{y_{2}}) \right\rangle}
\label{ncorb}
\end{eqnarray} 
where $ \ast_{\bold{y}} = e^{\frac{i}{2}\left( \overleftarrow{\partial_{y}}\right)_{\mu}\theta^{\mu\nu}\left( \overrightarrow{\partial_{y}}\right)_{\nu}}$. 

In rest of the chapter we restrict to only space-space noncommutativity \footnote{We have assumed only space-space noncommutativity for calculational simplicity but one can 
do similar analysis with both space-time as well as space-space noncommutativity} i.e. we take $\theta_{0i} = \theta_{i0} = 0$. As in commutative case, we take the incoming beam
 as a Gaussian wavepacket $ f(\bold{p}) = N e^{- \alpha(\bold{p}- \bold{p_{0}})^{2}}$ centered around some mean momentum $\bold{p_{0}}$ and restrict ourselves to only coincidence
 measurements. $C^{(B)}_{\theta}$ then turns out to be 
\begin{eqnarray}
C^{(B)}_{\theta} & = & 1 + \frac{4\alpha}{4\alpha^{2}+\lambda^{2}} \, \exp{\left[-\frac{4\alpha^{2}\bold{y}^{2} - (\bold{y}\cdot\bold{{\lambda}})^{2}}{2\alpha(4\alpha^{2}+\lambda^{2})}\right]}
\, \exp{\left[-\frac{2\alpha \{\bold{p_{0}}^{2}\bold{\lambda}^{2}-(\bold{p_{0}}\cdot\bold{\lambda})^{2}\}}{4\alpha^{2}+\lambda^{2}}\right]} \nonumber \\
& & \exp{\left[-\frac{4\alpha\bold{y}\cdot(\bold{p_{0}}\times\bold{\lambda})}{4\alpha^{2}+\lambda^{2}}\right]}
\label{gcor}
\end{eqnarray}
where we have defined $ \theta_{ij} = \varepsilon_{ijk}\lambda_{k} $. In getting (\ref{gcor}) we have used the standard result that, for $n$-dim column matrices X and  B and a 
$n\times n$ positive definite, symmetric square matrix A , we have 
\begin{eqnarray}
\int d^{n}X_{i} e^{-X^{T}AX + B^{T}X}= \left(\frac{\pi^{n}}{\det{A}}\right)^{\frac{1}{2}}e^{\frac{1}{4}B^{T}A^{-1}B}
\end{eqnarray}
We see that (\ref{gcor}) not only depends on $\bold{y}$ (the separation between detectors) but 
also on the mean momentum $\bold{p_{0}}$, the noncommutative length scale $\bold{\lambda}$, as well as on the angles between $\bold{\lambda}$ and $\bold{y}$ and 
between $\bold{\lambda}$ and $\bold{p_{0}}$. Moreover, as a check we can see that, in the limit $\bold{\lambda}\rightarrow 0$ we get back (\ref{corgaus}).

Similarly, for twisted (chiral) fermionic particles the noncommutative HBT correlation function is given by 
\begin{eqnarray}
\hspace{-.6cm} 
C^{(F)}_{\theta} = \frac{\left\langle \overline{\psi}^{(-)}(\bold{y_{1}})\gamma_{0} \left \{\overline{\psi}^{(-)}(\bold{y_{2}})\gamma_{0} 
\ast_{\bold{y_{2}}}\frac{1}{2}(1\pm \gamma_{5})\psi^{(+)}(\bold{y_{2}}) \right \}\ast_{\bold{y_{1}}} \frac{1}{2}(1\pm \gamma_{5})\psi^{(+)}(\bold{y_{1}})  \right\rangle}
{\left\langle\overline{\psi}^{(-)}(\bold{y_{1}})\gamma_{0}\ast_{\bold{y_{1}}} \frac{1}{2}(1\pm \gamma_{5})\psi^{(+)}(\bold{y_{1}}) \right\rangle
 \left\langle \overline{\psi}^{(-)}(\bold{y_{2}})\gamma_{0} \ast_{\bold{y_{2}}}\frac{1}{2}(1\pm \gamma_{5})\psi^{(+)}(\bold{y_{2}}) \right\rangle}
\label{ncor}
\end{eqnarray} 
Again restricting ourselves to only space-space noncommutativity, using Gaussian wavepackets and considering only coincidence measurements, $C^{(F)}_{\theta}$  can be written as
\begin{eqnarray}
& & C^{(F)}_{\theta}  =  1 - \frac{1}{2} e^{-2\alpha\bold{p_{0}}^{2}} \left[ e^{-\frac{\bold{z_{1}}^{2}}{4\alpha}} e^{i\left( \overleftarrow{\partial_{z_{1}}}\right)_{i}\theta^{ij}
\left( \overrightarrow{\partial_{z_{2}}}\right)_{j}} e^{-\frac{\bold{z_{2}}^{2}}{4\alpha}} \right] - \left( \frac{2 }{9\pi\alpha} \right) e^{-2\alpha\bold{p_{0}}^{2}}  \nonumber \\ 
& & 
\left[\left\{ \leftidx{_1}{F}{_1}\left( 2, \frac{5}{2}, -\frac{\bold{z_{1}}^{2}}{4\alpha} \right) (z_{1})_{a}\right\}
e^{i\left( \overleftarrow{\partial_{z_{1}}}\right)_{i}\theta^{ij}\left( \overrightarrow{\partial_{z_{2}}}\right)_{j}}
 \left\{ \leftidx{_1}{F}{_1}\left( 2, \frac{5}{2}, -\frac{\bold{z_{2}}^{2}}{4\alpha} \right) (z_{2})_{a}\right\} \right ]
\label{gennoncor}
\end{eqnarray} 
where $\bold{z_{1}} = \bold{y} - 2i\alpha\bold{p_{0}} $ and $\bold{z_{2}} = \bold{y} + 2i\alpha\bold{p_{0}} $.

Expanding this in $ \theta$ and taking terms upto second order we get
\begin{eqnarray}
& & C^{(F)}_{\theta} = 1 -  \frac{1}{2}  e^{-\frac{\bold{y}^{2}}{2\alpha}}  \left[1 + \frac{\bold{p_{0}}\cdot(\bold{y}\times\bold{\lambda})}{\alpha}
 - \frac{\bold{\lambda}^{2}}{4\alpha^{2}} + \frac{\bold{y}^{2}\bold{\lambda}^{2}}{8\alpha^{3}} - \frac{(\bold{y}\cdot\bold{\lambda})^{2}}{8\alpha^{3}} 
 -\frac{\bold{p_{0}}^{2}\bold{\lambda}^{2}}{2\alpha} +  \frac{(\bold{p_{0}}\cdot\bold{\lambda})^{2}}{2\alpha} \right. \nonumber \\ 
& &  \left.  + \frac{[\bold{p_{0}}\cdot(\bold{y}\times\bold{\lambda})]^{2}}{2\alpha^{2}} \right] \, - \,  \left( \frac{2 }{9\pi\alpha} \right)  e^{-2\alpha\bold{p_{0}}^{2}} 
\left[  \leftidx{_1}{F}{_1}\left( 2; \frac{5}{2};-\frac{1}{4\alpha}\left( \bold{y}^{2} - 4\alpha^{2}\bold{p_{0}}^{2} 
- 4i\alpha\bold{y}\cdot\bold{p_{0}} \right) \right) \right. \nonumber \\ 
& &  \left.  \leftidx{_1}{F}{_1}\left( 2; \frac{5}{2};-\frac{1}{4\alpha}\left( \bold{y}^{2} - 4\alpha^{2}\bold{p_{0}}^{2}
 + 4i\alpha\bold{y}\cdot\bold{p_{0}} \right) \right) \right.  \left( \bold{y}^{2} + 4\alpha^{2}\bold{p_{0}}^{2} \right)\nonumber \\
& &  +  \frac{4}{25\alpha^{2}} \left\{ ( 4 \alpha\bold{p_{0}}\cdot(\bold{y}\times\bold{\lambda}) 
- 2\bold{\lambda}^{2} ) \left( \bold{y}^{2} + 4\alpha^{2}\bold{p_{0}}^{2} \right)   + (\bold{y}\cdot\bold{\lambda})^{2} 
+ 4\alpha^{2}(\bold{p_{0}}\cdot\bold{\lambda})^{2} \right\} \nonumber \\
& &  \leftidx{_1}{F}{_1}\left( 3; \frac{7}{2};-\frac{1}{4\alpha}\left( \bold{y}^{2} - 4\alpha^{2}\bold{p_{0}}^{2} - 4i\alpha\bold{y}\cdot\bold{p_{0}} \right) \right) 
 \leftidx{_1}{F}{_1}\left( 3; \frac{7}{2};-\frac{1}{4\alpha}\left( \bold{y}^{2} - 4\alpha^{2}\bold{p_{0}}^{2} + 4i\alpha\bold{y}\cdot\bold{p_{0}} \right) \right) \nonumber \\
& & + \frac{6}{175 \alpha^{3}} \left\{ \bold{y}^{2}\bold{\lambda}^{2} - 4\alpha^{2}\bold{p_{0}}^{2}\bold{\lambda}^{2}  - 4i\alpha(\bold{p_{0}}\cdot\bold{y})\bold{\lambda}^{2}
 - (\bold{y}\cdot\bold{\lambda})^{2} + 4\alpha^{2}(\bold{p_{0}}\cdot\bold{\lambda})^{2} \right. \nonumber \\ 
& &  \left. + 4i\alpha(\bold{p_{0}}\cdot\bold{\lambda})(\bold{y}\cdot\bold{\lambda}) \right\} ( \bold{y}^{2} + 4\alpha^{2}\bold{p_{0}}^{2} )
\leftidx{_1}{F}{_1}\left( 4; \frac{9}{2};-\frac{1}{4\alpha}\left( \bold{y}^{2} - 4\alpha^{2}\bold{p_{0}}^{2} - 4i\alpha\bold{y}\cdot\bold{p_{0}} \right) \right) \nonumber \\ 
& & \leftidx{_1}{F}{_1}\left( 3; \frac{7}{2};-\frac{1}{4\alpha}\left( \bold{y}^{2} - 4\alpha^{2}\bold{p_{0}}^{2} + 4i\alpha\bold{y}\cdot\bold{p_{0}} \right) \right) 
\,  + \, \frac{6}{175 \alpha^{3}} \{ \bold{y}^{2}\bold{\lambda}^{2} - 4\alpha^{2}\bold{p_{0}}^{2}\bold{\lambda}^{2}   \nonumber \\ 
& &  + 4i\alpha(\bold{p_{0}}\cdot\bold{y})\bold{\lambda}^{2}  - (\bold{y}\cdot\bold{\lambda})^{2} + 4\alpha^{2}(\bold{p_{0}}\cdot\bold{\lambda})^{2} 
- 4i\alpha(\bold{p_{0}}\cdot\bold{\lambda})(\bold{y}\cdot\bold{\lambda}) \} \nonumber \\
 & &  ( \bold{y}^{2} + 4\alpha^{2}\bold{p_{0}}^{2} )  \leftidx{_1}{F}{_1}\left( 3; \frac{7}{2};-\frac{1}{4\alpha}\left( \bold{y}^{2} - 4\alpha^{2}\bold{p_{0}}^{2}
 - 4i\alpha\bold{y}\cdot\bold{p_{0}} \right) \right) \nonumber \\
& &  \leftidx{_1}{F}{_1}\left( 4; \frac{9}{2};-\frac{1}{4\alpha}\left( \bold{y}^{2} - 4\alpha^{2}\bold{p_{0}}^{2} + 4i\alpha\bold{y}\cdot\bold{p_{0}} \right) \right) 
- \frac{288}{1225\alpha^{2}} \{ \bold{p_{0}}\cdot(\bold{y}\times\bold{\lambda}) \}^{2}  ( \bold{y}^{2} + 4\alpha^{2}\bold{p_{0}}^{2} ) \nonumber \\
& & \left. \leftidx{_1}{F}{_1}\left( 4; \frac{9}{2};-\frac{1}{4\alpha}\left( \bold{y}^{2} - 4\alpha^{2}\bold{p_{0}}^{2} - 4i\alpha\bold{y}\cdot\bold{p_{0}} \right) \right)  
\leftidx{_1}{F}{_1}\left( 4; \frac{9}{2};-\frac{1}{4\alpha}\left( \bold{y}^{2} - 4\alpha^{2}\bold{p_{0}}^{2} + 4i\alpha\bold{y}\cdot\bold{p_{0}} \right) \right) \right] \nonumber \\
& & + \; O(\lambda^{3})
\label{noncor}
\end{eqnarray}
where again in writing (\ref{noncor}) we have defined $ \theta_{ij} = \varepsilon_{ijk}\lambda_{k} $.

As can be clearly seen from (\ref{noncor}), the noncommutative HBT correlation function not only depends on $\bold{y} $ and $ \bold{p_{0}}$ but also on $\bold{\lambda}$ 
and various angles that $\bold{\lambda}$ makes with $\bold{y} $ and $ \bold{p_{0}}$. 

A comparison of (\ref{corgaus})-(\ref{gcor}) and (\ref{comcor})-(\ref{noncor}) tells us that the HBT correlation function for twisted bosons/fermions are different from 
those for ordinary bosons/fermions and become more pronounced with increasing momenta of incoming particles. Since $\bold{\lambda}$ is expected to be a very small quantity,
the deviations will be highly suppressed and one has to look at particles with high enough energy-momentum (w.r.t noncommutative scale), so that the deviations get sufficiently 
enhanced to be detectable. Hence, as stated in beginning of the chapter, the best place to look for the signatures of twisted statistics, is to look at particles 
in L.H.C or in Ultra High Energy Cosmic Rays (UHECRs). Moreover, since HBT correlations are essentially quantum correlations between free identical particles and are manifestations 
of the particular statistics followed by the identical particles of a given beam, they are sensitive only to the statistics obeyed by these particles. Therefore despite our 
present lack of detailed knowledge about production mechanism and source of origin of UHECRs, a study of HBT correlations may still be able to provide unambiguous signatures, 
of underlying noncommutative nature of spacetime.

%
 \vspace{10cm}
The graphs shown below highlight the difference between $C^{(B)}_{0}$ - $C^{(B)}_{\theta}$ and $C^{(F)}_{0}$ - $C^{(F)}_{\theta}$ \footnote{The persistence of correlations
 to large distances is attributed to the large uncertainties in our present determination of energy-momentum of UHECRs. With better and more precise knowledge of the energy-momentum,
 the correlation will decrease significantly resulting in better bounds on noncommutative deviations}.

\begin{figure} [h!]
\centering
\subfloat []{\includegraphics[angle=0,width=2.7in]{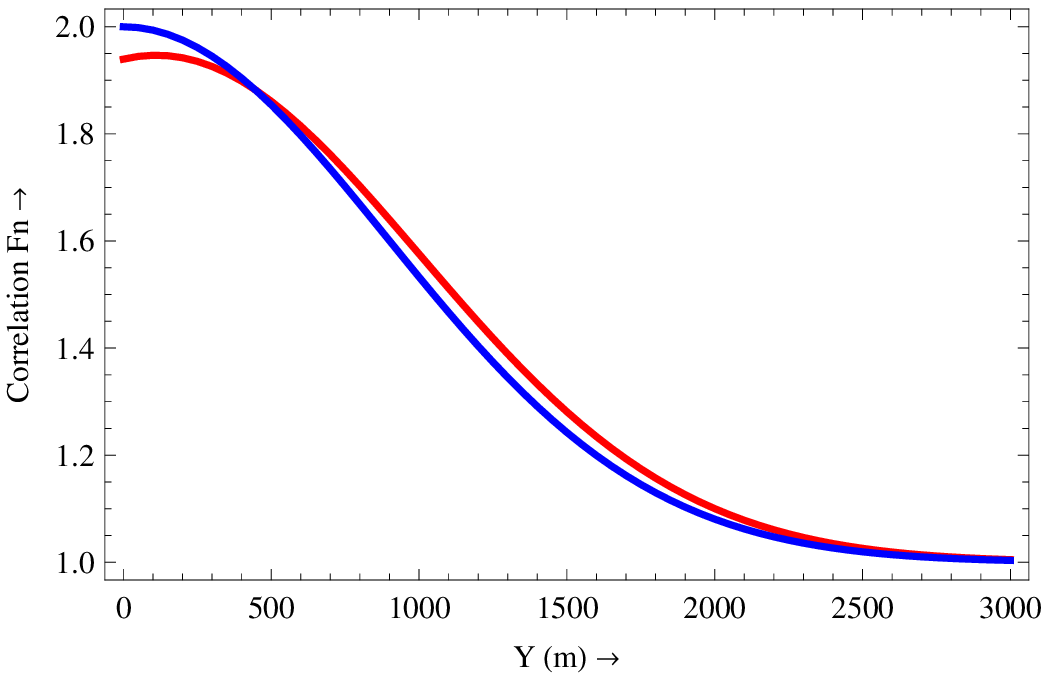} } 
 \hspace{.75cm}\vspace{.25cm}
\subfloat []{\includegraphics[angle=0,width=2.7in]{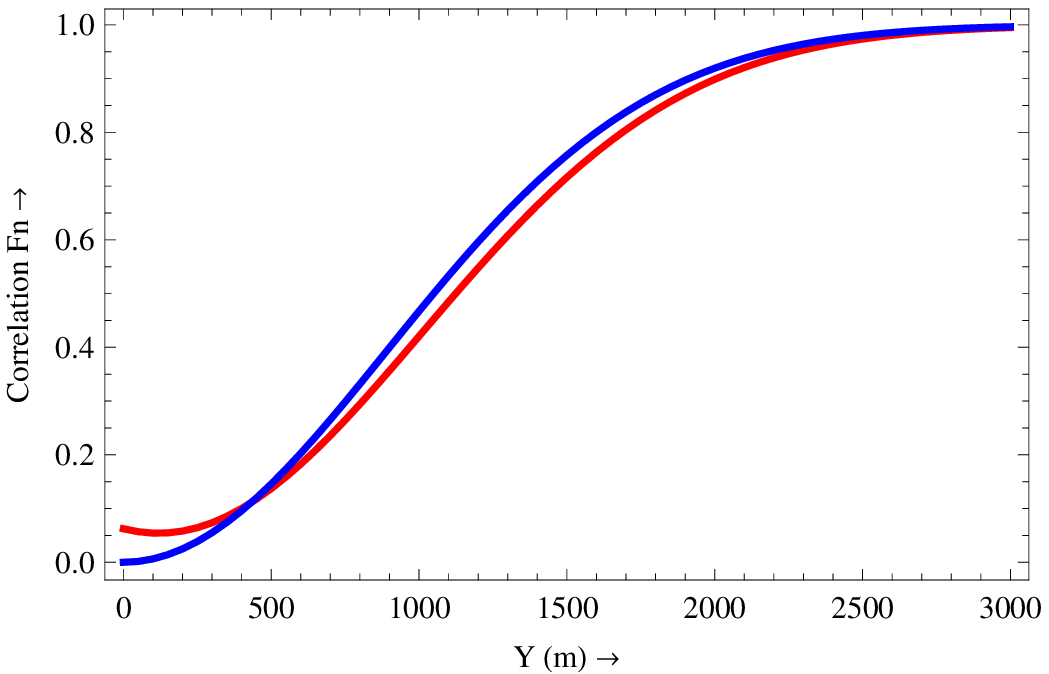}} \\
\subfloat []{ \includegraphics[angle=0,width=2.7in]{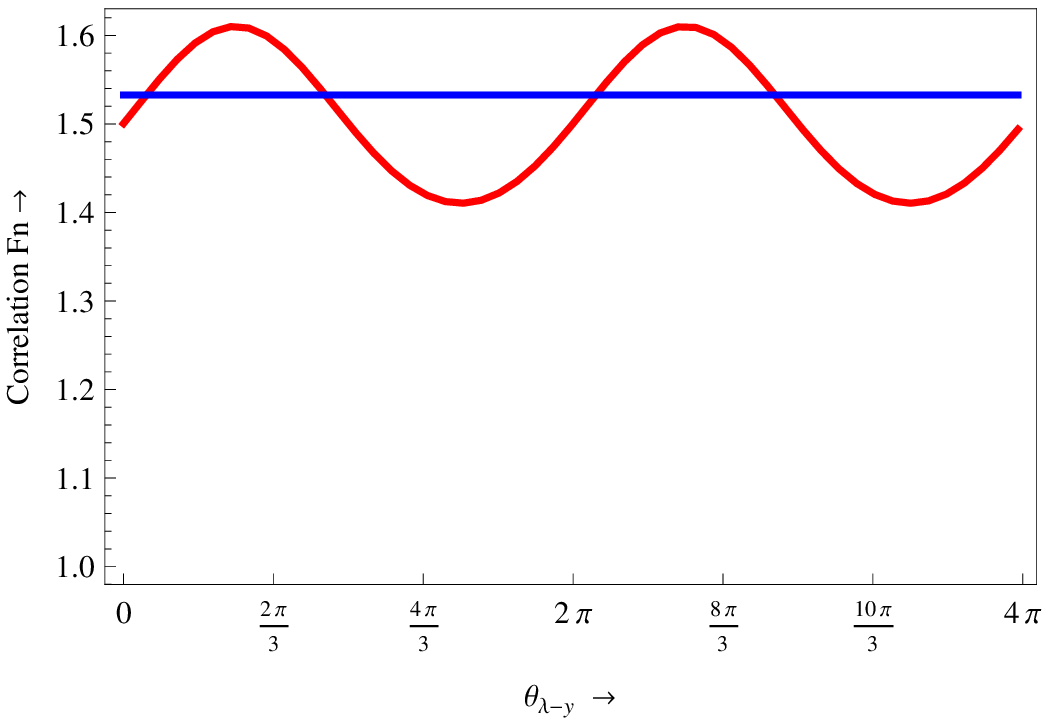}}
 \hspace{.75cm}\vspace{.25cm}
\subfloat []{\includegraphics[angle=0,width=2.7in]{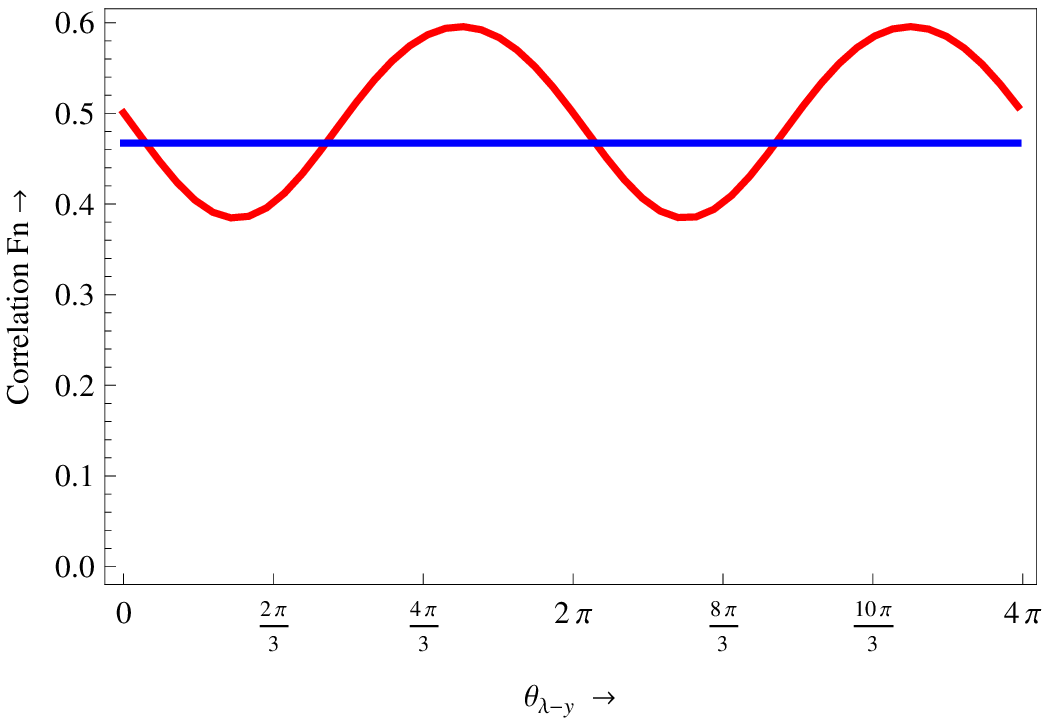}}
\caption{ In the above figures, the blue and red line represent the commutative and noncommutative HBT correlation functions respectively. 
The figures (a), (c) are plotted for ordinary and twisted bosons and (b), (d) are plotted for ordinary and twisted fermions.
The figures (a), (b) are plotted with $|\bold{p_{0}}| = 6 \times 10^{19} $eV (i.e. at G.Z.K cutoff),  $\alpha = 2.04 \times 10^{19}\text{eV}^{-2} $ (i.e taken numerically same
 as the present error in estimation of energy of UHECRs) \cite{nag}, $|\bold{\lambda}| =  1.47 \times 10^{-24}\text{m}^{2}$  and taken along z-axis and the angles as 
 $\theta_{\lambda - y}= \frac{\pi}{4}$, $\theta_{\lambda - p_{0}}= \frac{\pi}{4}$ , $\phi_{\lambda - y}= \frac{\pi}{3}$ and $ \phi_{\lambda - p_{0}}= \frac{\pi}{6} $. 
The figures (c), (d) are plotted with $|\bold{y}|= 1000$m  and same values for rest of the parameters.     }

\end{figure}

Of particular interest are the various angular dependences of noncommutative correlation functions (\ref{gcor}) and (\ref{noncor}) which are completely absent in commutative 
correlation functions (\ref{corgaus}) and (\ref{comcor}). Due to rotation and revolution of earth, the noncommutative correlation functions will show periodic oscillations completely
absent in commutative correlations and hence will perhaps provide the best and most unambiguous signal for underlying noncommutative structure of spacetime. Therefore, as claimed earlier, 
the information about noncommutative structure of spacetime, can be extracted out, from observing the nature of variation of HBT correlation functions (in particular by looking at 
angular variations) with varying certain experimentally measurable quantities, in a very unambiguous way. 

Also it is worth noting that if we take noncommutative length scale same as Planck scale i.e. $10^{-35}$m then due to an upper limit on momenta of UHECRs (GZK cutoff), the deviations 
will turn out to be O($10^{-50}$) which are too small to give any unambiguous signatures of it. So the noncommutative deviations are detectable only if the effective 
noncommutative scale is much larger than Planck scale, which is likely to happen in presence of large extra dimensions. Thus the noncommutative deviations in HBT correlations 
effectively provide us signatures of large extra dimensions.  

This chapter is based on the work published in \cite{hbt}.


\chapter{Summary}
\label{chap:chap6}

In this thesis we studied field theories written on a particular model of noncommutative spacetime, the Groenewold-Moyal (GM) plane. We started with briefly reviewing the novel features
of field theories on GM plane e.g. the $\ast$-product, restoration of Poincar\'e-Hopf symmetry and twisted commutation relations. After this brief review, in next chapter we discussed 
our work on renormalization of field theories on GM plane. We first gave a review of the noncommutative interaction picture, the LSZ reduction formula and showed their equivalence. 
We then took up the problem of renormalization of noncommutative theories involving only matter fields. We showed that any generic noncommutative theory involving pure matter fields is 
a renormalizable theory if the analogous commutative theory is renormalizable.
We further showed that all such noncommutative theories have same fixed points and $\beta$-functions for the couplings, as that of the analogous commutative theory. 
The unique feature of these field theories was the emergence of twisted statistics of the particles. Motivated by it, in the third chapter, we looked at 
the possibility of twisted statistics by deforming internal symmetries instead of spacetime symmetries. We constructed two different twisted theories which can be viewed as internal 
symmetry analogue of the GM plane and dipole field theories. We further studied their various properties like the issue of causality and the scattering formalism. 
Having studied the mathematical properties of noncommutative and twisted internal symmetries we moved on to discuss their potential phenomenological signatures. We first discussed the 
noncommutative thermal correlation functions and show that because of the twisted statistics, all correlation functions except two-point function get modified. 
Finally we discussed the modifications in Hanbury-Brown Twiss (HBT) correlation functions due to twisted statistics on GM plane 
and the potential of observing signatures of noncommutativity by doing a HBT correlation experiment with Ultra High Energy Cosmic Rays (UHECRs).

In the second chapter we presented a complete and comprehensive treatment of noncommutative theories involving only matter fields. We showed first for real scalar fields having a 
$\phi^4_{\theta,\ast}$ interaction and then for generic theories that the noncommutative $\hat{S}_{\theta}$ is the same as $\hat{S}_{0}$ and that the $S$-matrix elements only have 
an overall phase dependence on the noncommutativity scale $\theta$. We also argued that since there is only an overall phase dependence on the noncommutativity scale $\theta$, 
the physical observables like scattering cross-sections and decay rates do not depend on $\theta$ and there is no UV/IR mixing in any such theory. 

We further showed that all such theories are renormalizable if and only if the corresponding commutative theories are renormalizable. The usual commutative techniques for 
renormalization can be used to renormalize such theories. Moreover, we showed by explicit calculations for $\phi^4_{\theta,\ast}$ case and argued for generic case, that for 
all such theories the $\beta$-functions, RG flow of couplings or the fixed points are all same as those of the analogous commutative theory. 

The equivalence of physical observables like scattering cross-sections, decay rates and also that of $\beta$-functions, RG flow and fixed points with those of the corresponding 
commutative theories does not mean that all such noncommutative theories are one and the same as their commutative counterparts. One can always construct, even for free theories, 
appropriate observables which unambiguously distinguish between a noncommutative and commutative theory \cite{rahul-hbt}.

The discussion of the chapter was limited only to matter fields and interaction terms constructed out of only matter fields. Noncommutative field theories involving nonabelian gauge 
fields violate twisted Poincar\'e invariance and are know to suffer from UV/IR mixing \cite{amilcar-pinzul}. They require special treatment which is outside the scope of present 
work.

In the third chapter we discussed the possibility of having twisted statistics by deformation of internal symmetries. We constructed two such deformed statistics and discussed field 
theories for such deformed fields. We showed that both type of twisted quantum fields discussed in this chapter, satisfy commutation relations different from the usual bosonic/fermionic 
commutation relations. Such twisted fields by construction (and in view of CPT theorem) are nonlocal in nature. We showed that inspite of the basic ingredient fields being nonlocal,  
it is possible to construct local interaction Hamiltonians which satisfy cluster decomposition principle and are Lorentz invariant.

We first discussed a specific type of twist called ``antisymmetric twist''. This kind of twist is quite similar in spirit to the twisted noncommutative field theories. 
The formalism developed for antisymmetric twists was analogous (with appropriate generalizations and modifications) to the formalism of twisted noncommutative theories. 
We then constructed interaction terms using such twisted fields and discussed the scattering problem for such theories. We found that the twisted $SU(N)$ invariant interaction
Hamiltonian as well as $S-$matrix elements are identical to their untwisted analogues and hence by doing a scattering experiment it is rather difficult to distinguish between 
a twisted and untwisted theory. We further showed that relaxing the demand of $SU(N)$ invariance leads to differences between the two theories and for certain interaction 
terms the twisted theory is nonlocal although its analogous untwisted theory is local. As an interesting application of these ideas we showed that the marginal ($\beta$-)
deformations of the scalar matter sector of $N=4$ SUSY Hamiltonian density can be described in terms of a twisted interaction Hamiltonian density and hence the twisted internal
symmetries can be used to significantly simplify the discussion of such theories.
 
We then constructed more general twisted statistics which can be viewed as internal symmetry analogue of dipole theories. We also discussed the construction of interaction 
terms and scattering formalism for it. The main results for general twists are same as those for the antisymmetric twist.

We ended the chapter with discussion of causality of such twisted field theories. We showed that the twisted fields are noncausal and hence a generic observable constructed out 
of them is also noncausal. Inspite of this it is possible to construct certain interaction hamiltonians, e.g. the $SU(N)$ invariant interaction hamiltonian, which are causal and 
satisfy cluster decomposition principle.

In the fourth chapter we discussed the formalism of Green's functions to compute correlation functions and adapt it to the noncommutative case. We showed that due to twisted commutation 
relations satisfied by the fields on GM plane, all correlation functions apart from two-point correlation function get modified.

In the fifth chapter we looked at probable signatures of noncommutativity in Ultra High Energy Cosmic Rays (UHECRs). We looked at the modifications in noncommutative HBT correlation function
due to the twisted statistics. We showed that the commutative and noncommutative HBT correlation functions differ from each other and the difference gets more and more pronounced as 
we go to higher and higher energies. Hence an HBT experiment with UHECRs can provide us potential signatures of large scale noncommutativity.






\end{document}